\documentclass[a4paper,11pt]{article}

\usepackage{jheppub}
\usepackage{xcolor,colortbl}
\allowdisplaybreaks

\usepackage{physics}
\usepackage{mathtools}
\usepackage{slashed}
\usepackage{subfigure}
\usepackage{dsfont}

\def\as{{\alpha_s}}
\def\nn{\nonumber}
\def\ep{\epsilon}
\def\g{\gamma}

\title{NNLO QCD corrections to leptonic observables in top-quark pair production and decay}

\author[a]{Micha\l{} Czakon}
\author[b]{Alexander Mitov}
\author[b]{and Rene Poncelet}
\affiliation[a]{Institut f\"ur Theoretische Teilchenphysik und Kosmologie, RWTH Aachen University,\\ D-52056 Aachen, Germany}
\affiliation[b]{Cavendish Laboratory, University of Cambridge, Cambridge CB3 0HE, UK}
\emailAdd{mczakon@physik.rwth-aachen.de}
\emailAdd{adm74@cam.ac.uk}
\emailAdd{poncelet@hep.phy.cam.ac.uk}

\abstract{
We calculate a comprehensive set of spin correlations and differential distributions in top-quark pair production and decay to dilepton final states. Our calculation is performed in the Narrow Width Approximation. This is the first time such a complete study is performed at next-to-next-to leading order in QCD. Both inclusive and fiducial distributions are presented and analyzed. Good agreement between NNLO QCD predictions and data is found. We demonstrate that it is possible to perform high-precision comparisons of fixed-order calculations with fiducial-level data. Subtleties of the top quark definition are raised and clarified. Some of those are found to have a very significant impact on top-quark pair production at absolute threshold. 
}

\dedicated{\rm TTK-20-25, P3H-20-043, Cavendish-HEP-20/07}

\begin{document}
\maketitle
\flushbottom

\section{Introduction}

The hadroproduction of top-quark pairs is among the cornerstone processes for the LHC and future hadron colliders. The reason for this is twofold. On one hand, a massive experimental program is underway at the LHC both verifying the Standard Model (SM) aspects of top physics and conducting searches for physics beyond the SM. The corresponding boundaries for both these aspects are being pushed to unprecedented reach and precision. On the other hand, top quark production is central to the development of theoretical tools and techniques for collider processes and has been instrumental in advancing the high-precision physics program at hadron colliders.

Many SM measurements in the top-quark sector are currently systematics dominated. This trend will only accelerate in the future as more LHC data gets accumulated. A major contributor to the said systematics is the precise modeling of $t\bar t$ final states. The main goal of this work is to extend the already established approach of using Next-to-Next-to Leading Order (NNLO) predictions for stable top quark pair production to the description of the more realistic final state
\begin{equation}
  p p \to t\bar t \to b\bar{b}\ell^+\ell^-\nu\bar{\nu}+X\;,
  \label{eq:tt+decay}
\end{equation}
resulting from the decay of the top quark and antiquark into states containing two leptons.

The process (\ref{eq:tt+decay}) has been studied extensively at NLO in QCD. Initially, the calculations have been performed \cite{Bernreuther:2004jv,Melnikov:2009dn,Bernreuther:2010ny,Biswas:2010sa,Melnikov:2011qx,Campbell:2012uf} in the Narrow-Width Approximation (NWA) \cite{Denner:1999gp,Denner:2005fg} (i.e. in the limit $\Gamma_t/m_t \to 0$). The NWA approach has been extended to approximate NNLO in top-quark production and full NNLO in the decay \cite{Gao:2017goi}. Full NNLO precision, both in top production and decay, has recently been achieved for certain spin-correlation observables \cite{Behring:2019iiv}. Advances and automation in one-loop calculations eventually made it possible to account at NLO for both off-shell effects and non-doubly resonant contributions (i.e. dilepton final states not necessarily mediated by a pair of top quarks) \cite{Denner:2010jp,Bevilacqua:2010qb,Denner:2012yc,Frederix:2013gra,Heinrich:2017bqp,Jezo:2016ujg,Cascioli:2013wga}.

Although off-shell NLO calculations offer the most complete description for this process, they are very challenging computationally. Extending them to NNLO is not feasible at present, mainly due to the lack of two-loop amplitudes for such complicated multi-parton, multi-scale amplitudes. The computation of NNLO corrections to $t\bar t$ production and decay simplifies considerably in the NWA approach since it: a) factorizes radiative corrections into production of a top-quark pair and $t$ and $\bar t$ decays and, b), neglects the so-called non-resonant and single-resonant contributions. The accuracy of this approximation is parametrized by the ratio $\order{\Gamma_t/m_t}\approx 1\%$~\cite{Uhlemann:2008pm}. Dedicated phenomenological studies have confirmed this estimate for the total cross section as well as for many differential observables which are insensitive to the finite width of the top-quark. This indicates that as long as one does not look into phase space regions that are very sensitive to off-shell effects the NWA is a reliable approximation \cite{Bevilacqua:2010qb}. In particular observables which are sensitive to top-quark spin correlations can reliably be obtained within the NWA since the spin correlations between the top quark(s) production and their subsequent decay are taken into account exactly. The effects from a finite bottom quark mass are neglected throughout this work. This approximation has negligible impact on most of the observables studied, while exceptions are discussed explicitly.

The computation of any differential distribution with dilepton final states at NNLO in QCD offers many physics opportunities which we pursue in this work. Examples are high-precision comparisons of SM predictions with LHC data for the purpose of looking for deviations from the SM in the top quark sector; extraction of SM parameters like $m_t$ and pdf fits with improved precision; testing our ability to probe the SM at the few-percent level by verifying that theoretical predictions are not merely accurate enough but also closely match the experimental setups they are intended to be compared to, even though parton-shower and hadronization effects are neglected. The last point is particularly non-trivial since it requires comparing NNLO-accurate fixed-order predictions with data that has been unfolded using MC event generators of NLO precision. In fact, one of the main goals of the present work is to demonstrate that such a comparison is possible, albeit after dedicated efforts are made on both the theory and experimental sides. Such comparison is also a necessary step in the direction of producing and adopting NNLO-accurate MC event generators \cite{Mazzitelli:2020jio}, which is still years ahead in the future.

The content of the present work is as follows: in sec.~\ref{sec:theory} we specify our computational setup. Particular attention is paid to the implementation of NWA at NNLO. Section~\ref{sec:pheno} is devoted to phenomenological applications. We split them into inclusive and fiducial ones and a very thorough and comprehensive analysis of both is given. We present for the first time a complete set of NNLO-accurate predictions for all measured spin-density matrix and spin-correlation distributions as well as inclusive and fiducial one- and two-dimensional differential distributions of leptons, $b$-jets and top quarks. The impact of the definition of top quarks on the theory-data comparison is studied in sec.~\ref{sec:tops}. Our conclusions are summarized in sec.~\ref{sec:conclusions}.

\section{Computational setup}\label{sec:theory}

This work presents a comprehensive set of differential distributions in top-quark pair production and decay in NNLO QCD. It extends ref.~\cite{Behring:2019iiv} where leptonic angular observables sensitive to spin-correlations in top-quark pair production were studied.  This work utilizes the four-dimensional formulation of the sector-improved residue subtraction scheme {\sc Stripper} \cite{Czakon:2010td,Czakon:2014oma} implemented as in ref.~\cite{Czakon:2019tmo}. This framework has already been used for the calculation of NNLO QCD corrections to variety of processes like differential top-quark pair production \cite{Czakon:2014xsa,Czakon:2015owf,Czakon:2016ckf,Czakon:2016vfr,Czakon:2017dip,Czakon:2017wor,Czakon:2017lgo,Czakon:2019txp,Czakon:2019yrx}, inclusive jet production \cite{Czakon:2019tmo} and three-photon production \cite{Chawdhry:2019bji}, which was the first complete NNLO QCD calculation for a $2 \to 3$ process. Further technical details about our computation can be found in sec.~\ref{sec:amplitudes}. An exception is the treatment of a relatively minor NLO contribution which only enters the calculation at NNLO. For simplicity, it has been computed in the dipole subtraction formalism; see sec.~\ref{eq:nloxnlo} for details. A feature specific to top-quark production is the use of the Narrow-Width Approximation which we describe in detail starting in sec.~\ref{sec: NWA}.

\subsection{Implementation of the Narrow-Width Approximation}\label{sec: NWA}

\begin{figure}
\includegraphics[width=\textwidth]{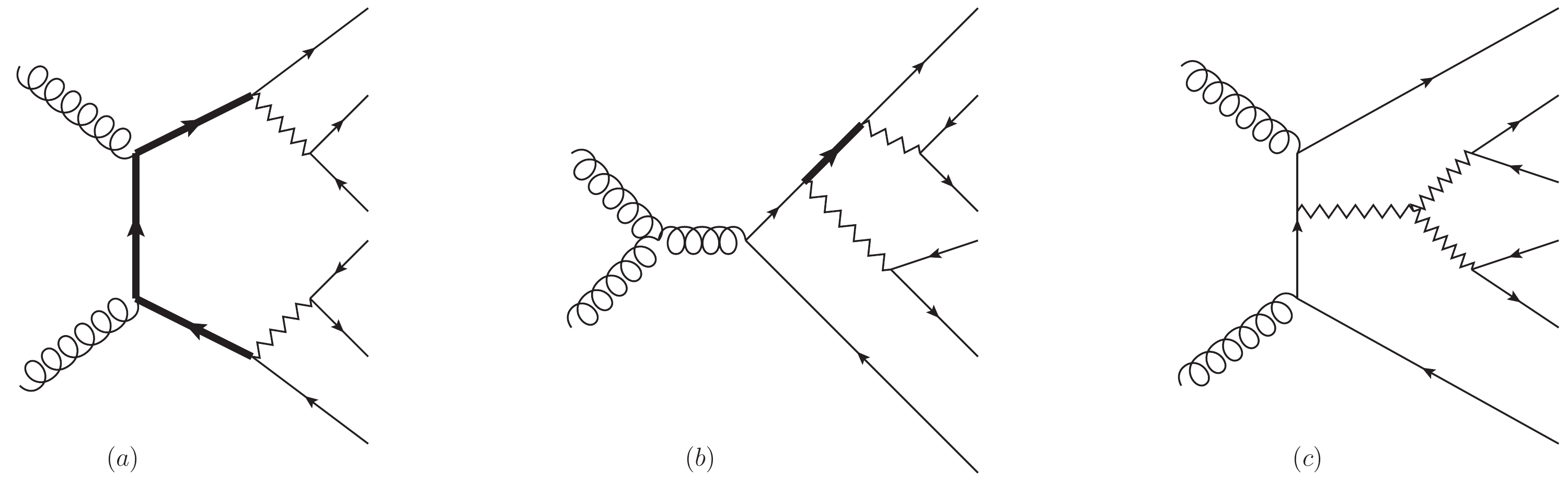}
\caption{Sample double- (a), single- (b) and non-resonant (c) diagrams contributing to the $pp\to b\bar{b}\ell^+\ell^-\nu\bar{\nu}$ process. Bold lines indicate top-quark propagators.}
\label{fig:diagrams}
\end{figure}

Three classes of diagrams contribute to the process eq.~(\ref{eq:tt+decay}). They are classified by the number of internal top-quark propagators into double-, single- and
non-resonant diagrams. Sample diagrams are shown in fig.~\ref{fig:diagrams}.

The factorization of the process (\ref{eq:tt+decay}) in the NWA into $t\bar t$ production and $t/\bar t$ decay works as follows. For simplicity we start at LO in perturbation theory. The (differential)  cross-section for the partonic reaction $q_1+q_2\to b(p_1)\ell^+(p_2)\nu(p_3)\bar{b}(p_4)\ell^-(p_5)\bar{\nu}(p_6)$ can be written as
\begin{equation}
 \hat{\sigma}^{(0)} = \frac{1}{2\hat{s}}\frac{1}{\mathcal{N}}
   \int \dd \Phi(P,\{p_i\}_{i=1}^6)
   \braket{\mathcal{M}}{\mathcal{M}} \,,
 \label{eq:nwa0}
\end{equation}
where $P = q_1+q_2$, $\sqrt{P^2} = \hat{s}$, ${\mathcal{M}}$ is the corresponding tree-level $2\to 6$ amplitude and $\mathcal{N}$ is a normalization factor. As mentioned above, some diagrams contributing to $\braket{\mathcal{M}}{\mathcal{M}}$ contain top-quark and top anti-quark propagators $T$ and $\overline T$. In the complex mass scheme, these propagators read
\begin{eqnarray}
T(p_t) &=& \frac{i}{\slashed{p}_t - \mu_t} \quad \text{with} \quad \mu_t^2 = m_t^2 - i m_t\Gamma_t \,,\\
\overline T(p_{\bar t}) &=& \frac{i}{\slashed{p}_{\bar t} - \mu_{\bar t}} \quad \text{with} \quad \mu_{\bar t}^2 = m_t^2 - i m_t\Gamma_{\bar t} \,.
\end{eqnarray} 
In the above equations $\Gamma_t$ and $\Gamma_{\bar t}$ are the inclusive widths for $t$ and $\bar t$ respectively. While the two widths are equal, we have labeled them this way since in the following we need to keep track of $t$ and $\bar t$ separately.

In each diagram the internal top-quark momentum $p_t$ is fixed by the external momenta via momentum conservation. When squaring the amplitude, interferences of diagrams containing a top-quark propagator with {\it the same} $p_t$ will contribute the following denominator
\begin{equation}
  |T(p_t)|^2 \sim \frac{1}{(p_t^2-m_t^2)^2+m_t^2\Gamma_t^2} \equiv T_D(p_t^2)\,.
\end{equation}
Same applies to the analogous denominator $\overline T_D(p_{\bar t}^2)$ corresponding to the $\bar t$ quark. The contributions to $\braket{\mathcal{M}}{\mathcal{M}}$ due to interferences of this kind will be denoted as
\begin{equation}
  T_D(p_t^2) \overline T_D(p_{\bar t}^2) \braket{\mathcal{M}_\text{res}}{\mathcal{M}_\text{res}}(p_t,p_{\bar t})\,,
\end{equation}
where $\mathcal{M}_\text{res}$, to be specified below, only receives contributions from the doubly-resonant part of the full amplitude $\mathcal{M}$.

The phase space integral in eq.~(\ref{eq:nwa0}) reads
\begin{equation}
 \dd \Phi(P,\{p_i\}_{i=1}^6) = (2\pi)^4\delta^{(4)}(P-\sum_{i=1}^6 p_i)
   \prod_{i=1}^6 \frac{\dd^3\vec{p}_i}{(2\pi)^3 2 E_i}\,.
\end{equation}
It can be rewritten {\it exactly} through the product of phase-spaces containing two intermediate states with invariant masses $p_t^2$ and $p_{\bar t}^2$ and two phase-spaces for the decays $p_t\to p_1+p_2+p_3$ and $p_{\bar t}\to p_4+p_5+p_6$:
\begin{equation}
 \dd \Phi(P,\{p_i\}_{i=0}^6) =
   \underbrace{\dd \Phi(P,p_t,p_{\bar{t}})}_{\equiv \dd \Phi_{t\bar{t}}}
               \frac{\dd p_t^2}{2\pi}\frac{\dd p_{\bar{t}}^2}{2\pi}
   \underbrace{\dd \Phi(p_t,\{p_k\}_{k=1}^3)}_{\equiv \dd \Phi_{\Gamma_t}}
   \underbrace{\dd \Phi(p_{\bar{t}},\{p_k\}_{k=4}^6)}_{\equiv \dd \Phi_{\Gamma_{\bar{t}}}}\,.
\end{equation}

The contribution to the (differential) cross-section from doubly-resonant diagrams reads
\begin{equation}
 \hat{\sigma}^{(0)}_{\rm res} = \frac{1}{2\hat{s}}\frac{1}{\mathcal{N}}
             \int_{q_\text{min}}^{q_\text{max}} \frac{\dd p_t^2}{2\pi} T_D(p_t^2)
             \int_{\bar{q}_\text{min}}^{\bar{q}_\text{max}} \frac{\dd p_{\bar{t}}^2}{2\pi} T_D(p_{\bar{t}}^2)
             \int \dd \Phi_{t\bar{t}}\, \dd\Phi_{\Gamma_t}\dd\Phi_{\Gamma_{\bar{t}}}
             \braket{\mathcal{M}_\text{res}}{\mathcal{M}_\text{res}}\;.
\end{equation}

In the NWA the above result simplifies as follows. First, the asymptotic behavior of the integral over $p_t$ in the limit $\Gamma_t/m_t \to 0$ reads
\begin{equation}
  \int_{-\infty}^\infty \frac{\dd p^2_t}{2\pi} T_D(p_t^2) \xrightarrow{\frac{\Gamma_t}{m_t}\to 0}
  \int_{-\infty}^\infty \dd p_t^2 \frac{\delta(p_t^2-m_t^2)}{2m_t\Gamma_t}\,.
\end{equation} 
An analogous result applies to the integration over $p_{\bar t}$. Second, the contribution to the cross-section from the region away from the resonance $1/|p_t^2-m_t^2| \approx 0$ is suppressed. For this reason one can extend the integration from $(q_\text{min},q_\text{max})$ to $(-\infty,+\infty)$. Third, the contributions from singly-resonant and non-resonant diagrams are suppressed, too, and can also be neglected. With these simplifications in mind we arrive at the NWA approximation for the partonic cross-section
\begin{equation}
  \hat{\sigma}^{(0)} \xrightarrow{\frac{\Gamma_t}{m_t}\to 0} \hat{\sigma}^{(0)}_{\rm NWA} \equiv
  \frac{1}{\hat{s}}\frac{1}{\mathcal{N}}
        \int \dd \Phi_{t\bar{t}}\, \dd\Phi_{\Gamma_t}\dd\Phi_{\Gamma_{\bar{t}}}
       \frac{
       \braket{\mathcal{M}_\text{res}}{\mathcal{M}_\text{res}}
       }{(2m_t\Gamma_t)^2} \bigg|_{p_t^2=m_t^2,\, p_{\bar{t}}^2=m_t^2}\,.
 \label{eq:NWA_LO}
\end{equation}

At this point we recall that the resonant amplitude $\ket{\mathcal{M}_{\text{res}}}$ still contains the numerator of the resonant top-quark propagator. Since, as follows from eq.~(\ref{eq:NWA_LO}), $t$ and $\bar t$ are on-shell one can make use of the polarization sums
\begin{eqnarray}
  \slashed{p}_t+m_t &=& \sum_h u(p_t,h) \bar{u}(p_t,h)\,,\\
  \slashed{p}_{\bar t}-m_t &=& \sum_{\bar h} v(p_{\bar t},{\bar h}) \bar{v}(p_{\bar t},{\bar h})\,,
\end{eqnarray}
and rewrite the resonant amplitude in terms of {\it polarized} on-shell $t\bar t$-production and $t/\bar t$-decay amplitudes
\begin{equation}
  \ket{\mathcal{M}_\text{res}} = \sum_{h,\bar h} \ket{\mathcal{M}_\text{prod}(h,\bar h)}\ket{\Gamma_t(h)}\ket{\Gamma_{\bar t}(\bar h)}\,.
\end{equation}
In the above equations $h$ (${\bar h}$) labels the helicity of the top quark (top antiquark). The squared matrix element thus takes the form of a spin-correlated product of squared production and decay matrix elements
\begin{equation}
\braket{\mathcal{M}_\text{res}}{\mathcal{M}_\text{res}} = 
\sum_{h,h',{\bar h},{\bar h}'} \braket{M_\text{prod}(h',{\bar h}')}{M_\text{prod}(h,{\bar h})} \braket{\Gamma_t(h')}{\Gamma_t(h)}\braket{\Gamma_{\bar t}({\bar h}')}{\Gamma_{\bar t}({\bar h})}\,.
\label{eq:spincorr}
\end{equation}

Combining eqs.~(\ref{eq:spincorr}) and (\ref{eq:NWA_LO}), the differential cross section in the NWA can be written in the following fully factorized form
\begin{equation}
  \dd \sigma = \dd \sigma_{t\bar{t}} \times \frac{\dd \Gamma_t}{\Gamma_t}
               \times \frac{\dd \Gamma_{\bar{t}}}{\Gamma_t}\,,
 \label{eq:width1}
\end{equation}
where $\dd\sigma_{t\bar{t}}$ denotes the differential cross-section for on-shell top-quark pair production and $\dd \Gamma_t$ is the differential top-quark decay rate to leptons. Spin correlations, explicitly given in eq.~(\ref{eq:spincorr}), are represented by the $\times$ symbol. Note that while eq.~(\ref{eq:width1}) was derived at LO in QCD it holds to all orders in $\as$. The explicit expansion through NNLO in QCD is given in sec.~\ref{sec:NWA-expand} below. 

Before closing this section we would like to comment on some features of $t\bar t$ production and decay in the NWA. As implied by eq.~(\ref{eq:spincorr}) the $t\bar t$ production and decay cross-section does not contain any interference effects between emissions from the $t\bar t$ production stage and the $t/\bar t$ decays or between the $t$ and $\bar t$ decays. Likewise, no loop diagrams are included that connect $t\bar t$ production and the $t/\bar t$ decays or the $t$ and $\bar t$ decays. 

All other loop effects and interferences contributing to a given order are included. In other words, the only interference and loop effects that are considered happen within either $t\bar t$ production, or $t$ decay or $\bar t$ decay. Effectively, $t\bar t$ production, $t$ and $\bar t$ decays are computed independently, the only connection between the three being the kinematics and spin of the top quark and antiquark which are passed correctly from the $t\bar t$ production to $t/\bar t$ decays. For example, at NNLO in QCD two-loop corrections are included in both the $t\bar t$ production cross-section and the decay rates. Similarly, at NNLO double real emissions, with their interferences, are included in both the $t\bar t$ production cross-section as well as in the two decay rates. An interesting example is the possibility at NNLO for emitting a $b\bar b$ pair which, in the calculation of the $t/\bar t$ decay, leads to final states with three $b/\bar b$-quarks. The interferences arising in such final states are properly taken into account.

\subsection{Matrix elements}\label{sec:amplitudes}

As follows from eq.~(\ref{eq:spincorr}) for the computation of the NNLO $t\bar t$ cross-section including top decay in the NWA, various polarized tree, one-loop and two-loop matrix
elements are required. 

All tree-level matrix elements are obtained from the AvH library \cite{Bury:2015dla,vanHameren:2007pt} which is the default tree-level matrix-element generator within the {\sc Stripper} c++ framework. The AvH library provides polarized matrix elements which makes it possible to compute all necessary spin- and/or color-correlated tree-level amplitudes required by the subtraction scheme. A number of cross-checks have been performed to ensure consistency of the polarization conventions. Transformation properties under discrete CPT transformations have been checked and agreement with the expectations has been found. The NWA matrix elements have been cross-checked against {MadGraph5\_aMC@NLO \cite{Alwall:2014hca} for several phase space points. Furthermore, we have checked that inclusive phase-space integrals reproduce the corresponding branching ratios.

Regarding loop corrections in $t\bar t$ production, the matrix elements for the following  processes are required at NNLO
\begin{eqnarray}
  &g g \to t \bar{t} \quad         &\text{at one- and two-loops}\,,\\
  &q \bar{q} \to t \bar{t} \quad   &\text{at one- and two-loops}\,,\\
  &g g \to t \bar{t} g \quad       &\text{at one-loop}\,,\\
  &q \bar{q} \to t \bar{t} g \quad &\text{at one-loop}\,,\\
  &g q \to t \bar{t} q \quad       &\text{at one-loop}\,.
\end{eqnarray}
The polarized matrix elements for $g g (q\bar{q}) \to t\bar{t}$ are taken from ref.~\cite{Chen:2017jvi}. Various checks of transformation properties and polarization sums of these matrix elements have been performed. The matrix elements for the real-virtual contribution are obtained from ref.~\cite{Denner:2002ii}. Combined with tree-level decays, these have been cross-checked at the phase-space point level against the {\sc OpenLoops} library \cite{Cascioli:2011va,Buccioni:2019sur}.

Regarding loop corrections in the $t/\bar t$ decay, one needs the following matrix elements
\begin{align}
 t \to b \ell^+ \nu \quad &\text{at one- and two-loop}\,,\\
 t \to b \ell^+ \nu g \quad &\text{at one-loop}\,,
\end{align}
as well as the corresponding charge conjugated processes. The matrix elements are available in the literature \cite{Lim:2018qiw,Bonciani:2008wf,Asatrian:2008uk,Beneke:2008ei}. To cross-check our implementation numerical results from \cite{Brucherseifer:2013iv} have been reproduced.

\subsection{Treatment of the top quark width and $W$ decay}\label{sec:top-width}

The top-quark width can be computed in perturbation theory
\begin{equation}
\Gamma_{t} = \Gamma_{t}^{(0)} + \as \Gamma_{t}^{(1)} +\as^2
\Gamma_{t}^{(2)} + \order{\as^3} \,. 
\label{eq:Gammatexp}
\end{equation}
It enters the NWA calculation as a parameter, through the denominator in eq.~(\ref{eq:NWA_LO}).

There are various possibilities for treating the $\frac{1}{\Gamma_t}$ factors. For example, one can set the value of $\Gamma_t$ to its numerical value corresponding to the perturbative order of the full calculation. Alternatively, one can formally expand eq.~(\ref{eq:width1}) in powers of $\as$ and keep only the terms consistent with the overall order of the full calculation \cite{Campbell:2012uf}. Both in this work and in ref.~\cite{Behring:2019iiv} we have employed the latter version since it leaves the normalization of the inclusive $t\bar{t}$ production cross-section unmodified.

In this work we consider only semileptonic decays of the $W$ boson. For this reason no QCD corrections are included i.e. we only consider the W decay at LO. For simplicity, we implement it also in the NWA approximation. Although the ratio $\Gamma_W/m_W$ is larger than the corresponding one for the top quark, it was shown in refs.~\cite{Denner:2010jp,Denner:2012yc} that the effects beyond NWA for the $W$ boson in $t\bar t$ production are very small. 

Treating the $W$-bosons in the NWA the (differential) top-quark decay rate can be written as
\begin{equation}
\dd\Gamma_t =\dd \Gamma( t \to b W^+)
             \sum_{ff'}\frac{\dd\Gamma(W^+\to ff')}{\Gamma_W}\,,
\end{equation}
assuming a diagonal CKM matrix. Assuming all $W$ decay channels are included, the integration of eq.~(\ref{eq:width1}), at LO, over the full phase space results in the total $t\bar{t}$ production cross section
\begin{equation}
  \sigma = \sigma_{t\bar{t}}\,.
\end{equation}
The preservation of the normalization also holds order-by-order as can also be seen in sec.~\ref{sec:NWA-expand}. However, one should keep in mind that this property may be altered if certain choices of scales are made; see sec.~\ref{sec:NWA-expand} for details. In the present work $W$ boson decays are restricted to $\mu$ and $e$ leptonic final states, i.e. no $W\to \tau$ decays are included
\footnote{Lepton flavor universality in $W\to\mu$ and $W\to\tau$ decays originating from $t\bar t$ events has been directly verified with high-precision at the LHC \cite{Aad:2020ayz}.}.
In such a setup the normalization of the $t\bar t$ production and decay cross-section reads
\begin{equation}
  \sigma = \sigma_{t\bar{t}} \left( \sum_{f,f' \in \{e,\mu\}} \text{BR} (W^+ \to
f^+ \nu_f) \text{BR} (W^- \to f^{'-} \bar{\nu}_f')\right)\;.
\label{eq:width2}
\end{equation}

\subsection{Explicit expressions for the NWA cross-section through NNLO}\label{sec:NWA-expand}

Each term appearing in the RHS of eq.~(\ref{eq:width1}) can be expanded systematically in $\as$
\begin{eqnarray}
 \dd \sigma_{t\bar{t}} &=& \dd \sigma_{t\bar{t}}^{(0)} + \as \dd
\sigma_{t\bar{t}}^{(1)} + \as^2 \dd \sigma_{t\bar{t}}^{(2)}\,,\\
  \dd \Gamma_{t(\bar{t})} &=& \dd \Gamma_{t(\bar{t})}^{(0)} + \as \dd
\Gamma_{t(\bar{t})}^{(1)} +\as^2 \dd \Gamma_{t(\bar{t})}^{(2)}\,.
\end{eqnarray}
Expanding eq.~(\ref{eq:width1}) up to second non-trivial order in $\as$ one obtains the following cross section decomposition
\begin{equation}
  \dd \sigma = \dd \sigma^{\text{LO}} + \as\dd \sigma^{\text{NLO}}
               +\as^2\dd \sigma^{\text{NNLO}}\,,
\label{eq:dsigma-expanded}
\end{equation}
where
\begin{eqnarray}
  \dd \sigma^{\text{LO}} &=& \sigma^{\text{LOxLO}}\,, \\
  \dd \sigma^{\text{NLO}} &=& \dd \sigma^{\text{NLOxLO}} +\dd
      \sigma^{\text{LOxNLO}} - \frac{2\Gamma_t^{(1)}}{\Gamma_t^{(0)}}
      \dd \sigma^{\text{LO}}\,,\\
  \dd \sigma^{\text{NNLO}} &=& \dd \sigma^{\text{NNLOxLO}} +\dd
      \sigma^{\text{NLOxNLO}}+\dd \sigma^{\text{LOxNNLO}}\nn\\
   & &- \frac{2\Gamma_t^{(1)}}{\Gamma_t^{(0)}}
      \dd \sigma^{\text{NLO}} - \left(\frac{\Gamma_t^{(1)2}}{ \Gamma_t^{(0)2}} 
      +\frac{2\Gamma_t^{(0)}\Gamma_t^{(2)}}{ \Gamma_t^{(0)2}} 
\right)\dd \sigma^{\text{LO}}\,.
\label{eq:width3}
\end{eqnarray}
As implied by the notation above, contributions at a given order are combined according to whether the corrections are
located in the production and/or decay:
\begin{eqnarray}
\dd\sigma^{\text{LOxLO}} &=& \dd \sigma_{t\bar{t}}^{(0)} \times \frac{\dd
\Gamma_t^{(0)}}{\Gamma_t^{(0)}}
               \times \frac{\dd \Gamma_{\bar{t}}^{(0)}}{\Gamma_t^{(0)}}\,, \\
\dd\sigma^{\text{NLOxLO}} &=& \dd \sigma_{t\bar{t}}^{(1)} \times \frac{\dd
\Gamma_t^{(0)}}{\Gamma_t^{(0)}}
               \times \frac{\dd \Gamma_{\bar{t}}^{(0)}}{\Gamma_t^{(0)}}\,, \\
\dd\sigma^{\text{NNLOxLO}} &=& \dd \sigma_{t\bar{t}}^{(2)} \times \frac{\dd
\Gamma_t^{(0)}}{\Gamma_t^{(0)}}
               \times \frac{\dd \Gamma_{\bar{t}}^{(0)}}{\Gamma_t^{(0)}}\,, \\
\dd\sigma^{\text{LOxNLO}} &=& \dd \sigma_{t\bar{t}}^{(0)} \times
 \left(\frac{\dd \Gamma_t^{(1)}}{\Gamma_t^{(0)}}\times
       \frac{\dd \Gamma_{\bar{t}}^{(0)}}{\Gamma_t^{(0)}}
       +\frac{\dd \Gamma_t^{(0)}}{\Gamma_t^{(0)}}\times
       \frac{\dd \Gamma_{\bar{t}}^{(1)}}{\Gamma_t^{(0)}} \right)\,, \\
\dd\sigma^{\text{LOxNNLO}} &=& \dd \sigma_{t\bar{t}}^{(0)} \times
 \left(\frac{\dd \Gamma_t^{(2)}}{\Gamma_t^{(0)}}\times
       \frac{\dd \Gamma_{\bar{t}}^{(0)}}{\Gamma_t^{(0)}}
       +\frac{\dd \Gamma_t^{(0)}}{\Gamma_t^{(0)}}\times
       \frac{\dd \Gamma_{\bar{t}}^{(2)}}{\Gamma_t^{(0)}}
       +\frac{\dd \Gamma_t^{(1)}}{\Gamma_t^{(0)}}\times
       \frac{\dd \Gamma_{\bar{t}}^{(1)}}{\Gamma_t^{(0)}}
\right)\,, \\
\dd\sigma^{\text{NLOxNLO}} &=& \dd \sigma_{t\bar{t}}^{(1)} \times
 \left(\frac{\dd \Gamma_t^{(1)}}{\Gamma_t^{(0)}}\times
       \frac{\dd \Gamma_{\bar{t}}^{(0)}}{\Gamma_t^{(0)}}
       +\frac{\dd \Gamma_t^{(0)}}{\Gamma_t^{(0)}}\times
       \frac{\dd \Gamma_{\bar{t}}^{(1)}}{\Gamma_t^{(0)}} \right)\,.
\label{eq:xsec_decomp}
\end{eqnarray}

We want to close the discussion of the cross-section decomposition in the NWA with a remark on its renormalization and factorization scale dependence. The factorization scale enters only the production cross-section $\dd\sigma^{(i)}_{t\bar{t}}$. For this reason the scale $\mu_F$ can be chosen based on past experience from stable top production. For the renormalization scale the situation is slightly more complicated since also the $t/\bar t$ decay rates depend on it. It is well-known that dynamical scale choices for $\mu_R$ lead to much better perturbative convergence for the production cross-section. Regarding decays, since the top-quarks are on-shell, one may expect that the top-quark mass is a good scale choice for $\mu_R$ in the inclusive and differential decay rates. While either scale choice can be implemented consistently, for reasons of simplicity we decided to evaluate the cross-section in a mixed scheme where all differential quantities (i.e. production and differential decay rates) are evaluated with the same scale (which itself could take dynamic or fixed values) while the inclusive decay rate $\Gamma^{(i)}$ appearing in eq.~(\ref{eq:dsigma-expanded}) is always evaluated at a fixed scale $\mu_R=m_t$. In particular, during scale variation, the scale in the inclusive rate is not varied. In practice this means that the inclusive decay rate in eq.~(\ref{eq:dsigma-expanded}) is treated as a fixed parameter independent of kinematics.

Due to our choice of scales, the normalization of the inclusive cross-section implied by eq.~(\ref{eq:width2}) is not going to be preserved at each order in $\as$. We have checked that the numerical impact due to this modification is negligible.

\subsection{Dipole subtraction for the $\dd\sigma^{\text{NLOxNLO}}$ contribution}\label{eq:nloxnlo}

With one exception, the present calculation is implemented in the four-dimensional formulation of the sector improved residue subtraction scheme. The one exception is the contribution
\begin{equation}
\dd\sigma^{\text{NLOxNLO}} = \dd \sigma_{t\bar{t}}^{(1)} \times
 \left(\frac{\dd \Gamma_t^{(1)}}{\Gamma_t^{(0)}}\times
       \frac{\dd \Gamma_{\bar{t}}^{(0)}}{\Gamma_t^{(0)}}
       +\frac{\dd \Gamma_t^{(0)}}{\Gamma_t^{(0)}}\times
       \frac{\dd \Gamma_{\bar{t}}^{(1)}}{\Gamma_t^{(0)}} \right)\,,
\end{equation}
in which two NLO computations have to be combined. For practical reasons, that have to do with our software implementation, it is not convenient to perform both in the sector improved scheme even though there is no conceptual problem in doing so. Instead we deal with the NLO corrections in the $t\to b+W$ decay within a modified Catani-Seymour subtraction \cite{Catani:1996jh,Catani:2002hc} as implemented in ref.~\cite{Campbell:2004ch}. We briefly review it in the following.

The NLO correction to the (differential) $t\to b+W$ top-quark width is given by
\begin{eqnarray}
  \Gamma_t^{(1)} &=& \Gamma_t^R + \Gamma_t^V \\
                 &=& \int \dd^d \Phi_3
                    \braket{\mathcal{M}^{(0)}_3} +
    \int \dd^d \Phi_2~ 2\Re\braket{\mathcal{M}^{(0)}_2}{\mathcal{M}^{(1)}_2}\,.
\end{eqnarray}
The infrared limits of the real radiation contribution is regulated in $d = 4-2\ep$ dimensions by subtracting the dipole
\begin{equation}
 D\left(p_t,p_W,p_b,p_g\right) = 4\pi C_F\left[
   \frac{1}{p_b\cdot p_g}\left(\frac{2}{1-z}-1-z-\ep(1-z) \right)
     -\frac{m_t^2}{(p_t\cdot p_g)^2}
  \right]\,,
\end{equation}
such that the expression
\begin{equation}
\int \dd^{d} \Phi_3 \Bigg\{ \braket{\mathcal{M}^{(0)}_3}(p_t,p_W,p_b,p_g) - D(p_t,p_W,p_b,p_g)
\braket{\mathcal{M}^{(0)}_2}(p_t,\tilde{p}_W,\tilde{p}_b) \Bigg\}
\end{equation}
is integrable in $d=4$. The momentum mapping necessary for the unresolved configuration is given by
\begin{equation}
 \tilde{p}_W = \frac{p_t^2-p_W^2}{2\sqrt{(p_t\cdot p_W)^2-p_W^2p_t^2}}\left(p_W-\frac{p_t\cdot p_W}{p_t^2}p_t\right) + \left(\frac{p_t^2+p_W^2}{2p_t^2}\right)p_t\,.
\end{equation}
The momentum $\tilde{p}_b$ is fixed through $\tilde{p}_b = p_t - \tilde{p}_W$. The integrated dipole is determined from 
\begin{equation}
 \dd \Gamma^I = \dd \Phi_3\, D\,\braket{\mathcal{M}^{(0)}_2} = \dd
\Phi_2(\tilde{p}_W,\tilde{p}_b) \braket{\mathcal{M}^{(0)}_2} \int \dd \mu_g(p_g)\, D \,.
\label{eq:Gamma-dipole}
\end{equation}
The above expression can be directly combined with the virtual contribution to cancel all $\ep$ poles. The explicit expression for the measure $\dd\mu_g$ appearing in eq.~(\ref{eq:Gamma-dipole}) can be found in ref.~\cite{Campbell:2004ch}.

\subsection{$\as$-expansion of normalized cross-sections}\label{sec:expand-ratio}

In collider physics one often encounters normalized differential distributions 
\begin{equation}
R \equiv \frac{1}{\sigma} \frac{\dd \sigma}{\dd X} \,,
\end{equation}
where $\dd\sigma/\dd X$ is some differential distribution and $\sigma$ is either the total cross-section (if it exists) or a suitably defined fiducial cross-section. Both have perturbative expansions
\begin{eqnarray}
\sigma &=& \sigma^{(0)} + \as \sigma^{(1)} + \as^2 \sigma^{(2)}\,,\\
\frac{\dd\sigma}{\dd X} &=& \frac{\dd\sigma^{(0)}}{\dd X}
                           + \as \frac{\dd\sigma^{(1)}}{\dd X}
                           + \as^2 \frac{\dd\sigma^{(2)}}{\dd X}\,.
\end{eqnarray}

Typically, normalized distributions are computed by simply dividing the differential distribution by the normalization factor which itself is computed numerically with the same precision as $\dd\sigma/\dd X$. However, sometimes one considers the formally equivalent definition when the normalized distribution itself is expanded in powers of $\as$; indeed, in the following we will consider one such example for dilepton $t\bar t$ final states. To introduce this definition we first introduce a formal perturbative expansion for $R$
\begin{equation}
R = R^{(0)} + \as R^{(1)} + \as^2 R^{(2)}\,.
\end{equation}
The perturbative coefficients in this expansion through NNLO read
\begin{eqnarray}
R^{(0)} &=& \frac{1}{\sigma^{(0)}} \frac{\dd \sigma^{(0)}}{\dd X}\;, \\
R^{(1)} &=& \frac{1}{\sigma^{(0)}}\frac{\dd \sigma^{(1)}}{\dd X} -
         \frac{\sigma^{(1)}}{\sigma^{(0)}}\frac{1}{\sigma^{(0)}} \frac{\dd \sigma^{(0)}}{\dd X}\;,\\
R^{(2)} &=& \frac{1}{\sigma^{(0)}}\frac{\dd \sigma^{(2)}}{\dd X} -
         \frac{\sigma^{(1)}}{\sigma^{(0)}}\frac{1}{\sigma^{(0)}} \frac{\dd \sigma^{(1)}}{\dd X} +
          \left(
            \left(\frac{\sigma^{(1)}}{\sigma^{(0)}}\right)^2-\frac{\sigma^{(2)}}{\sigma^{(0)}}
           \right)\frac{1}{\sigma^{(0)}} \frac{\dd \sigma^{(0)}}{\dd X}\,.
\end{eqnarray}

Finally we specify the order of parton distribution functions (pdf) we use in the calculation of $R$ at a given perturbative order
\begin{align}
  R^\text{LO} &= R^{(0)} &\text{all} \; \sigma^{(i)} \; \text{and} \; \dd\sigma^{(i)} \; \text{with LO pdf,}\\
  R^\text{NLO} &= R^{(0)}+\as R^{(1)} &\text{all} \; \sigma^{(i)} \; \text{and} \; \dd\sigma^{(i)} \; \text{with NLO pdf,}\\
  R^\text{NNLO} &= R^{(0)}+\as R^{(1)} + \as^2 R^{(2)} &\text{all} \; \sigma^{(i)} \; \text{and} \; \dd\sigma^{(i)} \; \text{with NNLO pdf.}
\end{align}

\section{Phenomenology}\label{sec:pheno}

The purpose of this section is twofold. First, we derive theoretical predictions for a variety of measurements by the ATLAS and CMS collaborations of $t\bar t$ final states in the dilepton channel. This includes one- and two-dimensional differential distributions as well as the complete set of observables needed for the reconstruction of the $t\bar t$ spin density matrix measured by CMS. Additional results - extending our previous work \cite{Behring:2019iiv} - on the $\Delta\phi(\ell\bar{\ell})$ distribution which is very sensitive to spin-correlations are also presented. All these results are accompanied by detailed theory/data comparison. Second, we analyze the ability of high-precision fixed-order calculations to describe experimental data at the particle level. As we will see in the following this is a highly-nontrivial task which requires refinement and better understanding of questions like jet composition, top-quark reconstruction and implementation of cuts. 

Before delving into the above mentioned predictions and lessons learned, we introduce the physical objects we work with together with some terminology which will be used extensively in the following. This terminology follows closely the experimental one and, hopefully, will make it easier to bridge the gap between theoretical and experimental jargon. 

The objects of interest are leptons, neutrinos, jets and $b$-jets. Since in our work no QED corrections are included, we do not impose any photon-lepton clustering requirement as is often done in LHC measurements. Both inclusive jets and $b$-jets are defined with the anti-$k_T$ algorithm \cite{Cacciari:2008gp} with size $R=0.4$. In our calculations $b$-quarks are always considered massless. Since at the second order in perturbation theory it is possible to radiate $b\bar b$ pairs off of any parton, including the $b$-quarks resulting from the decay of the top quark, we have to carefully consider the flavor content of jets. We define $b$-jet as a jet with {\it nonzero net bottomness}. For example, a jet containing a single $b$ quark (or $b$-antiquark) plus up to two quark(s) of different flavor and/or gluon(s) is considered to be a $b$-jet. Similarly, a jet containing $b\bar b b$ is also considered a $b$-jet as well as jets containing $bb$ of $\bar b\bar b$ plus up to one gluon. On the other hand a jet containing $b\bar b$ plus up to one gluon is not a $b$-jet.

It is well known \cite{Banfi:2006hf} that starting at NNLO, the anti-$k_T$ algorithm does not provide an infrared-safe definition of jet flavour. In the process under consideration this issue manifests itself through the emission of a soft $b\bar b$ pair. A proper treatment of jet flavour would require a modification of the jet-algorithm. To that end one could utilize the so-called flavour-$k_T$ algorithm \cite{Banfi:2006hf} which is infrared-safe, however, this would not allow for a direct comparison with the experimental measurements which use the anti-$k_T$ algorithm. In this work we adopt the following practical solution: since the numerical impact of a soft $b\bar b$ emission is small, see also refs.~\cite{Berger:2017zof,Catani:2020kkl,Campbell:2020fhf}, we use the standard anti-$k_T$ algorithm which in our calculation is automatically regulated by the technical cutoff defined along the lines of eq.~(118) in ref.~\cite{Czakon:2011ve}.

Next we explain what we mean by top quarks in our comparisons with data. Depending on the specific experimental analysis we will be considering either the {\it true} top or the so-called {\it reconstructed} top. By true top we mean the top quark (or antiquark) which was decayed. In practice the information about the true top is available to us in each event as a Monte Carlo truth and we do not need to reconstruct them from the momenta of their decay products. Alternatively, a reconstructed top is the (pseudo)object derived from an algorithm over the four-momenta of the final state objects (like jets, leptons and neutrinos) that pass the experimental cuts. The reconstruction algorithm is specific to a given analysis; here we only note that in general due to the incomplete knowledge of final state momenta, typically, there arises a difference between the four-momenta of the true and reconstructed tops. Quantitative analysis of this difference can be found in sec.~\ref{sec:pheno_fid}.

As it turns out, the best way to classify the possible analyses, is by the inclusiveness of observables with respect to hadronic radiation. Specifically, we will be considering analyses that are not fully inclusive in hadronic radiation. These are truly differential analyses where fiducial cuts are imposed on all objects like leptons and $b$-jets. Alternatively, many analyses exist which do not contain any cuts on jets. Clearly, this is possible because the differential data has been extrapolated to full phase-space with the help of Monte Carlo event generators. The details of this procedure are specific to each analysis. We would like to stress that both classes of distribution typically have additional fiducial cuts on the two leptons. This is the reason we do not speak of fully inclusive analyses but rather of analyses inclusive of hadronic radiation (although in the following we consider some fully inclusive analyses, too).

With the above qualifications in mind, in the following we will refer for short to the two types of analyses as {\it inclusive} and {\it fiducial} ones. Analyses of the inclusive type will be considered in sec.~\ref{sec:pheno_sdm} and in sec.~\ref{sec:pheno_inclusive}; a fiducial analysis is detailed in sec.~\ref{sec:pheno_fid}. 

Finally, let us specify the setup for the calculations of this work. Throughout we utilize the $G_F$ scheme with the following parameters
\begin{eqnarray}
  m_W &=& 80.385 \; \text{GeV} \\ \Gamma_W &=& 2.0928\;\text{GeV}\\
  m_Z &=& 91.1876 \; \text{GeV} \\ \Gamma_Z &=& 2.4952\;\text{GeV}\\
  G_F &=& 1.166379\cdot10^{-5} \;\text{GeV}^{-2} \\ \alpha &=&
       \frac{\sqrt{2} G_F}{\pi} m_W^2\left(1-(m_W/m_Z)^2\right)
\end{eqnarray}
The leading order top-quark width, entering the denominators in eq.~(\ref{eq:dsigma-expanded}), is computed from
\begin{equation}
\Gamma_t^{(0)} = G_F \frac{m_t^3}{8\pi\sqrt{2}}(1-x)^2(1+2 x) = 1.48063 \;\text{GeV}~ ({\rm for}~ m_t = 172.5\;\text{GeV})\,,
\end{equation}
where $x=(m_W/m_t)^2$. The default value of the top-quark mass is $m_t = 172.5$ GeV. The value for $\Gamma_t^{(0)}$ is always adapted, as needed, to the value of $m_t$. The order of the pdf sets is matched to the perturbative order of the calculation, i.e. we use a LO pdf set for a LO computation, NLO pdf set for a NLO computation and NNLO set for a NNLO computation. All results presented in this work are for the LHC at 13 TeV hadron-hadron center of mass energy. By default we use the NNPDF3.1 pdf set \cite{Ball:2017nwa} implemented through the LHAPDF interface \cite{Buckley:2014ana}. We consistently renormalize in a scheme with $n_f = 5$ active flavors and use pdf sets with five active flavors.

By default we use the dynamical central renormalization and factorization scale \cite{Czakon:2016dgf}
\begin{equation}
  \mu = \mu_R = \mu_F = \frac{H_T}{4} = \frac{1}{4}\sum_{i\in{t,\bar{t}}}
              \sqrt{m_t^2+p_{T,i}^2}\,.
\label{eq:scale}
\end{equation}
To estimate theory uncertainties due to missing higher-order terms we perform a standard independent 7-point $\mu_F,\mu_R$ variation by a factor of 2 around the central scale.
Monte Carlo integration uncertainties are usually small enough to be negligiable and are not stated explicitly. They can be accessed together with all other produced results in electronic format under \url{http://www.precision.hep.phy.cam.ac.uk/results/ttbar-decay/}.

\subsection{Top-pair spin correlations at the LHC}\label{sec:pheno_sdm}

Directly measuring the spin of the top quarks produced at hadron colliders is very challenging since top quarks are colored particles. There exists, however, the possibility to connect the spin of the top quark with its decay products because, due to top quark's short lifetime, hadronization cannot smear its spin. This offers the unique opportunity to study the spin of a bare quark indirectly through its decay products. While top quarks are produced un-polarized in hadron-hadron collisions, there is a correlation between the spins of the top and anti-top quarks in $t\bar t$ production. This correlation is imprinted in the $t$ and $\bar t$ decay products leading to a measurable effect on various angular observables \cite{Mahlon:1995zn,Mahlon:2010gw}. In this section two approaches for studying top-pair spin correlations are presented. The first, the so-called spin-density matrix formalism aims at reconstructing the complete set of spin correlations between the $t\bar t$ pair from the kinematics of the final state decay products. This approach requires the reconstruction of special reference frames connected to the top quark and antiquark, however, it presents certain experimental difficulties and tends to result in reduced precision. Alternatively, one may consider distributions of the $t\bar t$ decay products which are well defined and measured in the lab frame, and then try to disentangle the effects due to spin correlations from the ones due to kinematics. We will present examples of the first type in sec.~\ref{sec:spin-density-formalism} and sec.~\ref{sec:spin-density-pheno} while of the second type in sec.~\ref{sec:spin-correlations-angular}.

\subsubsection{Spin-density matrix: the formalism}\label{sec:spin-density-formalism}

A way of formalizing the $t\bar t$ spin correlations is by using the so-called spin-density matrix formalism \cite{Bernreuther:1993hq}. In the following we only emphasize the main points and refer the reader to the original literature for details. Starting with the LO cross-section, the matrix element for top quark pair production with subsequent leptonic decays is written as follows
\begin{equation}
 |\mathcal{M}(q\bar{q}/gg \to t\bar{t} \to \ell^+\ell^-\nu\bar{\nu}b\bar{b})|^2
   \sim \Tr{\rho \mathrm{R} \bar{\rho}}\,,
\label{eq:M-spin-corr}   
\end{equation}
where $\mathrm{R}$ is the spin density matrix describing the top-quark pair production and $\rho$ and $\bar{\rho}$ are the spin density matrices describing the $t$ and $\bar t$ decays, respectively. It is most convenient to define the matrix $\mathrm{R}$ in the rest frame of the $t\bar t$ system. For its definition one also needs two spin vectors $s_t$ and $s_{\bar{t}}$, which can be defined in the following way. One first constructs two unit 3-vectors $\hat s_t$ and $\hat s_{\bar{t}}$ defined in the $t$ and $\bar t$ rest frames, respectively. In turn, these 3-vectors naturally define two 4-vectors
\begin{equation}
s_t^r = \left(\begin{array}{c}0\\\hat{s}_t \end{array}\right)
\hspace{0.1\textwidth}\text{and}\hspace{0.1\textwidth}
s_{\bar{t}}^r = \left(\begin{array}{c}0\\\hat{s}_{\bar{t}}
  \end{array}\right)\,.
\end{equation}
The desired spin vectors $s_t$ and $s_{\bar{t}}$ are obtained from the 4-vectors $s_t^r$ and $s_{\bar{t}}^r$ by boosting them as appropriate from their respective $t$ and $\bar t$ rest frames to the $t\bar t$ rest frame: $s_t^r\to s_t$ and $s_{\bar{t}}^r\to s_{\bar{t}}$. The 4-vectors $s_t$ and $s_{\bar{t}}$ are valid spin 4-vectors since
\begin{equation}
 s_{t}^2 = s_{\bar{t}}^2 = -1 \quad \text{and}\quad p_t \cdot s_t = p_{\bar t}
 \cdot s_{\bar{t}} = 0 \;.
\end{equation}
The spin-correlated matrix element for $t\bar{t}$ production with spins parameterized by $s_t$
and $s_{\bar{t}}$ can be obtained by inserting the projectors
\begin{eqnarray}
  u(p_t, s_t)\bar{u}(p_t, s_t) &=&
\left(\slashed{p}_t+m\right)\frac{1}{2}\left(1+\g_5\slashed{s}_t\right)
  \,, \\
  v(p_{\bar t}, s_{\bar{t}})\bar{v}(p_{\bar t}, s_{\bar{t}})  &=&
\left(\slashed{p}_{\bar t}-m\right)\frac{1}{2}\left(1+\g_5\slashed{s}_{\bar{t}}
\right)\,,
\end{eqnarray}
into the squared matrix element in eq.~(\ref{eq:M-spin-corr}) and performing the required spin sums. The resulting spin-correlated $t\bar t$ production matrix element reads
\begin{equation}
|\mathcal{M}(s_{t},s_{\bar{t}})|^2 = \frac{1}{4} \Tr
\left[ 
\mathrm{R}(\mathbf{1}_{2\times 2}+\hat{s}_t\cdot\vec\sigma)\otimes(\mathbf{1}_{2\times 2}
      +\hat{s}_{\bar{t}}\cdot\vec\sigma)
\right] \;,
\label{eq:R-def}
\end{equation}
where $\sigma^i$ are the Pauli matrices and $\otimes$ is a tensor product in the spin indices of the $t$ and $\bar t$ quarks.

Eq.~(\ref{eq:R-def}) effectively serves as the definition of the matrix $\mathrm{R}$. Expanding the spin-density matrix in a basis of $\sigma$ matrices one arrives at the following decomposition
\begin{equation}
\mathrm{R} = \tilde{A}\; \mathds{1} \otimes \mathds{1} + \tilde{B}^+_i\sigma^i
  \otimes \mathds{1} + \tilde{B}_i^- \mathds{1} \otimes \sigma^i+
  \tilde{C}_{ij} \sigma^i \otimes \sigma^{j}\,.
  \label{eq:R-tilde}
\end{equation}
The coefficient functions $\tilde{A}$, $\tilde{B}^\pm_i$ and $\tilde{C}_{ij}$ defined by the above equation parameterize, respectively, the spin-summed matrix element, the top-quark/top-antiquark polarization and the spin correlation between $t$ and $\bar t$. Note that the dependence on the spin vectors $s_t$ and $s_{\bar{t}}$ is absorbed solely in the coefficients $\tilde{B}^\pm_i$ and $\tilde{C}_{ij}$.

To connect the above construction with experiment one needs to address several questions. First, one needs to promote the above considerations to the level of cross-section which is generalizable to inclusive observables at higher orders in perturbation theory. Second, one needs to find a convenient set of experimentally accessible angular variables in terms of which to express spin correlations. Finally, one has to overcome the problem that the spins of the top quark and antiquark are not accessible experimentally.

To overcome the above mentioned complications, one chooses an approach where a well-defined, experimentally measurable proxy for the spin-density matrix is introduced. To that end one starts by making a suitable choice of proxies for the two spin-vectors. A natural choice in the context of dilepton distributions is to use the 3-momenta of the leptons in the decay of the corresponding top quark/antiquark. In fact it can be shown that this is a very good choice. The differential decay rate of top-quark (in its rest frame) with respect to the angle $\chi_a$
between the momentum of the lepton $\ell$ and the top quark's spin vector is given by
\begin{equation}
\frac{1}{\Gamma}\frac{\dd\Gamma}{\dd\cos\chi_\ell} =
\frac{1}{2}(1+\kappa_\ell\cos\chi_\ell)\;.
\end{equation}
The parameter $\kappa_\ell$ is called spin analyzing power. As the name suggests, if $\kappa_\ell\ll 1$ then there is a very weak correlation between the kinematic variable $\chi_\ell$ and the top-quark spin, while the correlation is very strong if $\kappa_\ell$ is close to its maximal value $\kappa_\ell\approx 1$. With the help of direct calculations it has been verified \cite{Czarnecki:1990pe,Brandenburg:2002xr,Bernreuther:2008ju} that the spin analyzing power of the charged lepton momentum is nearly maximal. 

It is natural to express the $t\bar t$ spin correlations in terms of angles. To define those a suitable orthonormal basis $\{\hat{r},\hat{k},\hat{n}\}$ is needed and it can be constructed as follows \cite{CMS:2018jcg}. Considering the partonic kinematics in the $t\bar t$ rest frame, the vector $\hat{k}$ is given by the direction of the top quark 3-momentum. Together with the direction of the incoming partons $\hat{p}$ it defines the perpendicular direction to the top-quark scattering plane by $\hat{n} = \hat{p}\times\hat{k}/\sin\theta$, where $\theta$ is the top-quark scattering angle and $\times$ is the usual vector product. The remaining direction is given by $\hat{r} = (\hat{p}-\hat{k}\cos\theta)/\sin\theta$. In reality, at hadron colliders like the LHC the direction of the incoming partons in the $t\bar t$ rest frame is not necessarily known. For this reason the direction $\hat{p}$ is taken to be the direction of the beam axis, i.e. $\hat{p} = (0,0,1)$. As a last step the above construction is modified by 
\begin{equation}
  \{\hat{k},\hat{r},\hat{n}\} \to \{\hat{k},\text{sign}(\cos\theta)\hat{r},
                                        \text{sign}(\cos\theta)\hat{n}\}\,.
\end{equation}
The above replacement affects the C- and CP-transformation properties of amplitudes and as a result the $gg$-initiated contributions to CP-odd coefficients $\tilde{C}_{ij}$ are no longer vanishing.

Combining the above, one considers the following differential cross-section as a proxy for the spin-density matrix $\mathrm{R}$
\begin{equation}
\frac{1}{\sigma} \frac{\dd\sigma}{\dd \cos \theta^i_1 \dd \cos \theta^j_2}
 = \frac{1}{4}\left(1 + B^i_1 \cos\theta^i_1 +B^j_2 \cos\theta^j_2
                    - C_{ij} \cos\theta^i_1\cos\theta^j_2 \right)\;,
\label{eq:R-proxy}
\end{equation}
with new ``proxy" coefficients $B^i_1, B^i_1$ and $C_{ij}$. Note that while these coefficients are only proxies for the spin density components in eq.~(\ref{eq:R-tilde}) they inherit their P- and CP-transformation properties. The lepton momenta $\hat{p}_\ell$ define the angles
\begin{equation}
  \cos \theta^i_1 = \hat{p}_{\ell^+} \cdot \hat{i} \;,\quad
  \cos \theta^i_2 = \hat{p}_{\ell^-} \cdot \hat{i} \; \quad
  \text{with} \quad \hat{i} \in \{\hat{n},\hat{k},\hat{r}\}\;.
\end{equation}

The simplest way of extracting the coefficients $B^i_1$, $B^i_2$ and $C_{ij}$ is to reduce the cross-section (\ref{eq:R-proxy}) to several single-differential ones
\begin{eqnarray}
  \frac{1}{\sigma}\frac{\dd\sigma}{\dd\cos\theta^i_1}
     &=& \frac{1}{2} \left(1+B^i_1\cos\theta^i_1\right)\;,
     \label{eq:sdm_xsecs1}\\
  \frac{1}{\sigma}\frac{\dd\sigma}{\dd\cos\theta^i_2}
     &=& \frac{1}{2} \left(1+B^i_2\cos\theta^i_2\right)\;,
     \label{eq:sdm_xsecs2}\\
  \frac{1}{\sigma}\frac{\dd\sigma}{\dd\left(\cos\theta^i_1\cos\theta^i_2\right)}
     &=& \frac{1}{2} \left(1-C_{ii}\cos\theta^i_1\cos\theta^i_2\right)
          \ln\left(\frac{1}{|\cos\theta^i_1\cos\theta^i_2|}\right)\;,    
    \label{eq:sdm_xsecs3}\\
\frac{1}{\sigma} \frac{\dd\sigma}{\dd x_{ij}^\pm} &=& \frac{1}{2}
\left(1- \frac{C_{ij}\pm C_{ji}}{2}x_{ij}^\pm\right)\cos^{-1}\left(|x_{ij}^\pm|\right) ~(\text{for}~i\neq j)\,,
    \label{eq:sdm_xsecs4}
\end{eqnarray} 
where $x_{ij}^\pm = \cos\theta^i_1\cos\theta^j_2\pm \cos\theta^j_1\cos\theta^i_2$. 

With the above differential distributions one can determine the spin-correlation coefficients both experimentally and theoretically.

\subsubsection{Spin-density matrix: NNLO QCD predictions and comparison with data}\label{sec:spin-density-pheno}

In this section we present the results of our NNLO QCD calculation of the differential cross-sections eqs.~(\ref{eq:sdm_xsecs1}--\ref{eq:sdm_xsecs4}). The theoretical predictions are directly compared with experimental measurements published by the CMS Collaboration \cite{CMS:2018jcg}. The CMS analysis is inclusive in the sense defined in sec.~\ref{sec:pheno}. Since the analysis involves no fiducial phase-space cuts, our parton level predictions will be based on true tops and not on reconstructed tops (again, in the sense defined in sec.~\ref{sec:pheno}). The information about the true top-quark and top-antiquark momenta is derived directly from our Monte Carlo as MC truth, without the need for top-quark reconstruction. All input parameters in this calculation are set to the default values described in section \ref{sec:pheno}. Binnings for the distributions have been chosen in such a way that direct comparison of data against our parton-level results can be performed. 

\begin{figure}
\includegraphics[width=4.9cm]{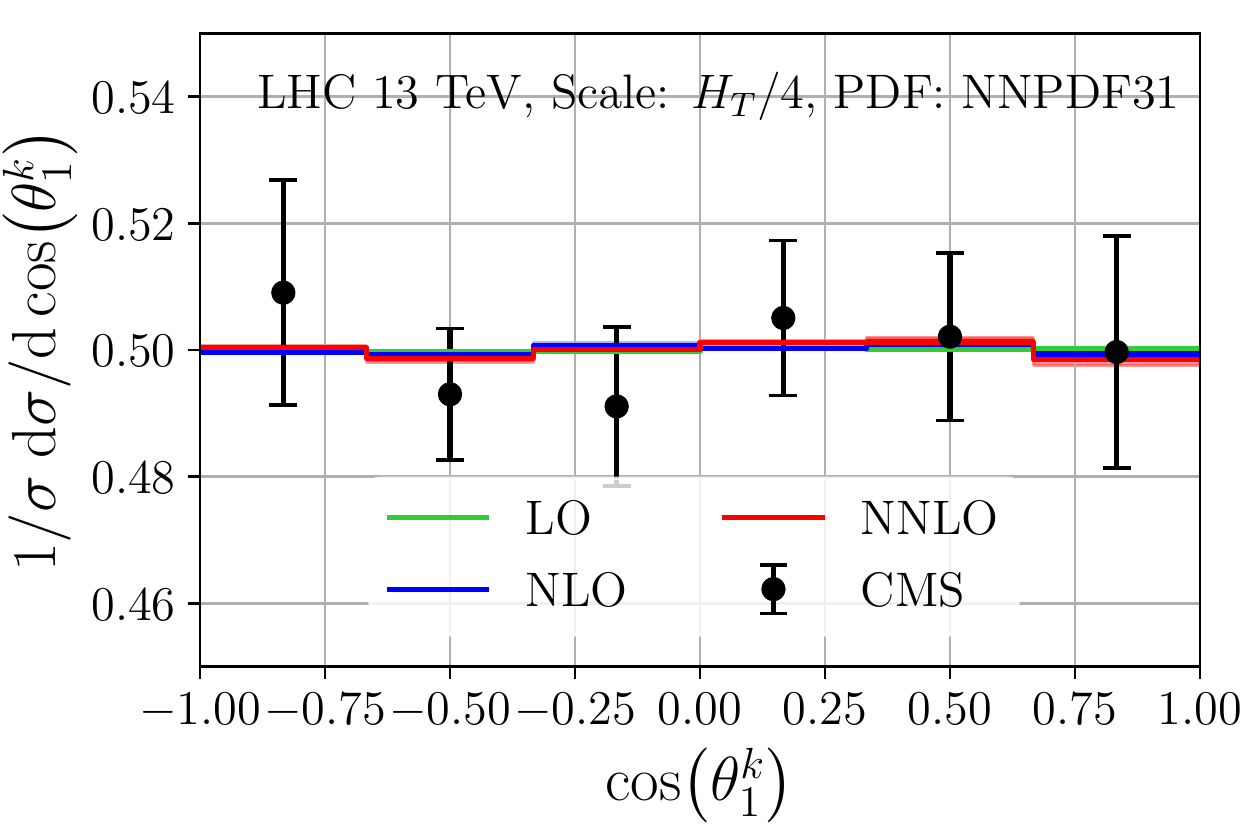}
\includegraphics[width=4.9cm]{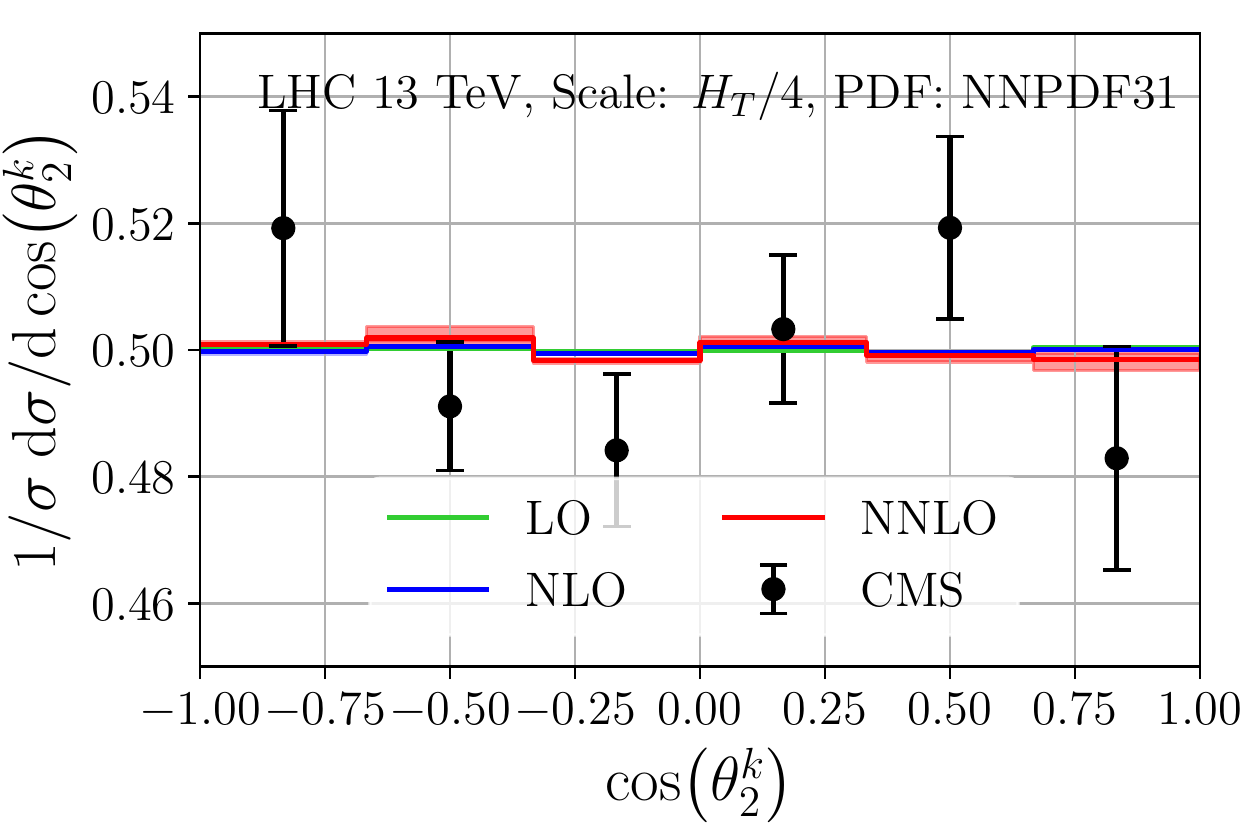}
\includegraphics[width=4.9cm]{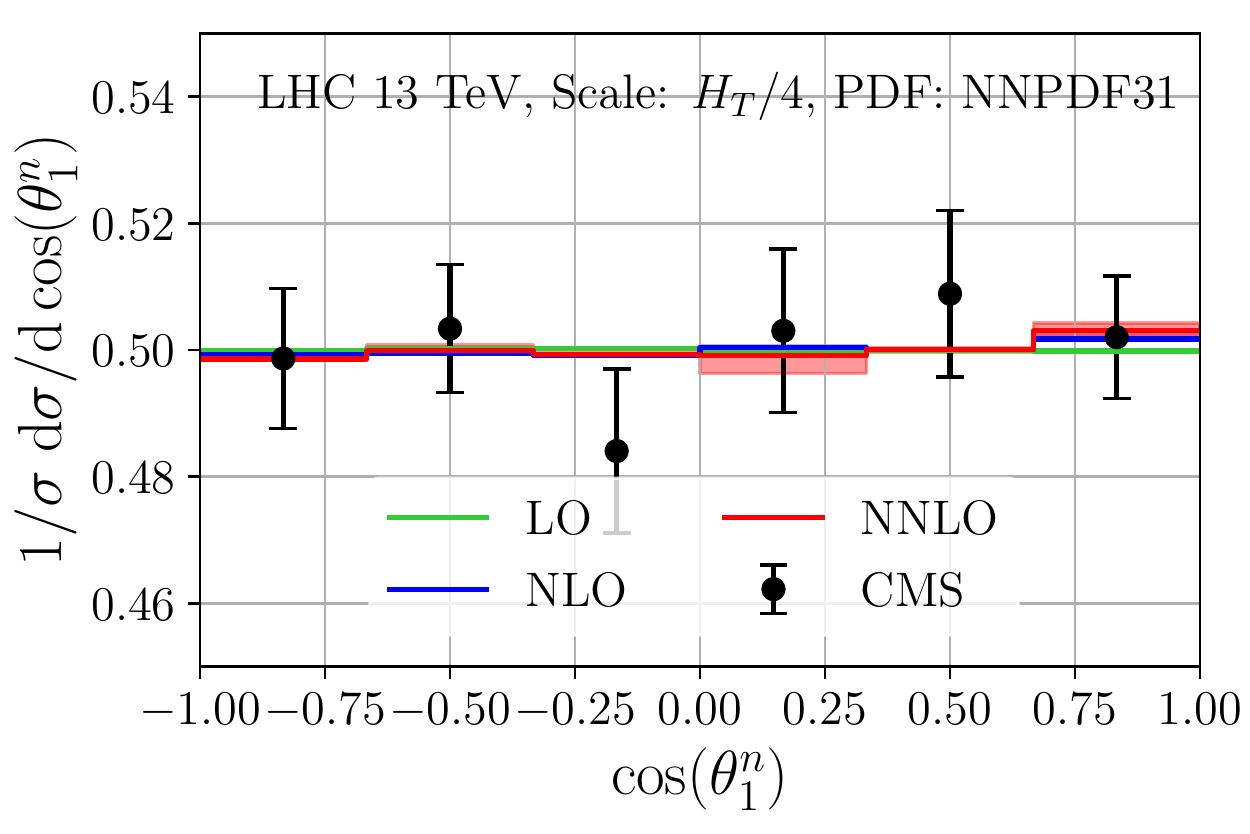}\\
\includegraphics[width=4.9cm]{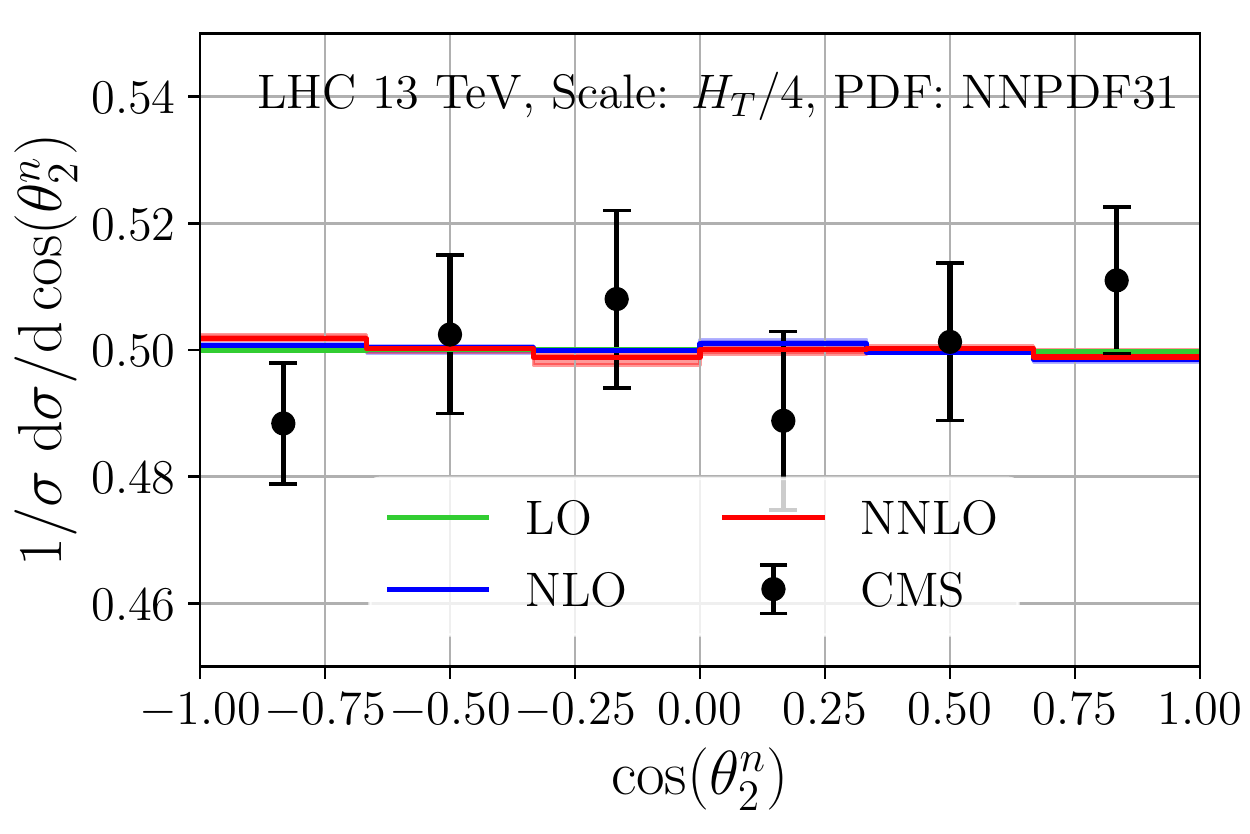}
\includegraphics[width=4.9cm]{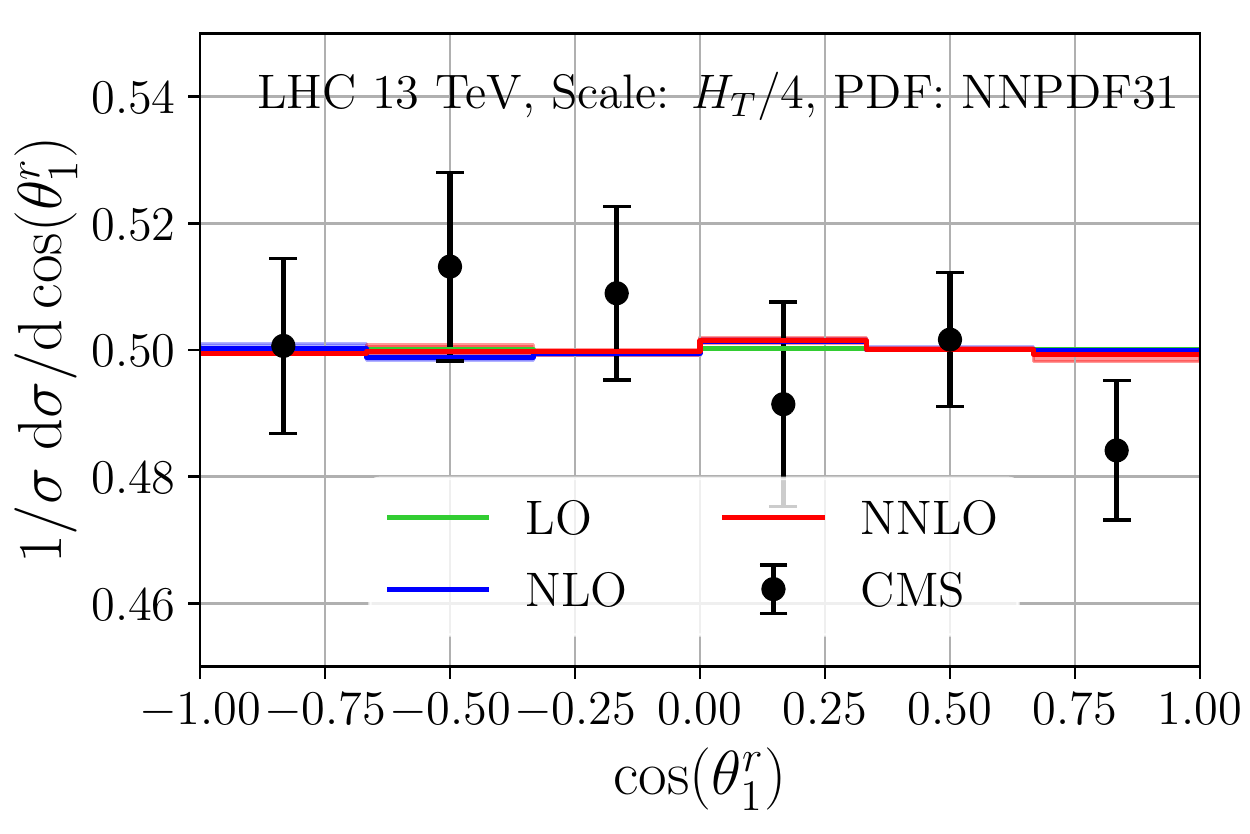}
\includegraphics[width=4.9cm]{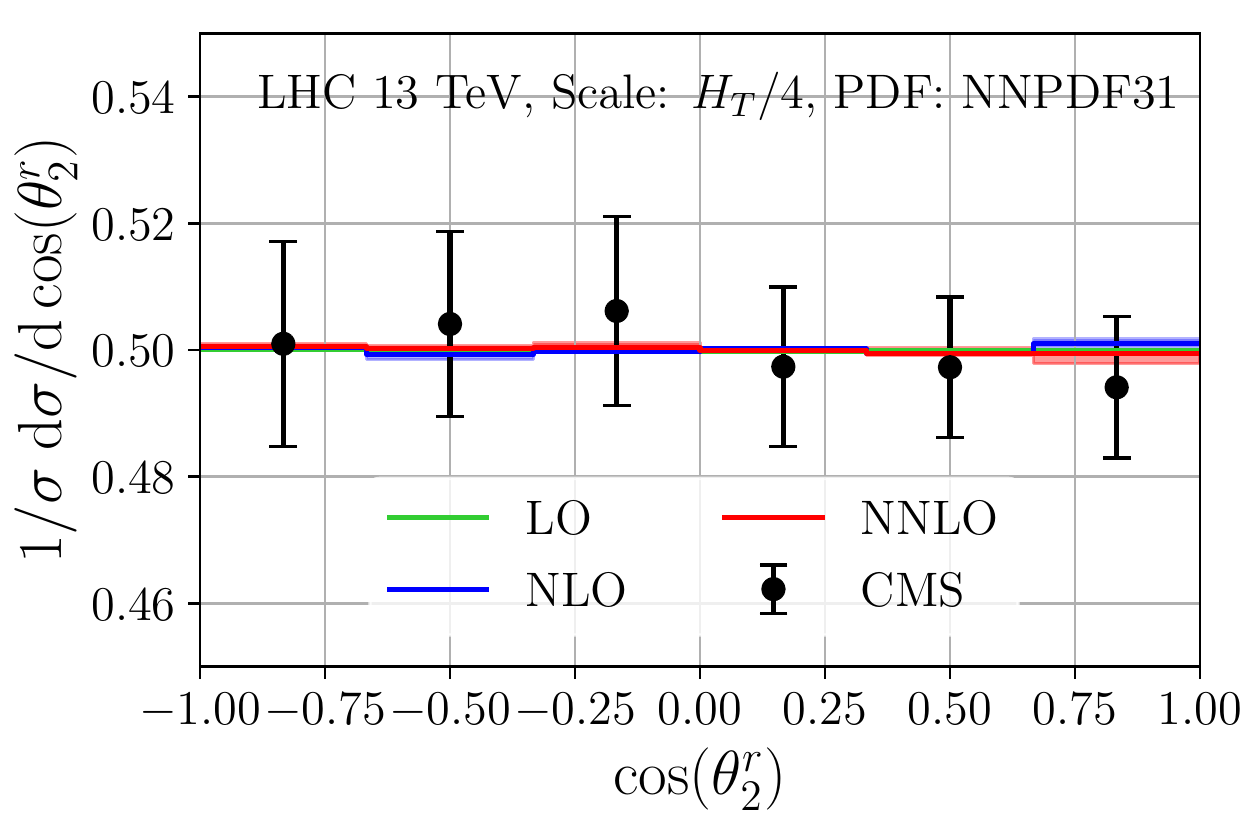}
\caption{LO, NLO and NNLO QCD predictions for the normalized differential cross section with respect to $\cos\theta^i_1$ and $\cos\theta^i_2$, $i=k, n, r$ versus CMS data \cite{CMS:2018jcg}. The error bands, tiny in size and barely visible, correspond to scale variation only.}
\label{fig:sdm_bi}
\end{figure}

We start by showing in fig.~\ref{fig:sdm_bi} the differential distributions (\ref{eq:sdm_xsecs1},\ref{eq:sdm_xsecs2}) which are sensitive to the polarization coefficients $B^i_1$ and $B^i_2$. The polarization is very small resulting in a very flat distribution around the value 0.5. The NNLO QCD corrections affect the distributions very little, i.e. these distributions appear to be very stable against radiative corrections. The fact that the scale variation is extremely small is mainly due to the normalization. The MC error of the calculation, not shown, is comparable to the scale variation. In fact, in few bins the MC error even exceeds the scale one. The exact estimate of the complete theory uncertainty is not too relevant since experimental uncertainties far exceed the theoretical ones. At any rate, within uncertainties, QCD predictions and data agree.

Next we consider the differential distribution in the product of angles $\cos\theta^i_1\cos\theta^i_2$ defined in eq.~(\ref{eq:sdm_xsecs3}); it is used for the extraction of the diagonal coefficients $C_{ii}$. The NNLO QCD results and corresponding comparison to CMS data are shown in fig.~\ref{fig:sdm_cii}. For this distribution, too, we observe negligible scale variation and tiny higher-order corrections. The experimental uncertainty dominates over the theory one and theory and data agree well in all bins. 

\begin{figure}
\includegraphics[width=4.9cm]{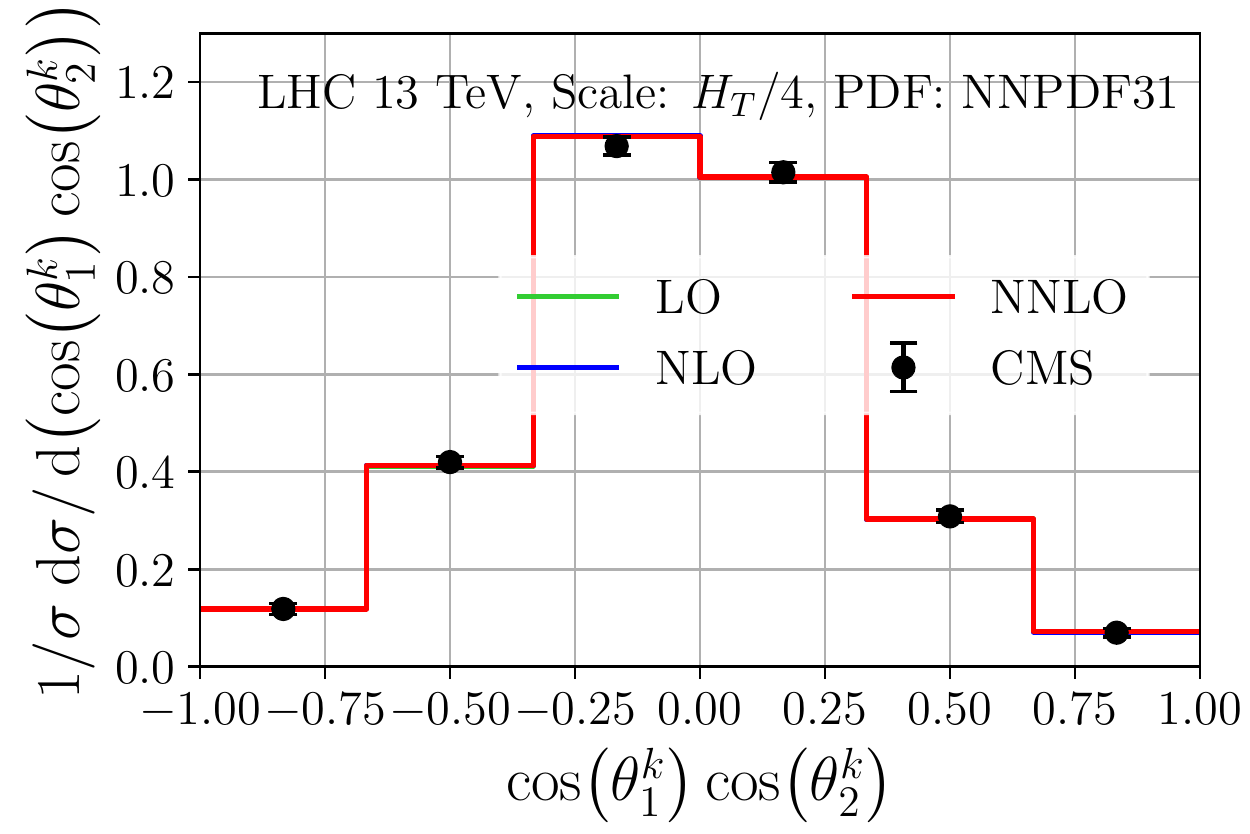}
\includegraphics[width=4.9cm]{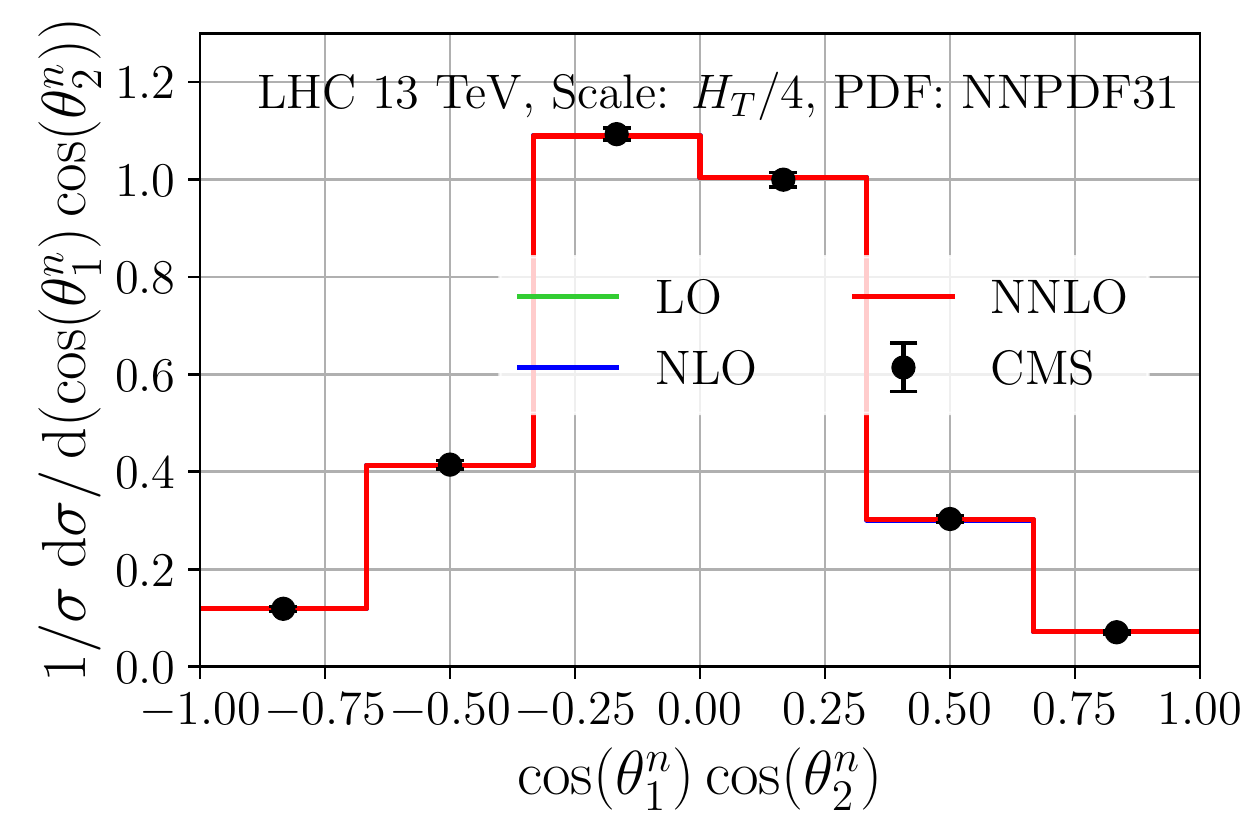}
\includegraphics[width=4.9cm]{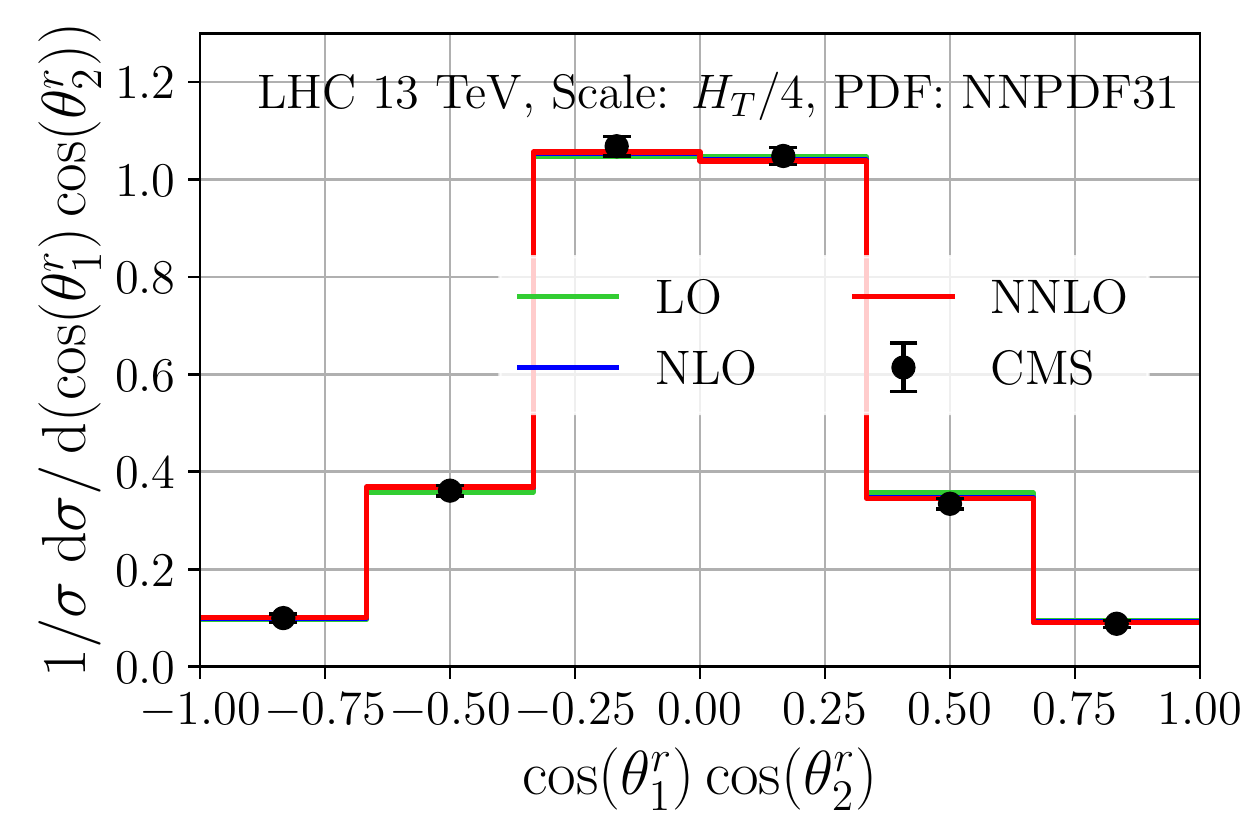}
\caption{As in fig.~\ref{fig:sdm_bi} but with respect to the variables $\cos\theta^i_1\cos\theta^i_2$.}
\label{fig:sdm_cii}
\end{figure}
\begin{figure}
\includegraphics[width=4.9cm]{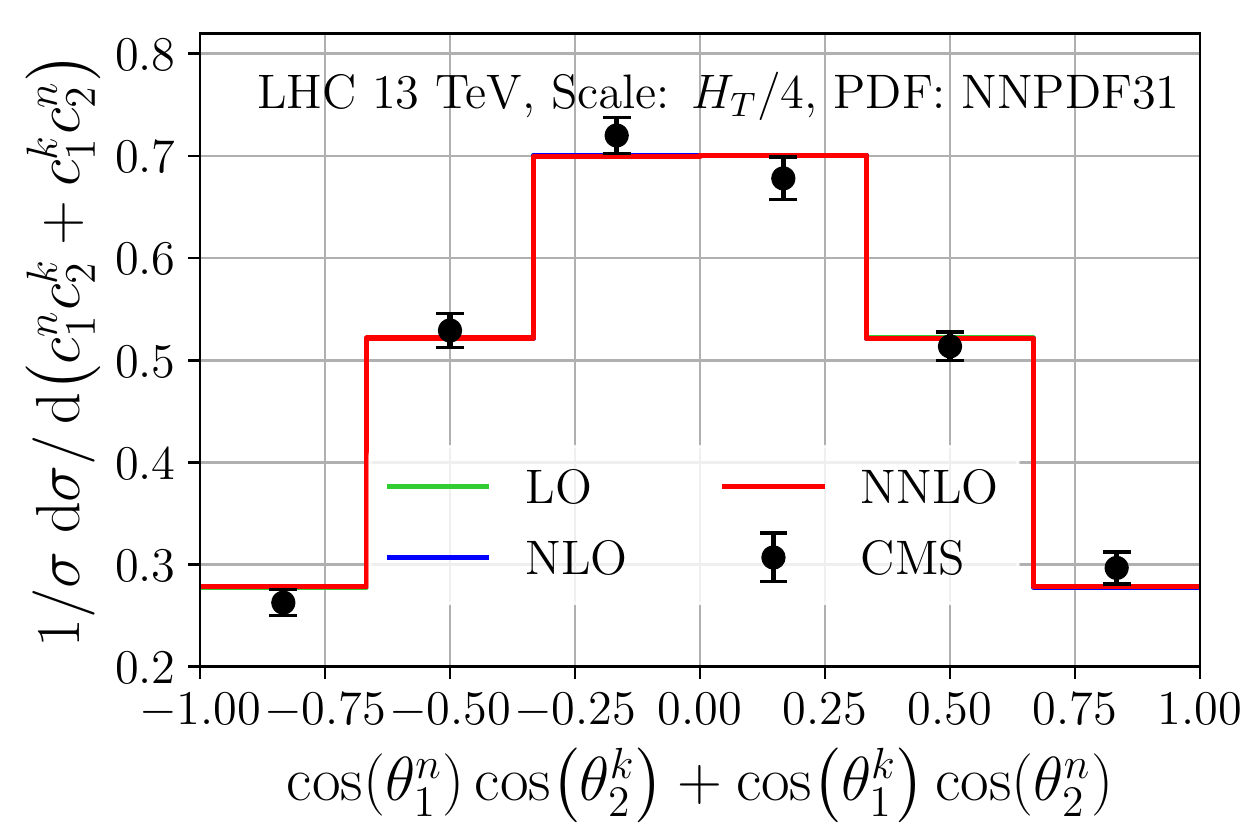}
\includegraphics[width=4.9cm]{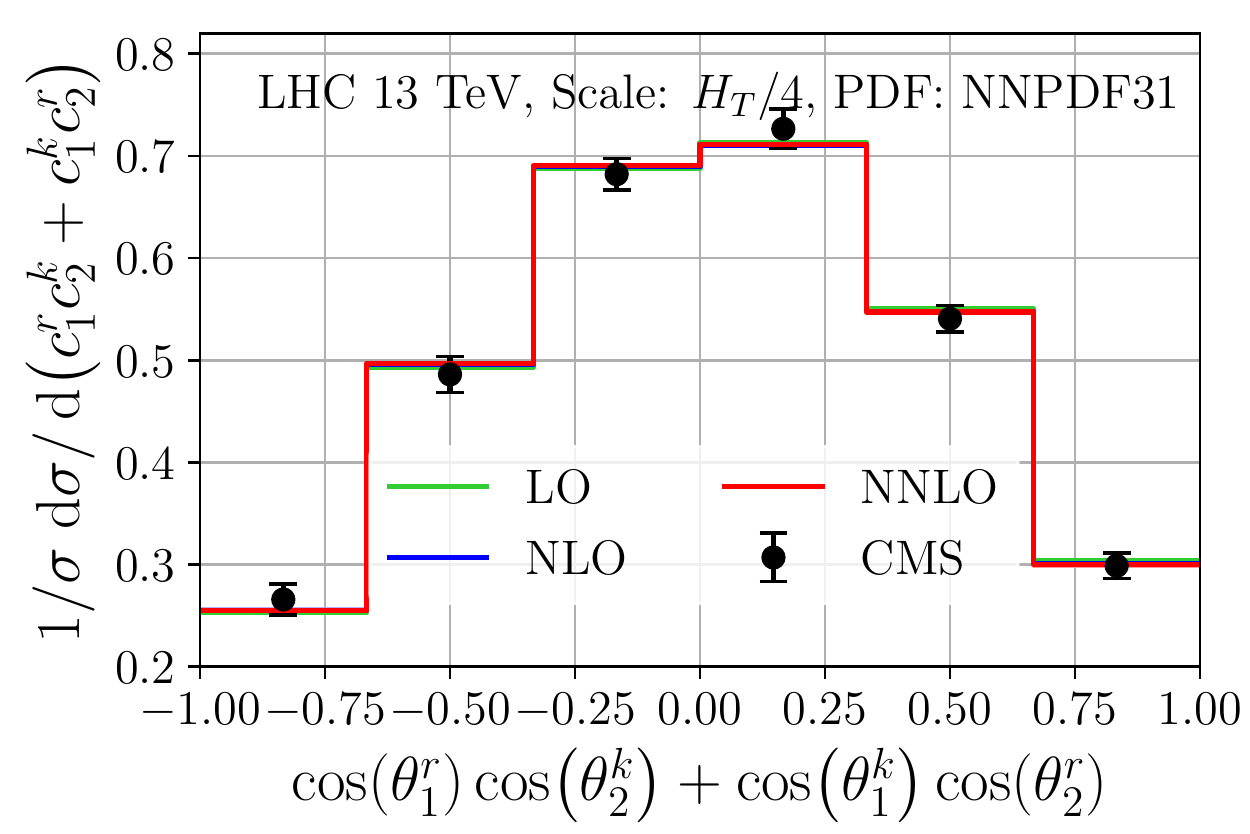}
\includegraphics[width=4.9cm]{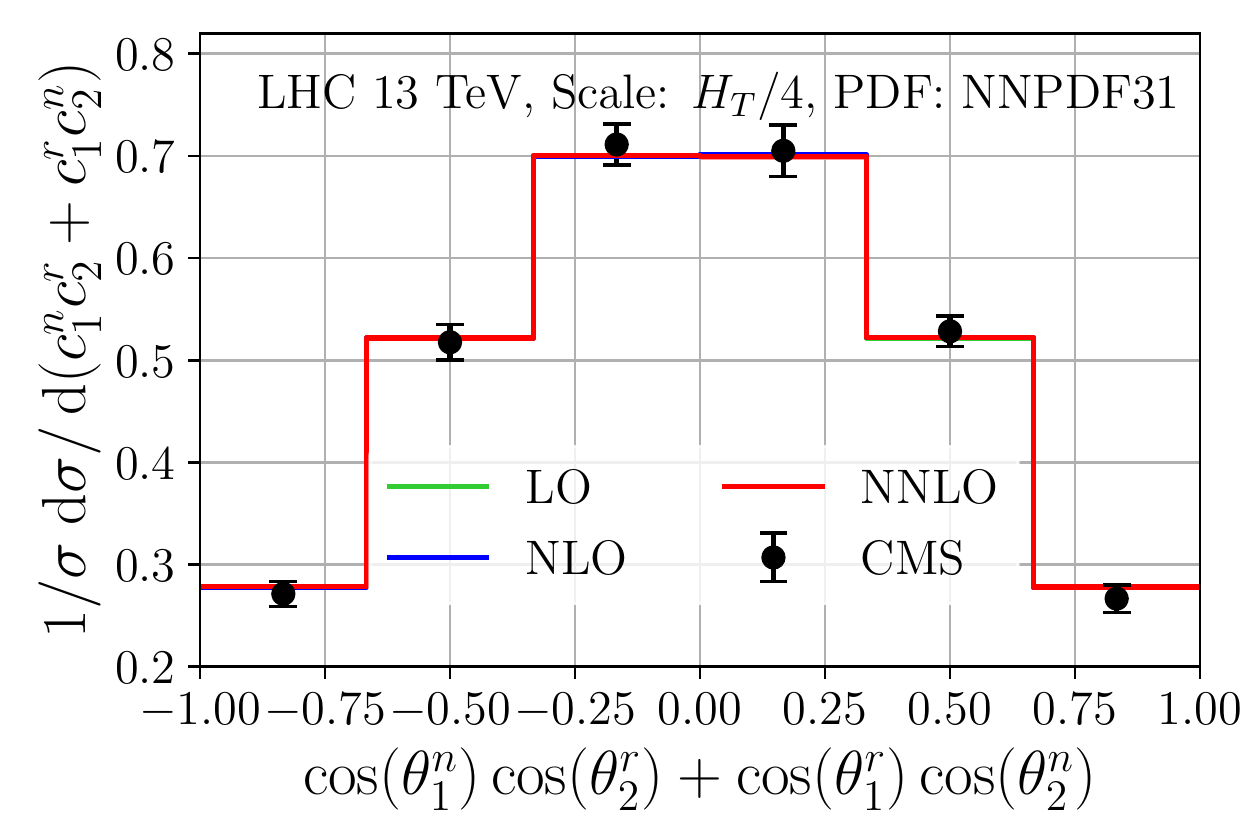}\\
\includegraphics[width=4.9cm]{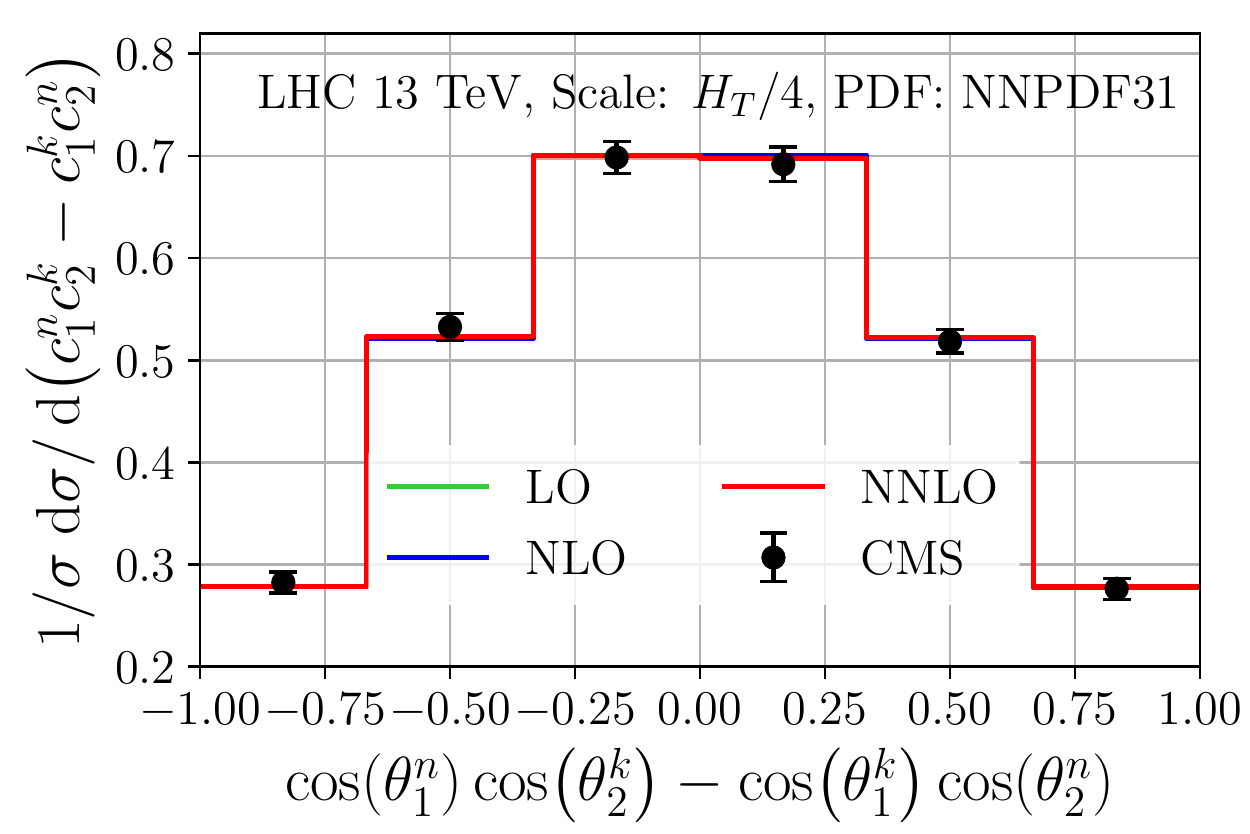}
\includegraphics[width=4.9cm]{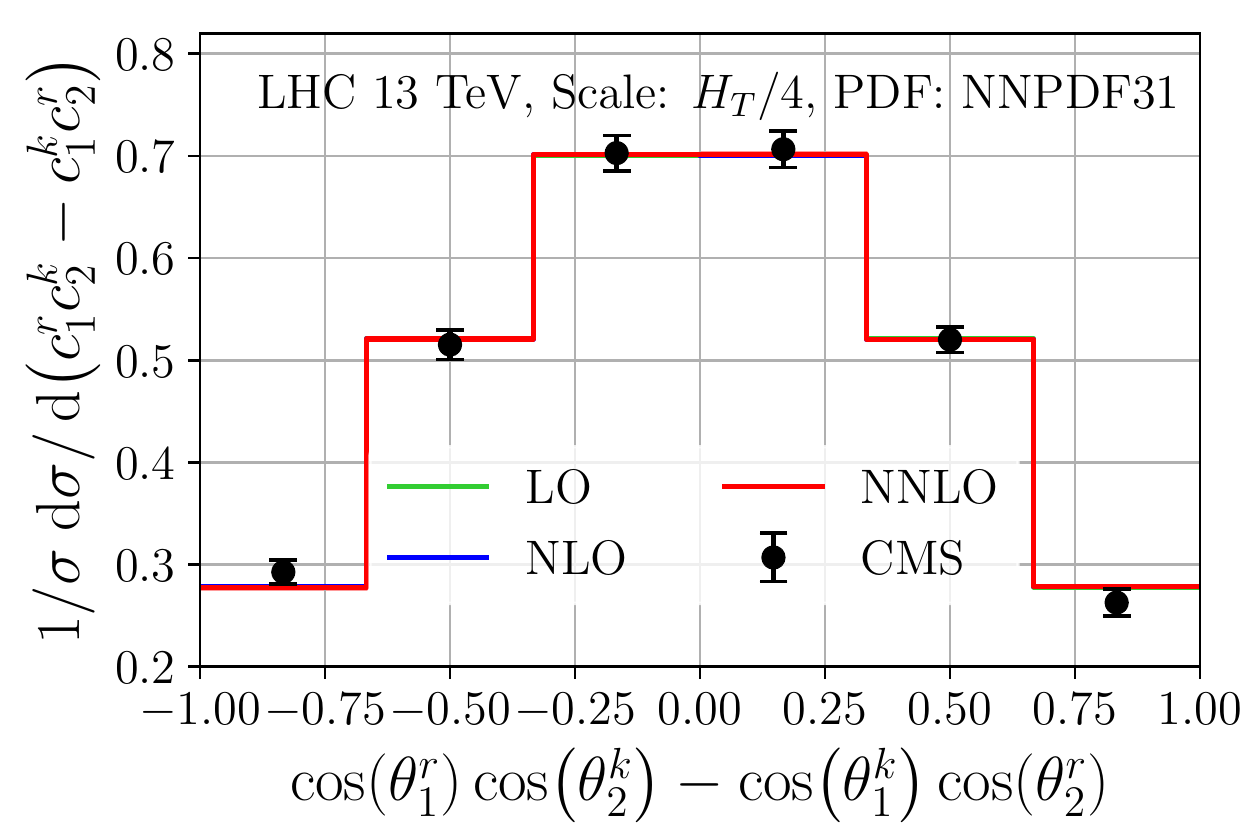}
\includegraphics[width=4.9cm]{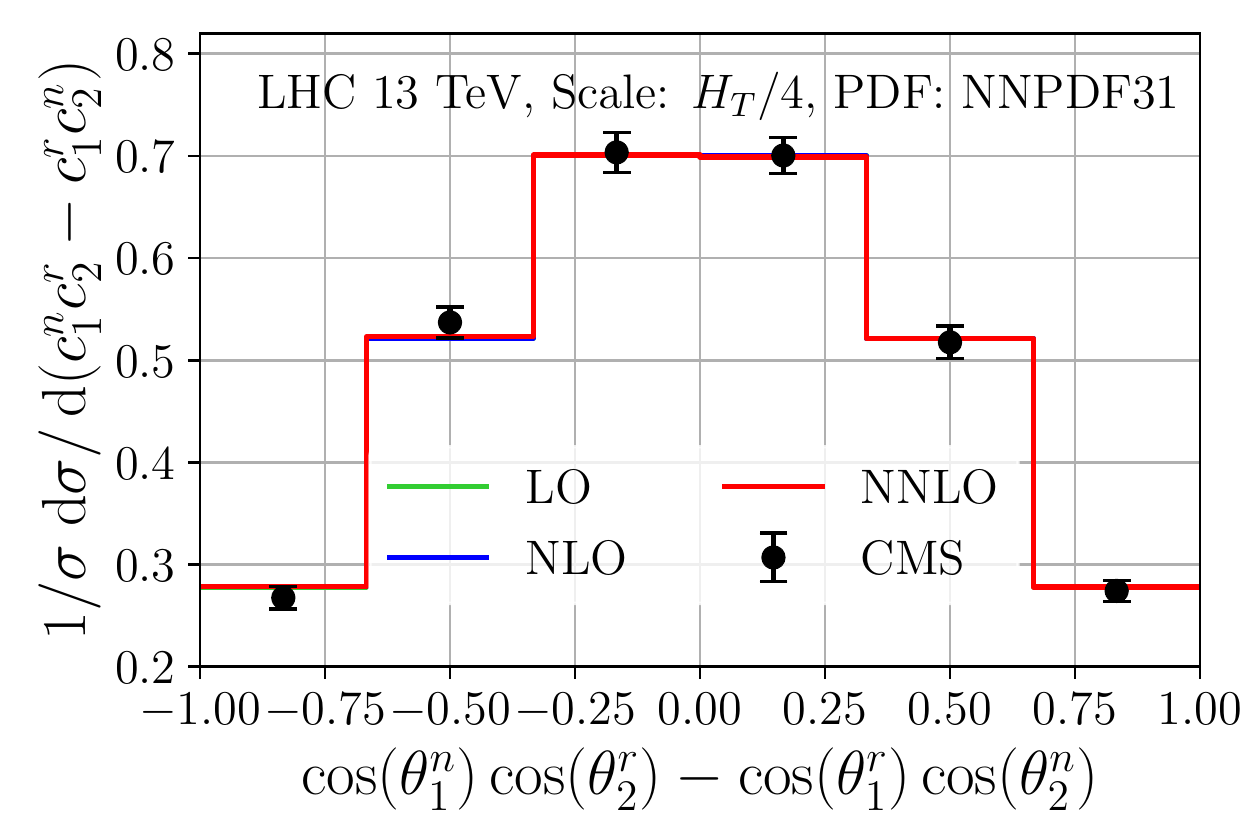}
\caption{As in fig.~\ref{fig:sdm_bi} but with respect to the variables $x_{ij}^\pm,~ i\neq j$.}
\label{fig:sdm_cij}
\end{figure}
The off-diagonal coefficients $C_{ij}$ are measured from the differential cross-section in the variables $x_{ij}^\pm$, see eq.~(\ref{eq:sdm_xsecs4}). The predictions for those distributions are shown in fig.~\ref{fig:sdm_cij}. For these distributions we also find tiny higher-order corrections and negligible theoretical uncertainties. The theory-data comparison is again dominated by the experimental uncertainties. An excellent agreement between theory and CMS data can be observed.

In summary, the power of this analysis is presently solely constrained by the size of the experimental uncertainties. Furthermore, through NNLO in QCD we observe no significant correction due to higher-order effects to the functional forms that define the spin-density coefficients $B^i_1, B^i_2$ and $C_{ij}$. This suggests that for the purpose of parametrizing spin correlations, the choice of the coefficients $B^i_1, B^i_2$ and $C_{ij}$ is optimal. The fact that the corrections to the spin-correlation coefficients are found to be rather small confirms the findings of our previous work on lepton angular distributions in the laboratory frame \cite{Behring:2019iiv}. There we managed to separate higher-order corrections to spin correlations from those due to recoil against additional radiation and demonstrated that the corrections to the spin correlation itself were small.

\begin{table}
\hskip-4mm
\begin{tabular}{c|r|r|r|r}
Coefficient & LO $(\times 10^{3})$ & NLO $(\times 10^{3})$& NNLO $(\times 10^{3})$& CMS $(\times 10^{3})$\\
\hline
$B^k_1$ & $1^{+0}_{-0}\,[{\rm sc}]\pm 1\,[{\rm mc}]$ & $1^{+0}_{-1}\,[{\rm sc}] \pm 2\,[{\rm mc}]$ & $-1^{+0}_{-1}\,[{\rm sc}] \pm 4\,[{\rm mc}]$ & $5 \pm 23$ \\
$B^r_1$ & $0^{+0}_{-0}\,[{\rm sc}] \pm 1\,[{\rm mc}]$ & $0^{+1}_{-0}\,[{\rm sc}] \pm 2\,[{\rm mc}]$ & $0^{+1}_{-2}\,[{\rm sc}] \pm 2\,[{\rm mc}]$ & $-23 \pm 17$ \\
$B^n_1$ & $0^{+0}_{-0}\,[{\rm sc}] \pm 1\,[{\rm mc}]$ & $3^{+1}_{-1}\,[{\rm sc}] \pm 1\,[{\rm mc}]$ & $4^{+1}_{-0}\,[{\rm sc}] \pm 3\,[{\rm mc}]$ & $6 \pm 13$ \\
$B^k_2$ & $0^{+0}_{-0}\,[{\rm sc}] \pm 1\,[{\rm mc}]$ & $0^{+0}_{-1}\,[{\rm sc}] \pm 1\,[{\rm mc}]$ & $-5^{+2}_{-3}\,[{\rm sc}] \pm 3\,[{\rm mc}]$ & $7 \pm 23$ \\
$B^r_2$ & $0^{+0}_{-0}\,[{\rm sc}] \pm 1\,[{\rm mc}]$ & $0^{+2}_{-0}\,[{\rm sc}] \pm 1\,[{\rm mc}]$ & $-2^{+0}_{-1}\,[{\rm sc}] \pm 2\,[{\rm mc}]$ & $-10 \pm 20$ \\
$B^n_2$ & $0^{+0}_{-0}\,[{\rm sc}] \pm 1\,[{\rm mc}]$ & $-2^{+0}_{-1}\,[{\rm sc}] \pm 1\,[{\rm mc}]$ & $-3^{+1}_{-0}\,[{\rm sc}] \pm 3\,[{\rm mc}]$ & $17 \pm 13$ \\
\hline
$C_{kk}$ & $324^{+7}_{-7}\,[{\rm sc}] \pm 1\,[{\rm mc}]$ & $330^{+2}_{-2}\,[{\rm sc}] \pm 3\,[{\rm mc}]$ & $323^{+2}_{-5}\,[{\rm sc}] \pm 6\,[{\rm mc}]$ & $300 \pm 38$ \\
$C_{rr}$ & $6^{+5}_{-5}\,[{\rm sc}] \pm 1\,[{\rm mc}]$ & $58^{+18}_{-12}\,[{\rm sc}] \pm 2\,[{\rm mc}]$ & $69^{+8}_{-7}\,[{\rm sc}] \pm 3\,[{\rm mc}]$ & $81 \pm 32$ \\
$C_{nn}$ & $332^{+1}_{-0}\,[{\rm sc}] \pm 1\,[{\rm mc}]$ & $330^{+1}_{-1}\,[{\rm sc}] \pm 2\,[{\rm mc}]$ & $326^{+1}_{-1}\,[{\rm sc}] \pm 4\,[{\rm mc}]$ & $329 \pm 20$ \\
\hline
$C_{nr}+C_{rn}$ & $1^{+0}_{-0}\,[{\rm sc}] \pm 1\,[{\rm mc}]$ & $-1^{+1}_{-0}\,[{\rm sc}] \pm 3\,[{\rm mc}]$ & $-4^{+4}_{-0}\,[{\rm sc}] \pm 6\,[{\rm mc}]$ & $-4 \pm 37$ \\
$C_{nr}-C_{rn}$ & $0^{+0}_{-1}\,[{\rm sc}] \pm 1\,[{\rm mc}]$ & $-1^{+1}_{-0}\,[{\rm sc}] \pm 2\,[{\rm mc}]$ & $2^{+4}_{-2}\,[{\rm sc}] \pm 8\,[{\rm mc}]$ & $-1 \pm 38$ \\
$C_{nk}+C_{kn}$ & $0^{+0}_{-0}\,[{\rm sc}] \pm 1\,[{\rm mc}]$ & $2^{+1}_{-0}\,[{\rm sc}] \pm 1\,[{\rm mc}]$ & $3^{+4}_{-1}\,[{\rm sc}] \pm 3\,[{\rm mc}]$ & $-43 \pm 41$ \\
$C_{nk}-C_{kn}$ & $1^{+0}_{-0}\,[{\rm sc}] \pm 1\,[{\rm mc}]$ & $1^{+1}_{-1}\,[{\rm sc}] \pm 2\,[{\rm mc}]$ & $6^{+0}_{-2}\,[{\rm sc}] \pm 7\,[{\rm mc}]$ & $40 \pm 29$ \\
$C_{rk}+C_{kr}$ & $-229^{+4}_{-4}\,[{\rm sc}] \pm 1\,[{\rm mc}]$ & $-203^{+9}_{-7}\,[{\rm sc}] \pm 2\,[{\rm mc}]$ & $-194^{+8}_{-6}\,[{\rm sc}] \pm 7\,[{\rm mc}]$ & $-193 \pm 64$ \\
$C_{rk}-C_{kr}$ & $1^{+0}_{-0}\,[{\rm sc}] \pm 1\,[{\rm mc}]$ & $1^{+0}_{-1}\,[{\rm sc}] \pm 4\,[{\rm mc}]$ & $-1^{+1}_{-3}\,[{\rm sc}] \pm 5\,[{\rm mc}]$ & $57 \pm 46$ 
 \end{tabular}
\caption{The extracted spin-density coefficients $B^i_1$, $B^i_2$ and $C_{ij}$ at various orders in perturbation theory. Estimates for the MC and scale theory uncertainties are shown separately; see sec.~\ref{sec:spin-density-pheno} for their definitions.}
\label{tab:sdm_coeffs}
\end{table}

To complete the discussion of the spin-density matrix, in addition to the differential distributions described above, we also extract the spin density matrix coefficients $B^i_1$, $B^i_2$ and $C_{ij}$. These coefficients are extracted one-by-one, from a fit to the corresponding binned differential distribution. The fit takes the functional form implied by eqs.~(\ref{eq:sdm_xsecs1}--\ref{eq:sdm_xsecs4}). A separate fit is performed for each one of the 7 scale combinations. We then construct seven $\chi^2$ functions (one for each scale combination) from the difference between the actual calculated bin values, see figs.~\ref{fig:sdm_bi},\ref{fig:sdm_cii},\ref{fig:sdm_cij}, and the values for each bin derived by integrating  the assumed functional form. The dependence on the unknown coefficients $B^i_1$, $B^i_2$ and $C_{ij}$ enters through the latter. That difference is divided by our estimate for the MC error of the calculation in a given bin. No cross-bin correlations are included. The value of the sought coefficient is determined by minimizing the $\chi^2$ function. The $\Delta\chi^2=1$ constraint on the central scale is interpreted as MC error on the extracted coefficient. The spread of the seven $\chi^2$ minima (one for each scale combination) with respect to the $\chi^2$ minimum corresponding to the central scale choice is then interpreted as a scale variation of the corresponding coefficient. 

We would like to emphasize that the procedure for estimating MC and scale errors of the spin-density coefficients is sensible but not perfect and should be considered as a guide as opposed to quantitive uncertainty estimate. The reason is that the procedure is very sensitive to the estimate of the MC error which is notoriously hard to pinpoint. In particular, in cases where the MC error dominates over the scale one (which is the case for most of the coefficients since their scale errors are tiny) the derived scale error is impacted by the MC error and thus not entirely independent from it. We believe that our estimate for the overall uncertainty (i.e. the combined scale and MC uncertainties) is correct as well as the two uncertainties individually in cases where the scale error is significantly larger than the MC one. 

The results for all spin-density coefficients $B^i_1$, $B^i_2$ and $C_{ij}$ are given in table \ref{tab:sdm_coeffs}. They are extracted at LO, NLO and NNLO in QCD. For comparison the corresponding CMS measurements are shown as well. It is easy to see that all measurements are compatible with the NNLO QCD results. Perhaps this comparison should not be over-interpreted since the experimental errors are still quite large, typically about 5 to 10 times larger than the combined scale and MC NNLO uncertainties. 

The behavior of the various spin-density coefficients at different orders of perturbation theory also deserves a comment. For almost all of the coefficients the estimated scale error is very small and is comparable to the MC one. This is the reason why in many cases it appears that the scale error at NNLO is larger than the NLO one. As explained above the estimated scale error is likely driven by the MC one and, as can be expected, the MC error at NNLO is much larger than the one at NLO. Only in the case of the coefficient $C_{rr}$ the scale error exceeds the MC one and for this coefficient one observes the expected pattern of decreasing scale uncertainty when going from NLO to NNLO. At present, given the dominant experimental uncertainties on the spin-density coefficients, the above calculations are quite satisfactory. They may need to be refined and recomputed with improved MC errors and fits if, in the future, the experimental measurements are improved by one order of magnitude.

\subsubsection{Spin correlations in dilepton angular distributions}\label{sec:spin-correlations-angular}

\begin{figure}
\includegraphics[width=7.7cm]{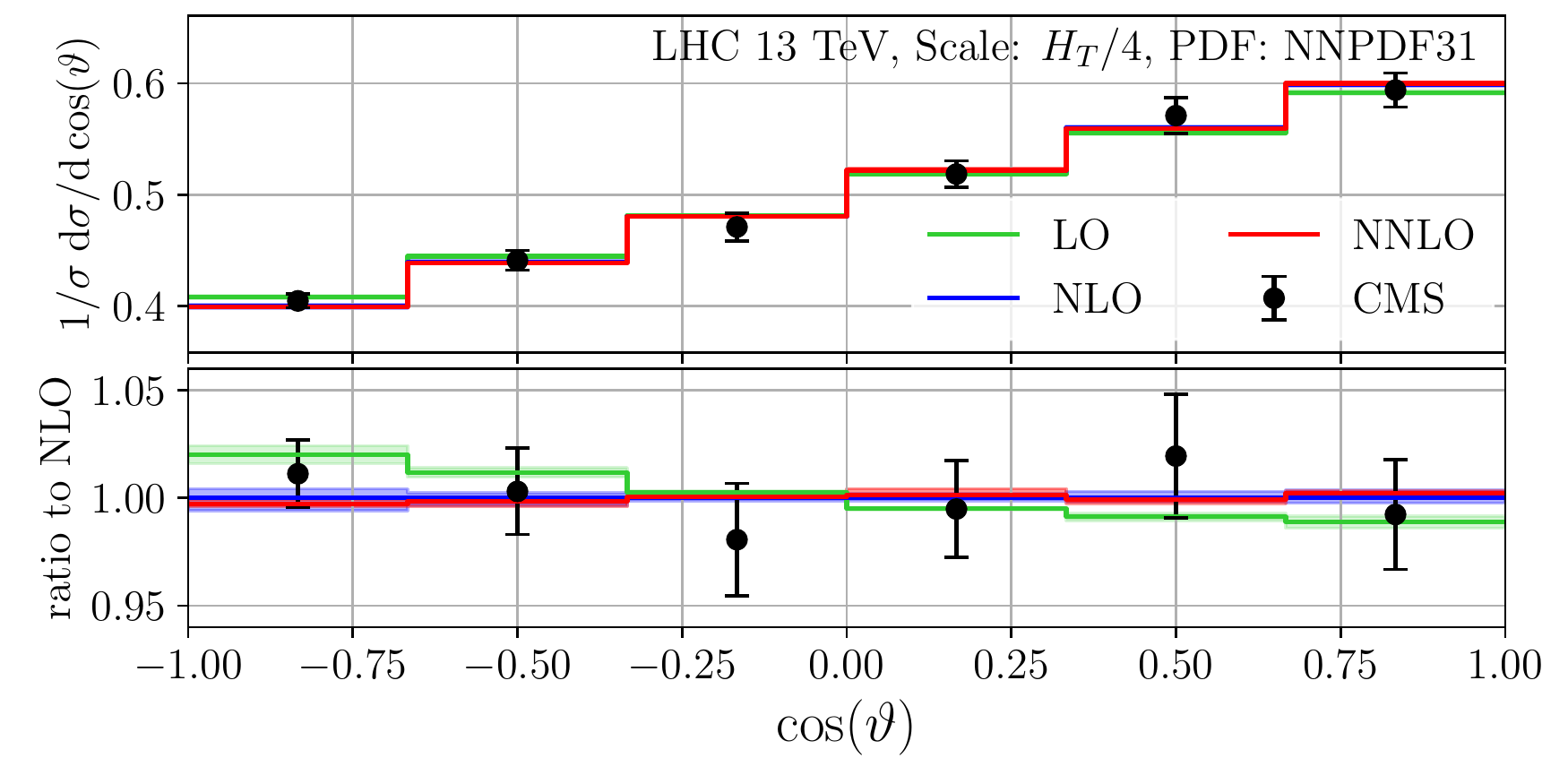}
\includegraphics[width=7.7cm]{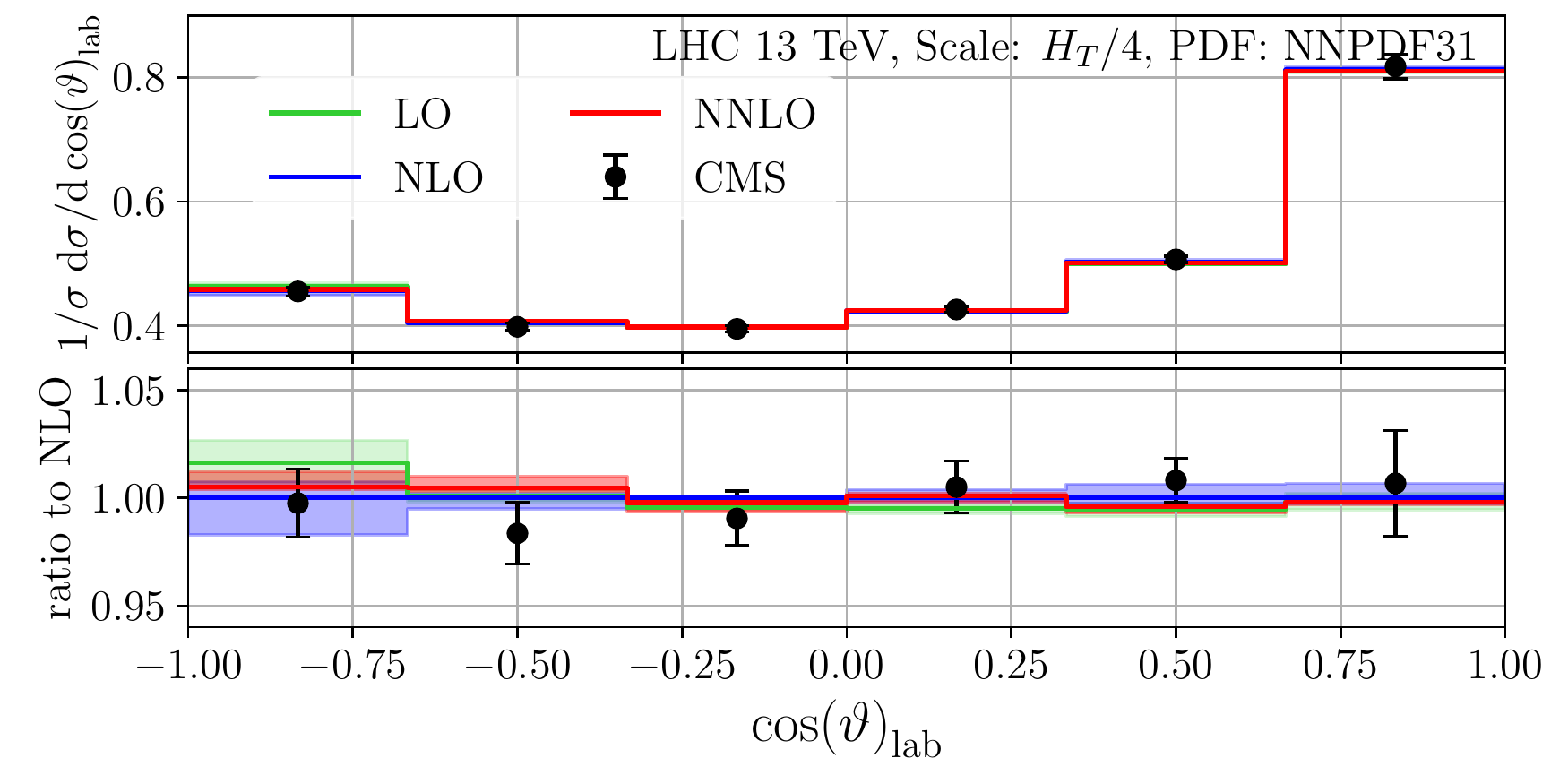}
\center
\includegraphics[width=7.7cm]{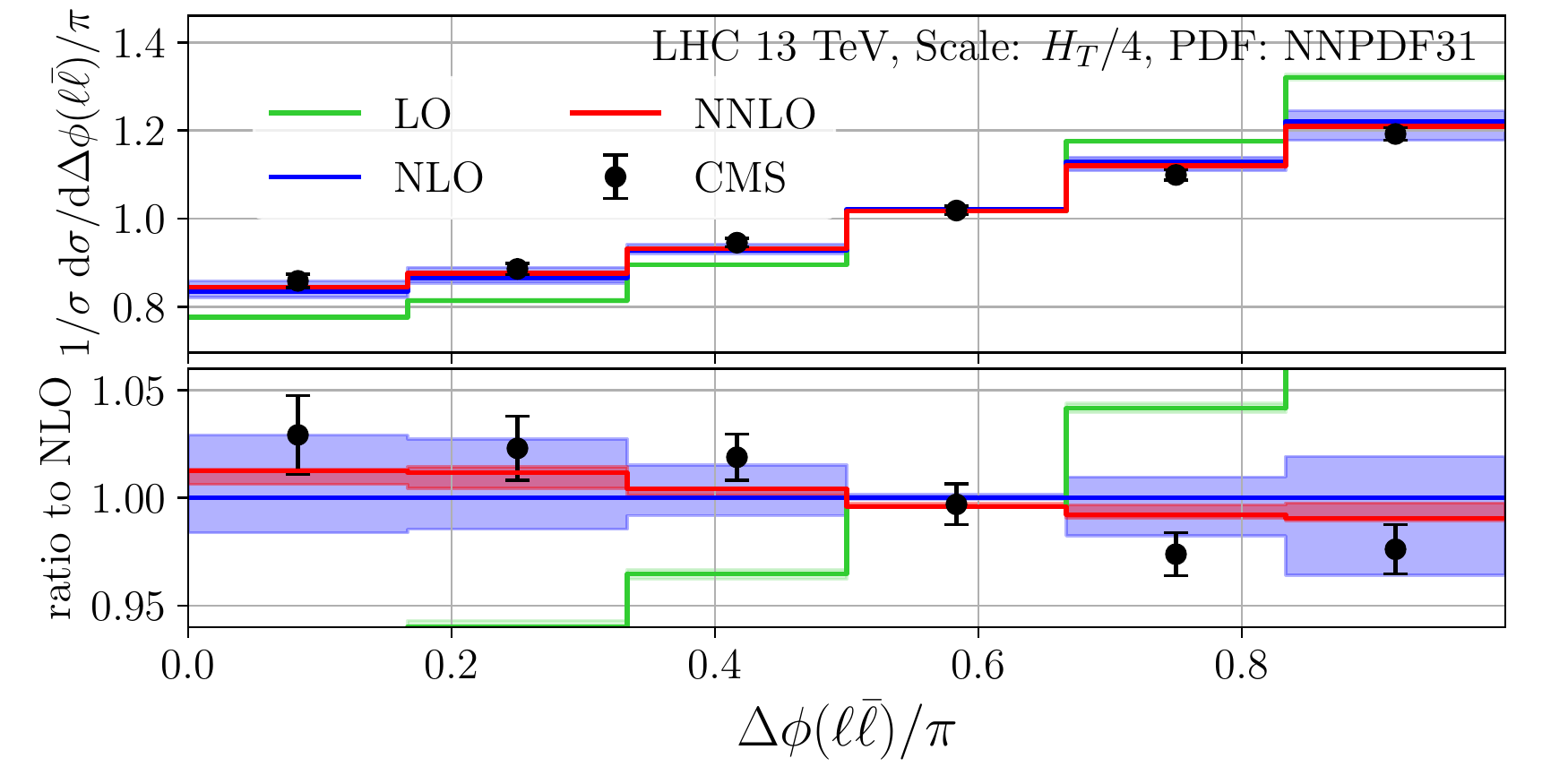}
\caption{As in fig.~\ref{fig:sdm_bi} but with respect to the variables $\cos\vartheta$, $\cos\vartheta_{\text{lab}}$ and $\Delta\phi(\ell\bar\ell)$.}
\label{fig:sdm_phi-theta}
\end{figure}

In this section we consider $t\bar t$ spin correlations in differential distributions measured directly in the lab frame. Although in this case spin correlations cannot be separated as cleanly from kinematics, the advantage of such measurements is that they can be both measured and computed with high precision. In the following we discuss two such sets of measurements: one by the ATLAS and one by the CMS Collaborations.

We start with the CMS publication \cite{CMS:2018jcg} which we already discussed in the previous section. In addition to the direct extraction of the spin-density matrix, the following set of spin-correlation sensitive variables has been measured by CMS and is computed by us:
\begin{equation}
\frac{1}{\sigma}\frac{\dd\sigma}{\dd\cos\vartheta}~~{\rm and}~~\frac{1}{\sigma}\frac{\dd\sigma}{\dd\cos\vartheta_{\text{lab}}}\,,
\end{equation}
where the angle $\vartheta$ is given by $\hat{p}_{\ell^+}\cdot\hat{p}_{\ell^-} = \cos\vartheta$, the angle $\vartheta_{\text{lab}}$ is defined as $\vartheta$ but in the laboratory frame.

The theory predictions through NNLO in QCD are shown in fig.~\ref{fig:sdm_phi-theta}. The main effect from the inclusion of the NNLO QCD corrections is to further reduce the size of the already small NLO scale uncertainty. The NNLO/NLO K-factor is rather small, much smaller than the NLO/LO one. This is along the lines of the pattern of higher-order corrections observed and discussed in sec.~\ref{sec:spin-density-pheno}. Good agreement between NNLO QCD and data is found although, due to the relatively large size of the experimental uncertainties, the current data cannot distinguish between the NLO and NNLO predictions.  

We next turn our attention to the normalized $\Delta\phi(\ell\bar\ell)$ distribution
\begin{equation}
\frac{1}{\sigma}\frac{\dd\sigma}{\dd\Delta\phi(\ell\bar\ell)}\,,
\label{eq:delta-phi}
\end{equation}
where $\Delta\phi(\ell\bar\ell)$ is the azimuthal (i.e. in the plane transverse to the beam) difference between the two charged leptons measured in the laboratory frame. The difference between the two angles is always taken in such a way as to ensure $0\leq \Delta\phi(\ell\bar\ell) \leq \pi$. We focus on the inclusive selection
\footnote{In fact this selection is fully inclusive since no cuts are imposed, not even on the leptons.}
which has received a lot of attention recently \cite{Aaboud:2019hwz,CMS:2018jcg,1802024}. We extend our previous calculation \cite{Behring:2019iiv} and provide predictions for the 6-bin CMS measurement \cite{CMS:2018jcg} together with expanded differential predictions (as described in sec.~\ref{sec:expand-ratio}) for the ATLAS measurement \cite{Aaboud:2019hwz}. To better assess the effect of the $\as$ expansion we compute the distribution with two different scale choices: the default dynamic scale eq.~(\ref{eq:scale}) as well as the fixed central scale $\mu_R=\mu_F=m_t$. 

\begin{figure}[t]
\includegraphics[width=7.5cm]{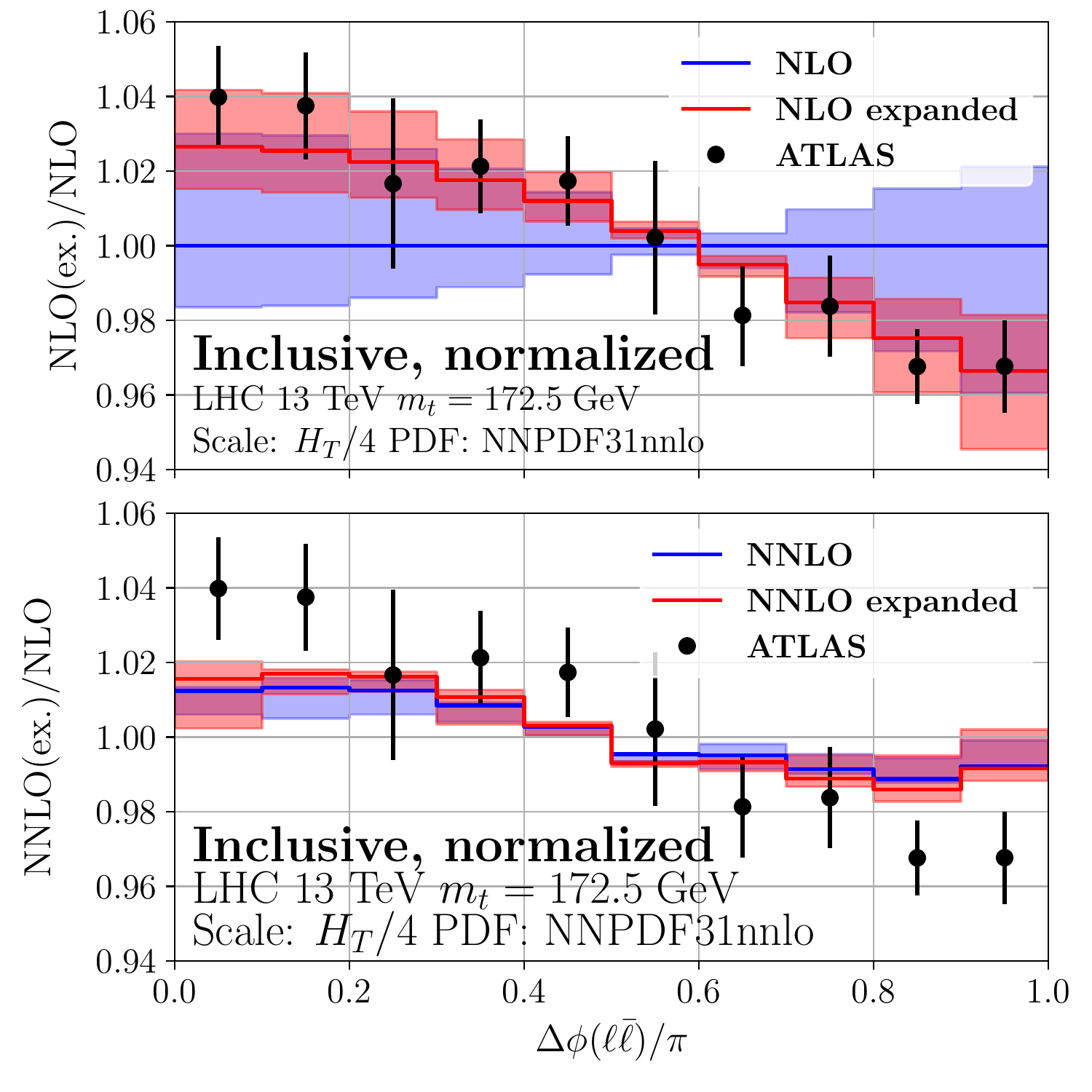}
\includegraphics[width=7.5cm]{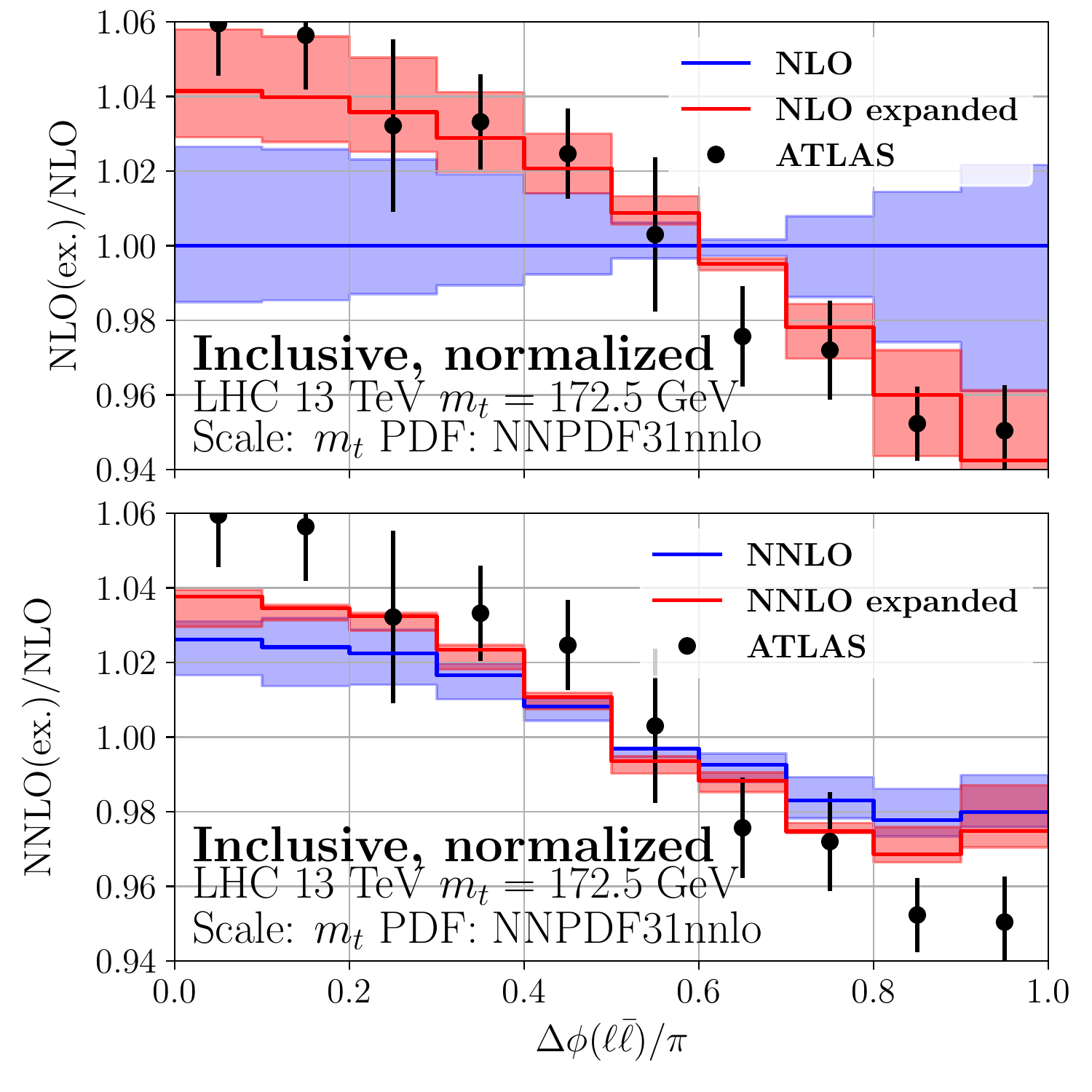}
\caption{Expanded (red) versus unexpanded (blue) predictions for the normalized $\Delta\phi(\ell\bar\ell)$ distribution at NLO (top) and NNLO (bottom) QCD. Predictions based on a dynamic (left) and static (right) scale are given. See sec.~\ref{sec:spin-correlations-angular} for details.}
\label{fig:dphi-expanded}
\end{figure}

The predictions, compared to CMS data, are shown in fig.~\ref{fig:sdm_phi-theta}. The expanded prediction is shown in fig.~\ref{fig:dphi-expanded} where it is compared with ATLAS data. The behavior of the unexpanded prediction is very similar to the one already discussed in detail in ref.~\cite{Behring:2019iiv}. This is not surprising since the only difference between the CMS and ATLAS \cite{Aaboud:2019hwz} measurements is the binning. It is interesting to note that the agreement with CMS closely mirrors the one found in ref.~\cite{Behring:2019iiv} with respect to ATLAS: NNLO QCD gets closer to data relative to NLO. It is also compatible with data within uncertainties in a per-bin basis, however, the difference in slopes seems to be significant. Regarding the compatibility of the ATLAS and CMS measurements, a tuned comparison of the two was recently presented in ref.~\cite{1802024}. A very good agreement between them was found.

The comparison between the expanded prediction and ATLAS data \cite{Aaboud:2019hwz} is presented in fig.~\ref{fig:dphi-expanded}; to the left we show the predictions based on a dynamic scale while to the right a fixed scale is used. The ratio between the expanded (in red) and unexpanded (in blue) definitions is shown at NLO (upper panel) and NNLO (lower panel). Only the scale variation is included in the corresponding uncertainty bands.

The behavior of the expanded prediction at higher orders is quite interesting. At LO the expanded and unexpanded definitions are identical. At NLO one observes a substantial difference between the two calculations. They are consistent within scale variation but hardly so when the fixed scale is used. In particular, the expanded definition agrees well with ATLAS data at NLO. The unexpanded prediction with dynamic scale agrees with data within errors (albeit with a different slope) while the prediction with a fixed scale does not agree with the data. This picture changes substantially once the NNLO QCD corrections are included. Unlike at NLO, we see that both the expanded and unexpanded predictions agree well with each other and both are somewhat discrepant with the data. The same conclusion holds for the predictions based on a fixed scale. The agreement of the two NNLO predictions is an additional indication that at NNLO in QCD the differential distribution in question is under good theoretical control and the error due to missing higher orders is small. It also confirms the conclusions of ref.~\cite{Behring:2019iiv} that NNLO predictions do not agree very well with measurements based on an inclusive selection.

\subsection{Leptonic differential distributions with inclusive selection}\label{sec:pheno_inclusive}

In this section we compare predictions for one- and two-dimensional leptonic differential distributions based on inclusive selection (in the sense of sec.~\ref{sec:pheno}) with ATLAS measurements  \cite{Aad:2019hzw}. The fiducial phase-space of the ATLAS analysis has only lepton requirements and no explicit cuts on jets. Two different measurements are presented in the ATLAS
publication \cite{Aad:2019hzw}: one that includes top-quark decays into $\tau$-leptons and one which does not. In this work we do not consider decays to $\tau$ leptons and therefore exclusively compare to the $\tau$-less measurement. For all distributions considered in this section we implement the following cuts designed to match the cuts for the 0-$b$-tagged jet region described in \cite{Aad:2019hzw}: required are two oppositely charged leptons with
\begin{itemize}
  \item $p_T(\ell) \geq 20\;\text{GeV}$\;,
  \item $|\eta(\ell)| \leq 2.5$\;.
\end{itemize}
No further requirements are imposed. In our computation we use the input parameters described in section \ref{sec:pheno} and the top quark momenta correspond to the true top-quarks. 

The measurement considers various differential distributions in the two charged leptons. The following one-dimensional distributions are studied
\begin{itemize}
  \item $|\eta(\ell)|$, the pseudo-rapidity distribution summed over both leptons,
  \item $p_T(\ell)$, the transverse momentum distribution summed over both leptons,
  \item $p_T(\ell) + p_T(\bar{\ell})$, the sum of the transverse momenta,
  \item $E(\ell) + E(\bar{\ell})$, the sum of the lepton energies,
  \item $m(\ell\bar{\ell})$, the invariant lepton pair mass,
  \item $|y(\ell\bar{\ell})|$, the rapidity of the lepton pair,
  \item $p_T(\ell\bar{\ell})$, the transverse momentum of the lepton pair,
  \item $\Delta\phi(\ell\bar{\ell})$, the azimuthal opening angle,
\end{itemize}
as well as the following two-dimensional distributions
\begin{itemize}
  \item $\Delta\phi(\ell\bar{\ell})$ in slices of $m(\ell\bar{\ell})$,
  \item $y(\ell\bar{\ell})$ in slices of $m(\ell\bar{\ell})$,
  \item $|\eta(\ell)|$ in slices of $m(\ell\bar{\ell})$.
\end{itemize}

For the above distributions we compute LO, NLO and NNLO QCD corrections in the NWA approximation. In order to elucidate the mass sensitivity of the various leptonic distributions we have produced two sets of predictions for two different values of the top-quark mass: $m_t = 171.5$ GeV as well as our default value $m_t = 172.5$ GeV. We hope that our study will aid future high-precision top-mass measurements along the lines of refs.~\cite{Frixione:2014ala,CMS:2016xfv,Aaboud:2017ujq}.

\begin{figure}
\includegraphics[width=7.5cm]{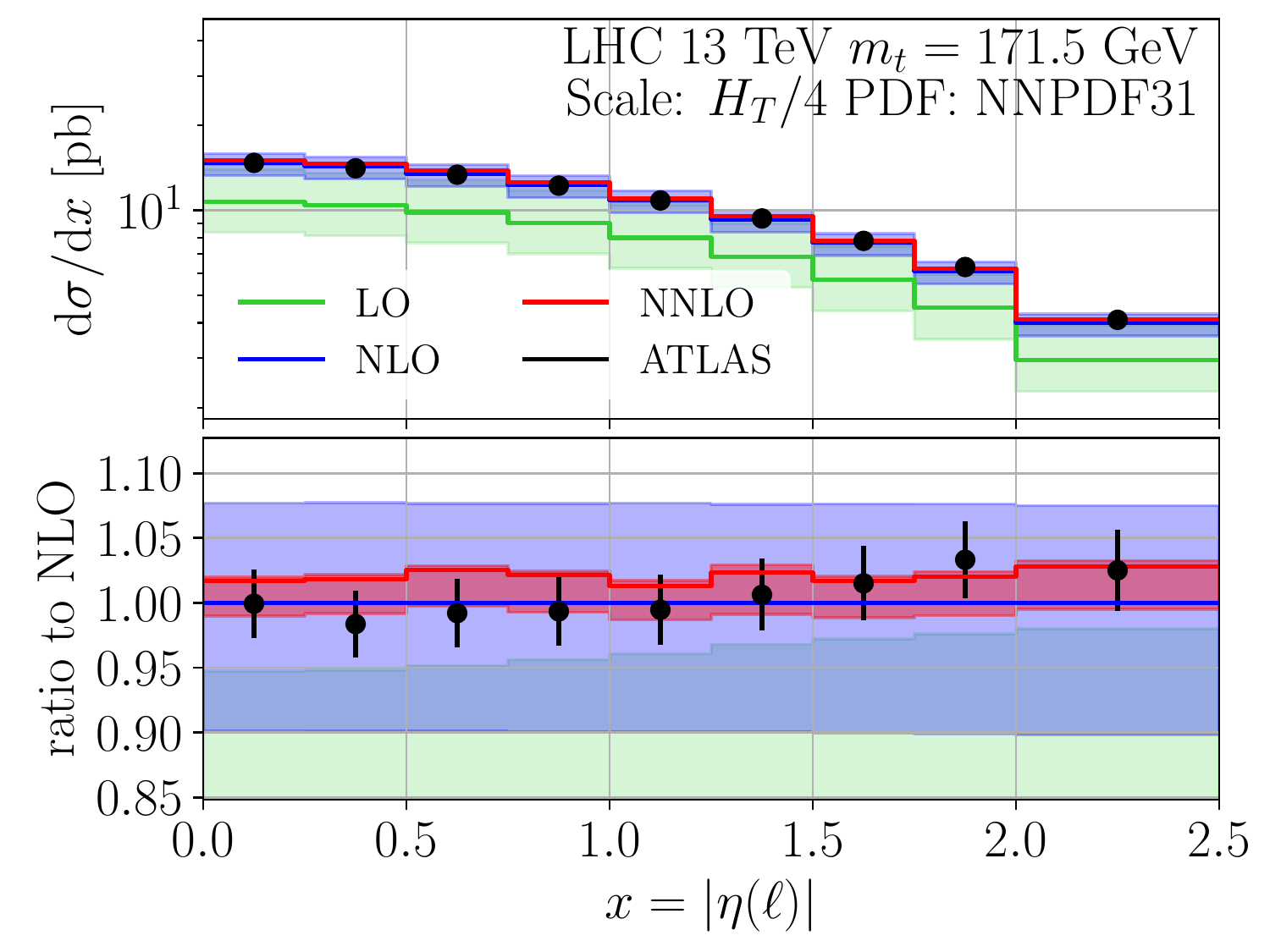}
\includegraphics[width=7.5cm]{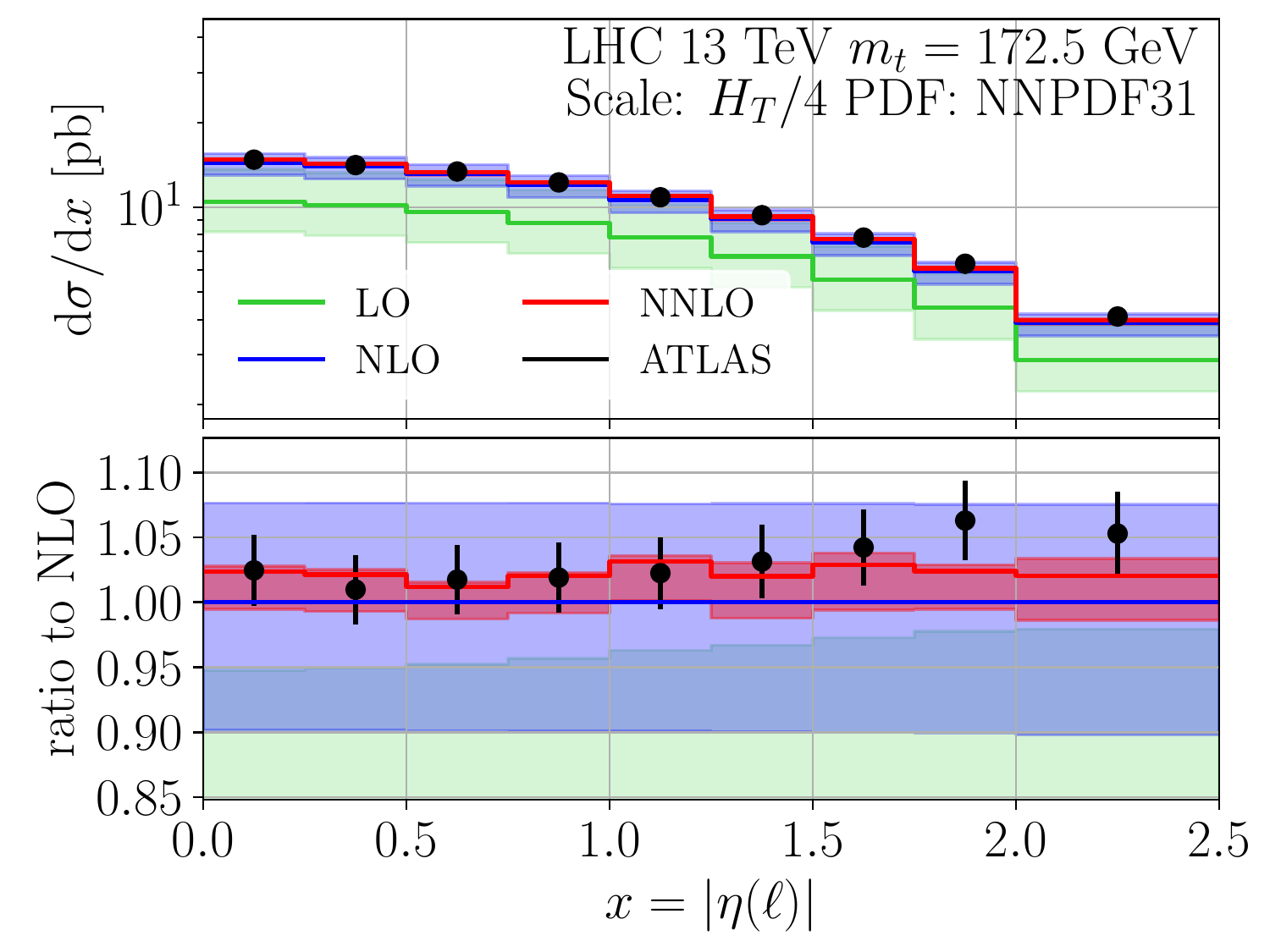}
\includegraphics[width=7.5cm]{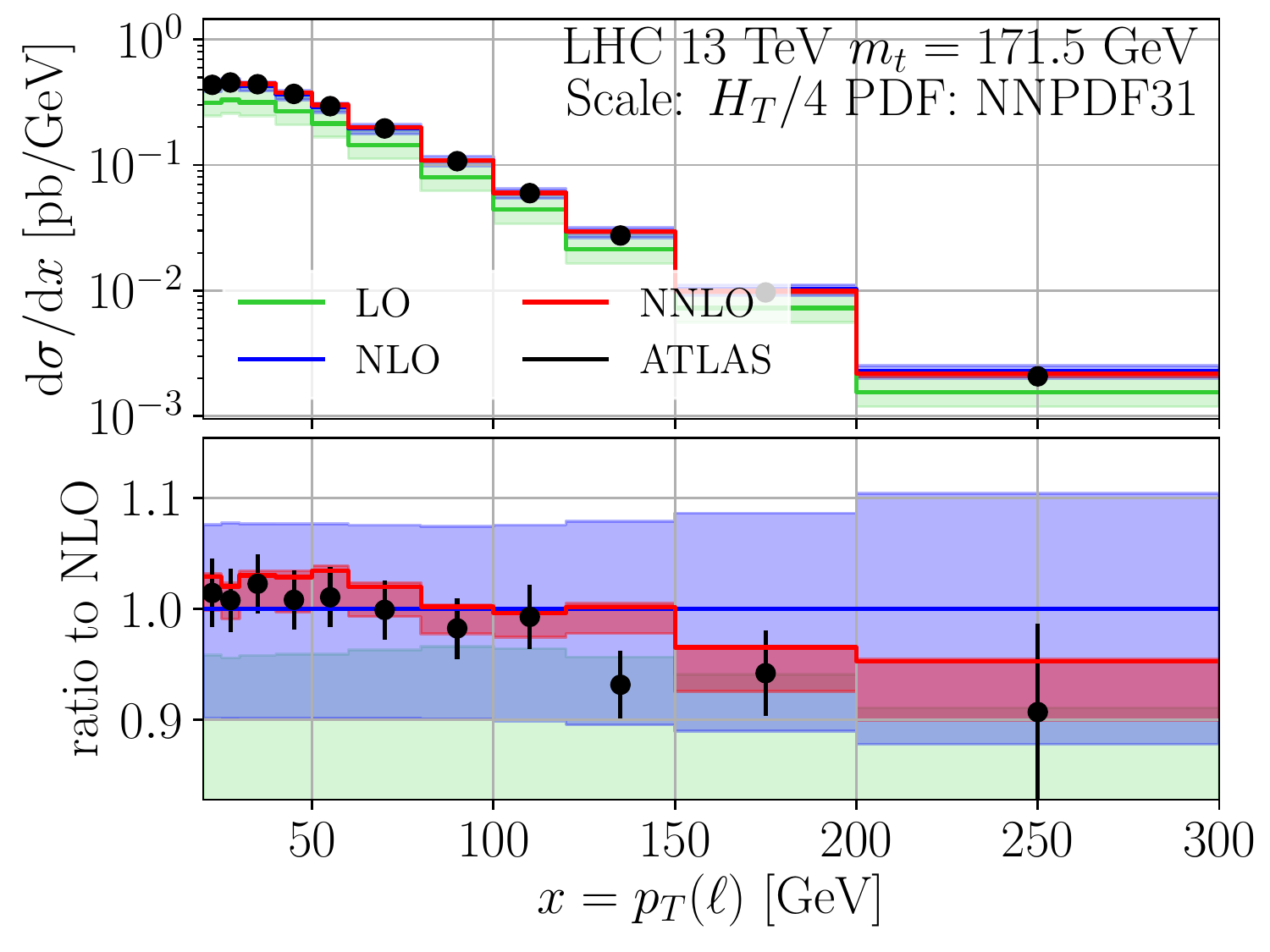}
\includegraphics[width=7.5cm]{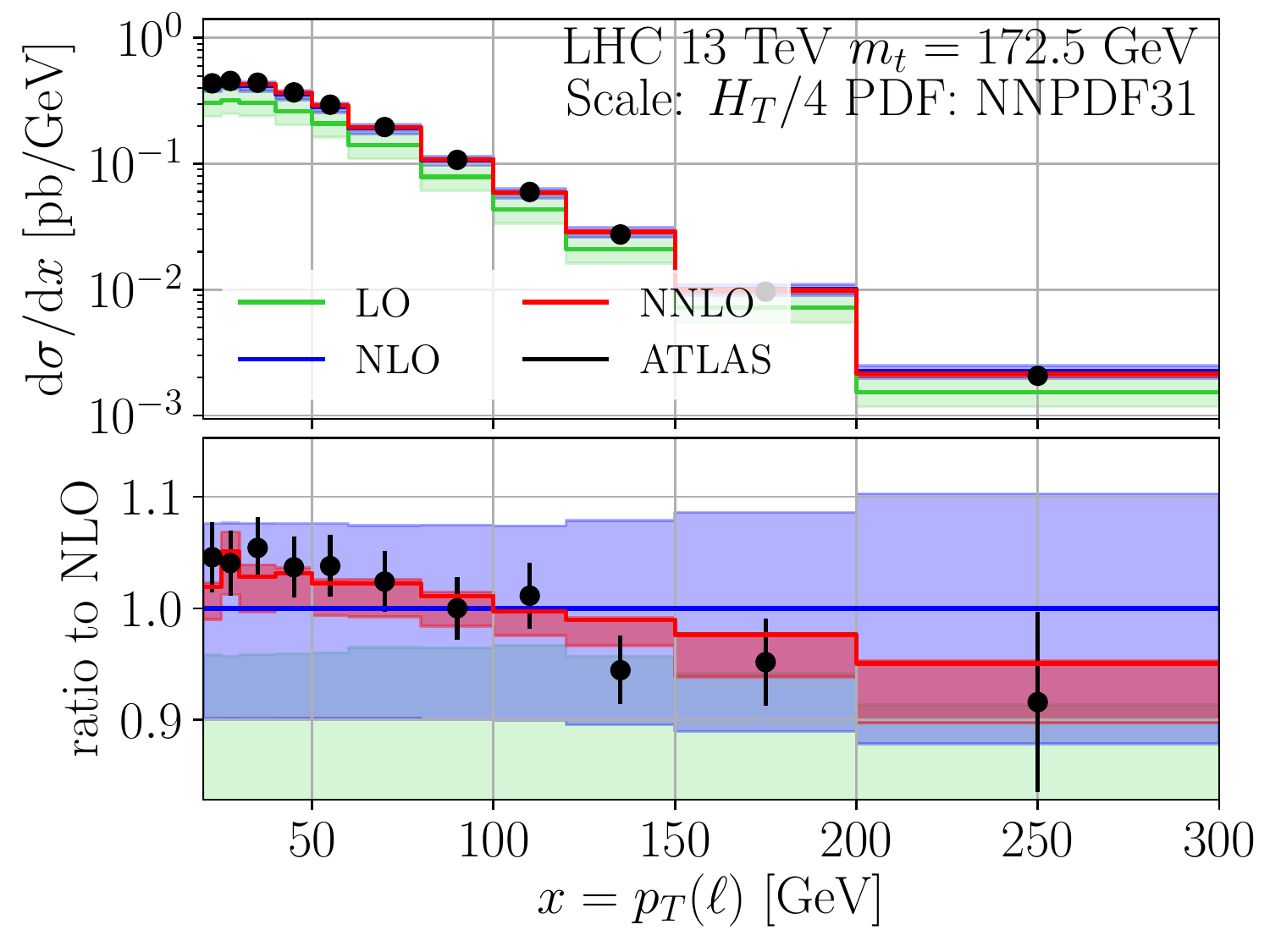}
\caption{Absolute differential distributions in $p_T(\ell)$ and $|\eta(\ell)|$ at LO (green), NLO (blue) and NNLO (red) versus ATLAS data \cite{Aad:2019hzw}. Uncertainty bands represent scale variation. Shown are two fixed-order predictions corresponding to $m_t=171.5$ GeV (left) and $m_t=172.5$ GeV (right).}
\label{fig:atlas_1}
\end{figure}

In fig.~\ref{fig:atlas_1} we show the single differential observables $p_T(\ell)$ and $|\eta(\ell)|$ related to single charged leptons. We observe that the inclusion of higher order QCD corrections leads to significantly reduced scale dependence which at NNLO is as low as $\order{1\%-5\%}$. We also notice that the size of the experimental uncertainty in most bins is significantly below the uncertainty of the NLO QCD predictions and is comparable in size to the NNLO one. Therefore, the inclusion of the NNLO QCD corrections is essential for any meaningful theory-data comparison.

Within uncertainties, data and NNLO QCD predictions agree very well. Besides reduced scale uncertainty the inclusion of the NNLO corrections has an important impact on the shape of the lepton $p_T$ spectrum and brings it closer to data. The 1 GeV change in $m_t$ affects mostly the normalization of the $|\eta(\ell)|$ distribution. For the lepton $p_T(\ell)$ spectrum it leads to a non-trivial change in its slope which, at low $p_T(\ell)$, results in a shift comparable in size with the experimental uncertainty.

\begin{figure}
\includegraphics[width=7.5cm]{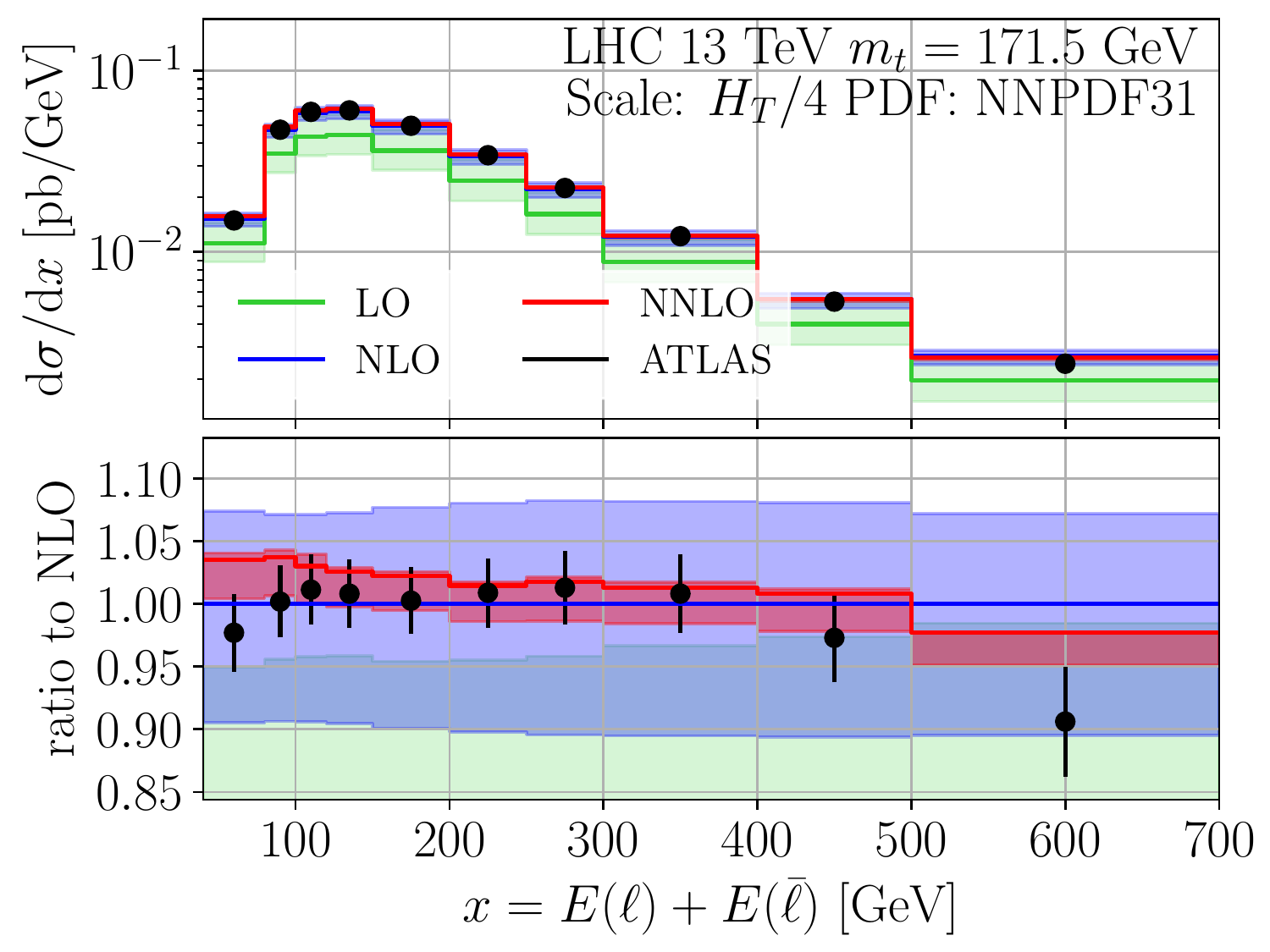}
\includegraphics[width=7.5cm]{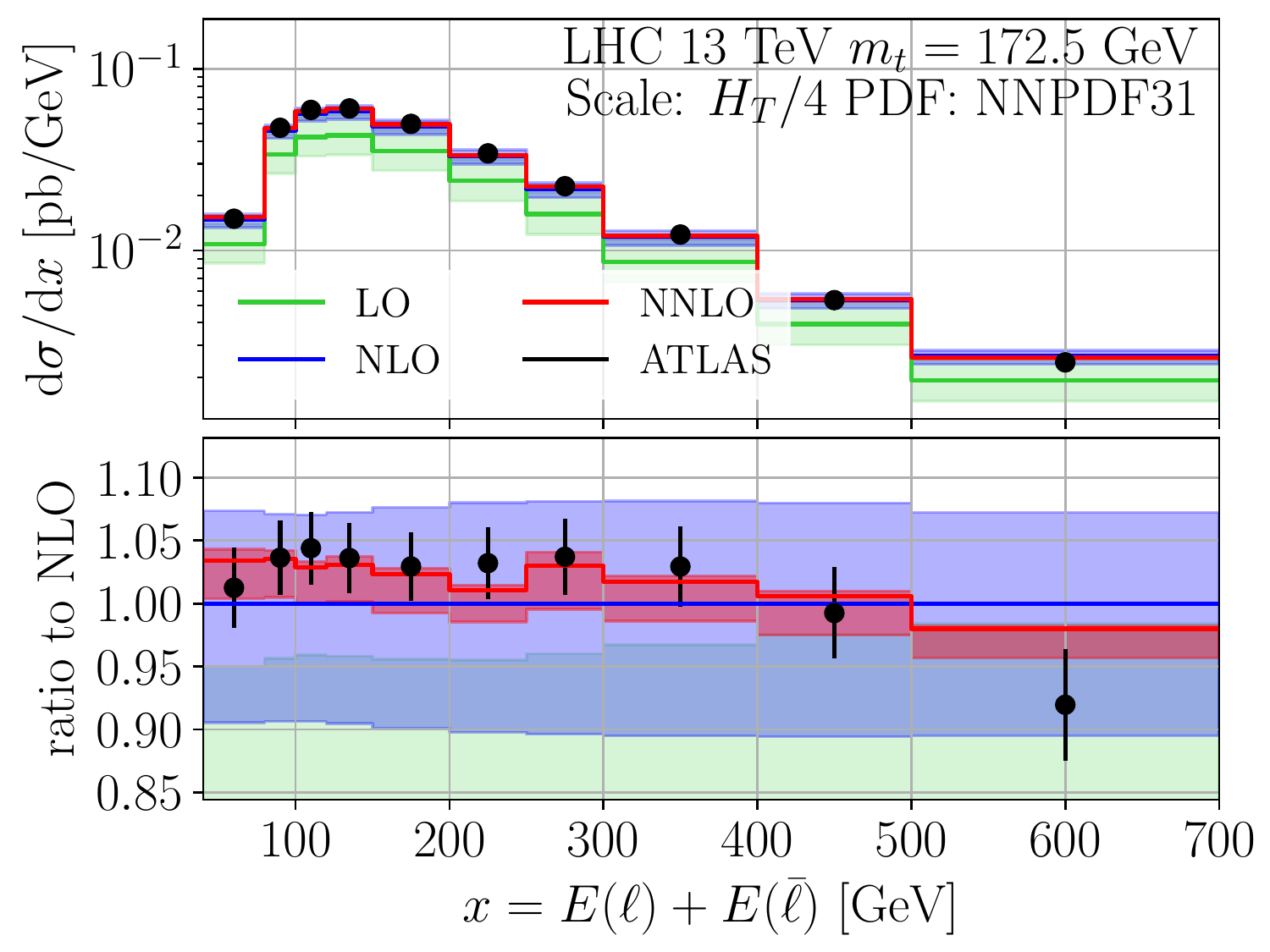}
\includegraphics[width=7.5cm]{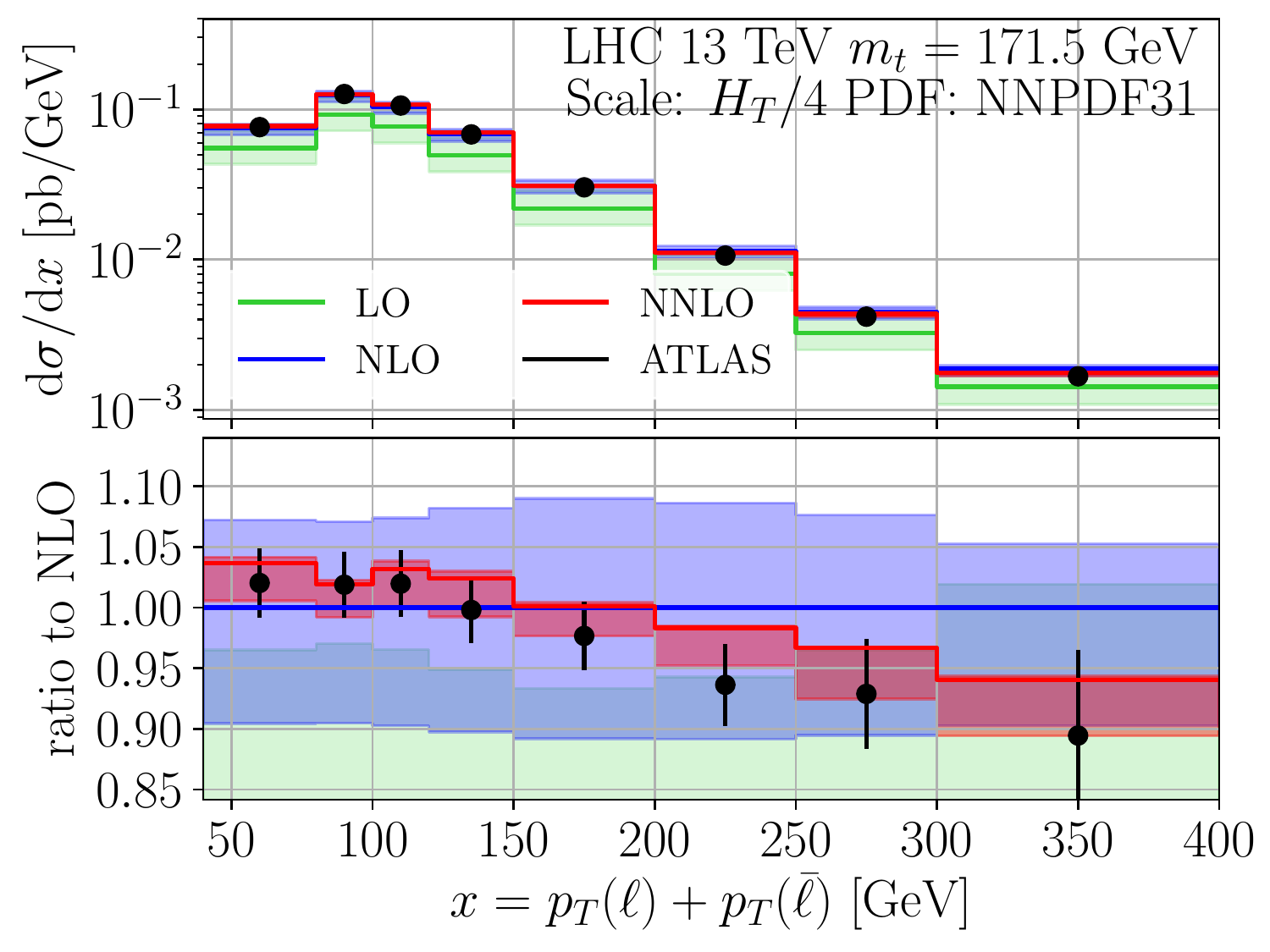}
\includegraphics[width=7.5cm]{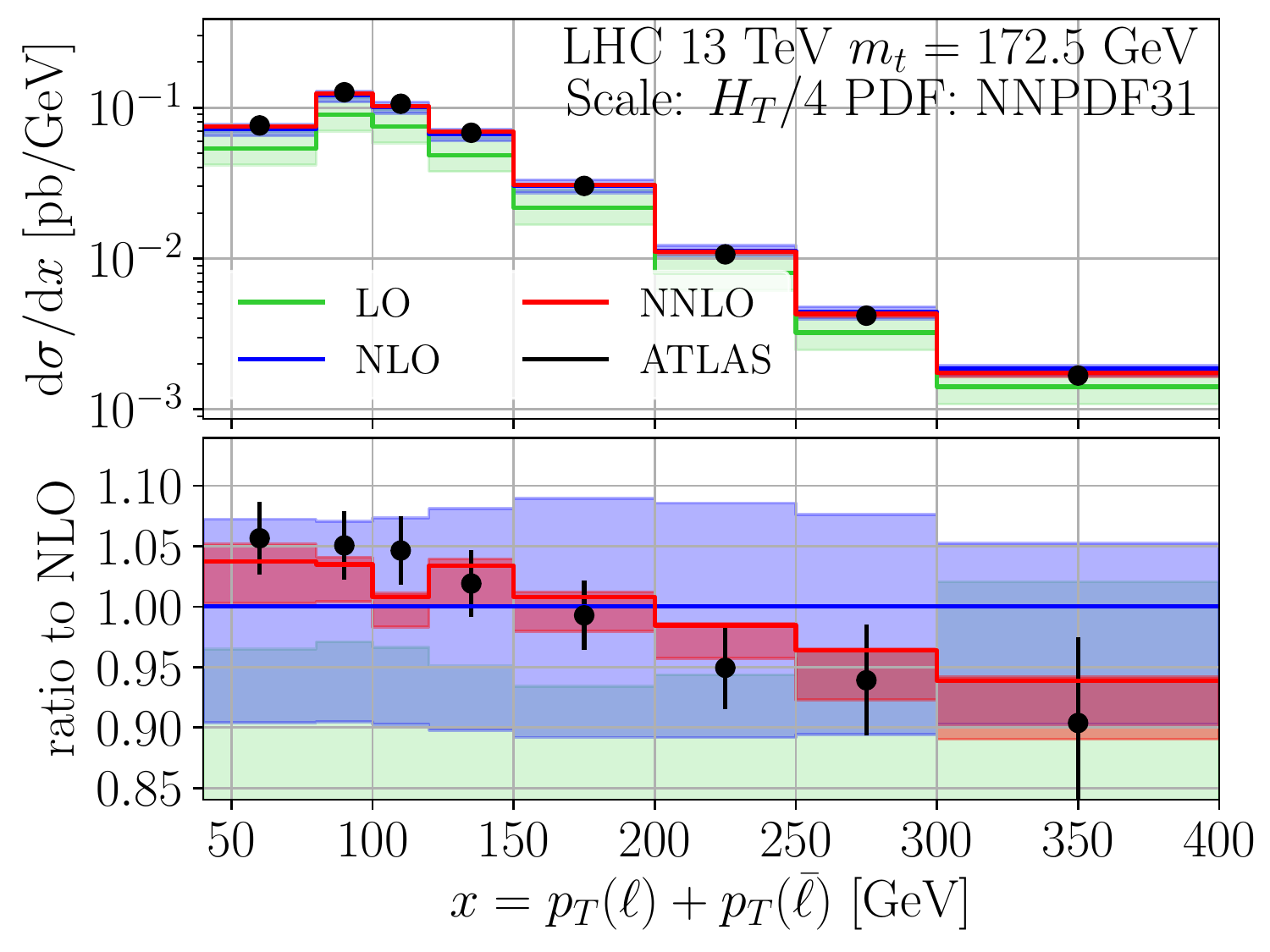}
\caption{As in fig.~\ref{fig:atlas_1} but for the $E(\ell)+E(\bar{\ell})$ and $p_T(\ell)+p_T(\bar{\ell})$ distributions.}
\label{fig:atlas_2}
\end{figure}
\begin{figure}
\includegraphics[width=7.5cm]{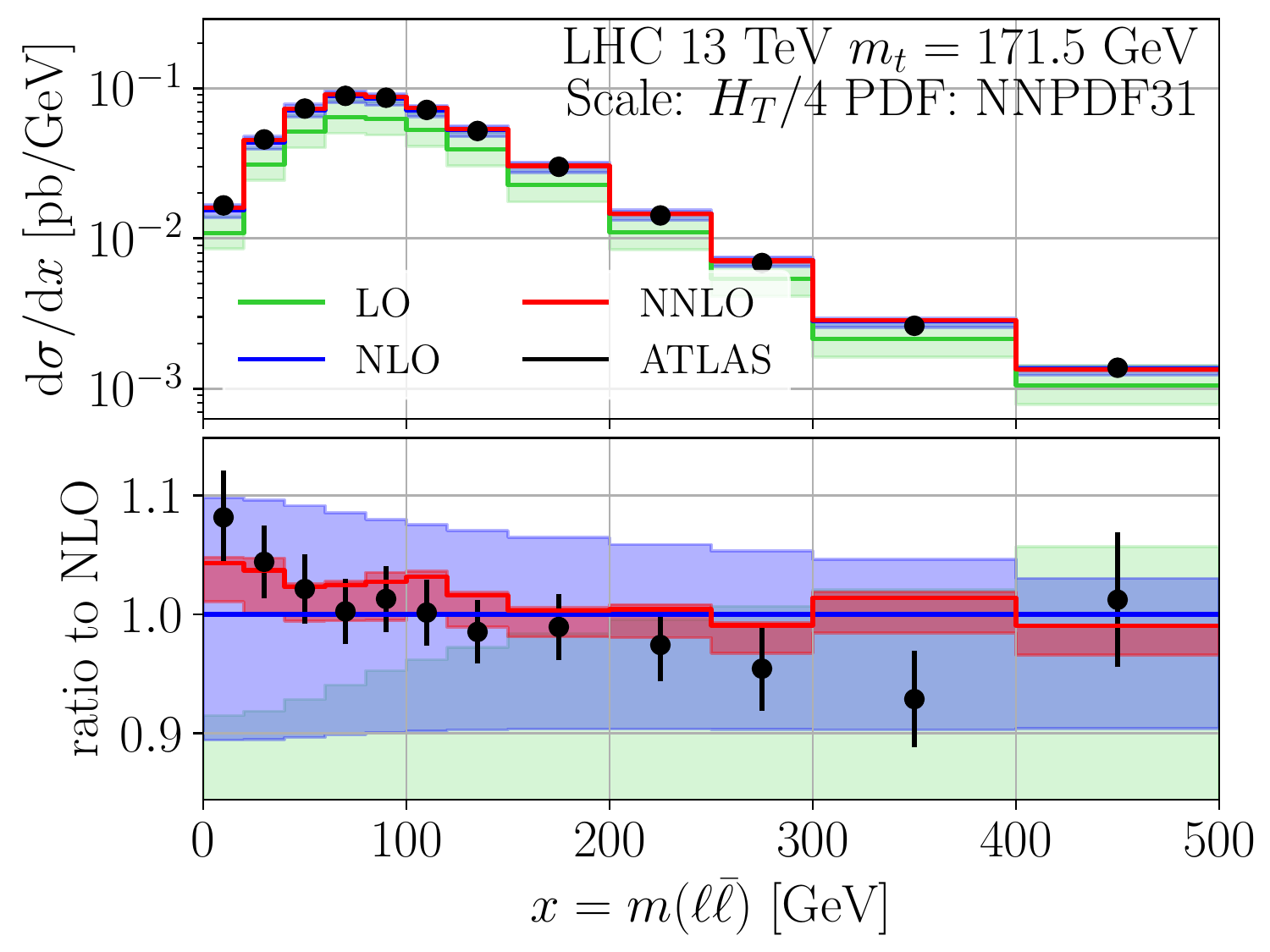}
\includegraphics[width=7.5cm]{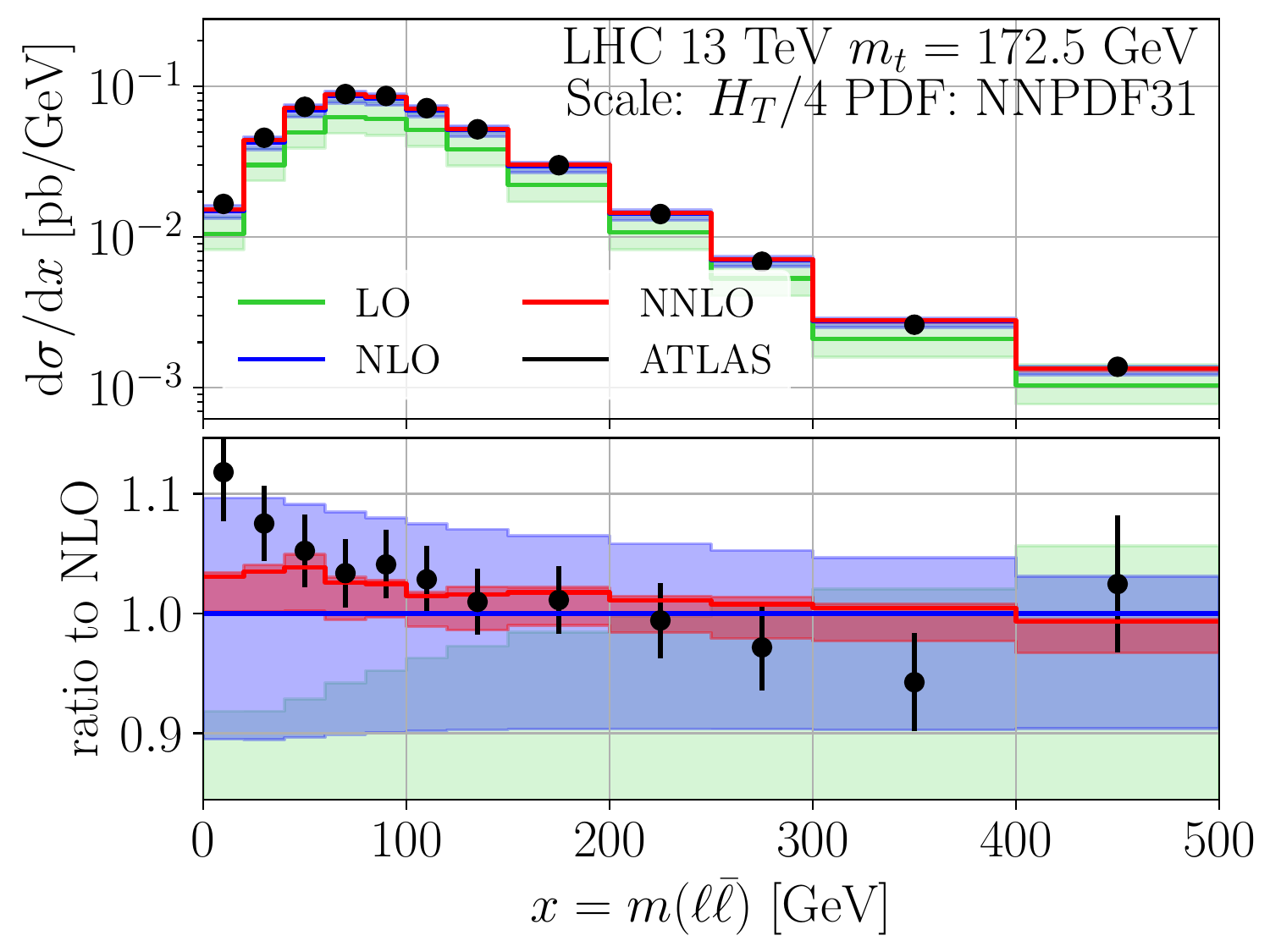}
\includegraphics[width=7.5cm]{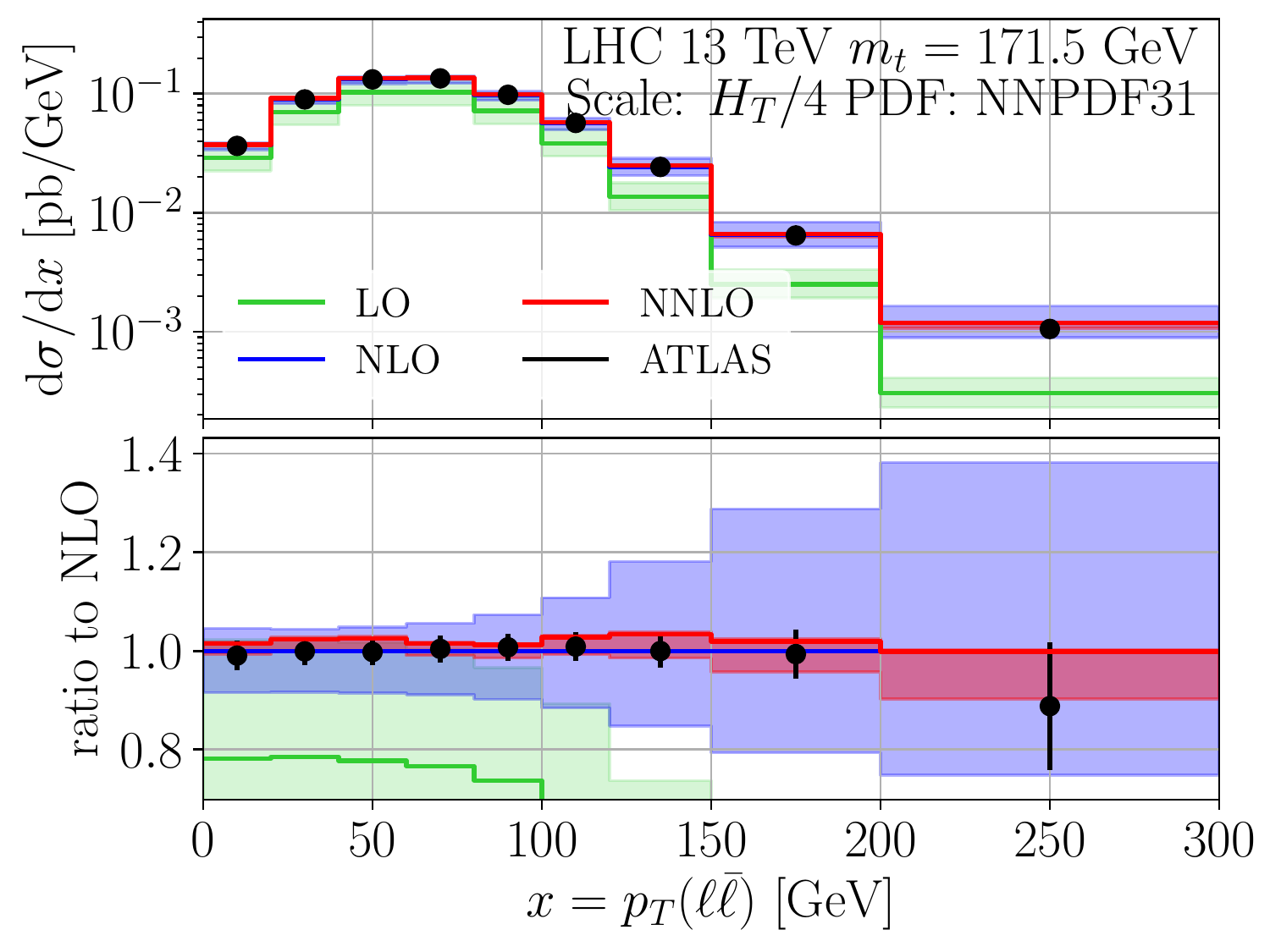}
\includegraphics[width=7.5cm]{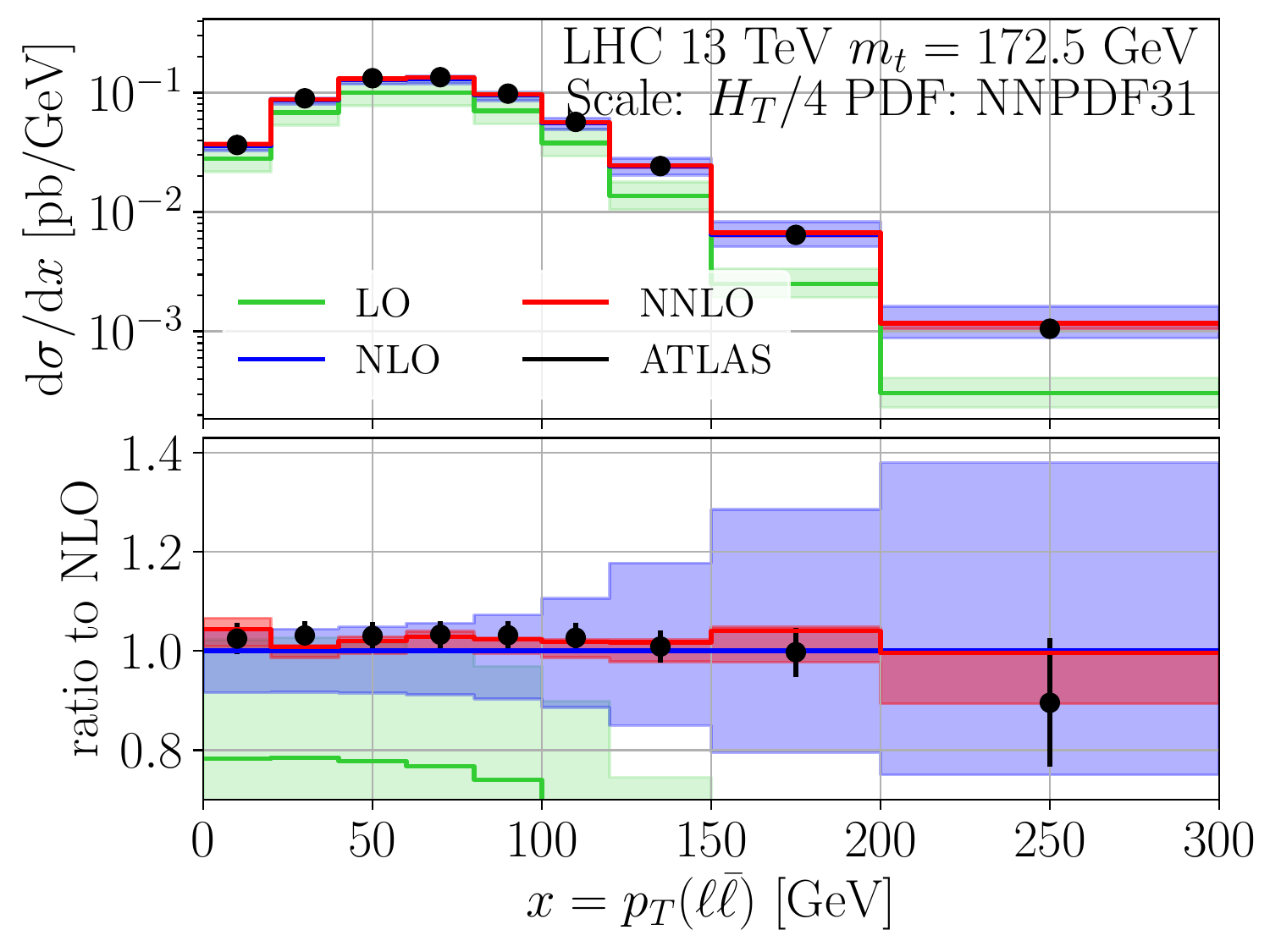}
\includegraphics[width=7.5cm]{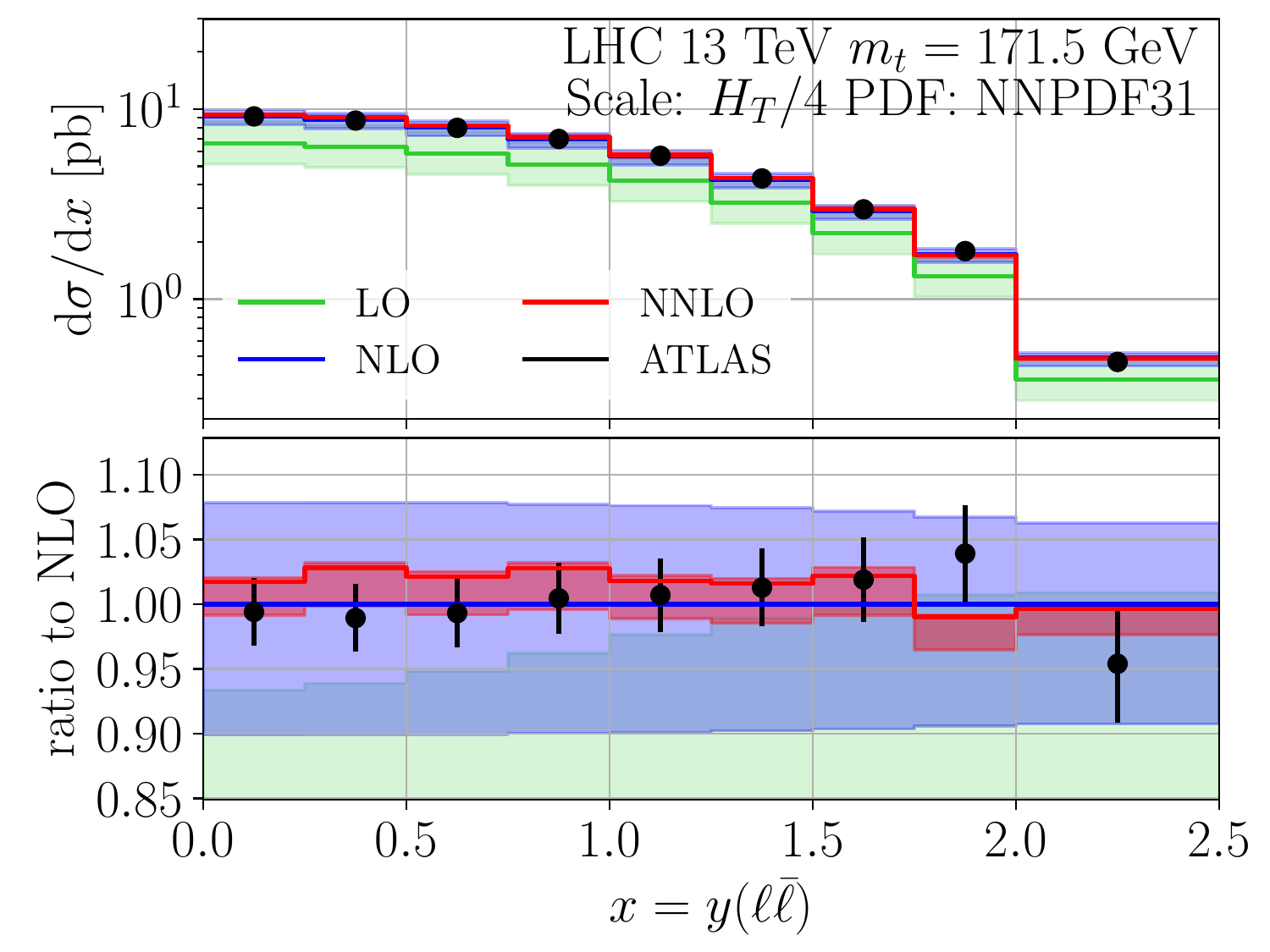}
\includegraphics[width=7.5cm]{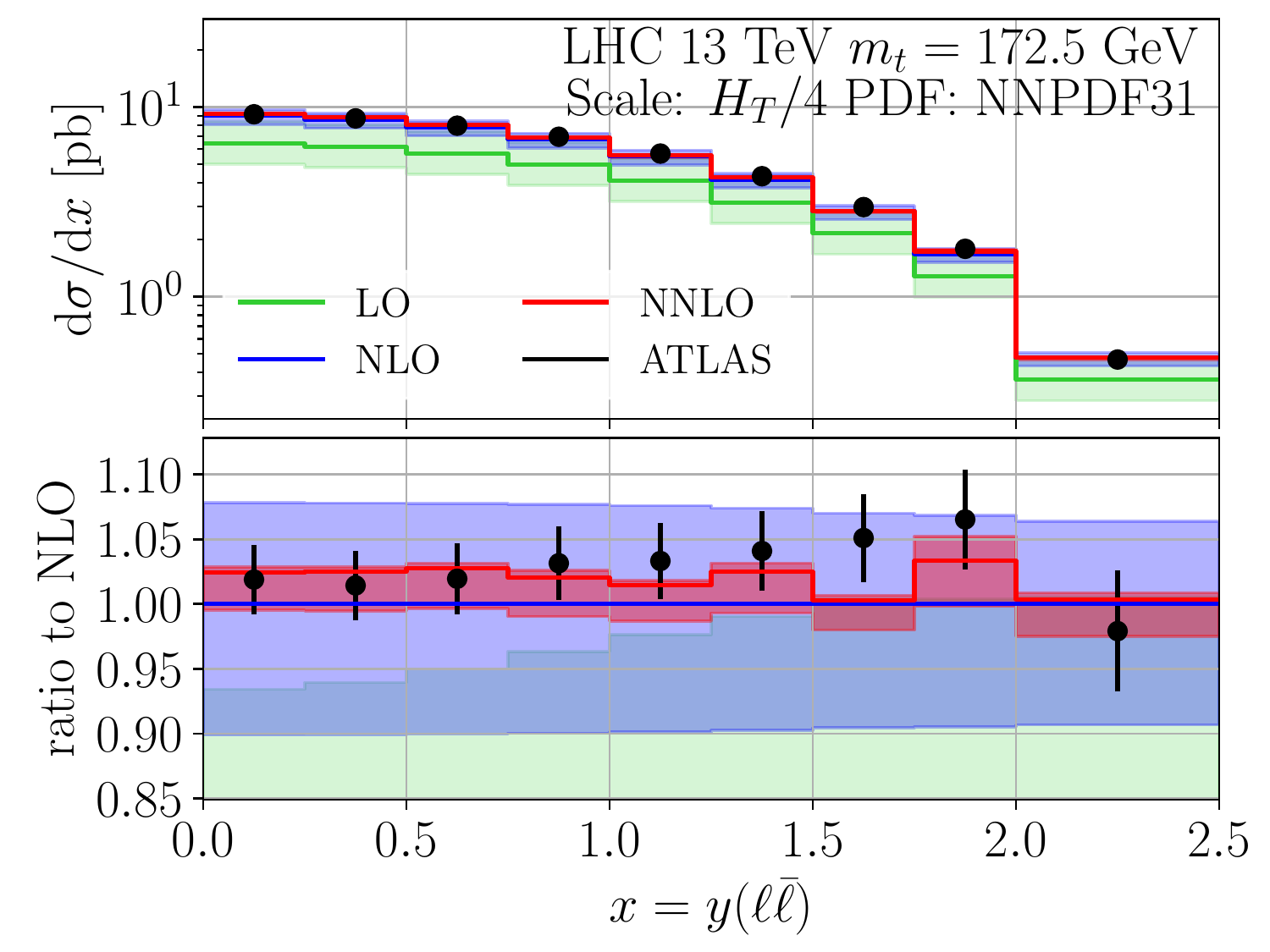}
\caption{As in fig.~\ref{fig:atlas_1} but for the $m(\ell\bar{\ell})$, $p_T(\ell\bar{\ell})$ and $y(\ell\bar{\ell})$ distributions.}
\label{fig:atlas_3}
\end{figure}

Qualitatively the same picture emerges from the differential distributions that depend on the combined kinematics of the lepton pair. In fig.~\ref{fig:atlas_2} we show the differential cross-sections related to the sum of the two lepton energies as well as the scalar sum of the lepton transverse momenta, while in fig.~\ref{fig:atlas_3} we show the $m(\ell\bar{\ell})$, $p_T(\ell\bar{\ell})$ and $y(\ell\bar{\ell})$ distributions. The general pattern of higher-order corrections described above is also evident in these lepton-pair spectra. A notable feature is the $m_t$ sensitivity of the $m(\ell\bar{\ell})$ distribution at low $m(\ell\bar{\ell})$. Unlike the $t\bar t$ threshold, this region can be measured precisely, mostly free from modeling ambiguities and the mass sensitivity is spread over several bins. 

It is also interesting to note a feature related to the pattern of higher order corrections. While in all cases the shift from LO to NLO is associated with a large K-factor, the corrections from NLO to NNLO are generally very small, typically only several percent. This is a sign of good perturbative convergence and confirms that the dynamic scale choice (\ref{eq:scale}) motivated in ref.~\cite{Czakon:2016dgf} works very well not only at the level of stable top quarks but also for leptonic differential distributions. Generally, from this one can conclude that missing yet-higher order corrections should be small.

\begin{figure}
\includegraphics[width=7.5cm]{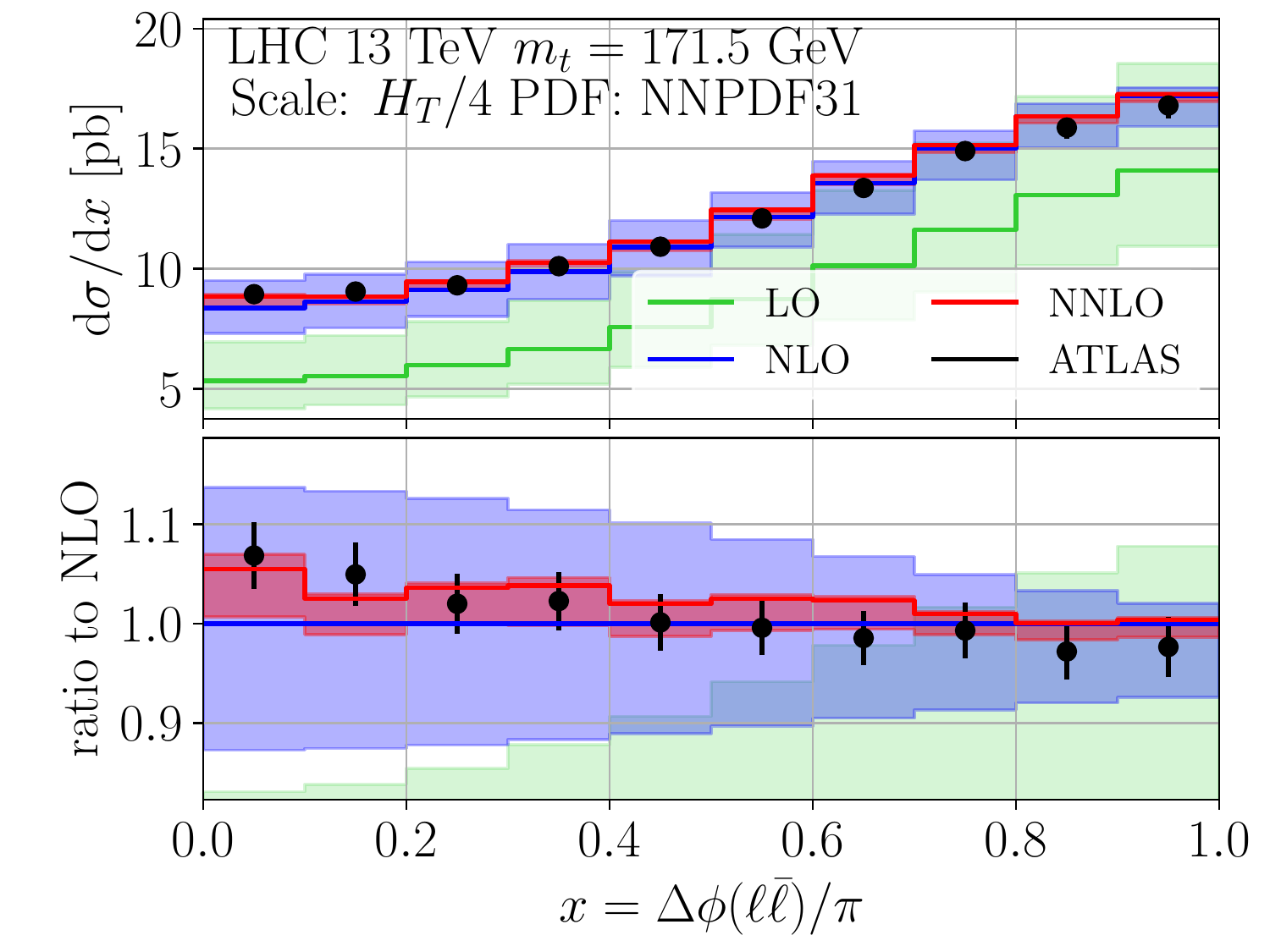}
\includegraphics[width=7.5cm]{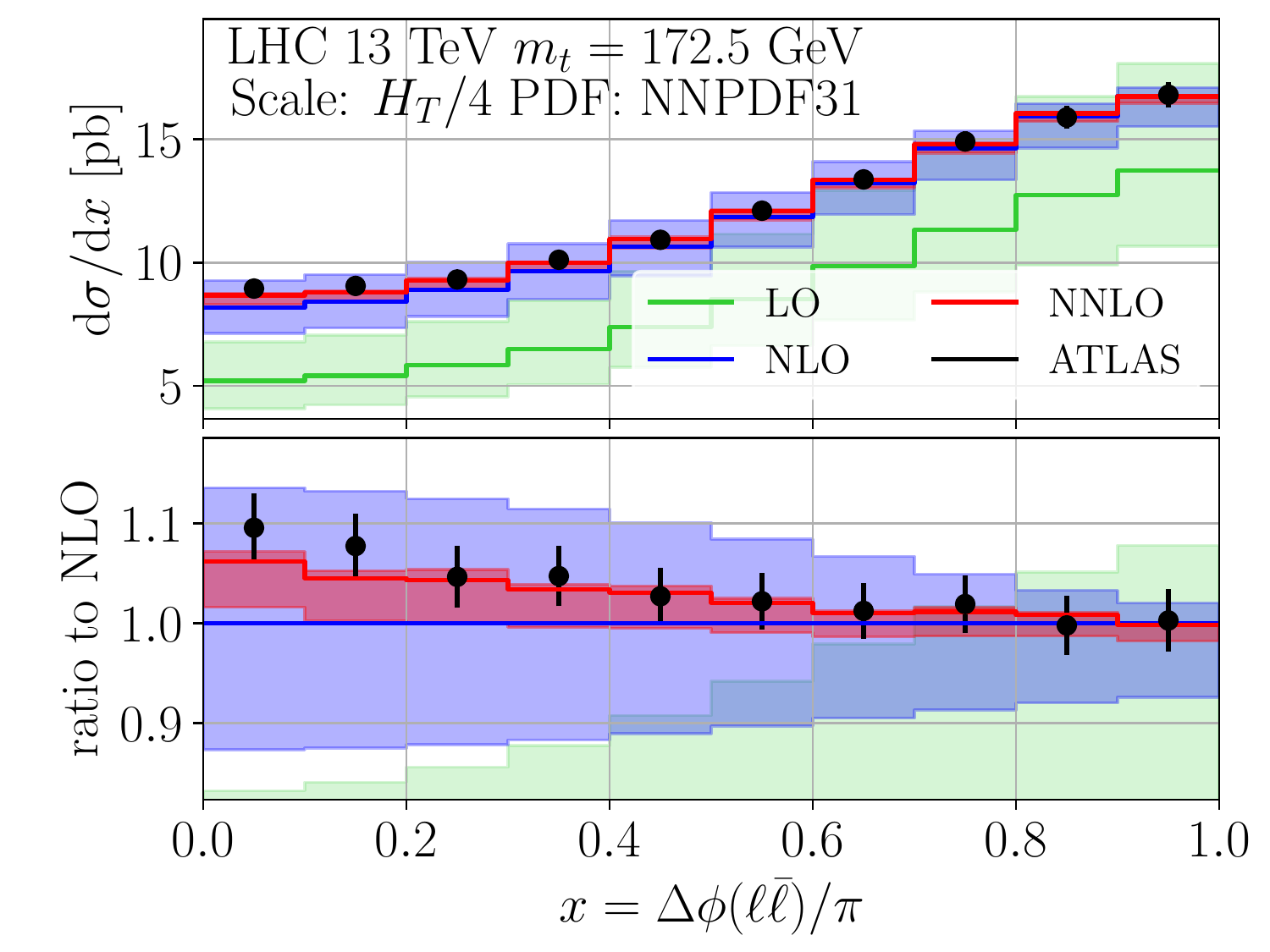}
\includegraphics[width=7.5cm]{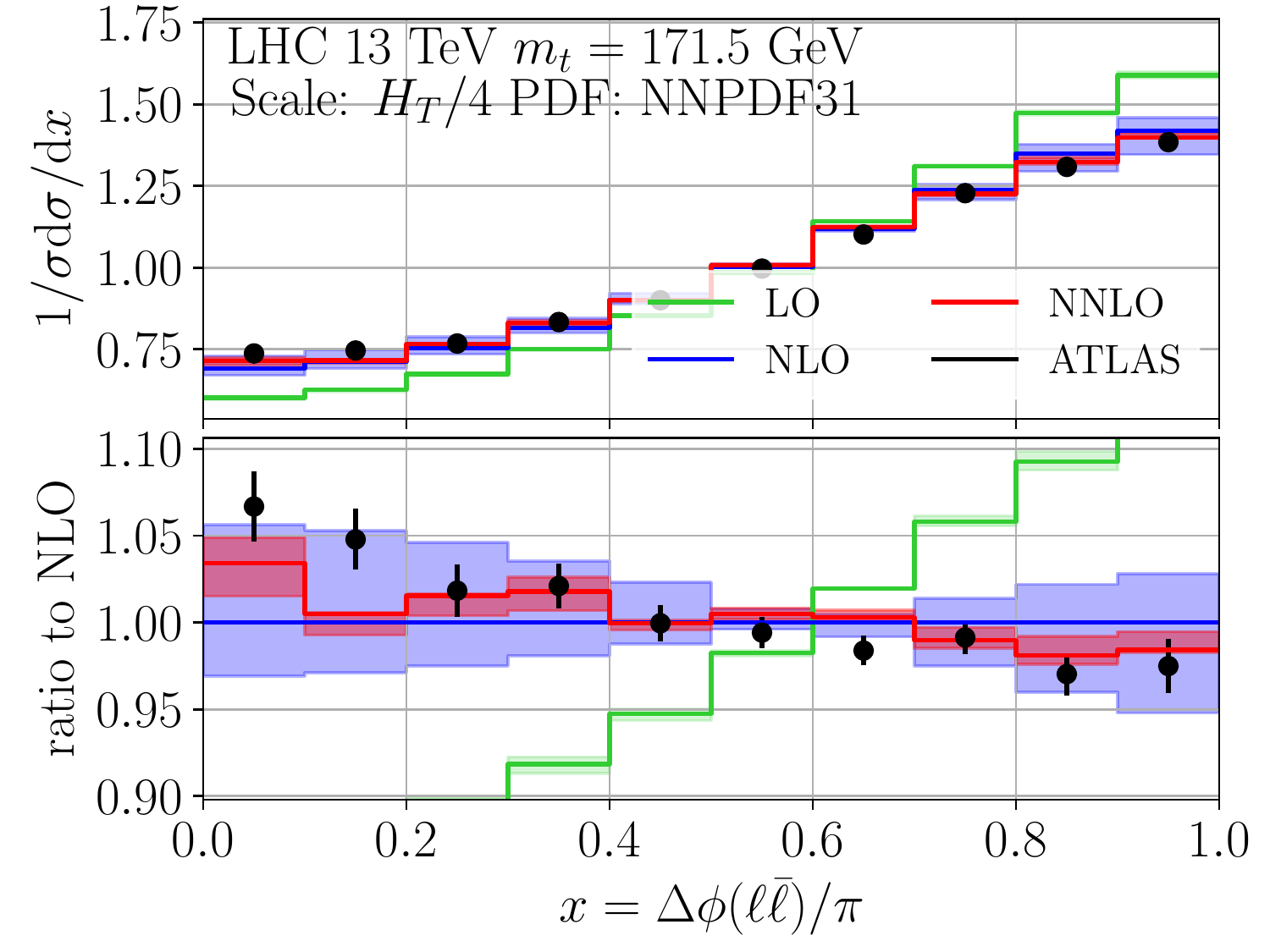}
\includegraphics[width=7.5cm]{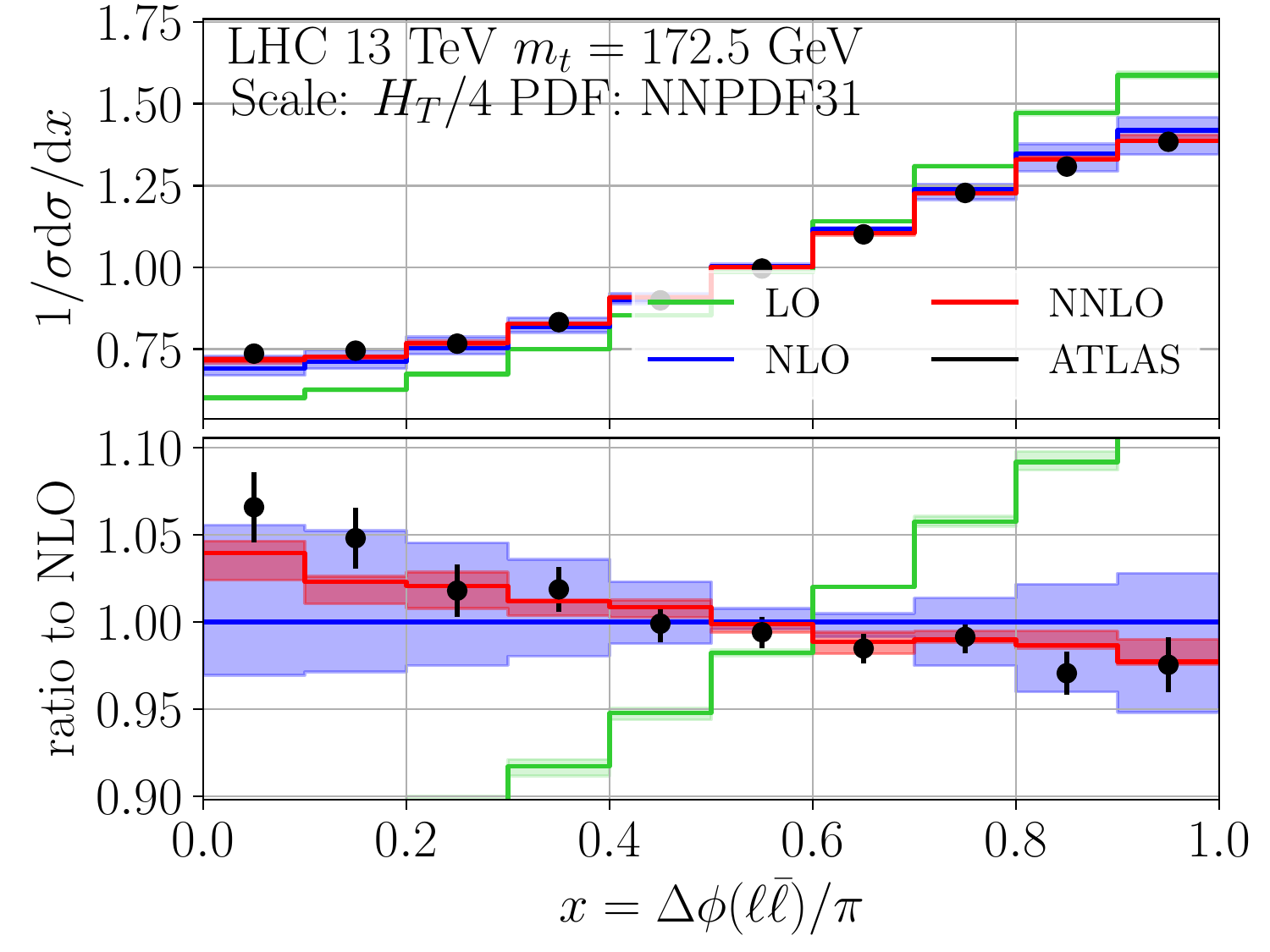}
\caption{As in fig.~\ref{fig:atlas_1} but for the absolute (top) and normalized (bottom) $\Delta\phi(\ell\bar{\ell})$ distribution.}
\label{fig:atlas_5}
\end{figure}

In fig.~\ref{fig:atlas_5} we show the differential distribution in the azimuthal opening angle $\Delta\phi(\ell\bar{\ell})$ between the two leptons. As already discussed in sec.~\ref{sec:spin-correlations-angular} this distribution is sensitive to $t\bar t$ spin correlations. We note that the $\Delta\phi(\ell\bar{\ell})$ distributions shown in sec.~\ref{sec:spin-correlations-angular} and here are not equivalent since in this section the $\Delta\phi(\ell\bar{\ell})$ distribution is subject to fiducial cuts on the two leptons. As already discussed in ref.~\cite{Behring:2019iiv} the presence of cuts on the final state leads to significant shape corrections. We show both the absolute and normalized distributions. As expected, normalizing the distribution results in significant reduction in scale uncertainty which is essential for precision comparison with data. In both the absolute and normalized distributions we observe very good agreement between NNLO QCD and data. The pattern of higher order corrections already noted for other differential distributions can also be observed here: one observes large NLO/LO K-factor while the NNLO corrections is rather mild relative to the NLO one. As expected from an angular distribution the top mass dependence is very small, mostly affecting the overall normalization.

\begin{figure}
\includegraphics[width=7.5cm,height=7.1cm]{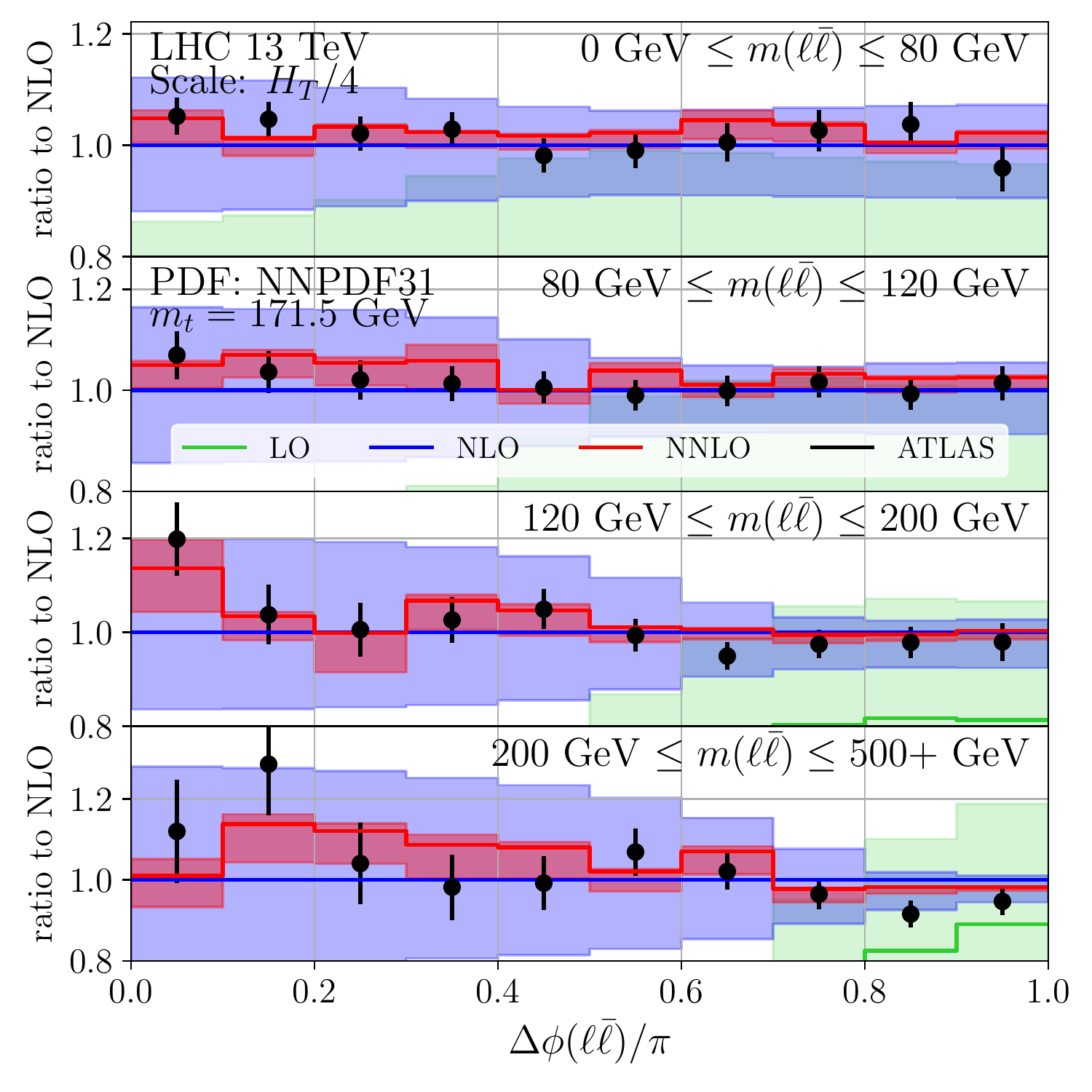}
\includegraphics[width=7.5cm,height=7.1cm]{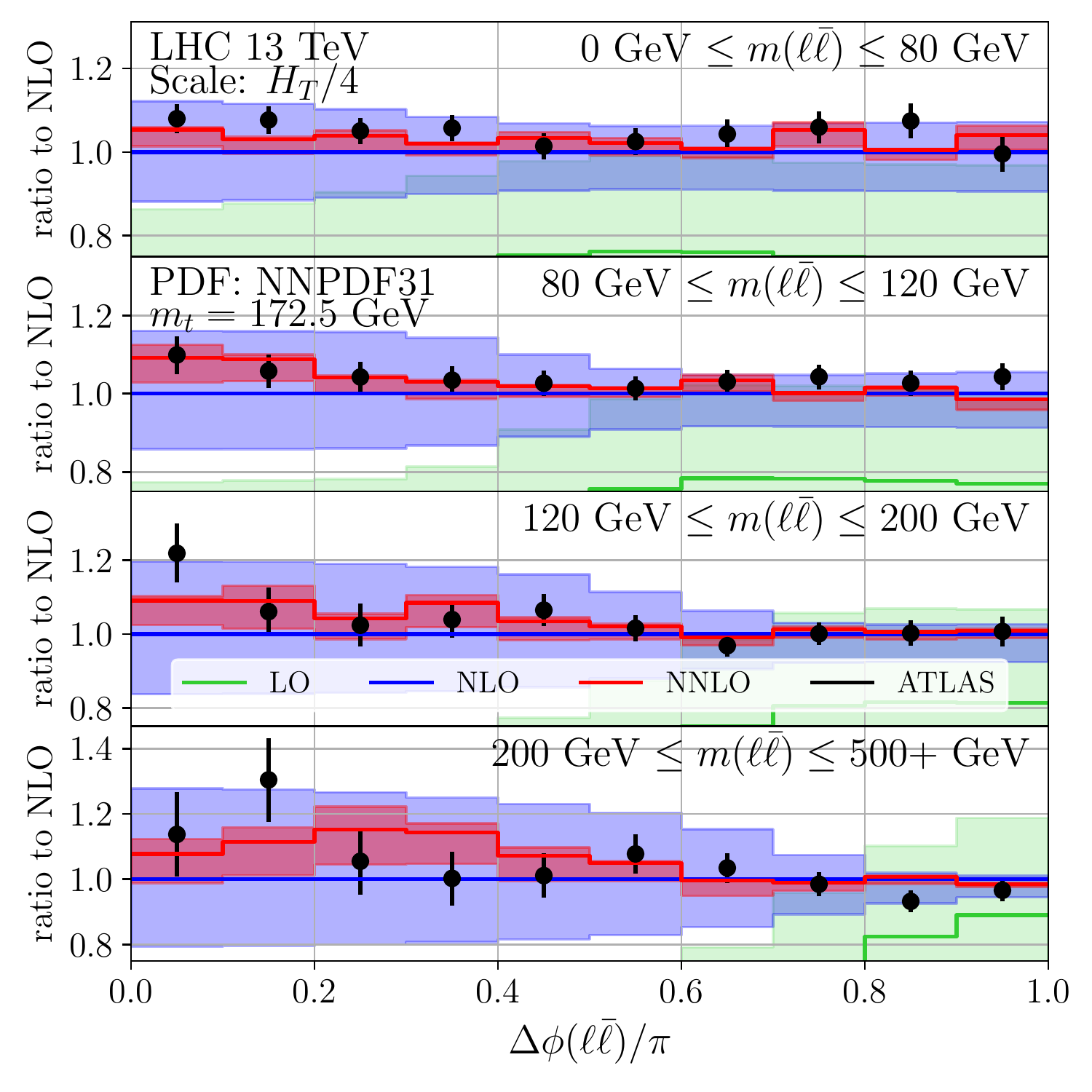}
\includegraphics[width=7.5cm,height=7.1cm]{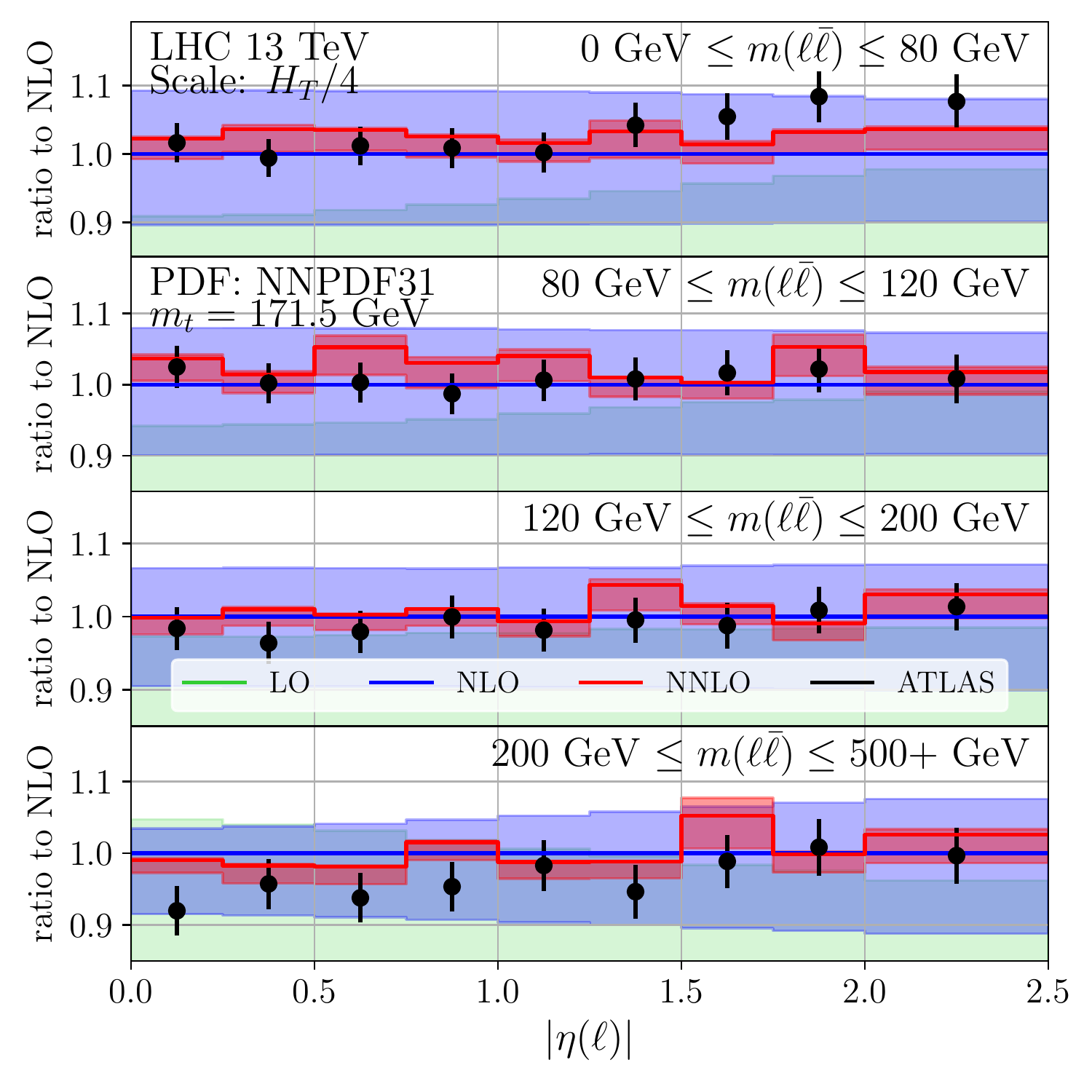}
\includegraphics[width=7.5cm,height=7.1cm]{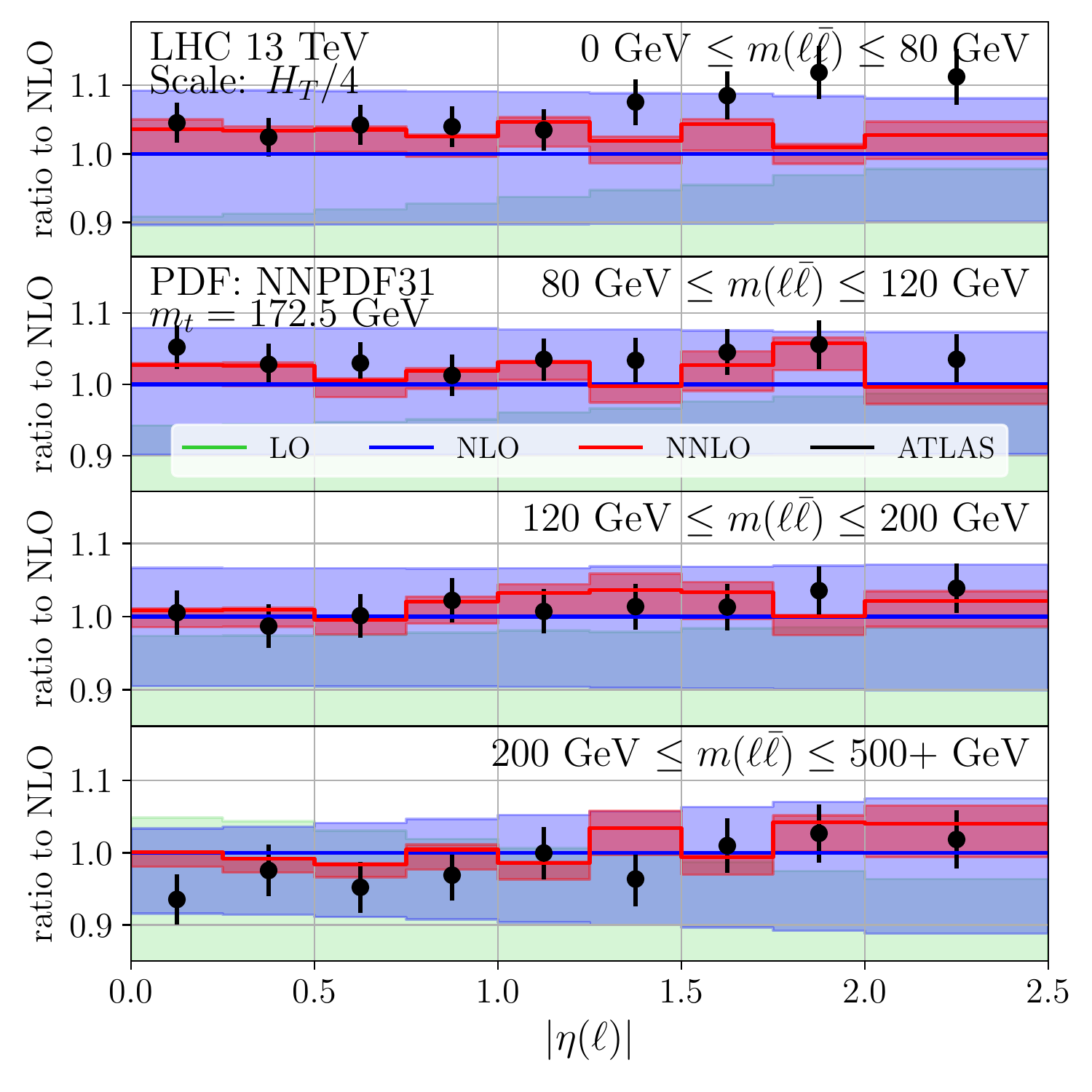}
\includegraphics[width=7.5cm,height=7.1cm]{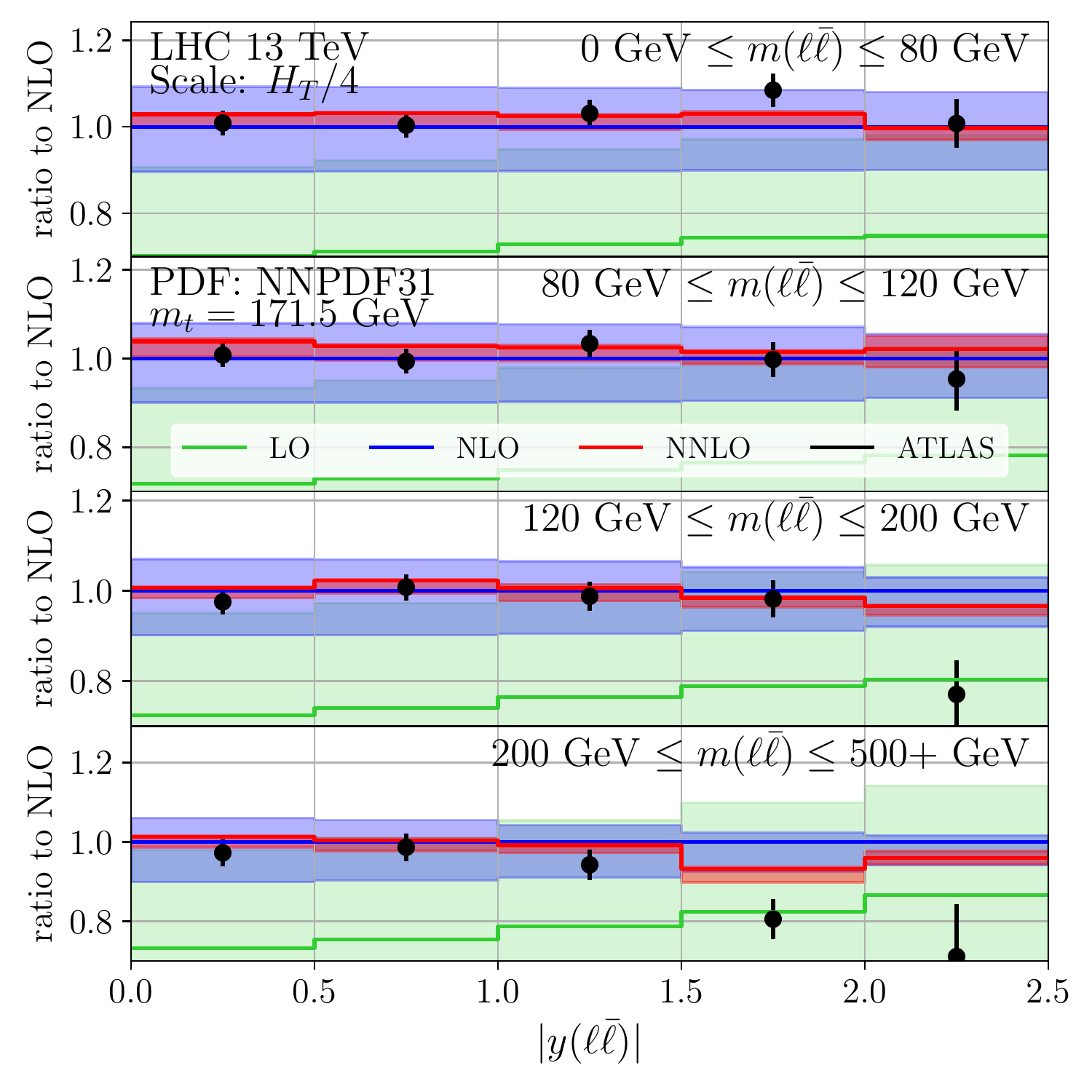}
\includegraphics[width=7.5cm,height=7.1cm]{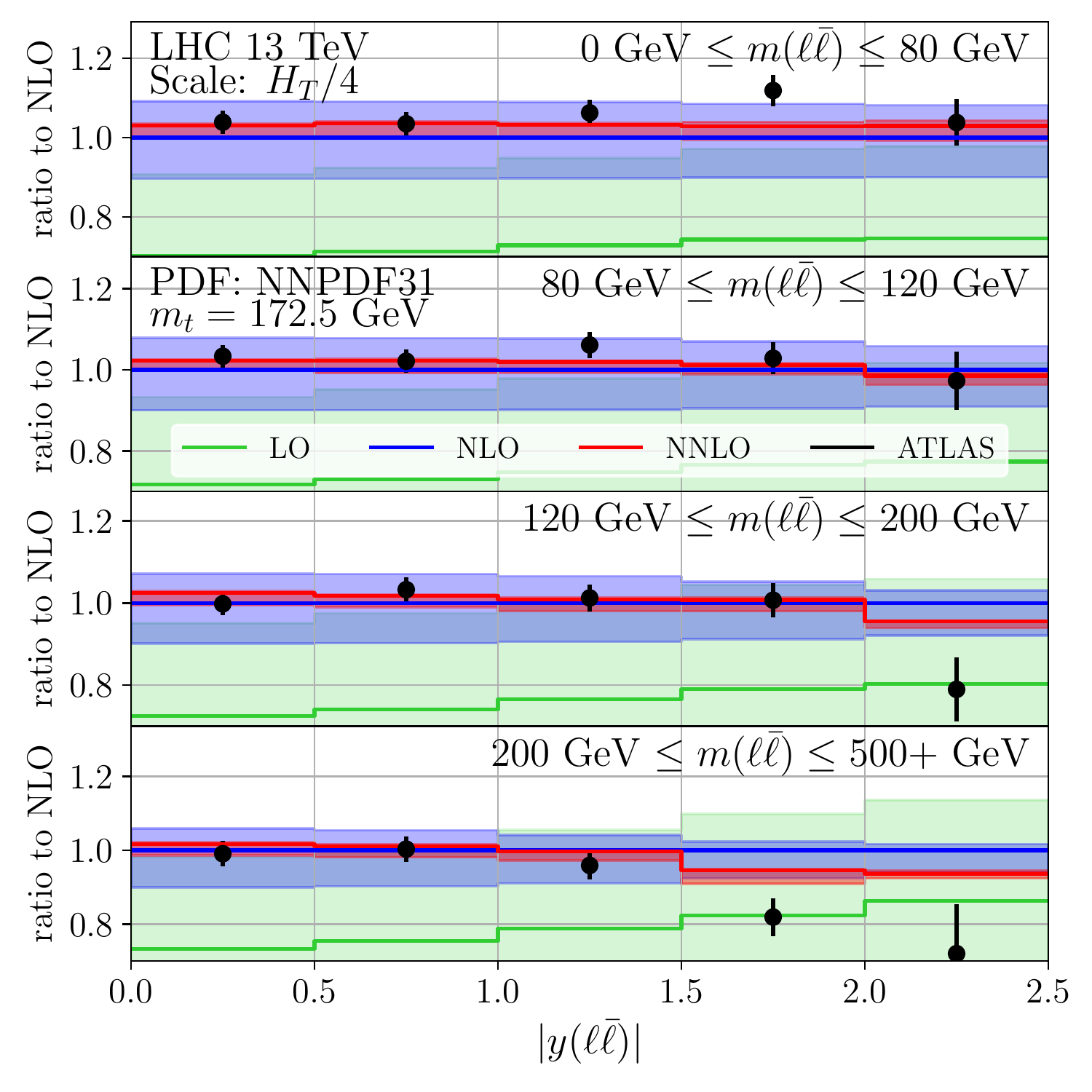}
\caption{Absolute double differential distributions in $m(\ell\bar{\ell})$ and $\Delta\phi(\ell\bar{\ell})$, $\eta(\ell)$ and $y(\ell\bar{\ell})$ versus ATLAS data \cite{Aad:2019hzw}. Uncertainty bands represent scale variation. Shown are fixed-order predictions at LO (green), NLO (blue) and NNLO (red) for two top quark masses $m_t=171.5$ GeV (left) and $m_t=172.5$ GeV (right).}
\label{fig:atlas_6}
\end{figure}

In the rest of this section we turn our attention to a set of two-dimensional distributions measured by the ATLAS Collaboration \cite{Aad:2019hzw}. Specifically, in fig.~\ref{fig:atlas_6} we show predictions for the single differential observables $\Delta\phi(\ell\bar{\ell})$, $\eta(\ell)$ and $y(\ell\bar{\ell})$ in slices of $m(\ell\bar{\ell})$. Just like for all one-dimensional distributions we compute predictions for two values of $m_t$. 

A number of features can be seen. The pattern of higher-order corrections for all cases is roughly similar to the one already seen for the 1-dim distributions: large NLO/LO K-factors while the NNLO corrections are small relative to the NLO ones and typically about several percent. The only exception is the $\Delta\phi(\ell\bar{\ell})$ distribution for large values of $m(\ell\bar{\ell})$ where the NNLO/NLO K-factor reaches 10\%-20\%. The size of the scale uncertainty at NNLO is much smaller than the NLO scale variation. In all bins the uncertainty of the NNLO theory prediction is comparable to or smaller than the experimental uncertainty.

In almost all cases we find very good description of data with NNLO QCD. Only in a few bins (with large $y(\ell\bar{\ell})$ and $m(\ell\bar{\ell})$), which are exhibiting relatively large experimental uncertainties, data lies outside of the scale uncertainty band. This difference cannot be attributed to the value of the top quark mass since these bins are insensitive to $m_t$.

It is interesting to note that the $\Delta\phi(\ell\bar{\ell}) \times m(\ell\bar{\ell})$ distribution exhibits notable sensitivity to the value of $m_t$ which in some slices results in a significant shape modification. The $\Delta\phi(\ell\bar{\ell})$ distribution, differential in $m(t\bar{t})$ has been proposed as a very sensitive probe for $t\bar t$ spin-correlations. Since such a measurement requires top quark reconstruction which reduces the precision of the measurement and thus limits the usefulness of this observable, the $\Delta\phi(\ell\bar{\ell}) \times m(\ell\bar{\ell})$ distribution may be seen as a proxy which can both be computed and measured very precisely and it may turn out to be a good place for studying spin correlations. 

In summary we note that the available NNLO QCD predictions for leptonic differential distributions match or exceed in precision the best available LHC measurements. This in turn offers the possibility for a detailed precision phenomenology in $t\bar t$ production with dilepton final states.

\subsection{Leptonic observables with fiducial selection}\label{sec:pheno_fid}

Up to this point we considered only observables inclusive in hadronic radiation. Such measurements necessarily involve extrapolations from the fiducial to the full phase-space and therefore contain some residual sensitivity to modeling. An alternative approach would be to compare fixed-order theory predictions with data directly at the fiducial level. Fiducial level comparisons also allow for interesting top-quark phenomenology since, due to the presence of $b$-jets, the top-quarks can be reconstructed and themselves studied within the fiducial phase space. 

While the fiducial approach is appealing it has drawbacks of its own. The main criticism towards fiducial comparisons of data with fixed-order calculations is related to the modeling of $b$-jets. An example is the ongoing discussion of the fiducial comparison for the  $\Delta\phi(\ell\bar{\ell})$ distribution of the NNLO computation in ref.~\cite{Behring:2019iiv} and the ATLAS measurement \cite{ATLAS:2018rgl}, see also ref.~\cite{Aaboud:2019hwz}. In the rest of this section we will demonstrate that while such differences do arise (and they can even be very significant) it is possible to have a meaningful fiducial theory-data comparison once the jets in data are modified so that they are sufficiently compatible with the ones in a fixed-order partonic calculation.

To be less abstract, in the following we focus on the CMS analysis \cite{Sirunyan:2018ucr} (although we believe the conclusions we reach are generic). The definition of the fiducial phase-space is as follows: required are two identified $b$-jets and
\begin{itemize}
\item $p_T(\ell) \geq 20 \;\text{GeV}$ and $|\eta(\ell)| \leq 2.4$ for both charged leptons\,,
\item $m(\ell\bar{\ell}) \geq 20\; \text{GeV}$\,,
\item 2 anti-$k_T$, $R = 0.4$ jets with $p_T \geq 30 \;\text{GeV}$
        and $|y| \leq 2.4$. Both jets are required to be $b$-tagged.
        In our fixed order computation a $b$-tag is implemented by counting the
        bottomness of a jet after clustering. Non-vanishing bottomness results in a $b$-tag\,.
\item Only jets that are well separated from the leptons, $\Delta R (j,\ell) \geq 0.4$, are taken into account.
\end{itemize}

In the following we will consider theory-data comparisons for leptonic, $b$-jet and top-quark fiducial differential distributions. It would be useful to first focus on the issues related to the definition of $b$-jets which affect the leptonic and $b$-jet distributions. Once this is clarified we will move to fiducial top-quark comparisons which, in turn, will require us to understand another layer of subtleties related to the definition of top quarks.

In the following we address the question {\it what is the content of a $b$-jet}? Although it may appear superfluous at first, this question turns out to have profound implications for precision theory-data comparisons. In a parton level calculation like ours jets are constructed by clustering partons, i.e. quarks and gluons. The clustering uses the same algorithm used in the experimental analysis. The measurement clusters particles which have resulted from the fragmentation of hard partons like the ones clustered in our fixed-order calculation. The idea is that although the two jets are built out of different objects, jets constructed in these two ways can be directly compared due to the jets' inclusiveness (within QCD). Basically, the hard partons clustered inside a partonic jet will undergo timelike fragmentation which, ultimately, will result in the production of a bunch of hadrons. The key argument is that the hadrons will tend to remain inside the same partonic jet. It is for this reason partonic and particle jets can be compared to each other. 

When does the above particle-parton jet equivalence argument fail? One possibility is out-of-cone radiation during the subsequent jet evolution. These effects are relatively suppressed and should not be a large contribution. In particular since a NNLO calculation like ours accounts exactly for up to two emitted (soft, collinear and/or hard) partons, such out-of-cone radiation is a N$^3$LO effect. A second reason that can lead to breaking of the partonic-particle jet equivalence is contributions from non-QCD effects. One such example are semileptonic $B$-decays which are mediated by the weak interaction. Our argument above states that the $B$-meson resulting from the in-jet fragmentation of a $b$-quark will be part of the final jet. However during semileptonic decays the $B$ meson decays to final states containing neutrinos and leptons. Even if the soft lepton is included in the jet the neutrino will not be since it is not registered by the detector. This leads to a difference between the partonic and particle jets. One may wonder if the numerical impact due to the loss of the neutrino momentum is significant. As it turns out, it is.

The original CMS publication \cite{Sirunyan:2018ucr} is based on jets that do not include the neutrinos from semileptonic $B$-meson decays. In order to compare with our parton level calculations, the CMS Collaboration has reanalyzed the data and has unfolded it with the help of Monte Carlo simulation to more inclusive jets that include the neutrino momentum. In the following we compare leptonic and $b$-jet distributions with this more inclusive CMS data \cite{CMS-fiducial-mod}. To demonstrate the numerical impact from the inclusion of the neutrino momentum we will also show in that comparison the original CMS data for jets that exclude neutrinos. 

The following set of leptonic and $b$-jet observables are computed within the fiducial phase space:
\begin{itemize}
  \item Leptonic observables, shown in figs.~\ref{fig:fid-lep-abs},\ref{fig:fid-lep-norm}:
    \begin{itemize}
      \item $p_T(\ell)$: transverse momentum of the negatively or positively charged lepton,
      \item $\eta(\ell)$: the pseudo rapidity of the negatively or positively charged lepton,
      \item $m(\ell\bar{\ell})$: lepton pair invariant mass,
      \item $p_T(\ell\bar{\ell})$: transverse momentum of the lepton pair,
      \item $\Delta\phi(\ell\bar{\ell})$: the angle defined in eq.~(\ref{eq:delta-phi}),
      \item $\Delta|\eta(\ell\bar{\ell})| \equiv |\eta(\bar{\ell})|-|\eta(\ell)|$: difference between the two absolute rapidities,
    \end{itemize}
  \item $b$-jet observables, shown in figs.~\ref{fig:fid-bjet-abs},\ref{fig:fid-bjet-norm} (only the two $b$-jets used in the top-quark reconstruction, see text around eq.~(\ref{eq:top-reconstruct}) for details, are used here):
    \begin{itemize}
      \item $p_T(b_1)$: transverse momentum of the leading $b$-jet,
      \item $p_T(b_2)$: transverse momentum of the sub-leading $b$-jet,
      \item $\eta(b_1)$: pseudo rapidity of the leading $b$-jet,
      \item $\eta(b_2)$: pseudo rapidity of the sub-leading $b$-jet,
      \item $m(b_1 b_2)$: invariant mass of the two leading $b$-jets,
      \item $p_T(b_1 b_2)$: transverse momentum of the two leading $b$-jets.
    \end{itemize}
\end{itemize}
\label{text:fid-plots-begin} 
The format of the plots of leptonic and $b$-jet spectra is as follows. The top panel in each plot shows the LO, NLO and NNLO QCD prediction for the corresponding observable. The bands around the theory predictions reflect the 7-point scale variation around the central scale eq.~(\ref{eq:scale}). Shown as black vertical bars is the modified CMS data \cite{CMS-fiducial-mod} which has $b$-jets including the neutrinos from semileptonic $B$-decays. Also shown as vertical grey bars is the original CMS data \cite{Sirunyan:2018ucr} which does not include neutrinos from semileptonic $B$-decays. For consistency theory predictions should be compared with the black data, while the data in grey is shown as an indication of the size of the effect of excluding the neutrinos from semileptonic decays.

The uncertainty on the original CMS data \cite{Sirunyan:2018ucr} (in grey) represents the full experimental uncertainty (statistics and systematics). Only statistical uncertainty estimate for the new CMS data \cite{CMS-fiducial-mod} (in black) has been made available. For this reason, with the black error bars we show a derived by us uncertainty {\it estimate} which is obtained by rescaling the systematics uncertainty of the grey data to the value of the black data and then adding it in quadrature to the statistical one. While this procedure is imperfect it is adequate for the qualitative theory-data comparison we perform here. 

The middle panel contains the same information as the top one, however, all curves and data are plotted relative to the NNLO QCD prediction. The bottom panel shows the size of the pdf uncertainty (shown as a dark grey band) for an alternative calculation using $m_t=171.5$ GeV. All curves in the middle and bottom panels are normalized to the central NNLO prediction with $m_t=172.5$ GeV.

Since the relative pdf error for a small change in $m_t$ remains roughly unchanged, we have computed the pdf uncertainty only for $m_t=171.5$ GeV and not for $m_t=172.5$ GeV. We believe this is sufficient for understanding the relative importance of pdf versus scale variation in all distributions considered in this section. When computing the pdf error we followed two approaches. One is the usual prescription appropriate for the pdf set we use, which utilizes the full set of pdf members. We have also computed the pdf uncertainty following the approach of reduced pdf sets proposed in ref.~\cite{Carrazza:2016htc}
\footnote{We would like to thank Zahari Kassabov for producing the reduced pdf sets used in this study.}. 
We find that the two ways of estimating the pdf error are extremely close for all distributions and cannot be distinguished within the MC error of the calculation. This is the first time reduced pdf sets have been checked in NNLO QCD for top quark production with decay and we find them very promising for future applications.

Besides the absolute distributions described above, we also show in separate plots the full set of normalized distributions. They are normalized to the fiducial cross-section which in some cases is different from the integral under a curve (for example, $p_T(\ell)$ is shown up to 400 GeV excluding the events with $p_T(\ell)>400$ GeV). 
\label{text:fid-plots-end} 

\subsubsection{Analysis of the lepton-only distributions}

\begin{figure}
\includegraphics[width=7.4cm]{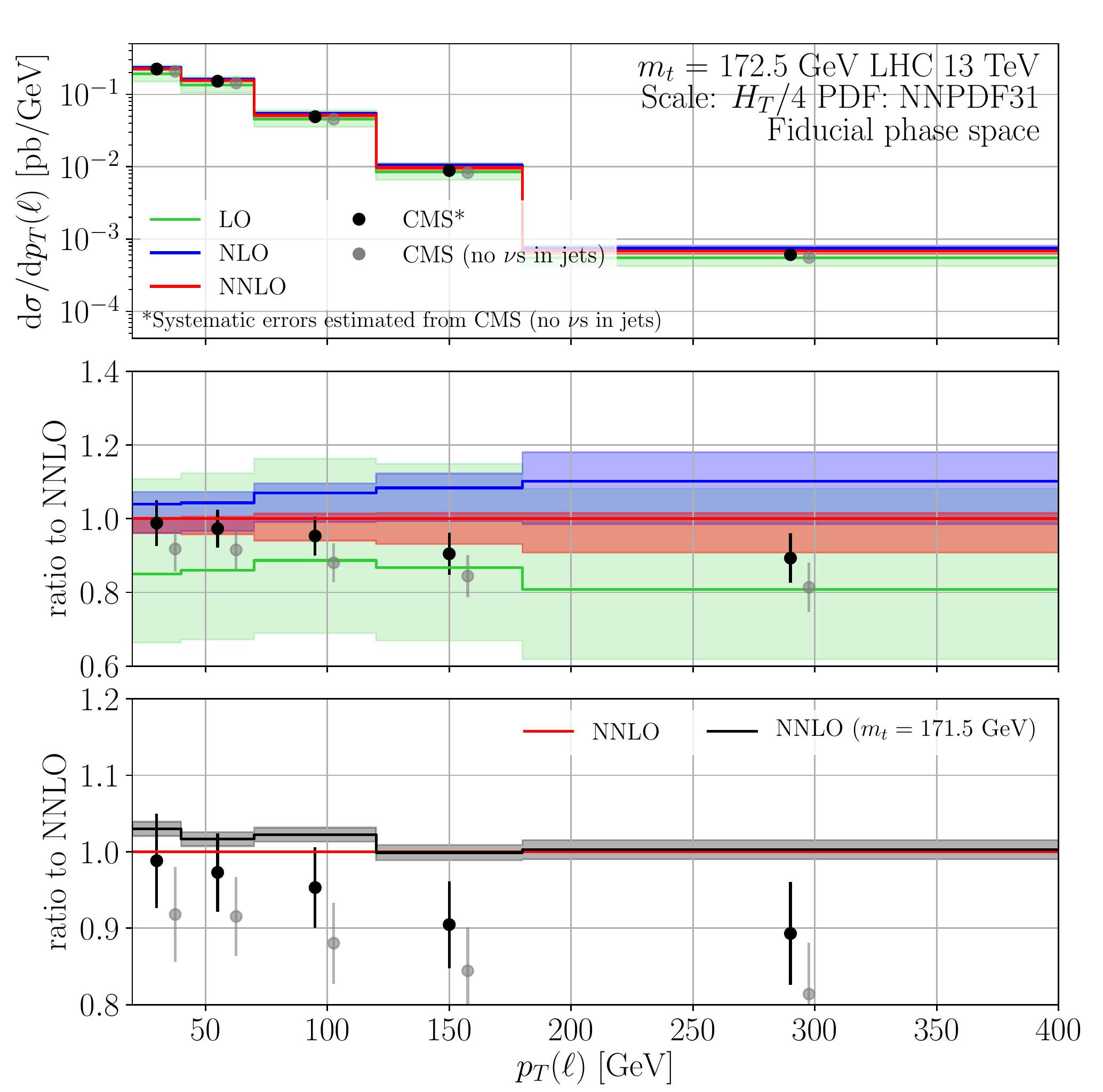}
\includegraphics[width=7.4cm]{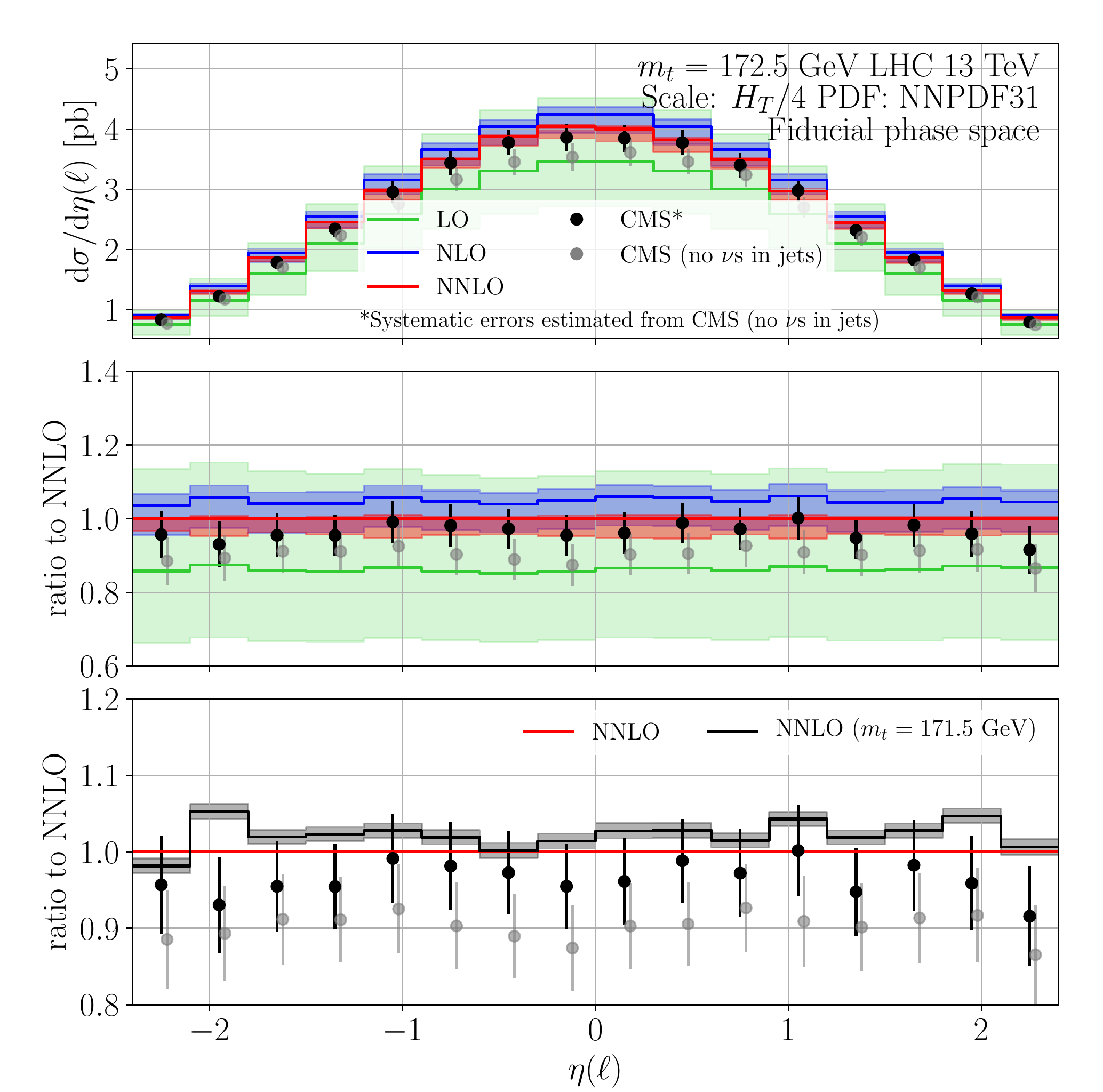}\\
\includegraphics[width=7.4cm]{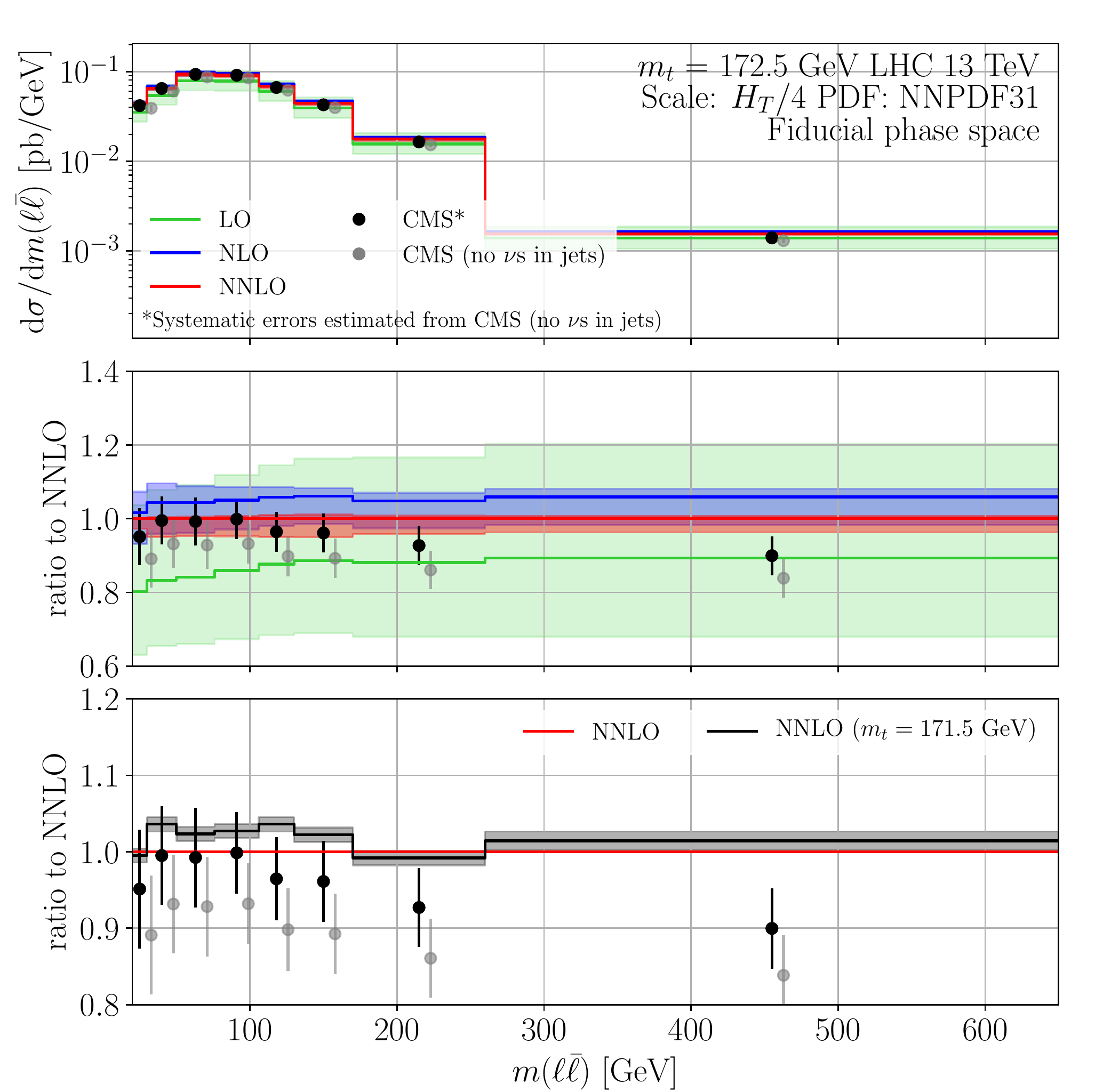}
\includegraphics[width=7.4cm]{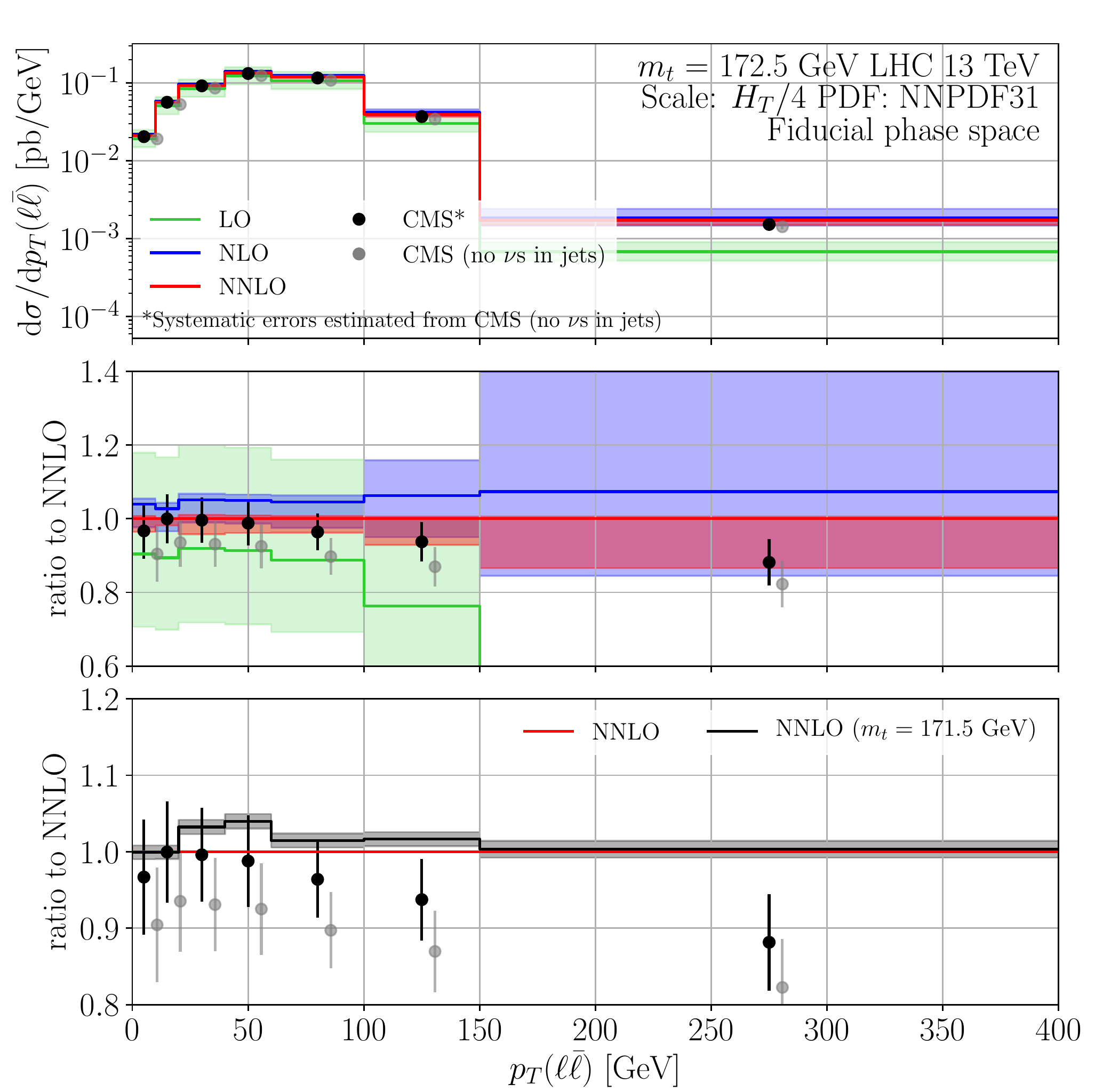}\\
\includegraphics[width=7.4cm]{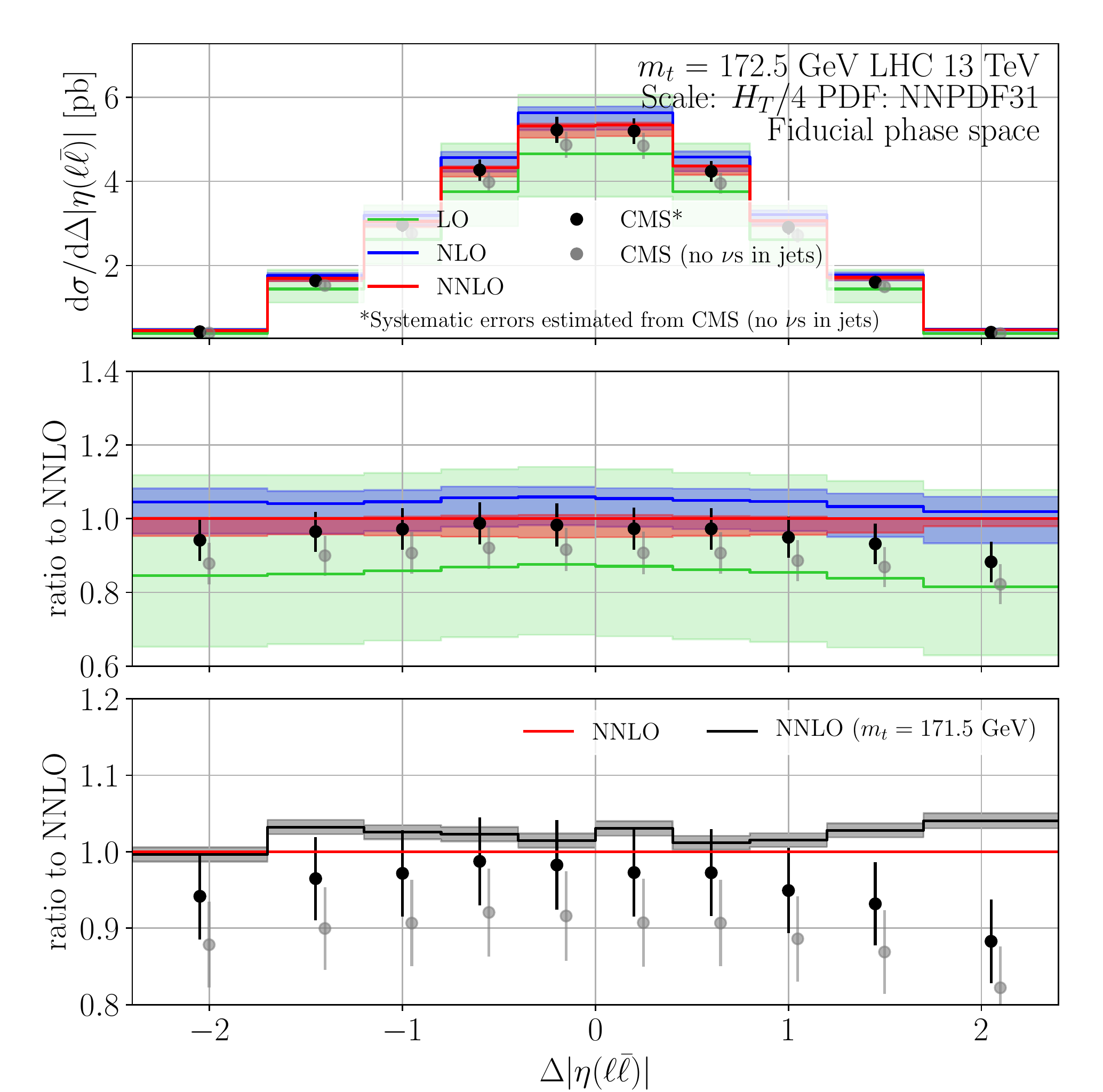}
\includegraphics[width=7.4cm]{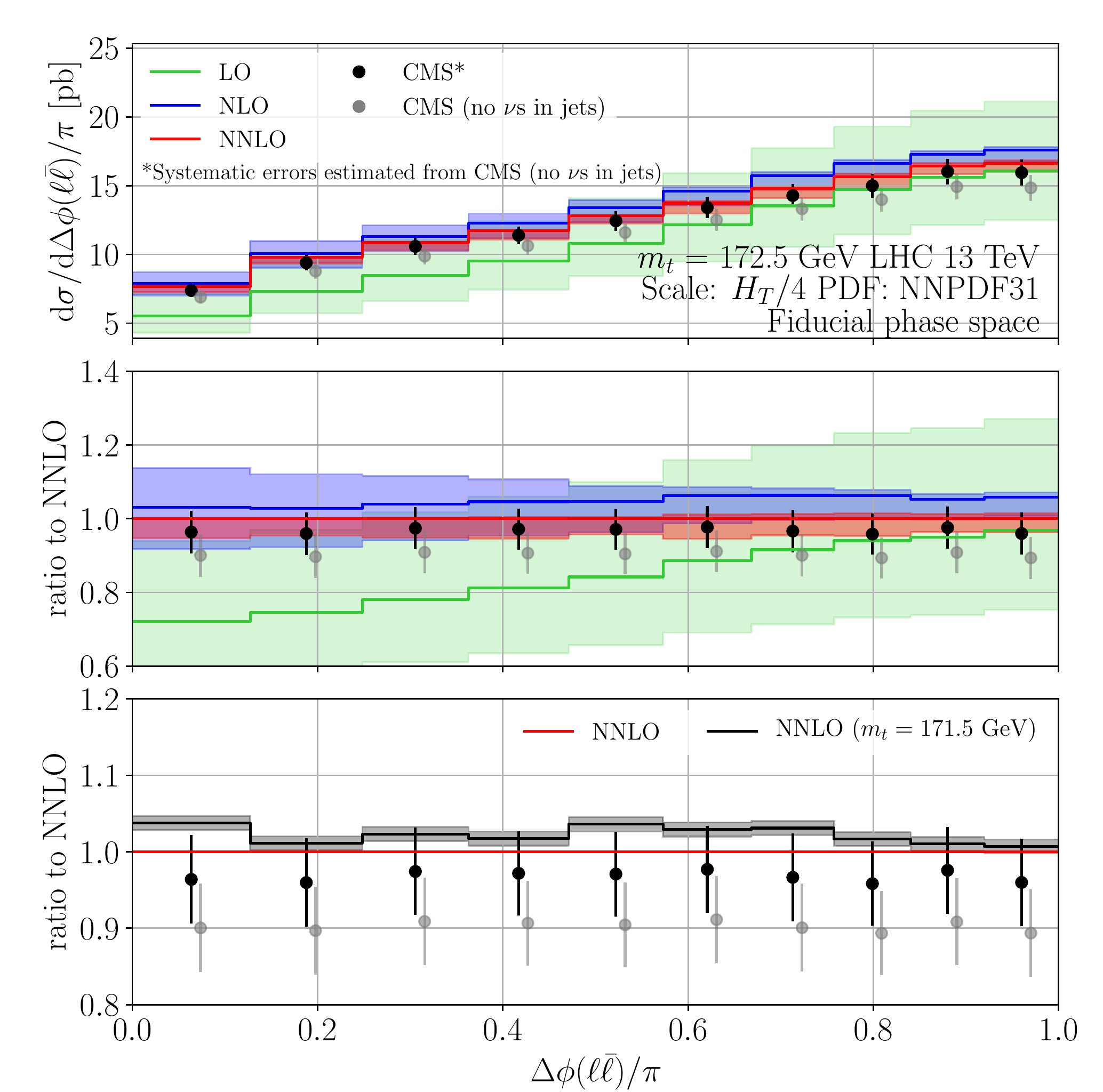}
\caption{Absolute fiducial leptonic distributions in LO, NLO and NNLO QCD versus CMS data without neutrinos \cite{Sirunyan:2018ucr} (in grey) and with neutrinos \cite{CMS-fiducial-mod} (in black). See pages~\pageref{text:fid-plots-begin}--\pageref{text:fid-plots-end} for details.}
\label{fig:fid-lep-abs}
\end{figure}
\begin{figure}
\includegraphics[width=7.4cm]{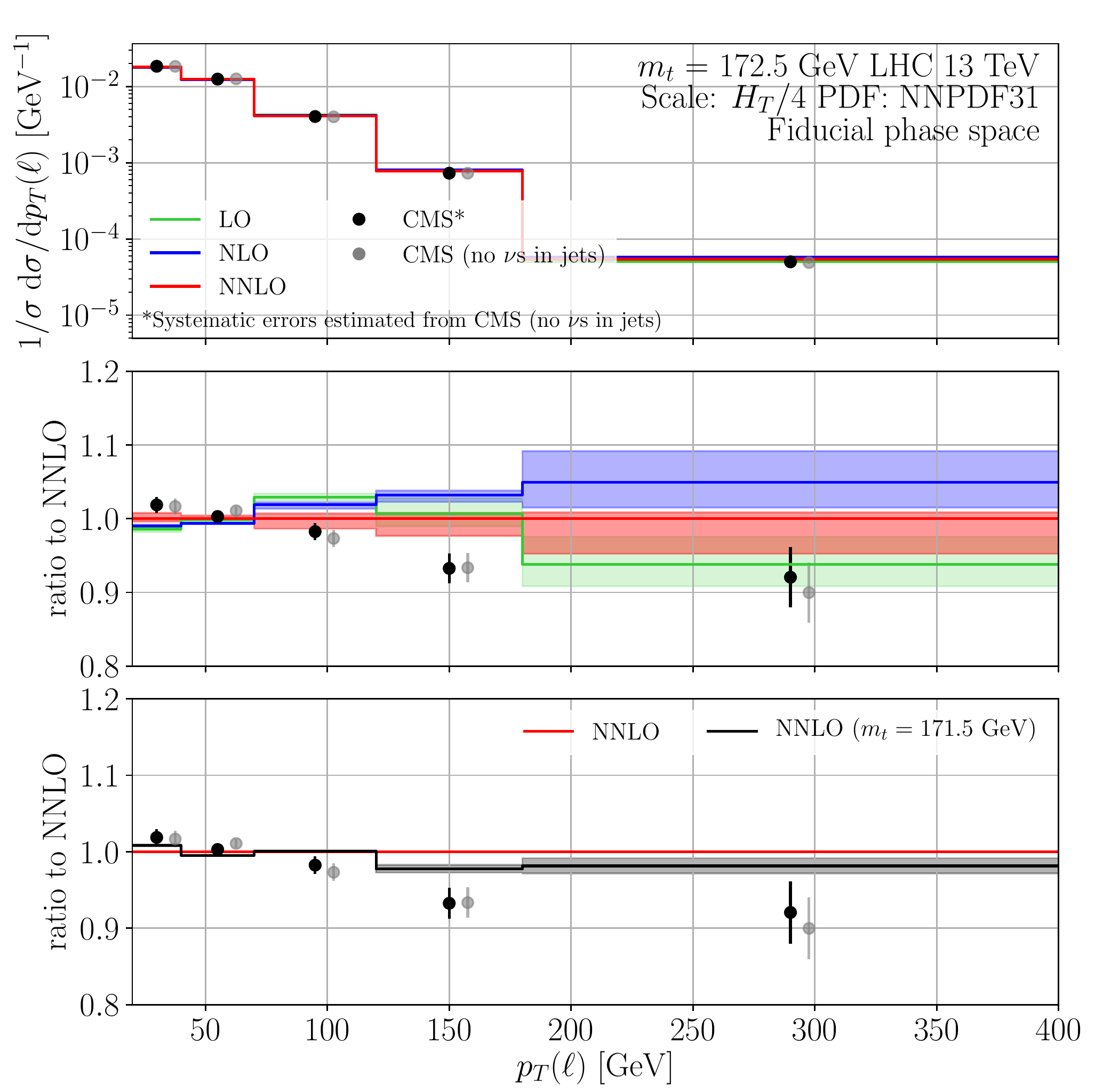}
\includegraphics[width=7.4cm]{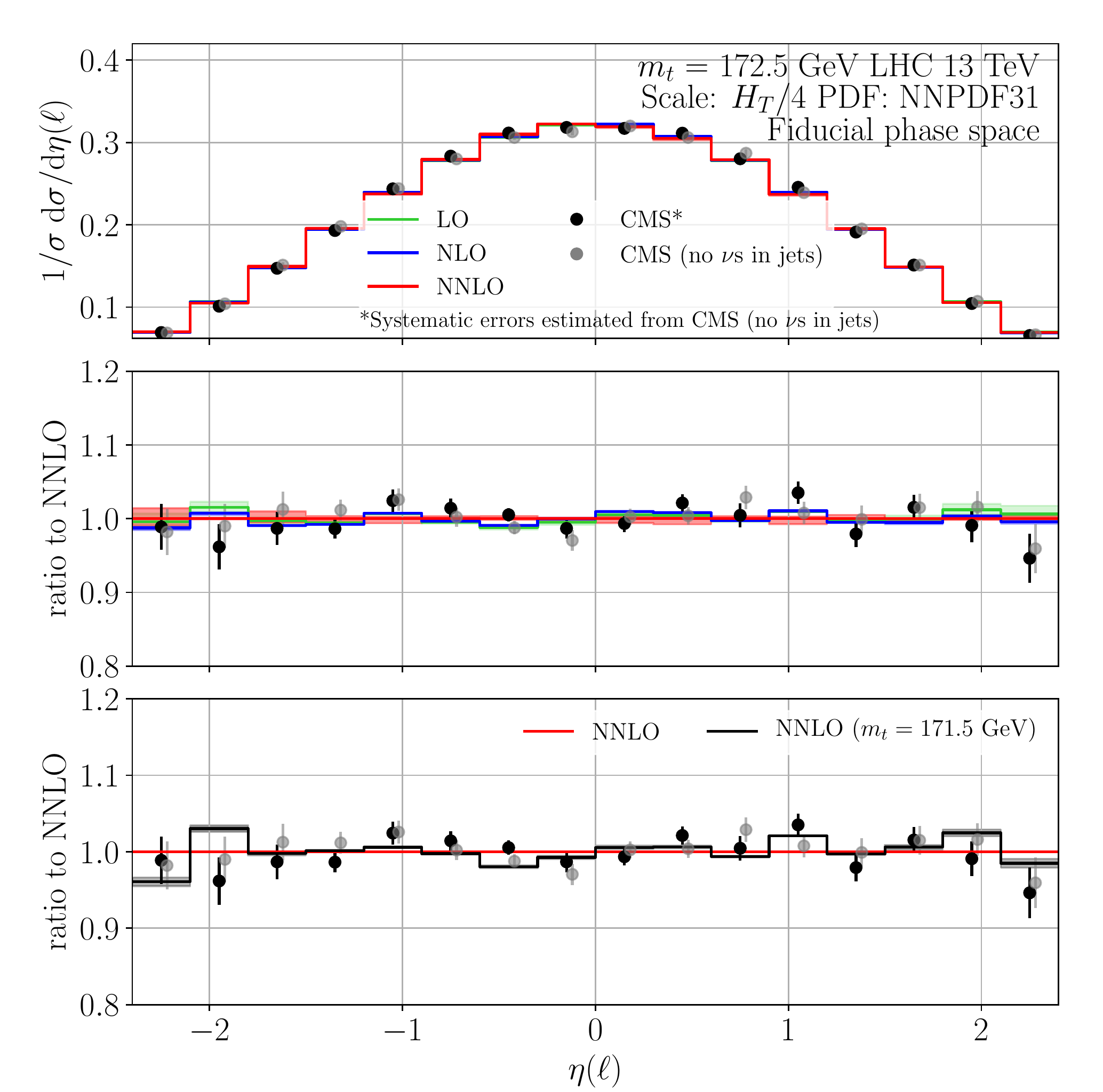}\\
\includegraphics[width=7.4cm]{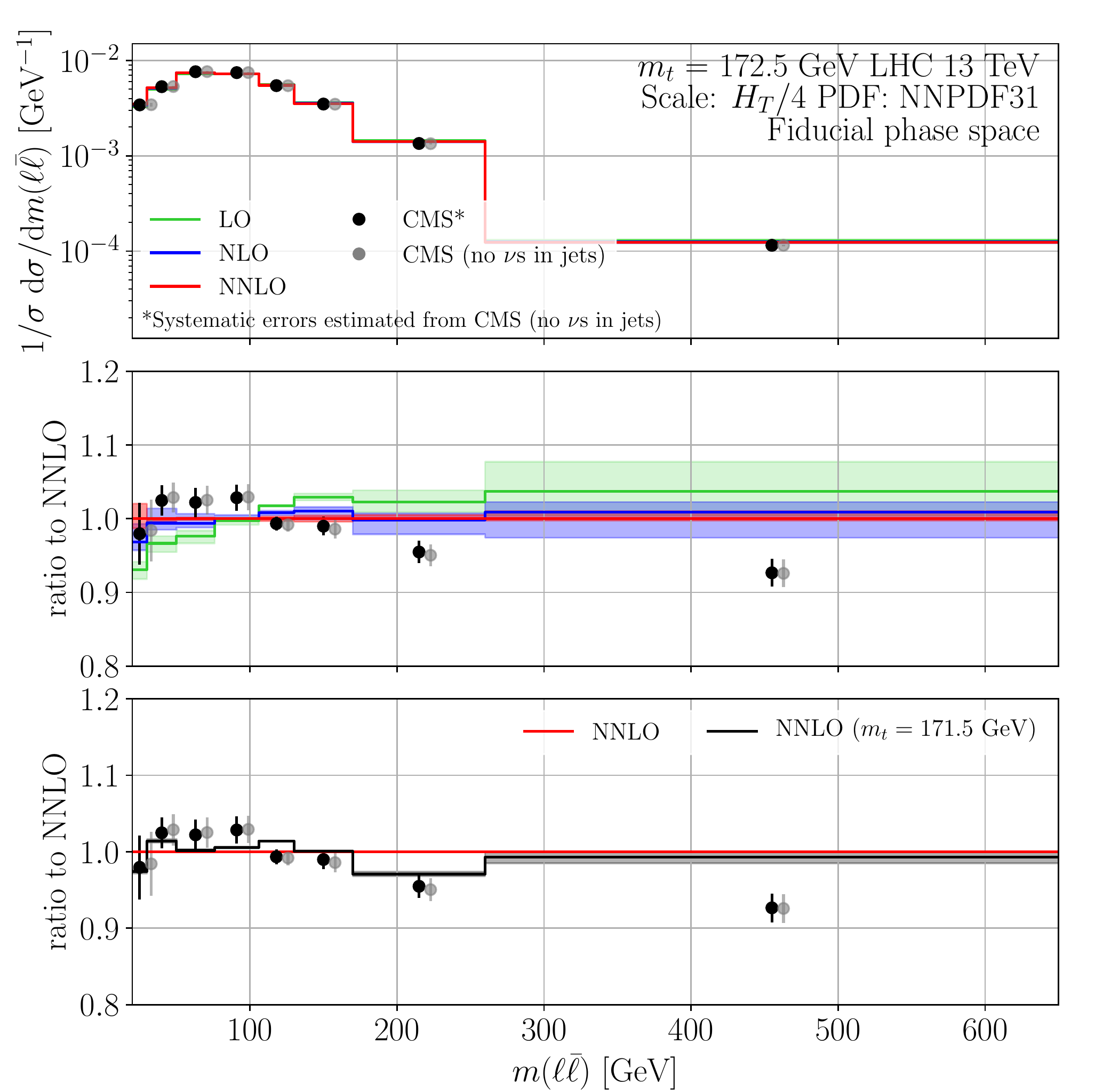}
\includegraphics[width=7.4cm]{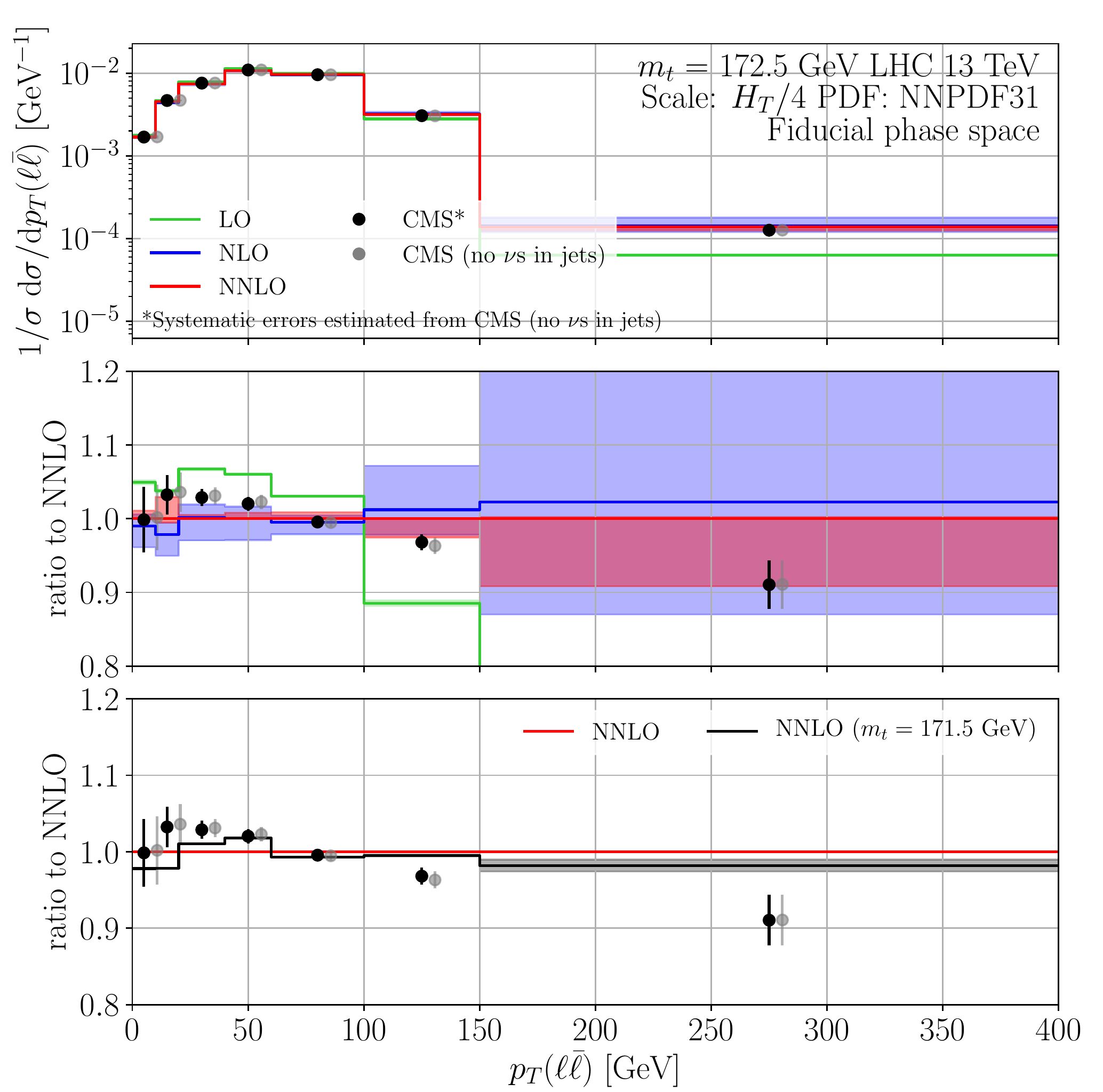}\\
\includegraphics[width=7.4cm]{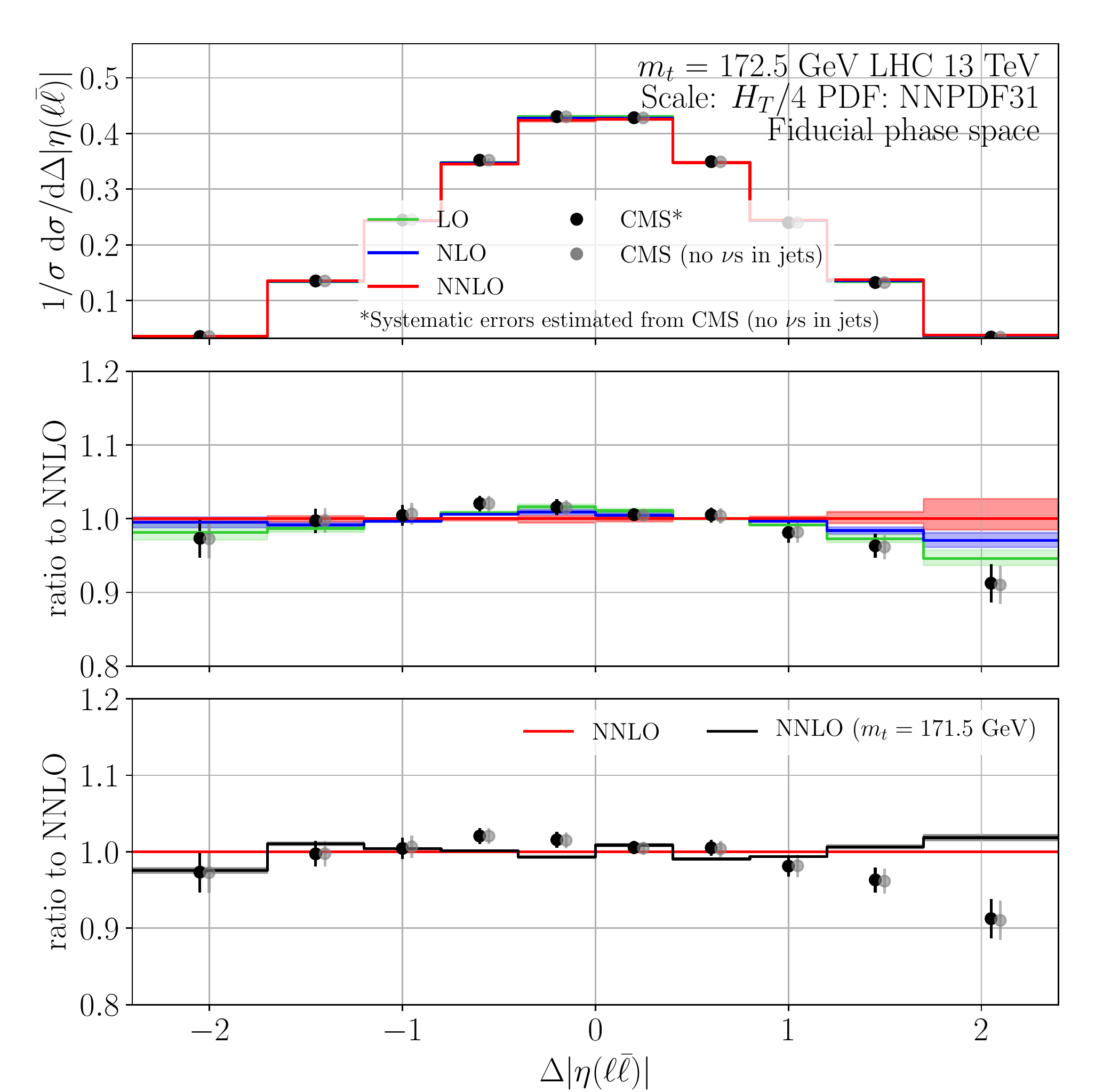}
\includegraphics[width=7.4cm]{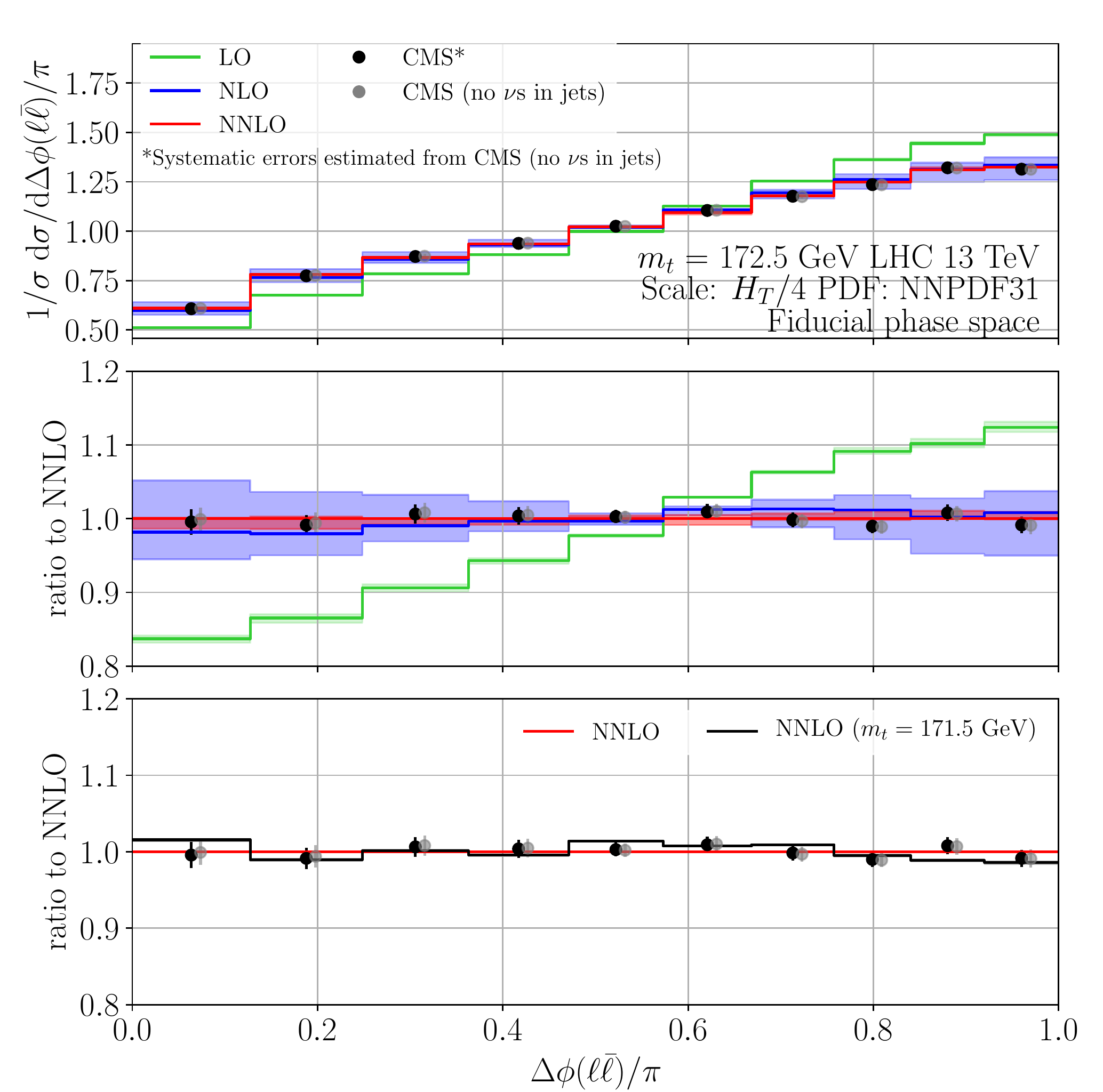}
\caption{As in fig.~\ref{fig:fid-lep-abs} but for the normalized leptonic distributions.}
\label{fig:fid-lep-norm}
\end{figure}

The lepton distributions specified in sec.~\ref{sec:pheno_fid} are shown in figs.~\ref{fig:fid-lep-abs},\ref{fig:fid-lep-norm}. A number of interesting features can be observed. The effect of the inclusion of neutrinos from semileptonic $B$-decays has significant impact on the data. For the absolute leptonic distributions discussed here the effect is mostly on the overall normalization with only a very small shape effect. While it may be counterintuitive at first, the reason leptonic distributions are affected by the definition of the $b$-jets is that a change in the $b$-jet $p_T$ affects the acceptance for jets and thus for the whole event. The effect on normalized distributions is marginal at most and is only noticeable for the $\eta(\ell)$ distribution. In a clear contrast, the data including neutrinos agrees much better with the NNLO predictions than the data excluding them.

The pattern of higher-order corrections in these fiducial distributions shares some features with the inclusive distributions discussed previously. It also shows some differences. Overall, the inclusion of higher order corrections leads to a substantial decrease in the scale variation of the theory predictions. Furthermore, the NNLO/NLO K-factor is substantially smaller than the NLO/LO one indicating good perturbative convergence of the scale (\ref{eq:scale}) also for the fiducial distributions. Unlike the inclusive case, however, for the fiducial distributions with this particular selection we observe a negative NNLO K-factor which is not atypical for fiducial distributions. 

The pdf uncertainty at NNLO is typically much smaller than the scale one. Only in few bins the two uncertainties are comparable. This implies that for the current generation of pdf sets the scale variation, alone, is a good indicator for the overall theoretical uncertainty at NNLO for fiducial distribution with non-extreme kinematics. This pattern applies to both absolute and normalized distributions; one exception appears in the rightmost bin of the $\Delta|\eta(\ell\bar\ell)|$ distribution where the scale NNLO variation appears smaller than the NLO one. This is likely not indicative of a departure from the pattern discussed above but is rather due to the larger MC error at NNLO which affects the scale variation error
\footnote{The two are not statistically independent and in cases where scale variation reduction occurs, as in normalized distribution, the effect of the MC error becomes more pronounced.}.

A 1 GeV change in $m_t$ shifts the spectra by an amount that is smaller, or at most comparable, to the size of the NNLO scale uncertainty band. This is the case for both the absolute and normalized distributions. The impact of the change in $m_t$ is typical: the dimensional distributions are affected towards small $p_T$ or $m(\ell\bar\ell)$ while the angular/rapidity distributions show less of a change in their shape than in their overall normalization. 

Overall, the agreement between NNLO QCD and data is quite good for the purely leptonic distributions considered here. A distribution that deserves particular attention is $\Delta\phi(\ell\bar{\ell})$ since it has received a lot of attention recently in the context of $t\bar t$ spin correlations. As can be seen from fig.~\ref{fig:fid-lep-norm} the agreement between NNLO QCD and data is quite impressive. This agrees with the findings of our prior work \cite{Behring:2019iiv} where we concluded that the fiducial prediction agrees with the fiducial ATLAS measurement \cite{ATLAS:2018rgl}. Such fiducial-level agreement is even more impressive given the CMS and ATLAS fiducial measurements are quite different in terms of event selection and phase-space extrapolation. We would like to stress that the agreement of the NNLO predictions with data at the fiducial level is unaffected by the issues related to $b$-jet definition discussed above and by the value of $m_t$, as can also be seen from fig.~\ref{fig:fid-lep-norm}. The picture that emerges from these two different fiducial-level comparisons is consistent as is the picture arising from the two (ATLAS and CMS) inclusive analyses. Based on this we think we have found arguments that further support our finding in ref.~\cite{Behring:2019iiv} that the $\Delta\phi(\ell\bar{\ell})$ distribution is well described by NNLO fixed order calculations in the fiducial volume while some disagreement remains at the inclusive level. Such a conclusion seems to be fairly robust and seems to suggest that LHC data and NNLO fixed-order calculations are already so precise that they are starting to be sensitive to secondary modeling effects within the LHC data.

\subsubsection{Analysis of the $b$-jet-only distributions}\label{sec:b-jets}

\begin{figure}
\includegraphics[width=7.4cm]{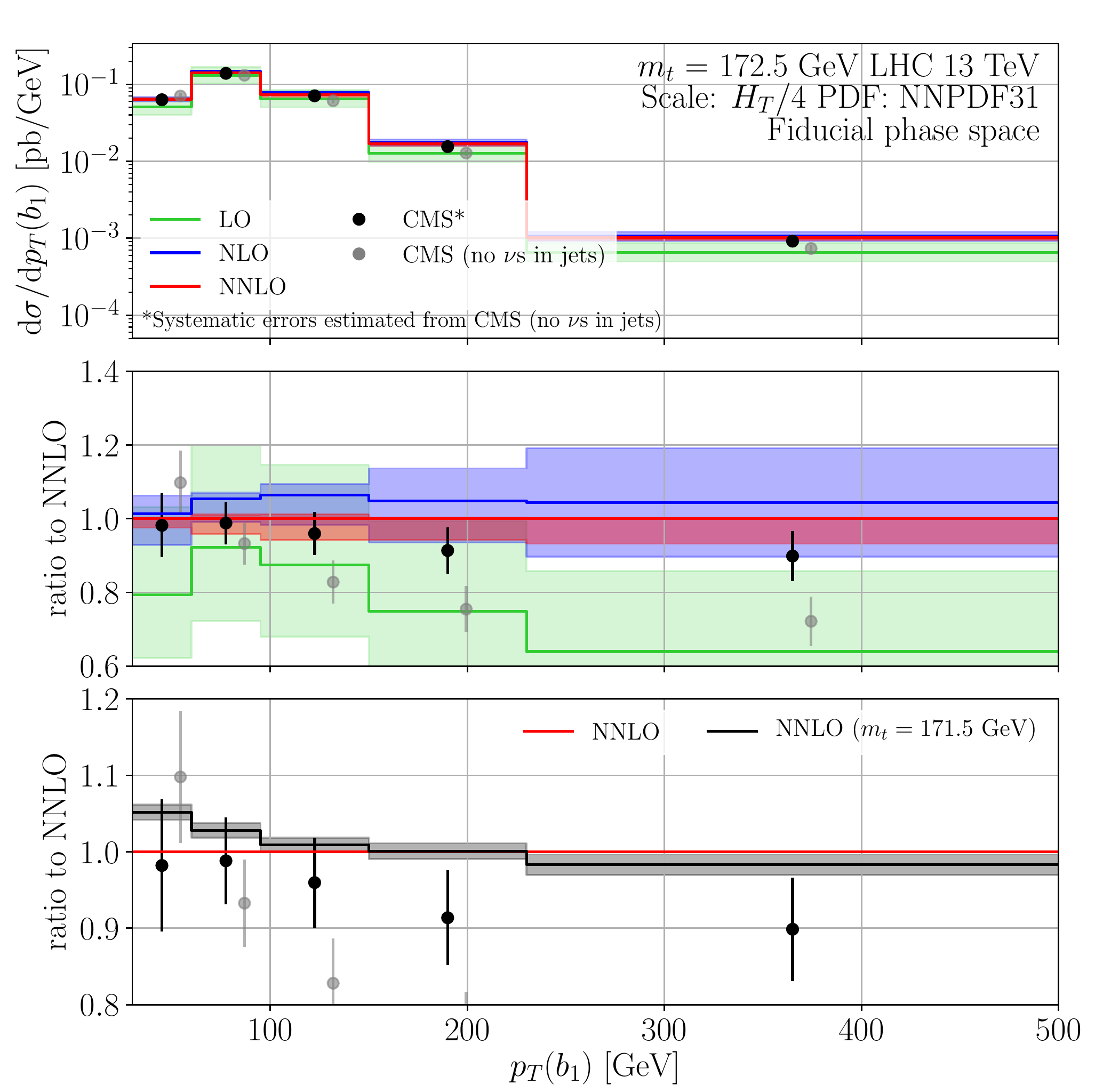}
\includegraphics[width=7.4cm]{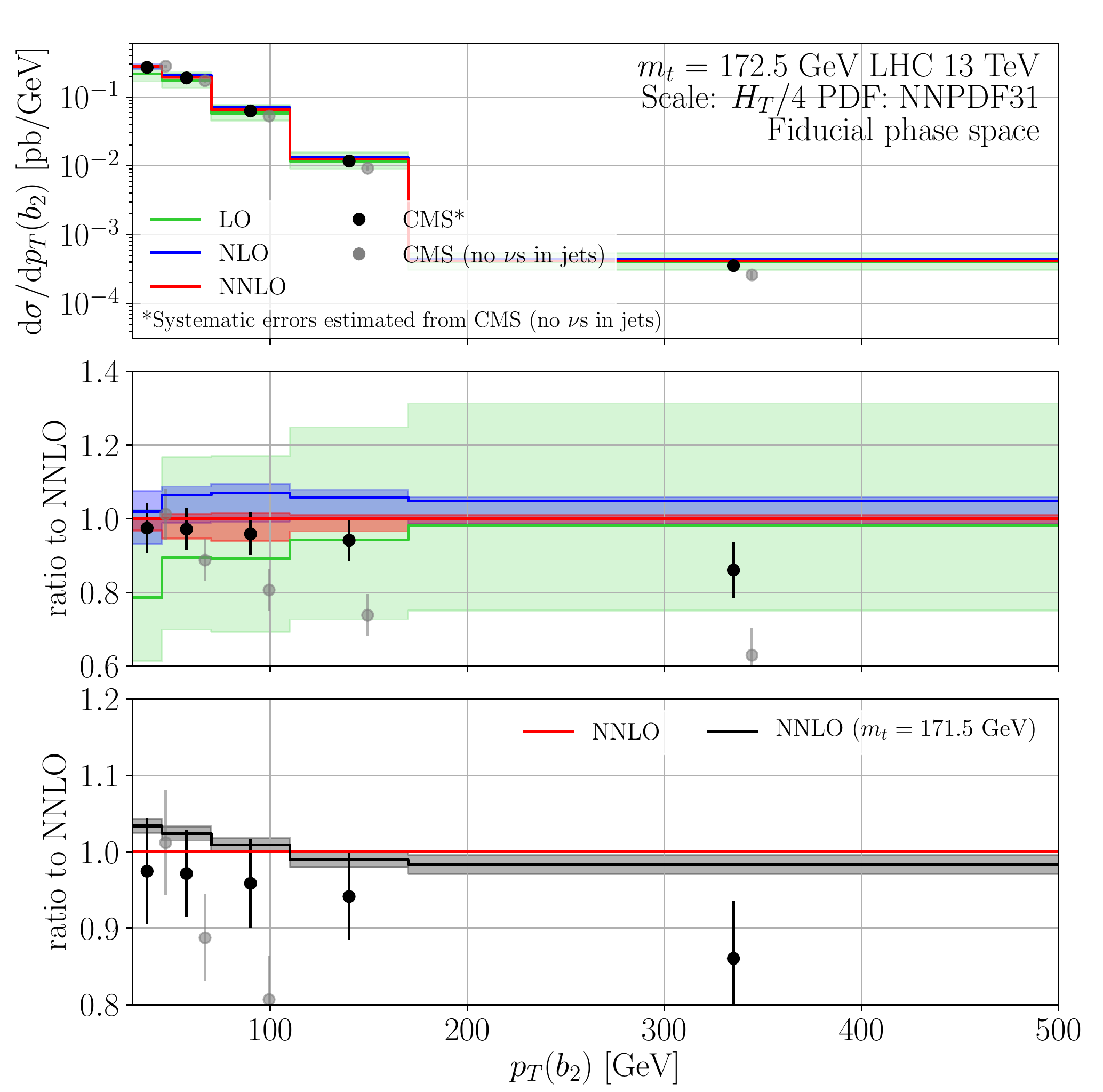}\\
\includegraphics[width=7.4cm]{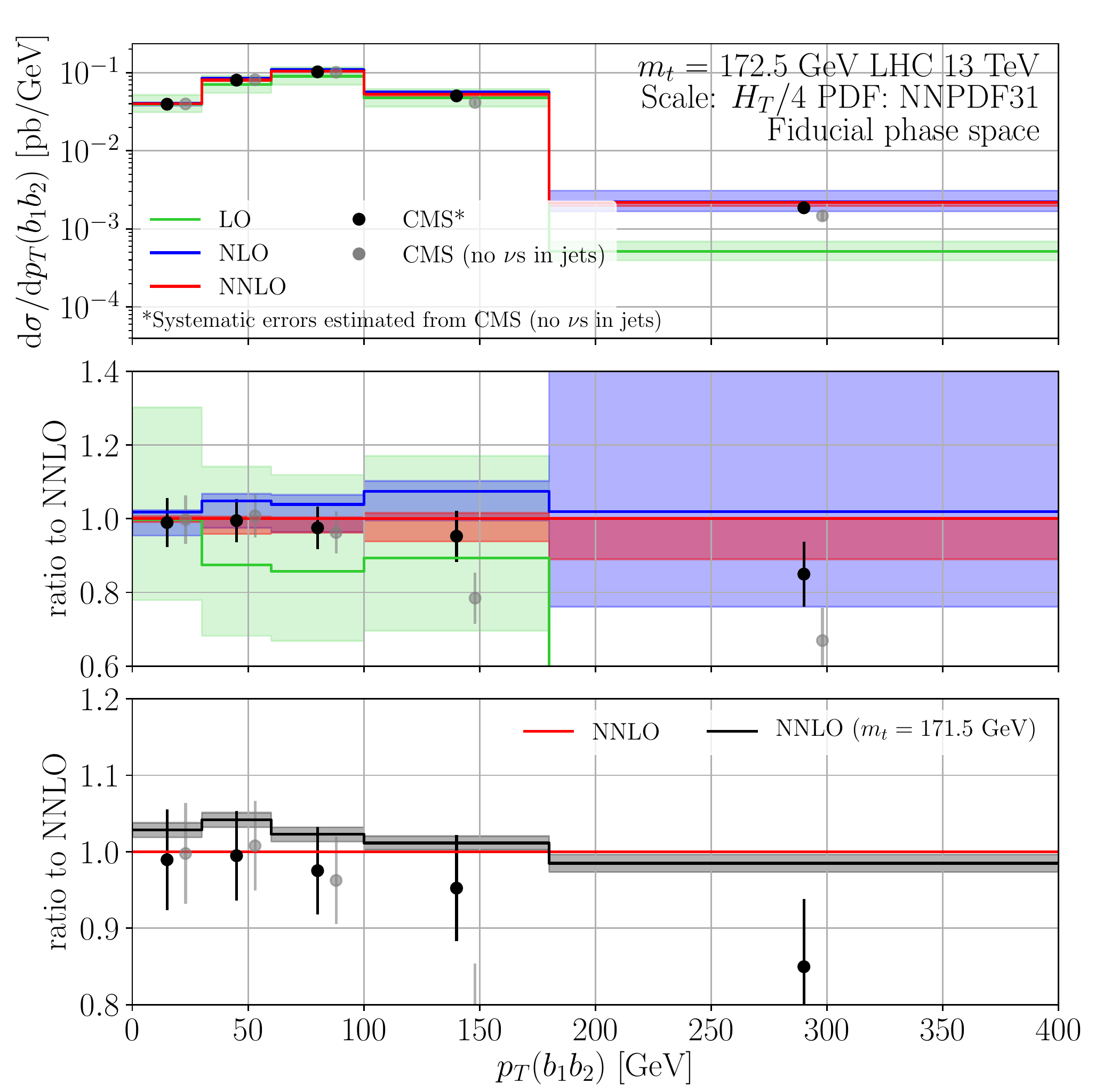}
\includegraphics[width=7.4cm]{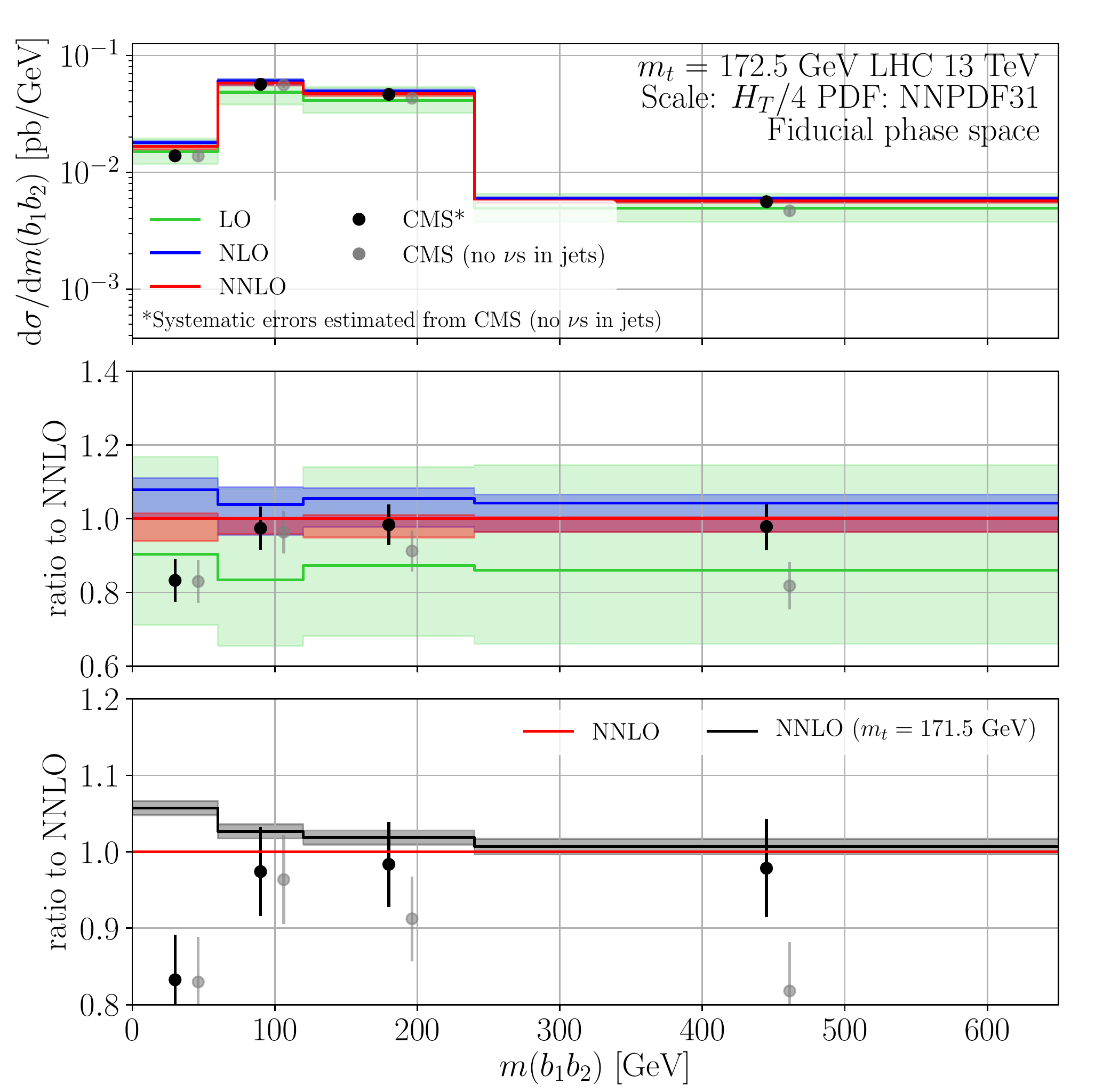}\\
\includegraphics[width=7.4cm]{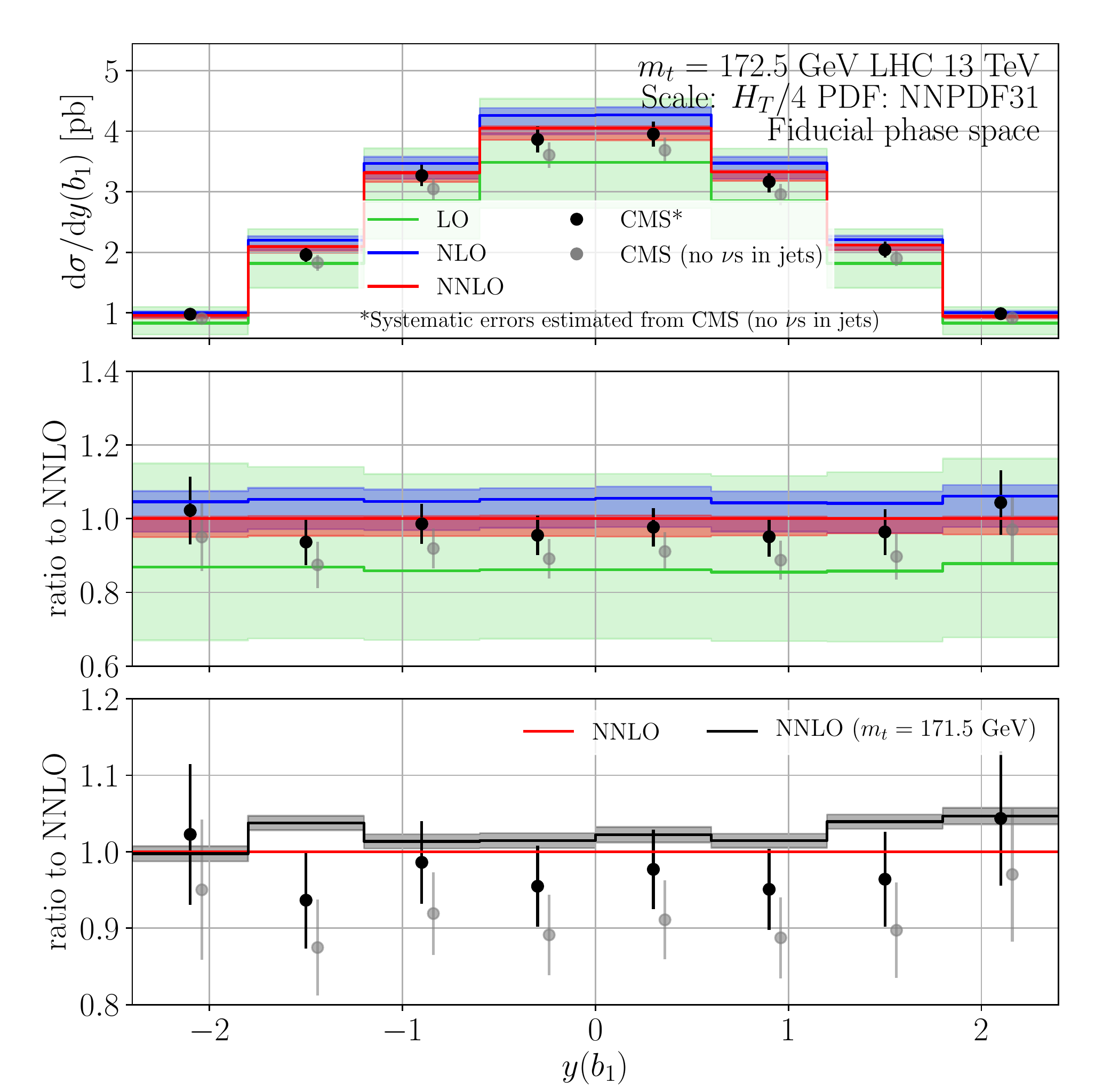}
\includegraphics[width=7.4cm]{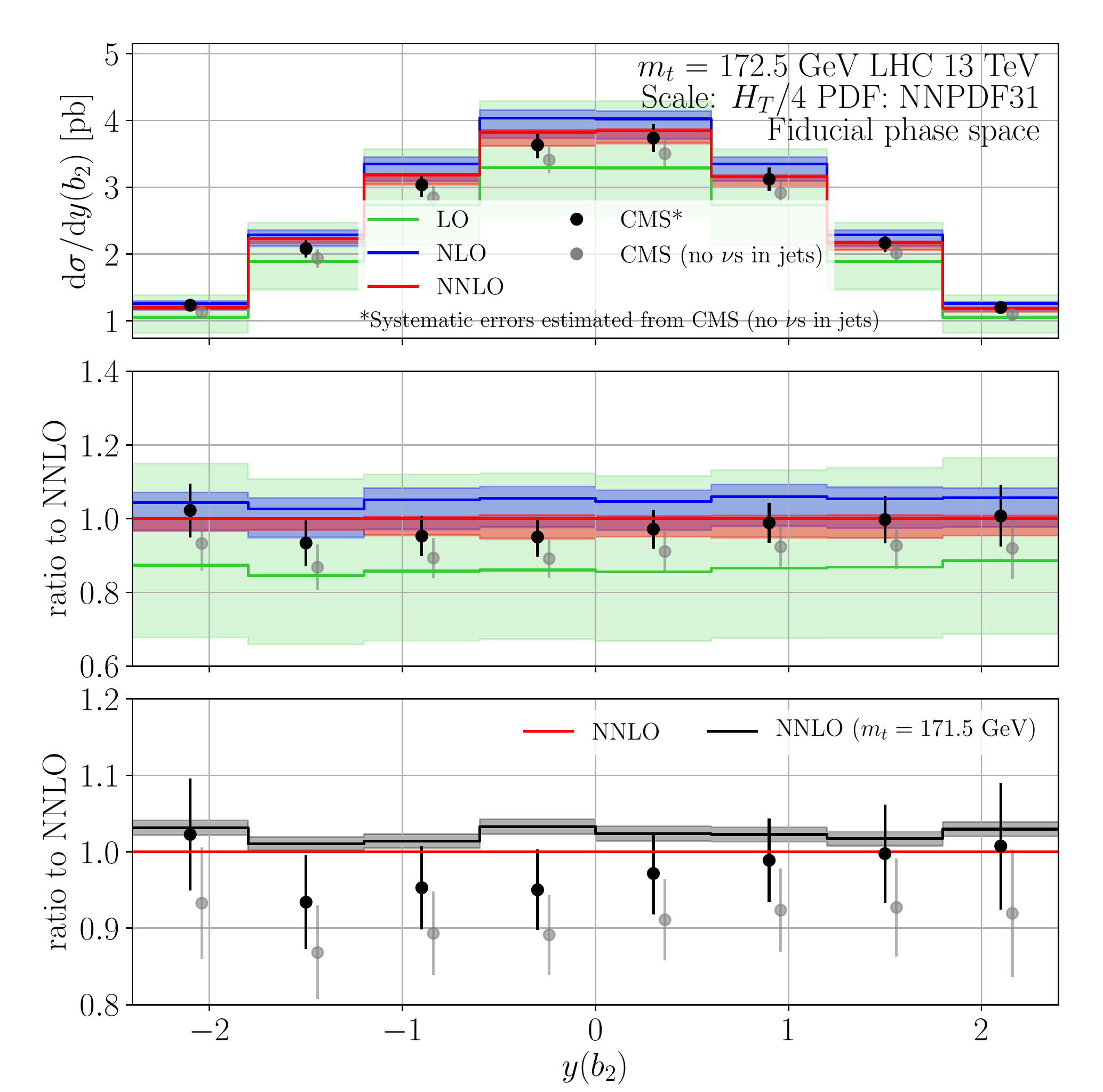}
\caption{As in fig.~\ref{fig:fid-lep-abs} but for the absolute $b$-jet distributions.}
\label{fig:fid-bjet-abs}
\end{figure}
\begin{figure}
\includegraphics[width=7.4cm]{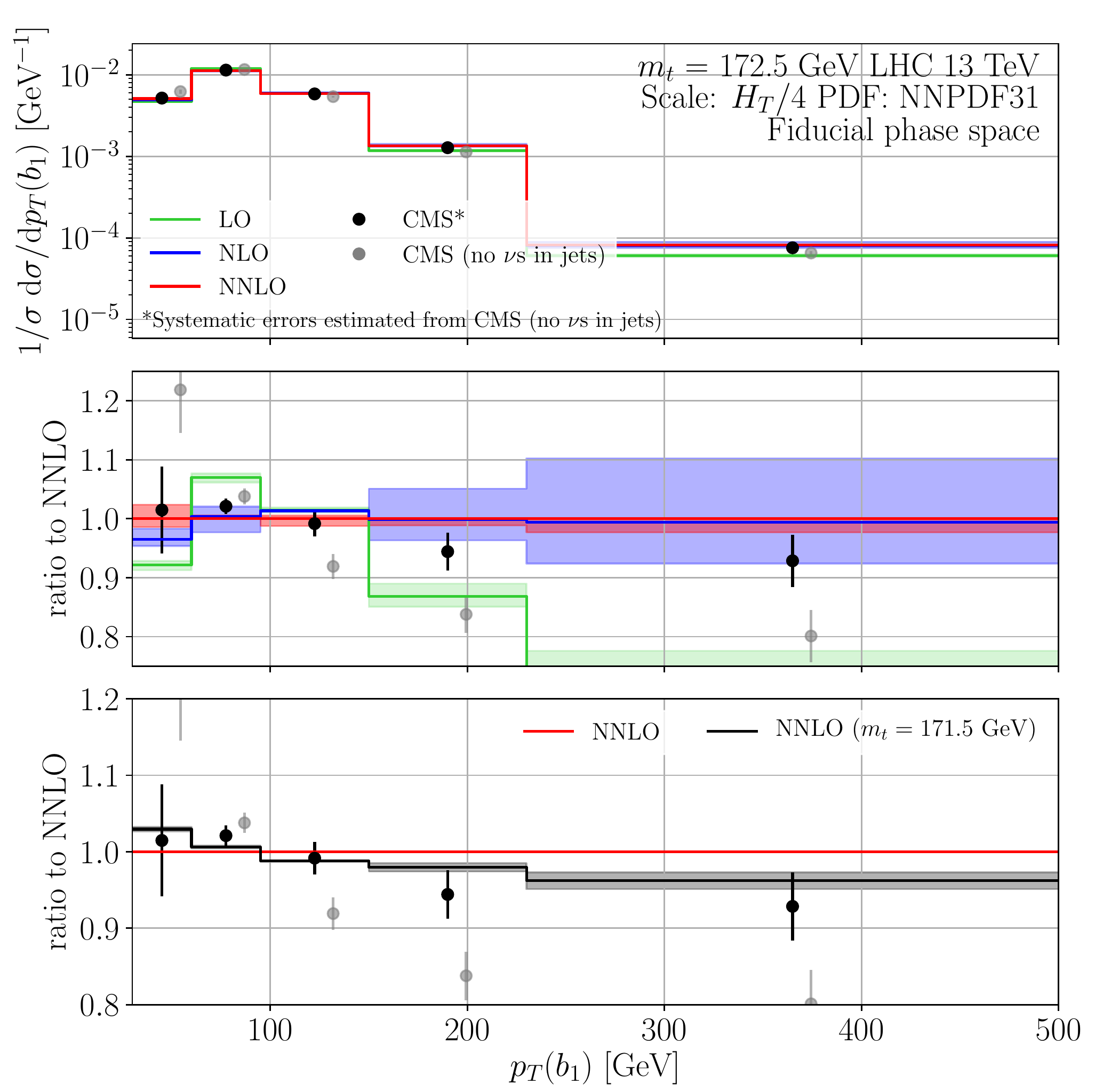}
\includegraphics[width=7.4cm]{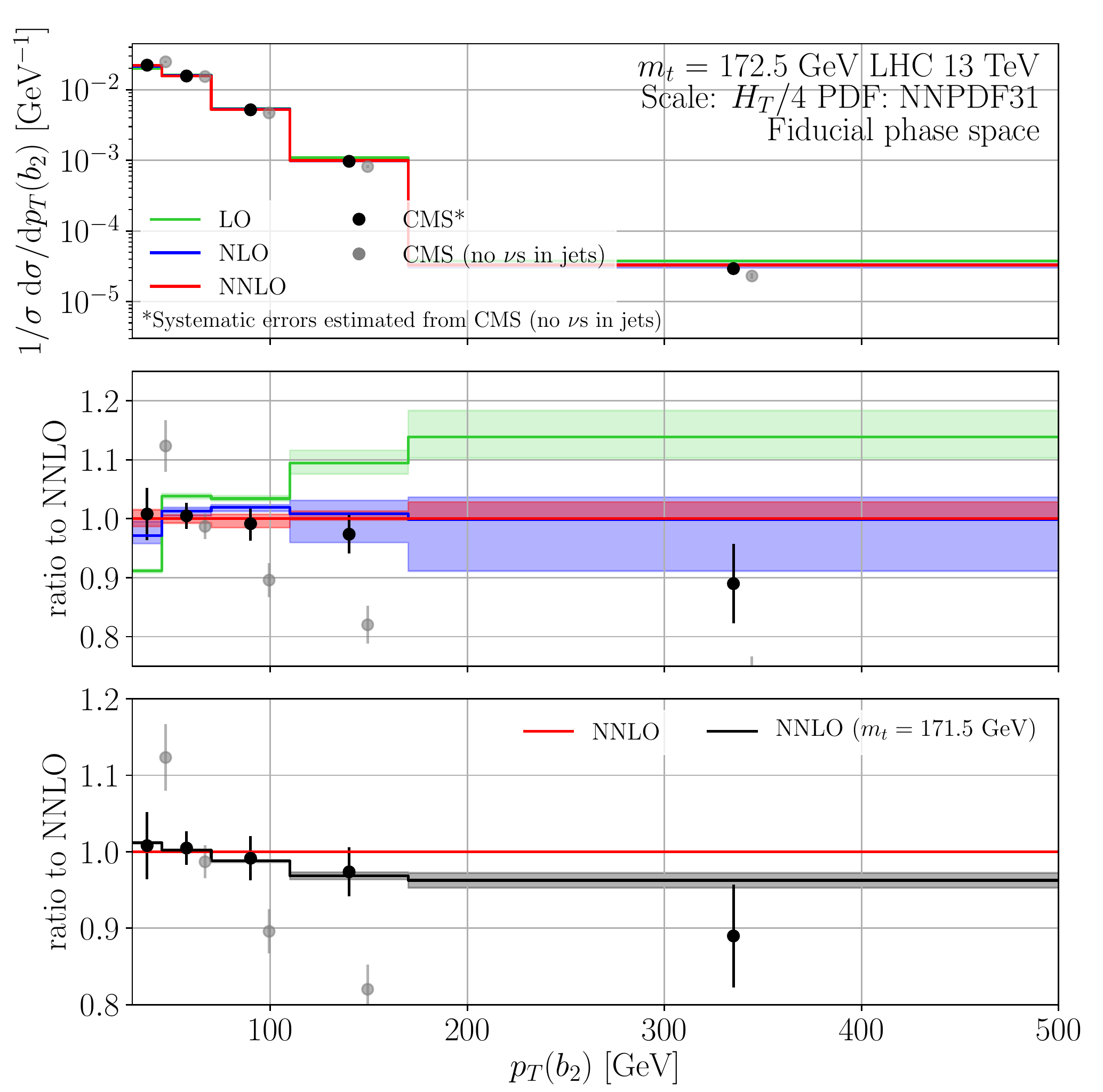}\\
\includegraphics[width=7.4cm]{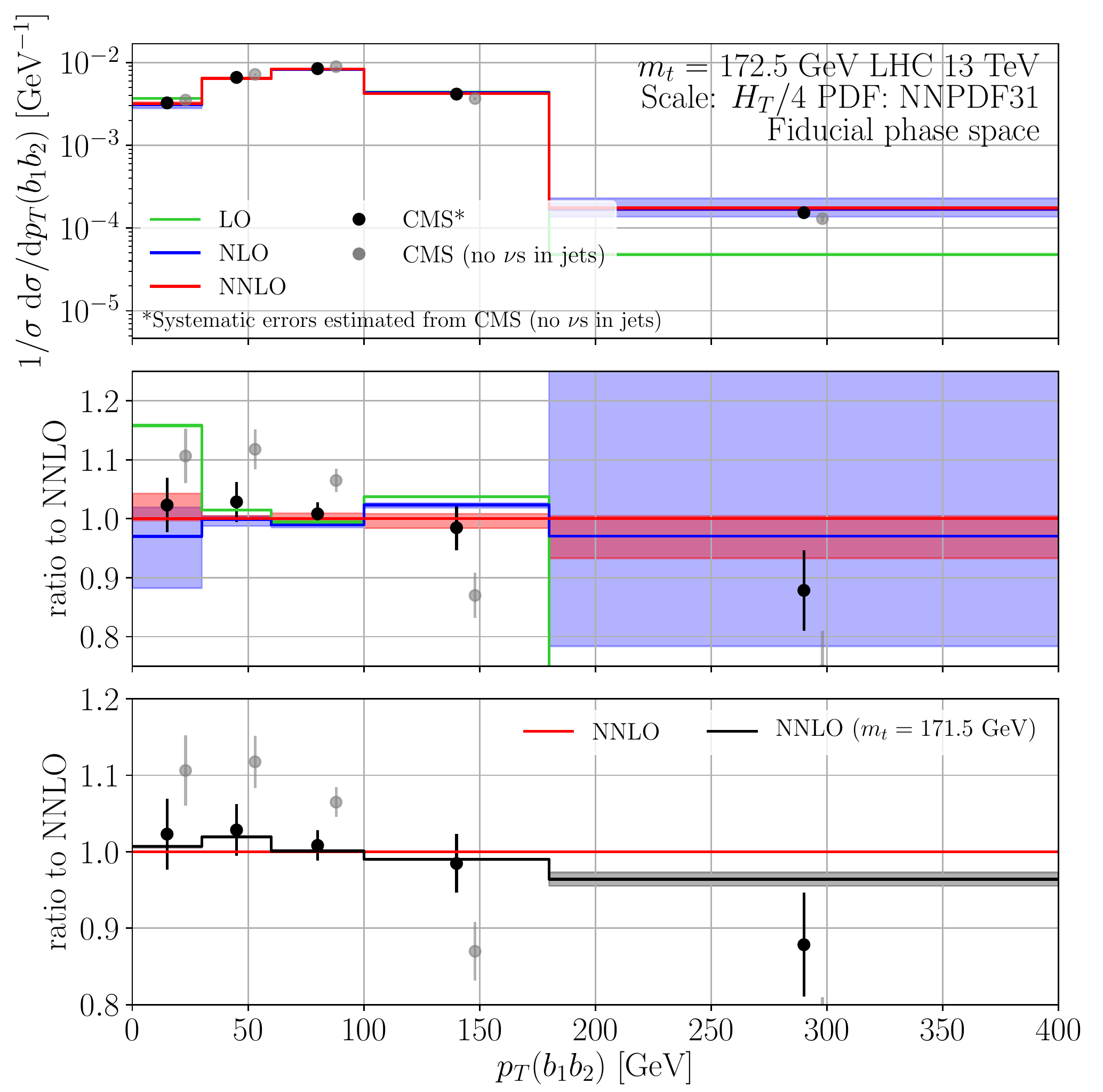}
\includegraphics[width=7.4cm]{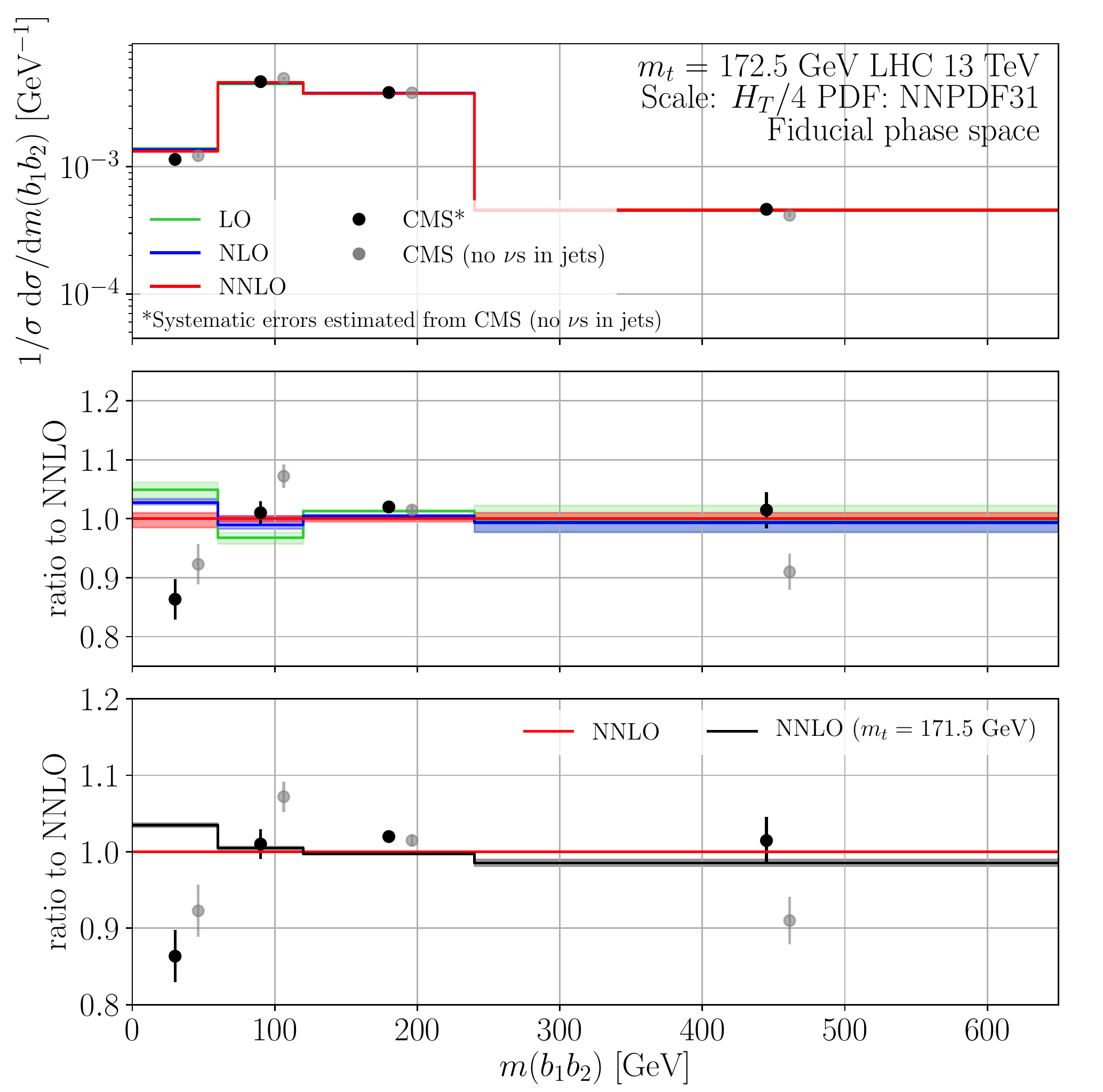}\\
\includegraphics[width=7.4cm]{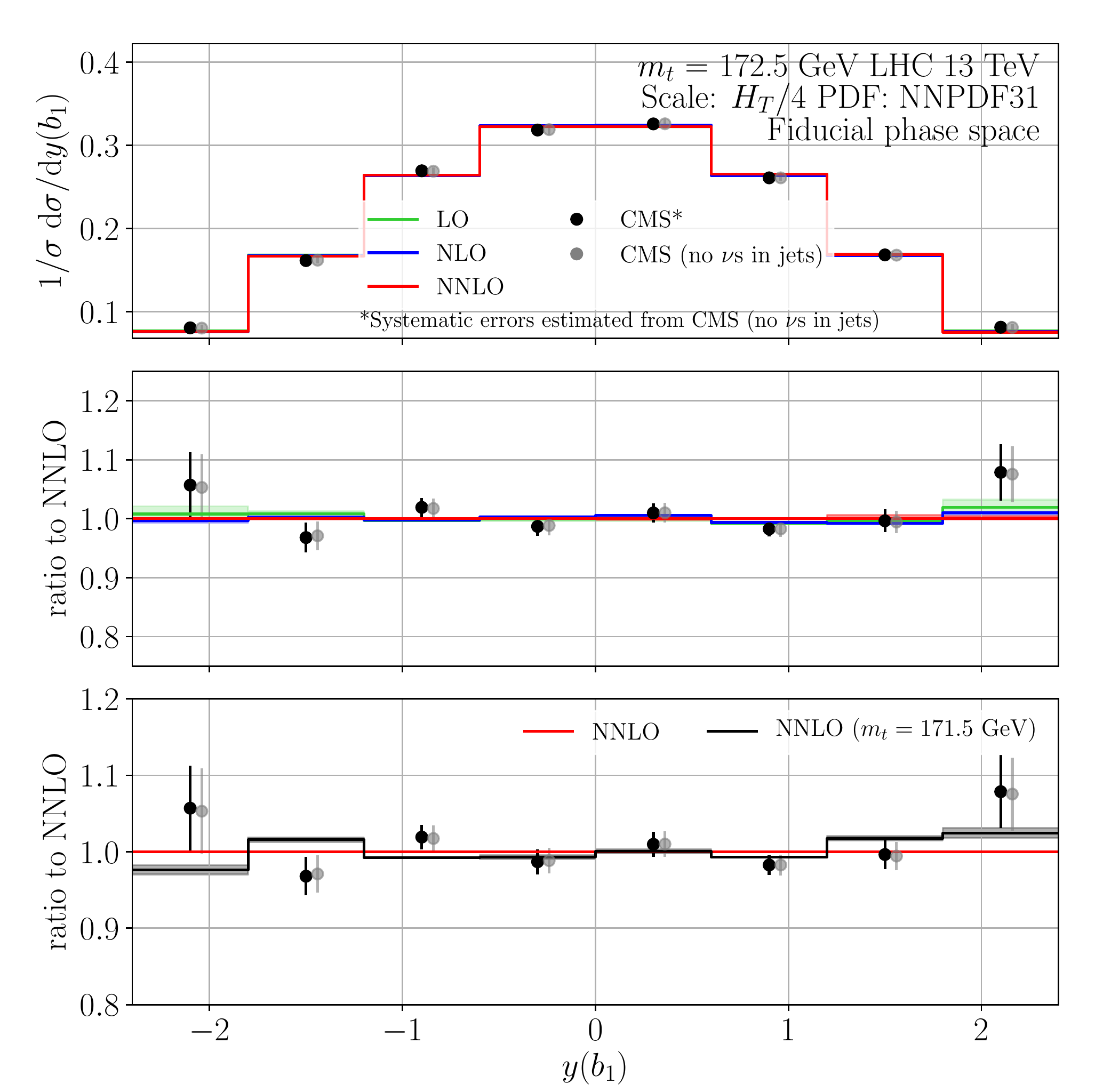}
\includegraphics[width=7.4cm]{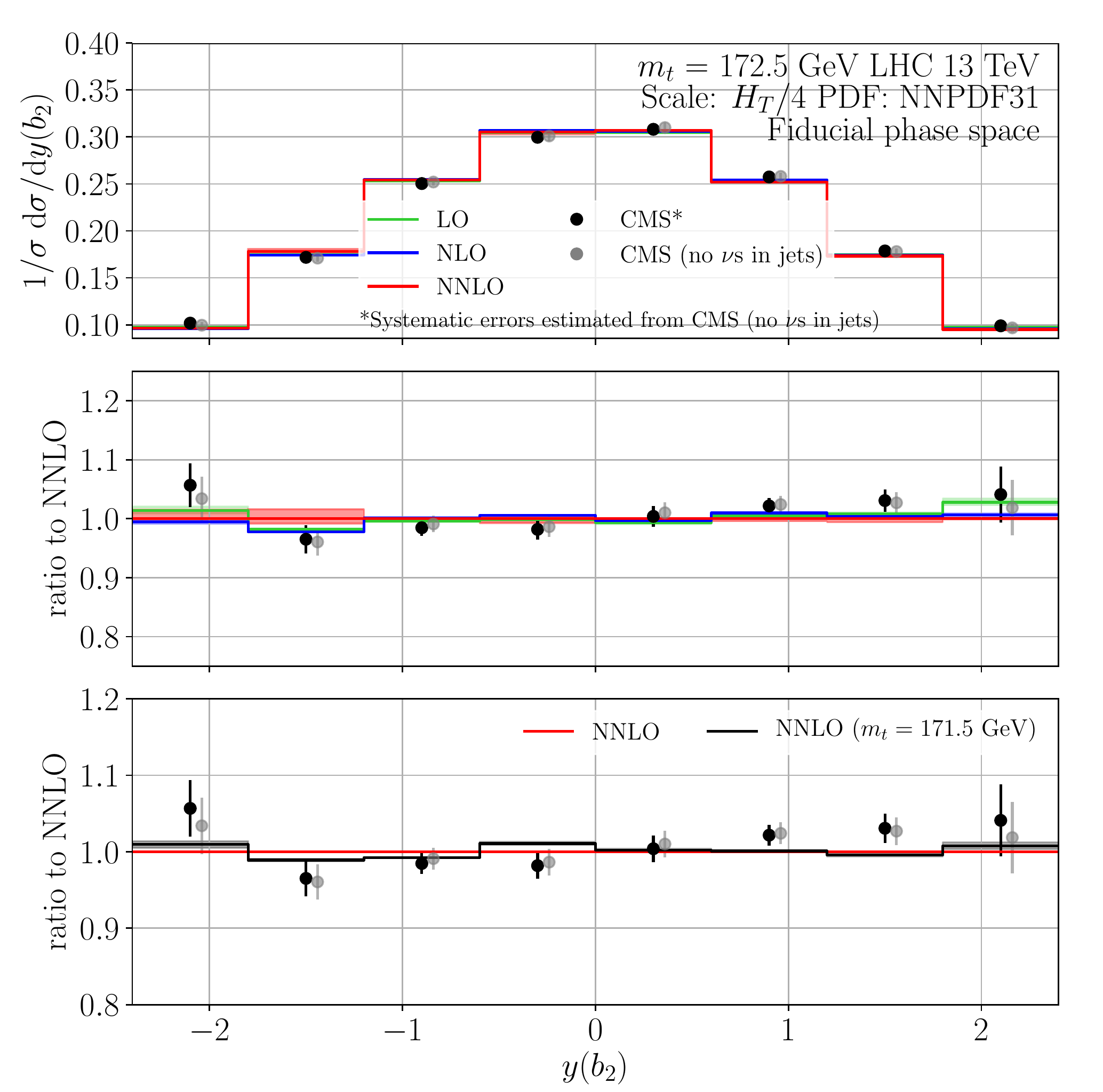}
\caption{As in fig.~\ref{fig:fid-lep-abs} but for the normalized $b$-jet distributions.}
\label{fig:fid-bjet-norm}
\end{figure}

The absolute and normalized $b$-jet distributions are shown in figs.~\ref{fig:fid-bjet-abs},\ref{fig:fid-bjet-norm}. As expected, the difference between the two data sets - the ones with and without neutrinos from semileptonic decays - is larger in the $b$-jet distributions than in the leptonic distributions. In addition to the overall normalization, the shape of the distributions is also strongly affected. With the exception of the first bin of the $m(b_1 b_2)$ distribution the data including neutrinos (black bars) agrees well with NNLO theory, unlike the data excluding neutrinos which clearly disagrees with NNLO QCD. 

What we mean by $b_1$ and $b_2$ is non-trivial and we refer to sec.~\ref{sec:tops} for a detailed explanation. In short, these are the two $b$-jets used in the reconstruction of the top quark and antiquark (there could be up to four $b$-jets in an event) and $b_1$ is the hardest among these two $b$-jets.

We have investigated the first bin of the $m(b_1 b_2)$ distribution; the unusually large data-theory discrepancy can be understood as a kinematics effect which is pronounced in fixed-order perturbation theory
\footnote{We thank Malgorzata Worek for an illuminating discussion on this point.}.
The $m(b_1 b_2)$ distribution cannot reach the point $m(b_1 b_2)=0$ since a) jets are massive and b) the two $b$-jets are never collinear to each other due to the jet clustering requirement. The minimal value $m(b_1 b_2)$ can take, however, depends on the order in perturbation theory, and whether $b$-quarks are treated as massive or massless. With respect to the $b$-mass it is clear that the mass of a realistic $b$-jet cannot be less than about 5 GeV which is about the mass of the $B$-meson resulting from the $b$-quark fragmentation. Since each $b$-jet has net bottomness the two jets should have a minimum invariant mass of about 10 GeV, irrespective of their kinematics. In our calculation $b$-quarks are massless and therefore the jet mass can be as low as zero. Separately from the issue of parton masses, the jet mass is given in terms of the invariant mass of all constituent partons. For realistic jets the jet mass is proportional to the jet $p_T$ (times a slowly varying function of $p_T$). However in fixed order perturbation theory the jet mass reaches this value more slowly. For example, at LO all partons are massless and therefore any jet will have zero mass at any $p_T$. At NLO at most one $b$-jet can have non-zero mass while NNLO is the first order in perturbation theory where both $b$-jets can have non-zero mass. While the above effects are small or negligible for a typical jet $p_T$ distribution with a typical jet $p_T$ cut they may have an outsized impact on the position of the end-point of the $m(b_1 b_2)$ distribution and on its behavior close to this end-point. The main consequence of the above effects is that at LO the $m(b_1 b_2)$ spectrum has a minimum value which is the lowest possible and this minimum value increase at higher orders of perturbation theory or when the $b$-quark is considered massive. As a result, smaller proportion of events will contribute to the first bin at higher perturbative orders. The effect on the absolute distribution is harder to predict because the relative depletion of the first bin can be compensated by the potential rise at higher orders of the number of events as a whole. The relative impact should be easier to see in the normalized distribution.

The above picture is consistent with the pattern of scale variation and higher-order corrections through NNLO exhibited by the first $m(b_1 b_2)$ bin as can be best observed in the normalized distribution shown in fig.~\ref{fig:fid-bjet-norm}. Indeed, once the effect of the overall normalization has been removed, we see that at higher orders the QCD corrections systematically decrease the predicted value in this bin while the scale variation is not decreasing as much as for the rest of the $m(b_1 b_2)$ bins. The interpretation that the first $m(b_1 b_2)$ bin is affected by higher-order corrections due to modified kinematics is also consistent with what can be observed in figs.~204 and 221 of ref.~\cite{CMS-fiducial-mod}: the default MC event generator used by CMS for this analysis is able to describe this bin of the absolute distribution within uncertainties. The MC generator however does not describe well this bin of the normalized data (it undershoots it). This behavior is consistent with our picture presented above: due to its more complete description of the jet mass, the MC generator predicts smaller cross-section in the first $m(b_1 b_2)$ bin than our fixed order calculation. This results from MC generator's inclusion of multiple soft and/or collinear emissions and correctly including the kinematic effect of the nonzero $B$-meson mass.

To complete this discussion we would like to mention that a number of potential sources of discrepancy in the first $m(b_1 b_2)$ bin can be excluded. For example, the MC error of the theory calculation is fairly small in this bin and is unlikely to be the reason behind this discrepancy. Since our calculation is performed in the NWA approximation one may worry that NWA-breaking effects are affecting this bin. NLO calculations of $t\bar t$ production including all corrections beyond NWA exist \cite{Denner:2010jp,Bevilacqua:2010qb,Denner:2012yc,Frederix:2013gra,Heinrich:2017bqp,Jezo:2016ujg}, however, we have not been able to find setting similar to ours that has estimated the size of the beyond-NWA corrections. Similar setup exists \cite{Bevilacqua:2019quz} for the related \cite{Bevilacqua:2018dny} process $pp\to t\bar t\gamma$. Ref.~\cite{Bevilacqua:2019quz} has specifically studied the quality of the NWA approximation and has found no noticeable corrections to that approximation. For this reason we suspect that the quality of the NWA is not responsible for the discrepancy observed in the first $m(b_1 b_2)$ bin. From figs.~\ref{fig:fid-bjet-abs},\ref{fig:fid-bjet-norm} one can also conclude that the value of $m_t$ is unlikely to be responsible for this discrepancy. While the first bin of the $m(b_1 b_2)$ distribution is fairly sensitive to the value of $m_t$ the top quark mass needs to be at least 2 GeV larger than the default value 172.5 GeV in order to account for this discrepancy.

All other $b$-jet distributions are described quite well by NNLO QCD. In particular as can be clearly observed for the normalized distributions in fig.~\ref{fig:fid-bjet-norm} the theory-data comparison is sensitive to the value of $m_t$. As can be observed from this figure the agreement seems to be improved for $m_t=171.5$ GeV compared to $m_t=172.5$ GeV. Although we do not advocate here that this simple comparison should be considered as a measurement of the top quark mass, it is clear that these distributions are sensitive to the value of $m_t$ and may offer the possibility of measuring this parameter.

\subsubsection{Analysis of the top quark distributions}\label{sec:tops}

As promised in sec.~\ref{sec:pheno_fid}, we next turn our attention to top-quark-level observables with fiducial selection. Such a comparison will require us to address the question: {\it what is meant by top quarks when fiducial selection is imposed on their decay products}? There are two top-quark concepts one can apply: the first one is the so-called true top quark which is simply the intermediate top quark generated in our Monte Carlo program and then subsequently decayed. It directly corresponds to the top quark field in the SM Lagrangian. The true top quark is related to the fiducial volume as follows. If in a given event the decay products of a true top quark (or top antiquark, as appropriate) pass the fiducial selection requirements, then the true top quark itself is used in the binned distribution. 

The second concept is that of the so-called reconstructed top. The reconstructed top quark is merely a proxy for the true top quark; its introduction is necessitated by experimental realities: no top quark can be observed directly, and only its decay products are measurable. Typically, experimental analyses rely on predefined algorithms that produce, for a given final state, a four-momentum which is interpreted as a proxy for the top (anti-)quark. In this work, this proxy will be called reconstructed top. While the top reconstruction algorithms are designed to produce reconstructed top which is close to the true top, differences between the two are present. Quantifying the size of those differences is one of the main goals of this section. 

In this discussion we consider the following top-quark level distributions:
\begin{itemize}
      \item $p_T(t)$: transverse momentum of the top-quark,
      \item $y(t)$: rapidity of the top-quark,
      \item $m(t\bar{t})$: top-quark pair invariant mass,
      \item $p_T(t\bar{t})$: transverse momentum of the top-quark pair,
      \item $y(t\bar{t})$: top-quark pair rapidity.
\end{itemize}
\begin{figure}
\includegraphics[width=7.4cm]{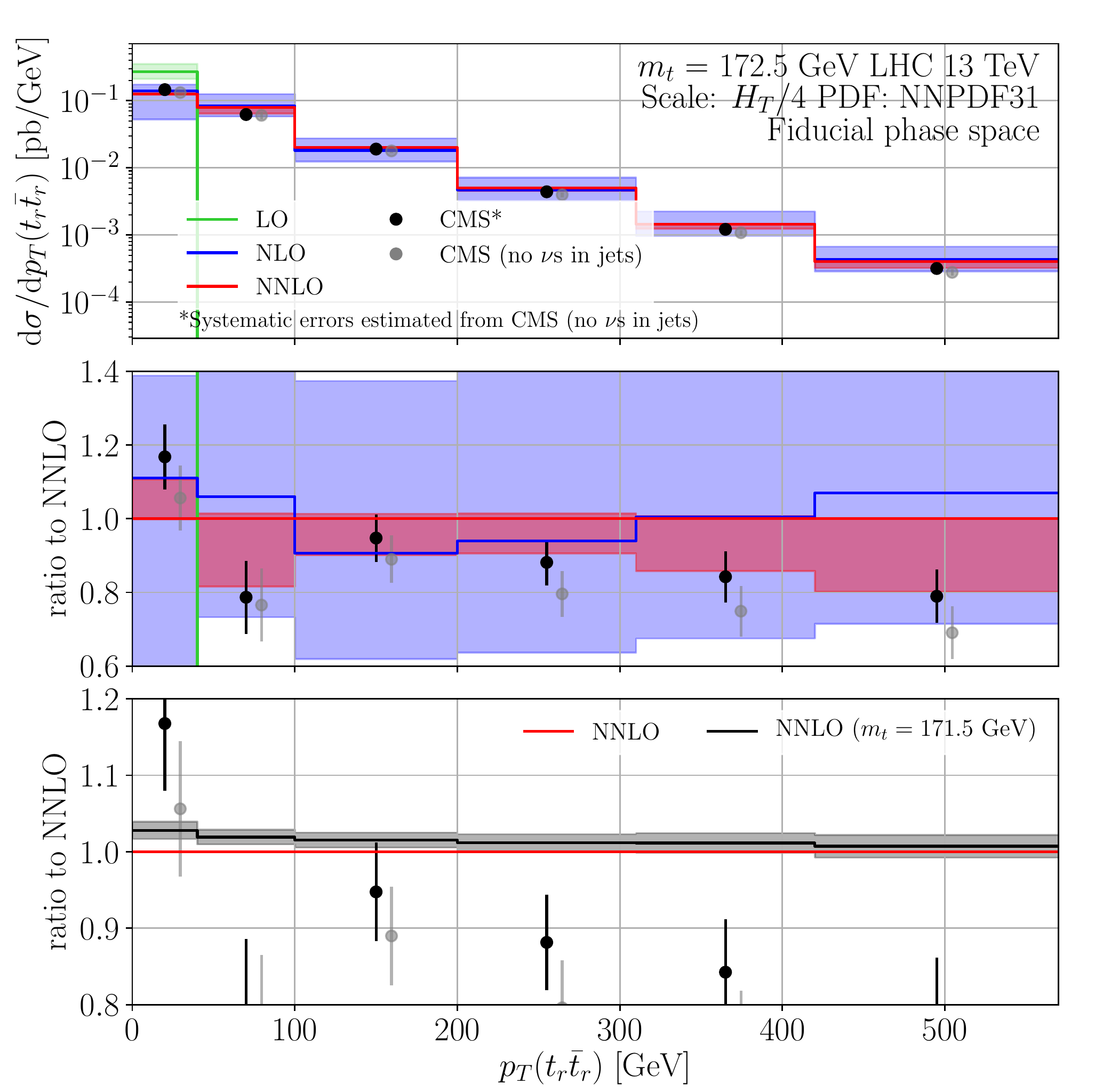}
\includegraphics[width=7.4cm]{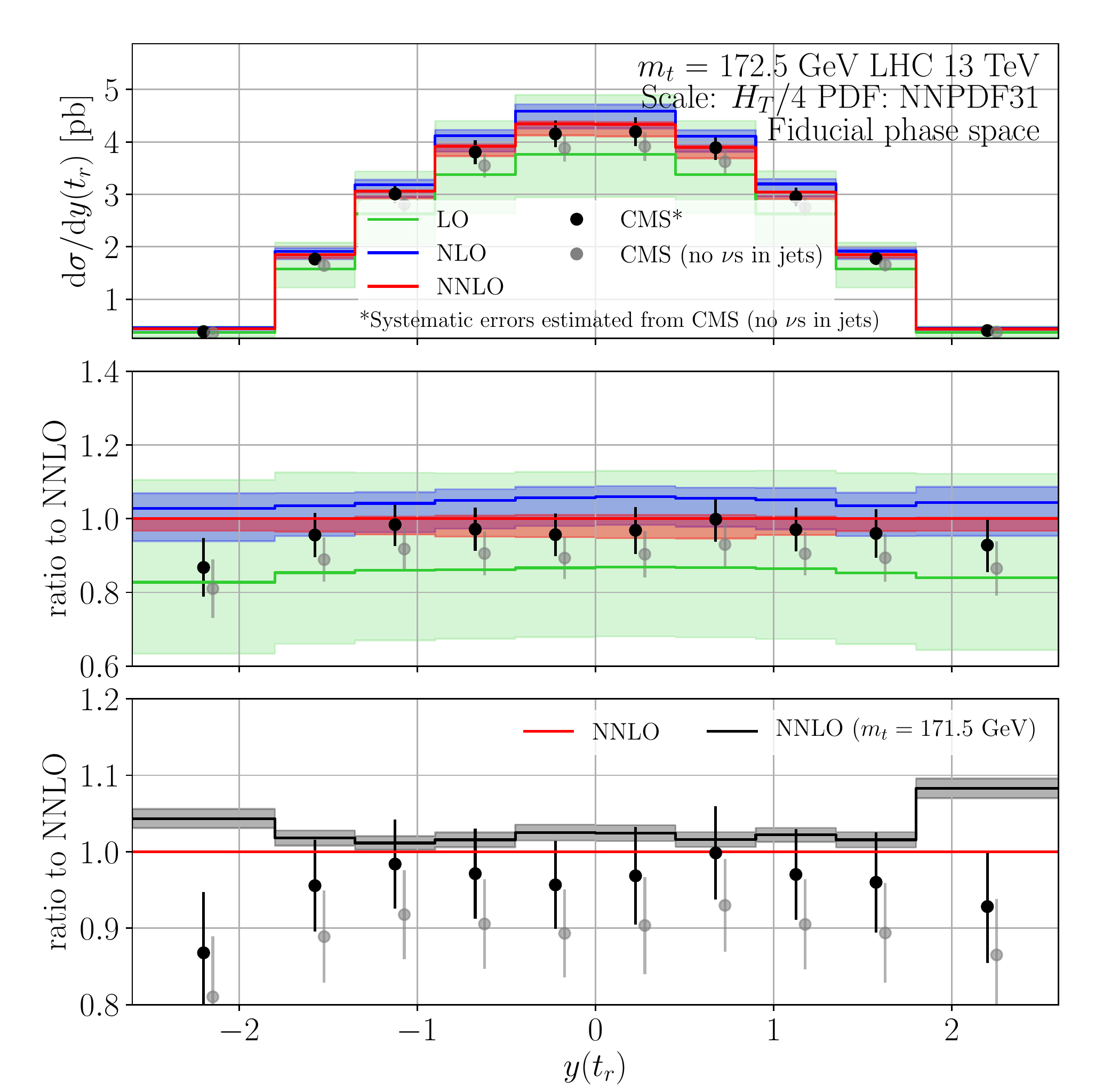}\\
\includegraphics[width=7.4cm]{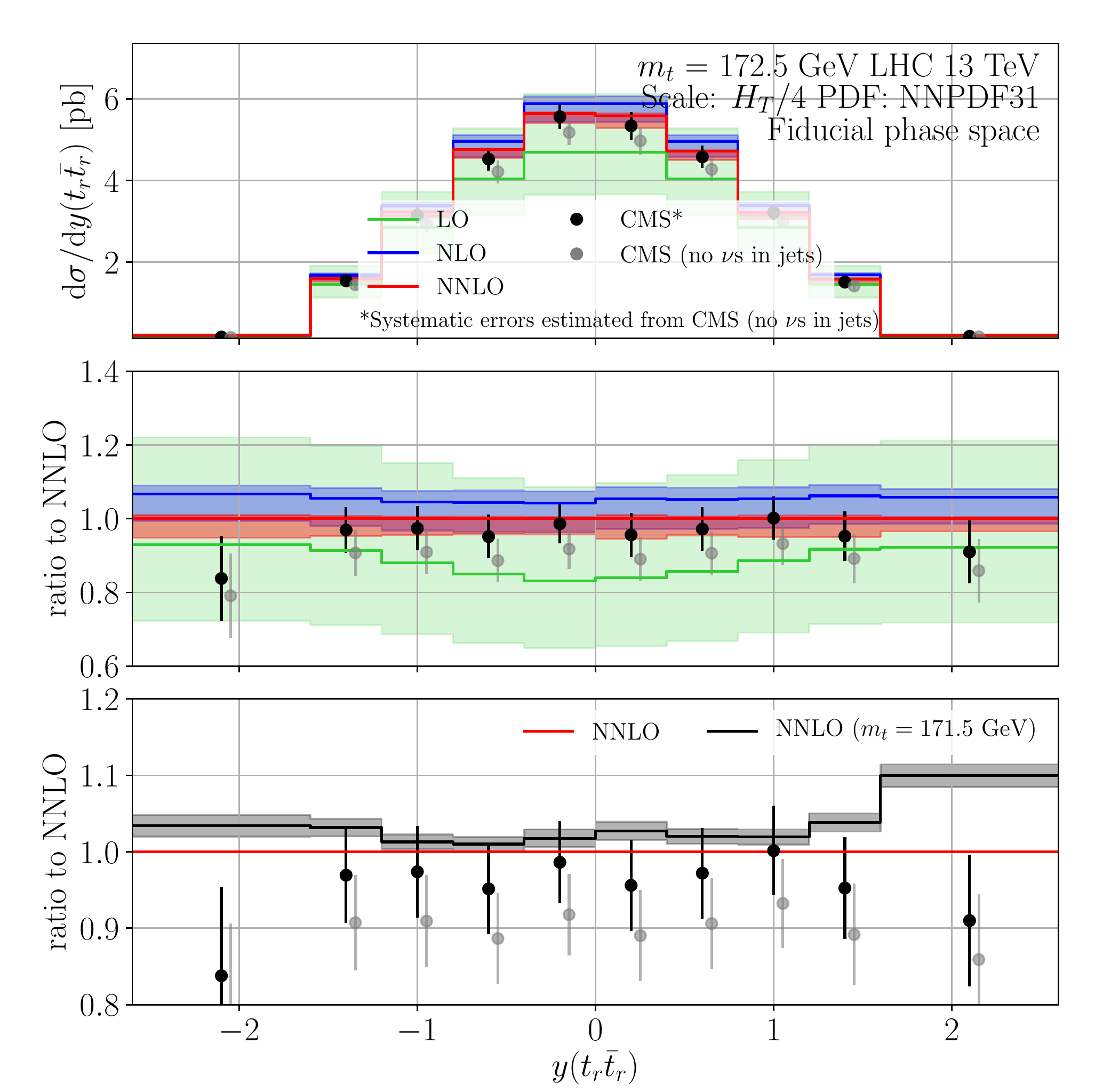}
\includegraphics[width=7.4cm]{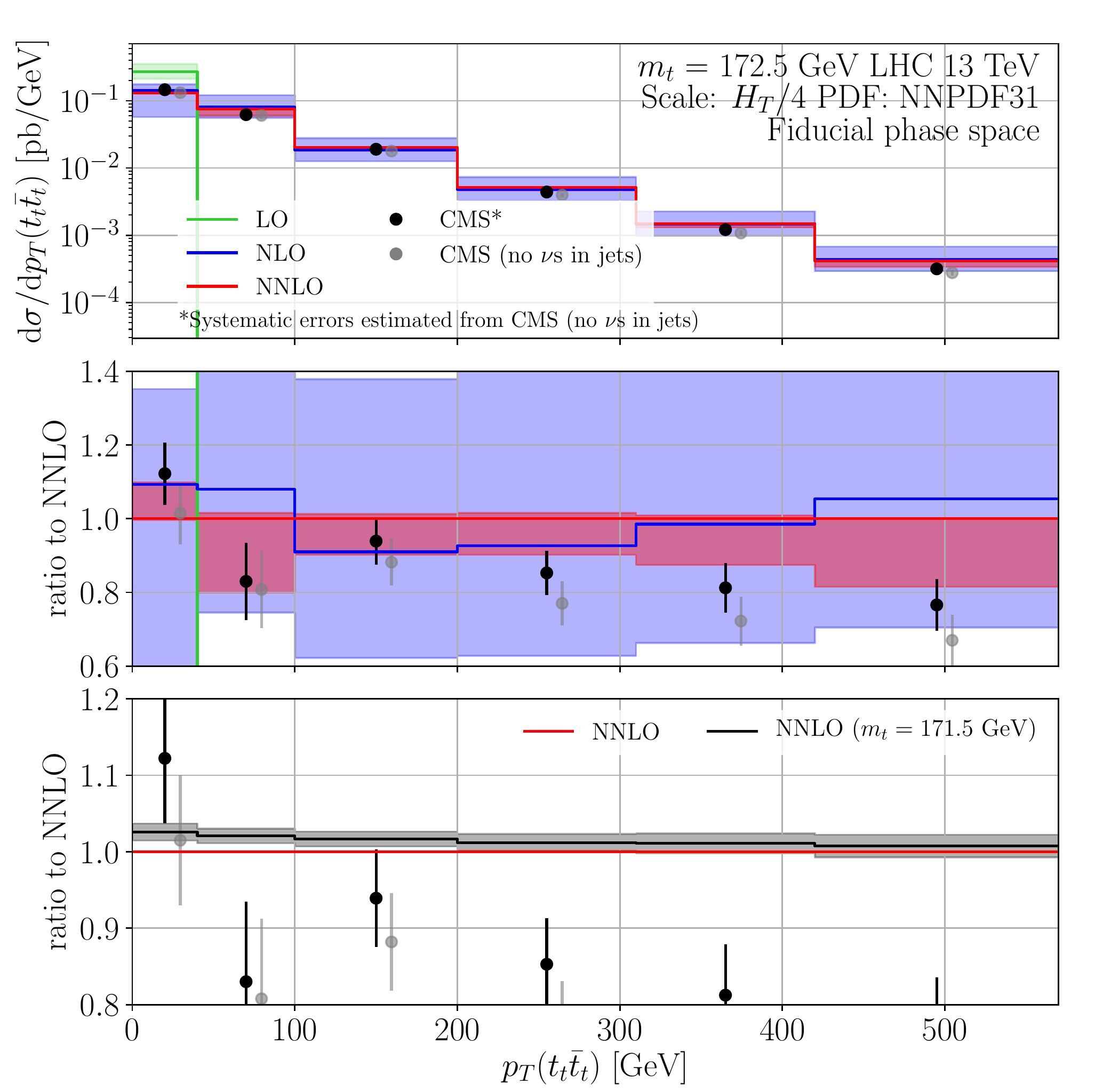}\\
\includegraphics[width=7.4cm]{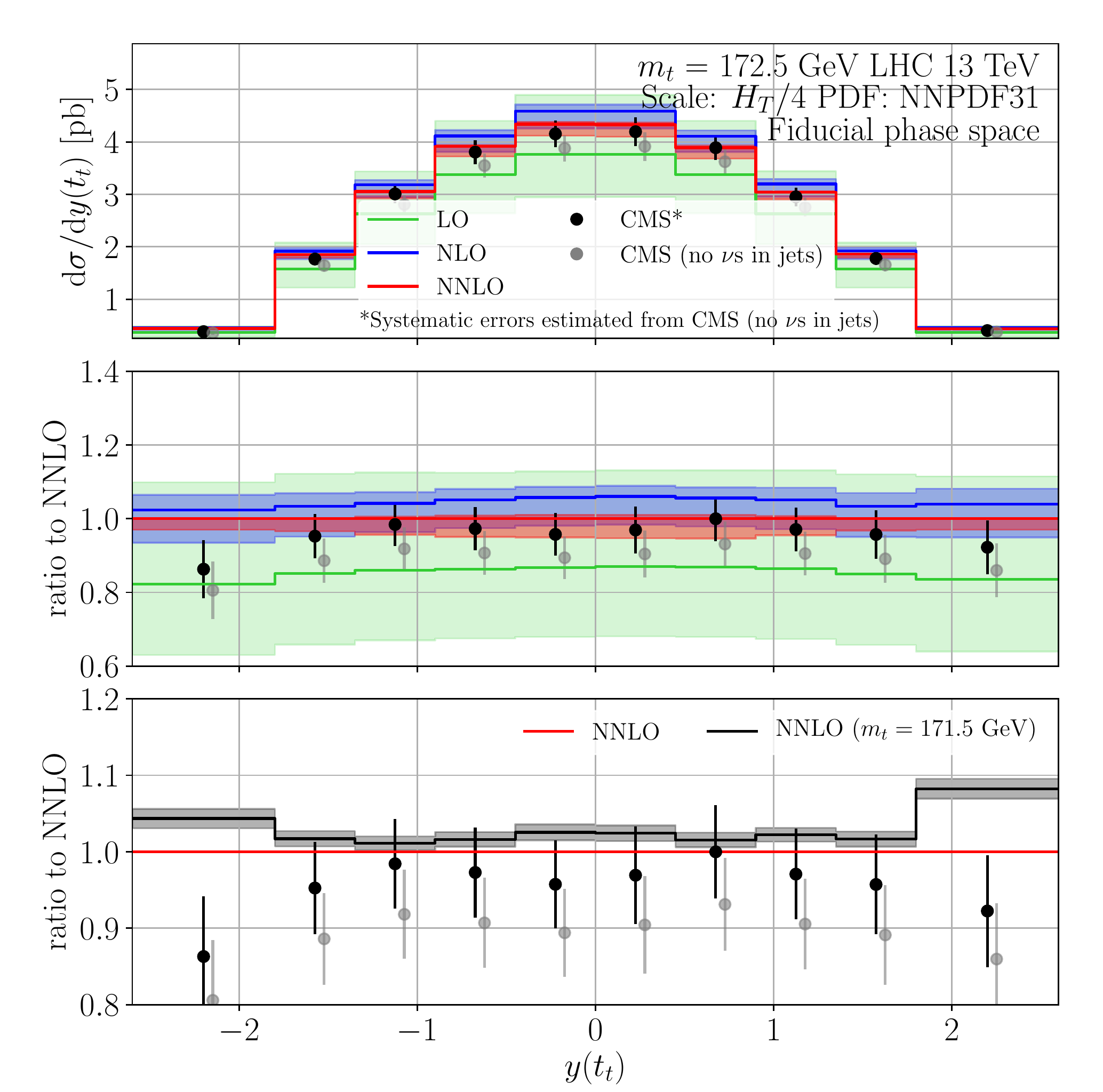}
\includegraphics[width=7.4cm]{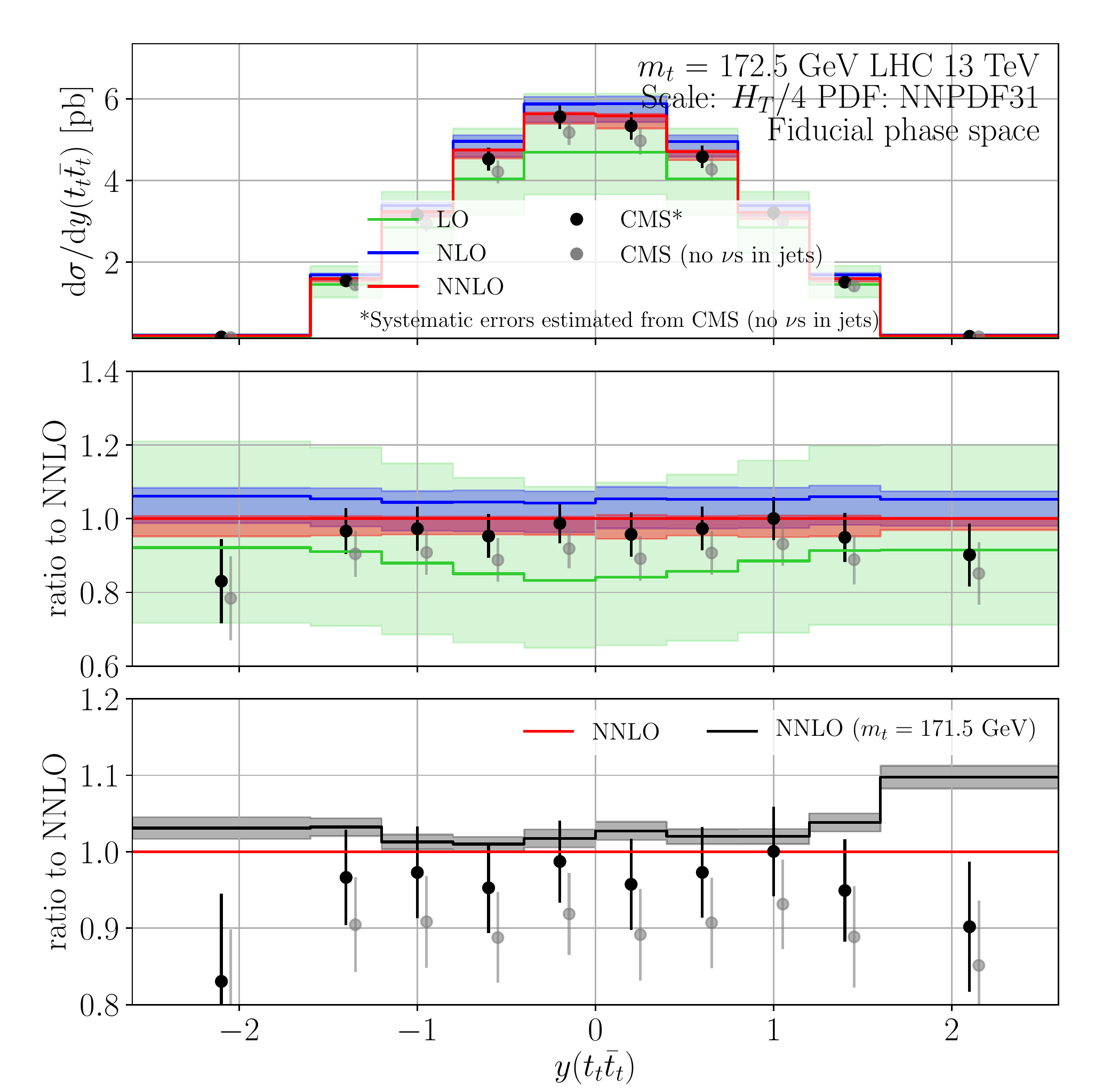}
\caption{As in fig.~\ref{fig:fid-lep-abs} but for the reconstructed (top) and true (bottom) top quark absolute $p_T(t\bar{t}), y(t)$ and $y(t\bar{t})$ distributions.}
\label{fig:fid-top1-abs}
\end{figure}

The above distributions are computed for both the true and reconstructed top quarks and are compared to CMS data without neutrinos \cite{Sirunyan:2018ucr} (in grey) and with neutrinos \cite{CMS-fiducial-mod} (in black). Both absolute and normalized distributions are shown, see figs.~\ref{fig:fid-top1-abs},\ref{fig:fid-top1-norm},\ref{fig:fid-top2-abs},\ref{fig:fid-top2-norm}. The format of the plots is the same as the ones for the leptonic and $b$-jet distributions described on pages~\pageref{text:fid-plots-begin}--\pageref{text:fid-plots-end}. 

The reconstruction algorithm used by CMS in ref.~\cite{Sirunyan:2018ucr} has been implemented by us in three steps. Step one: the two neutrino momenta are extracted from our MC. Step two: the two $W$ bosons are reconstructed from the two leptons and the two neutrinos. Since we work in the NWA for the $W$ bosons this step is unambiguous for us; it is also helped by the fact that we always have exactly two leptons in our events and we do not consider QED radiation. Step three: the $t$ and $\bar t$ quarks are reconstructed from the two $W$'s constructed in step two and from two $b$-jets. Since in our calculation we can have up to four $b$-jets this reconstruction step is ambiguous. The ambiguity is resolved by choosing the two $b$-jets (among all $b$-jets that pass the fiducial requirements) which minimize the difference:
\begin{equation}
|m_{Wb_1}-m_t| + |m_{Wb_2}-m_t|\,.
\label{eq:top-reconstruct}
\end{equation}
We stress that when $b$-jets are mentioned in this section as well as in sec.~\ref{sec:b-jets}, $b_1$ and $b_2$ refer to the hardest and subleading $b$-jets, respectively, among the two $b$-jets used in the $t$ and $\bar t$ reconstruction.

\begin{figure}
\includegraphics[width=7.4cm]{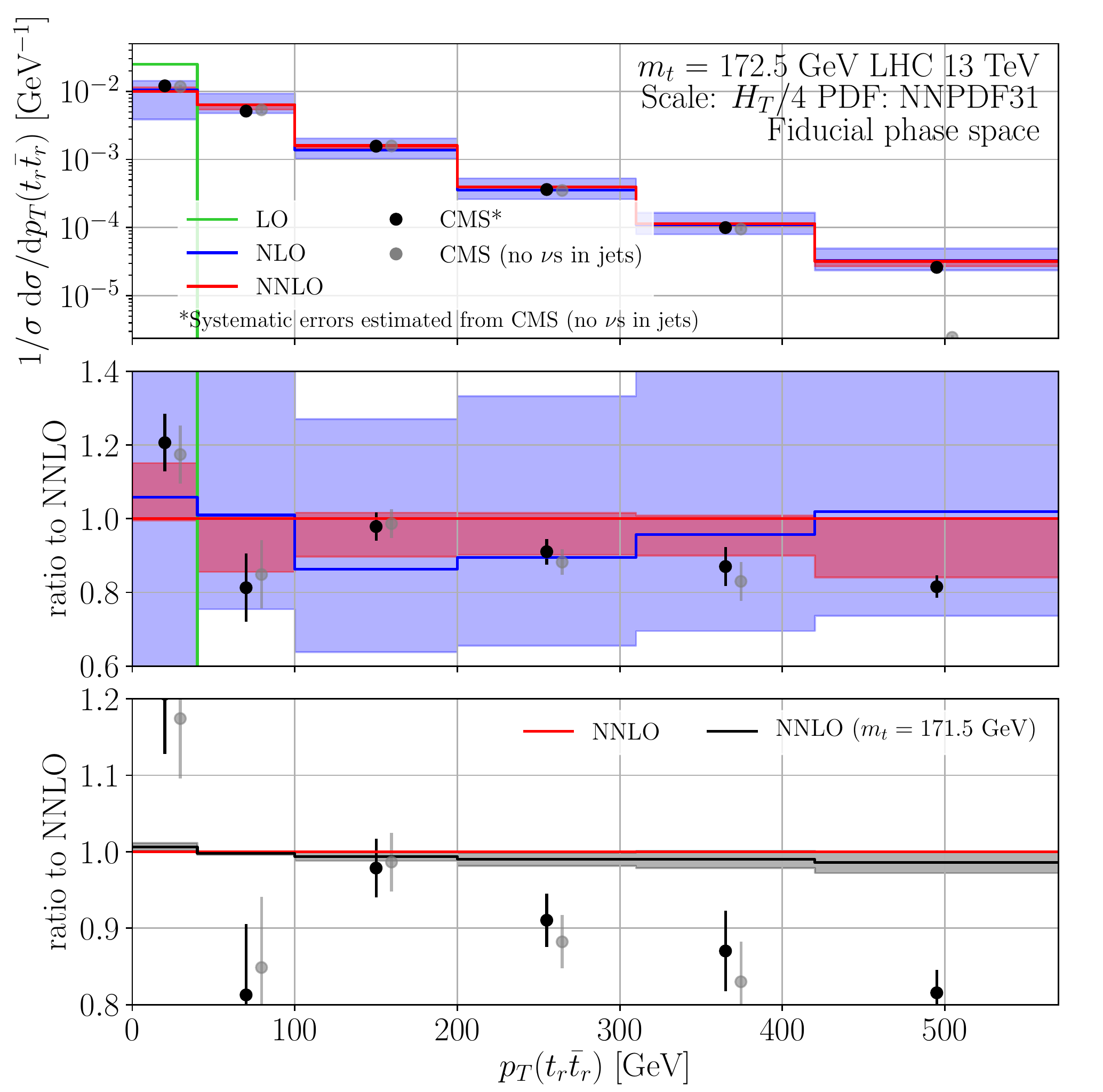}
\includegraphics[width=7.4cm]{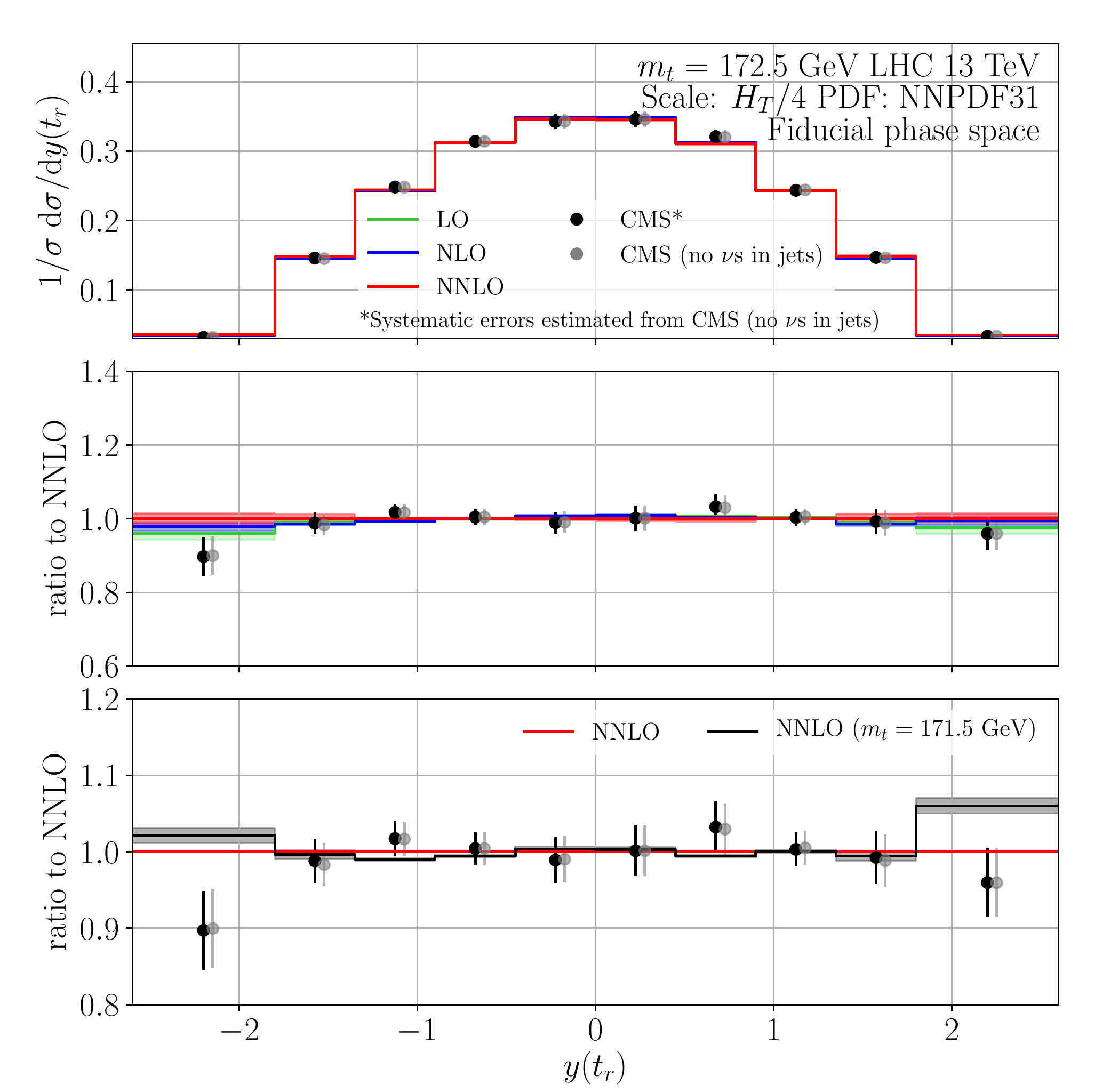}\\
\includegraphics[width=7.4cm]{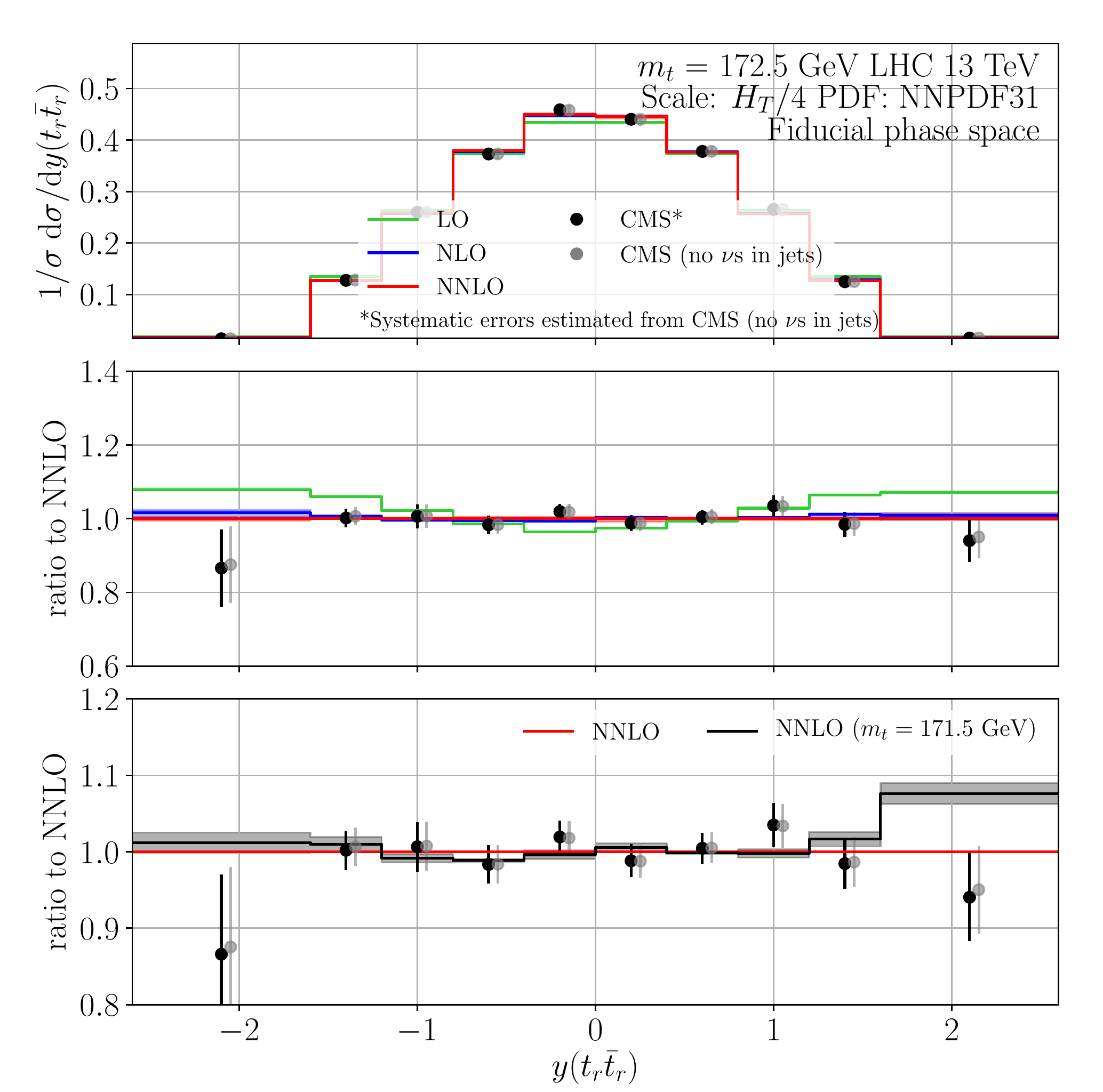}
\includegraphics[width=7.4cm]{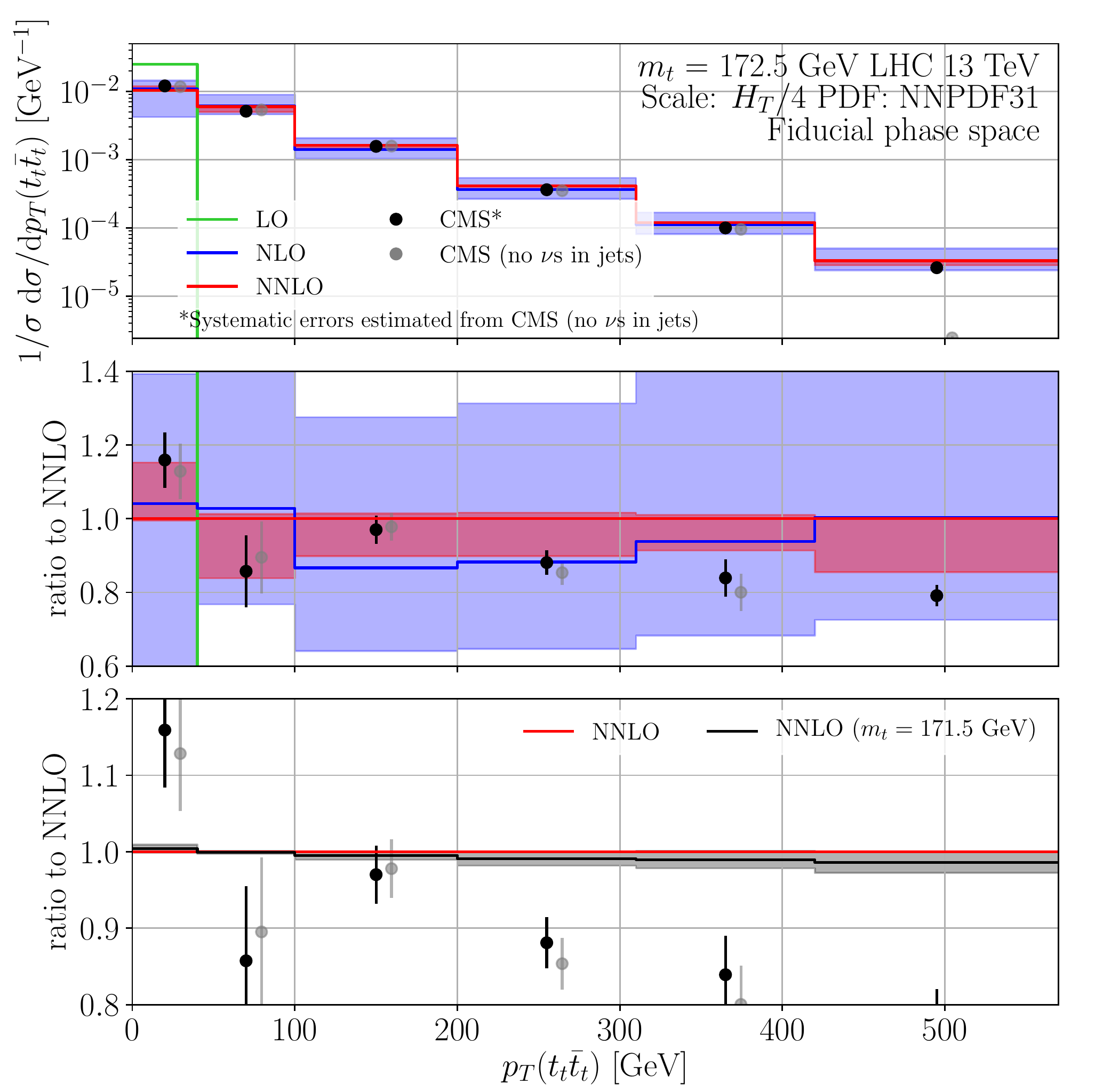}\\
\includegraphics[width=7.4cm]{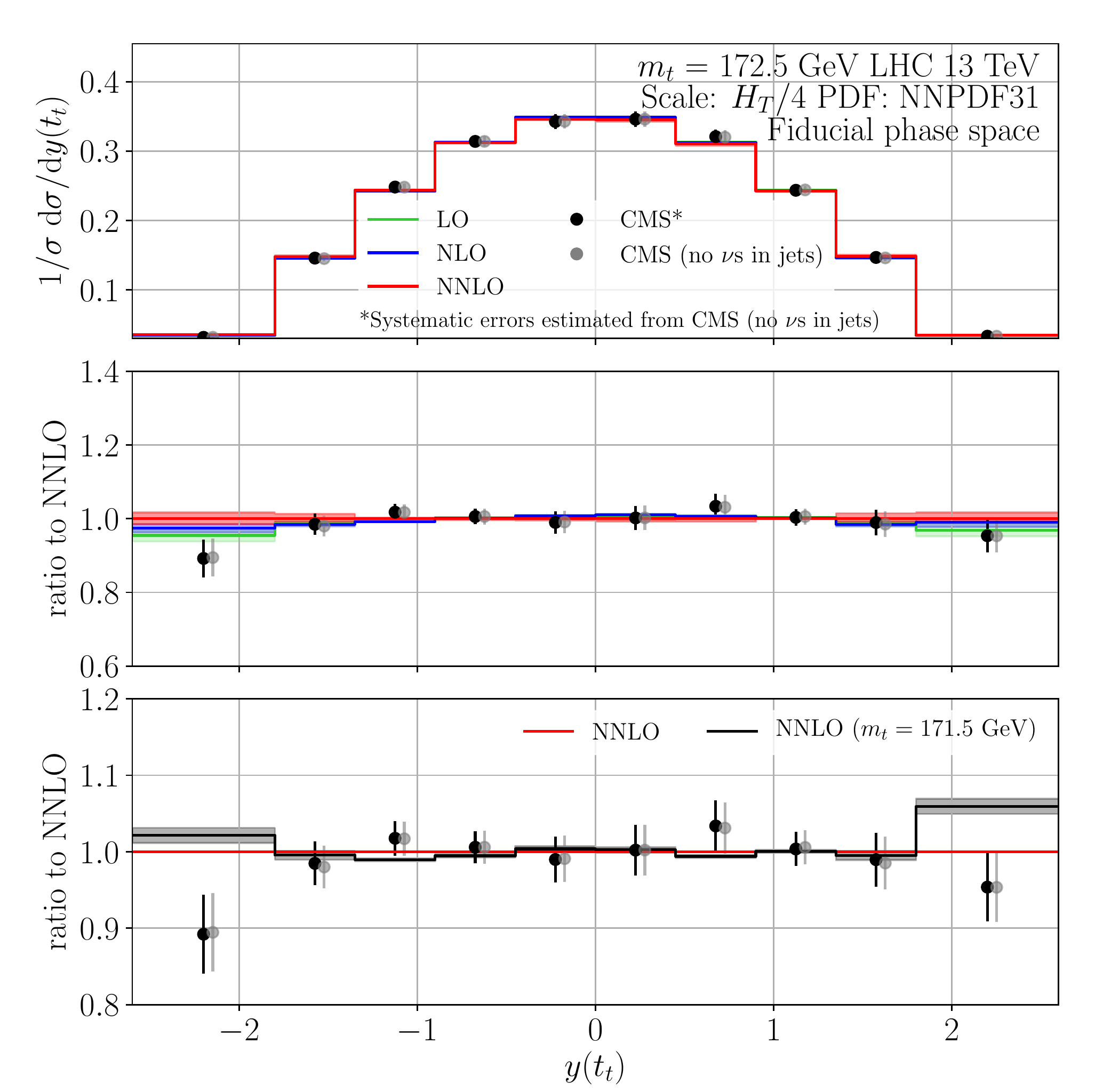}
\includegraphics[width=7.4cm]{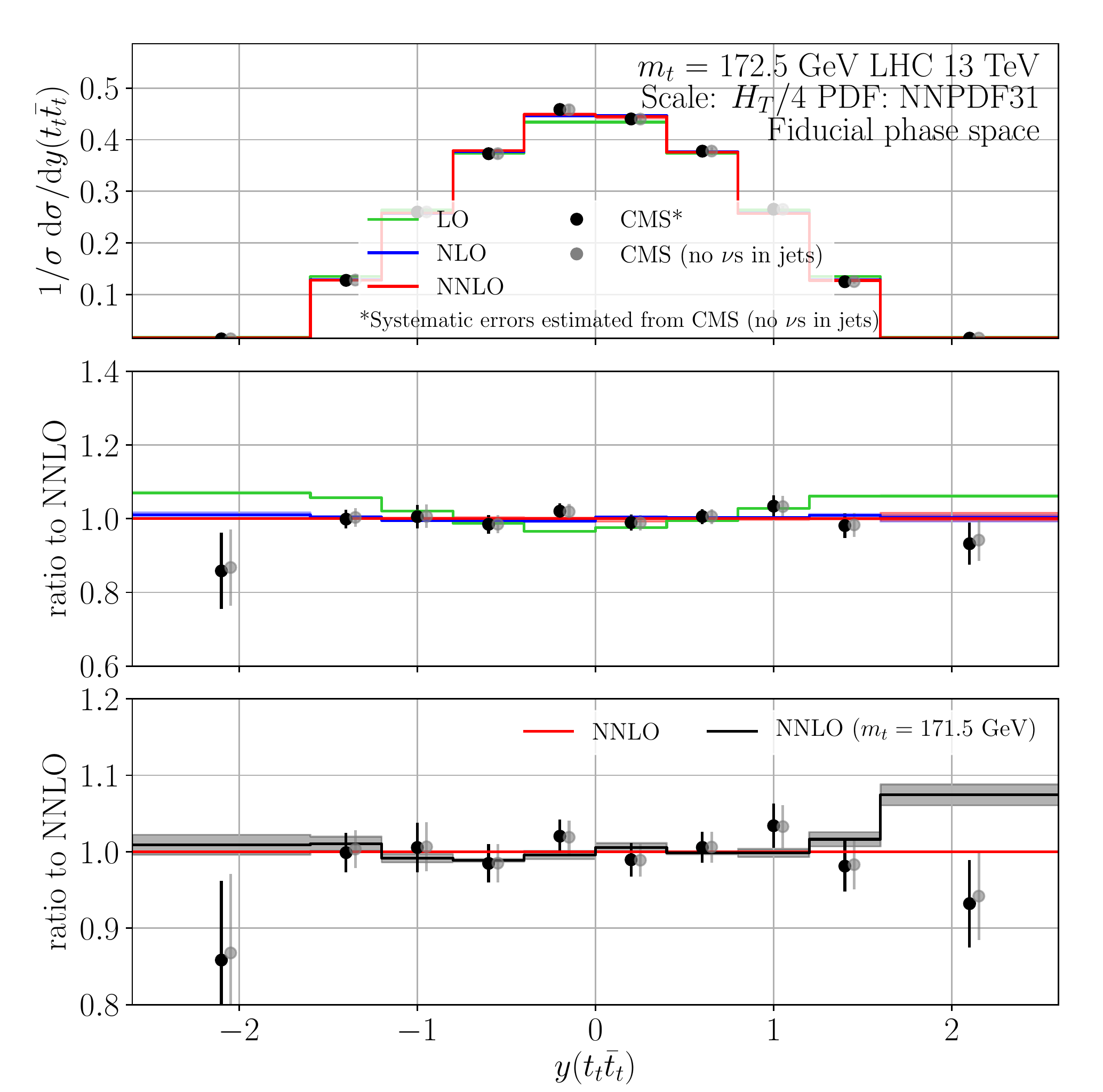}
\caption{As in fig.~\ref{fig:fid-top1-abs} but for the normalized distributions.}
\label{fig:fid-top1-norm}
\end{figure}
\begin{figure}
\center
\includegraphics[width=6.5cm]{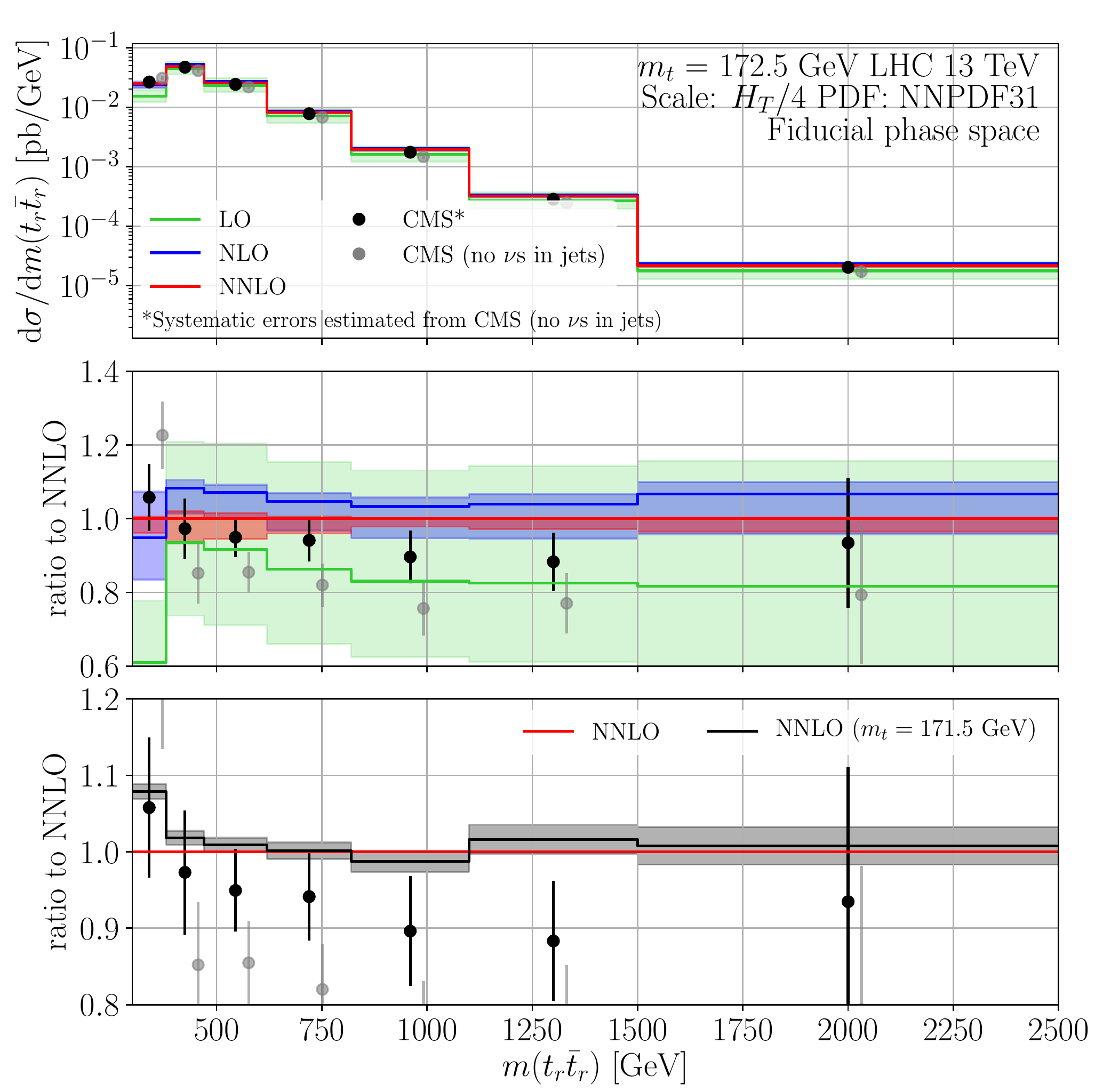}
\includegraphics[width=6.5cm]{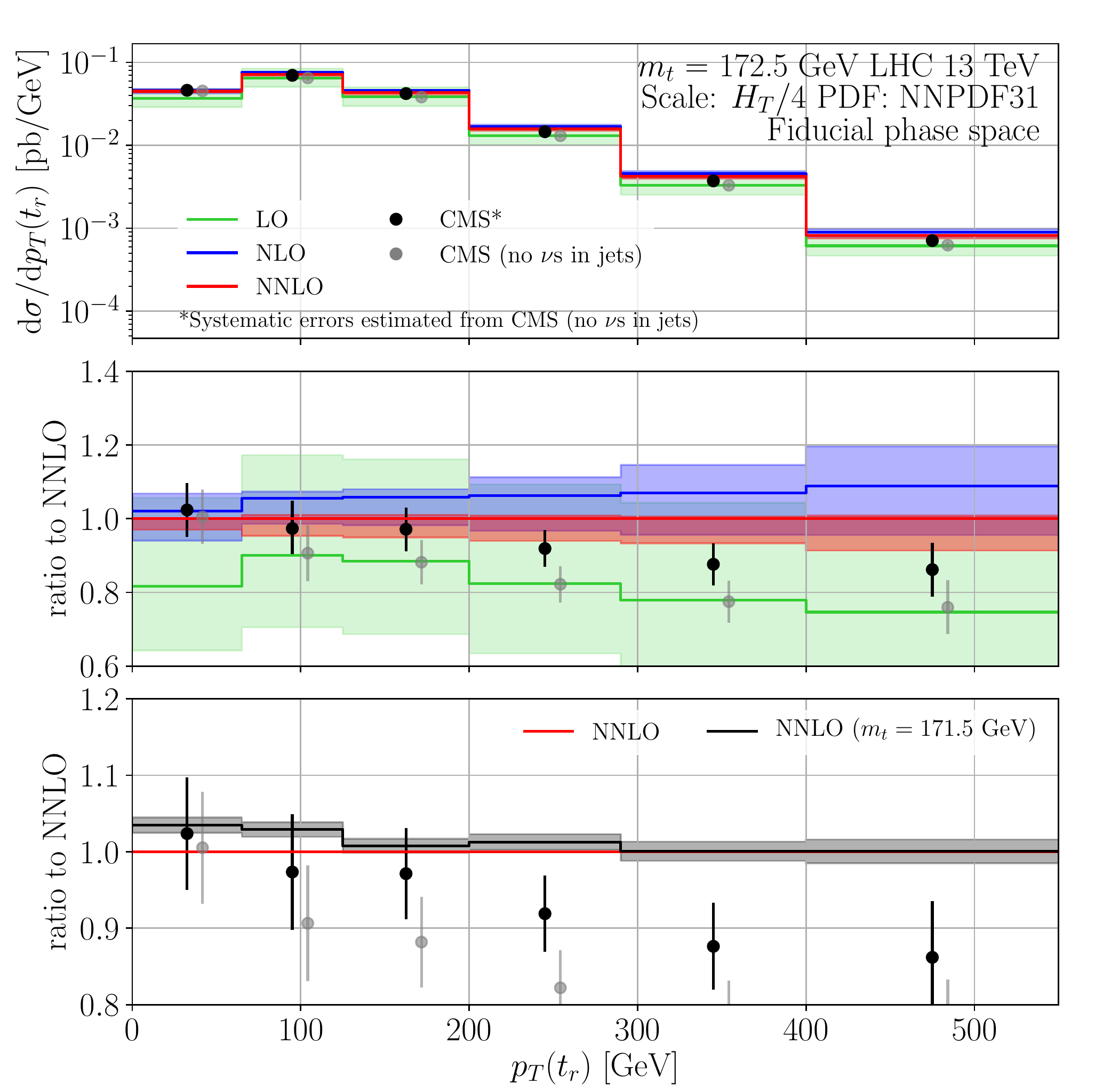}\\
\includegraphics[width=6.5cm]{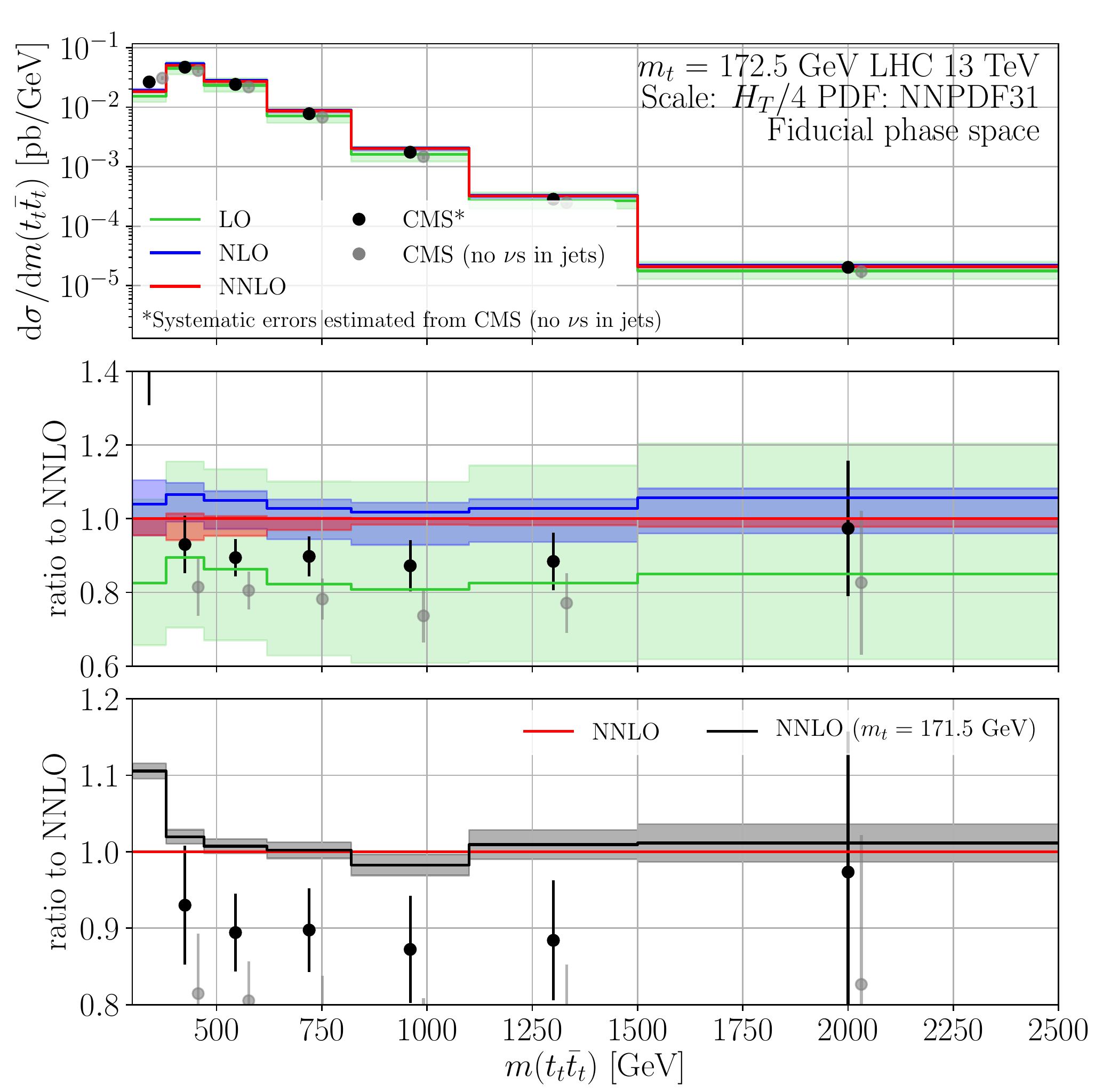}
\includegraphics[width=6.5cm]{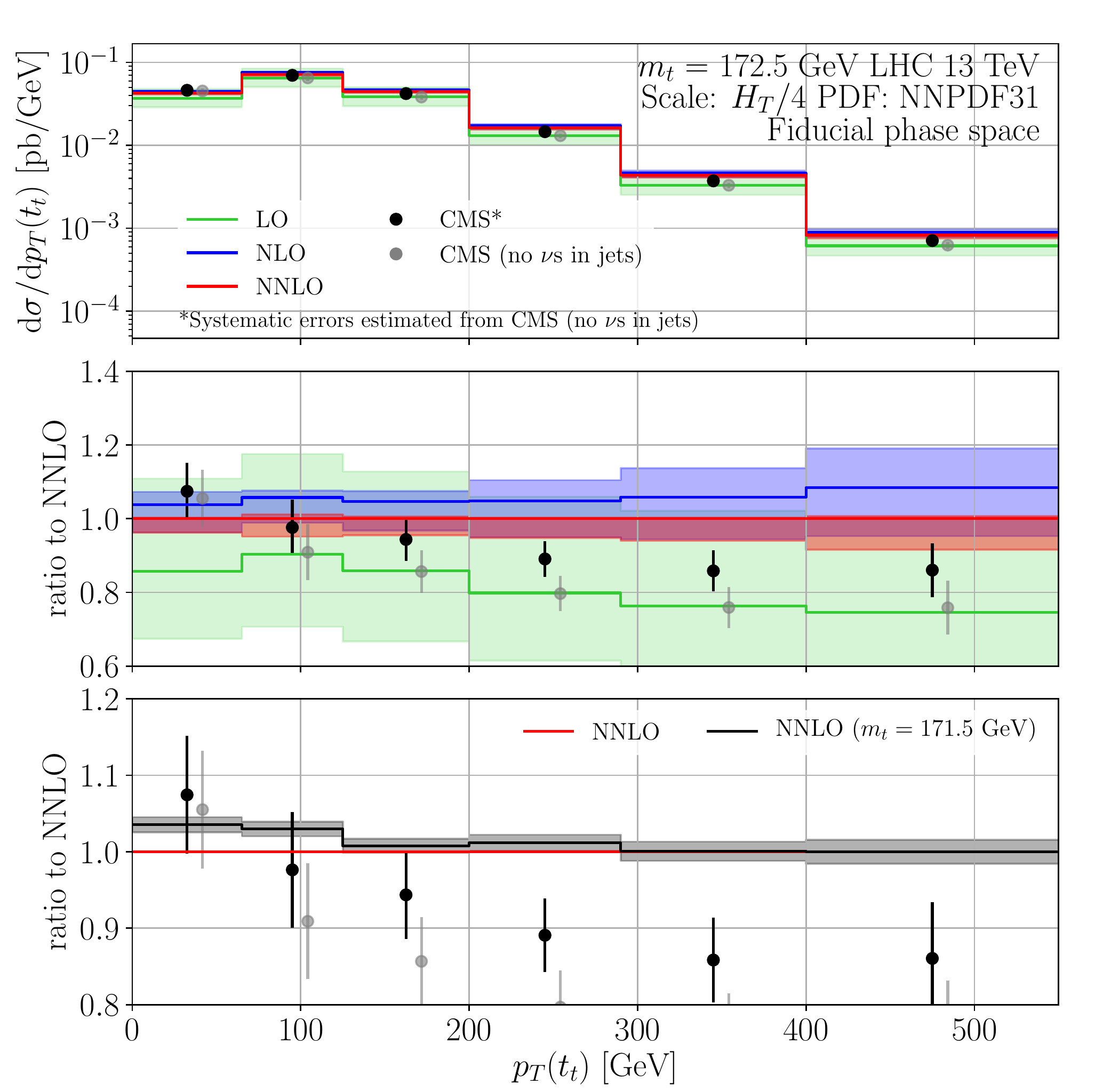}
\caption{As in fig.~\ref{fig:fid-lep-abs} but for the reconstructed (top) and true (bottom) top quark absolute $m(t\bar{t})$ and $p_T(t)$ distributions.}
\label{fig:fid-top2-abs}
\end{figure}
\begin{figure}
\center
\includegraphics[width=6.5cm]{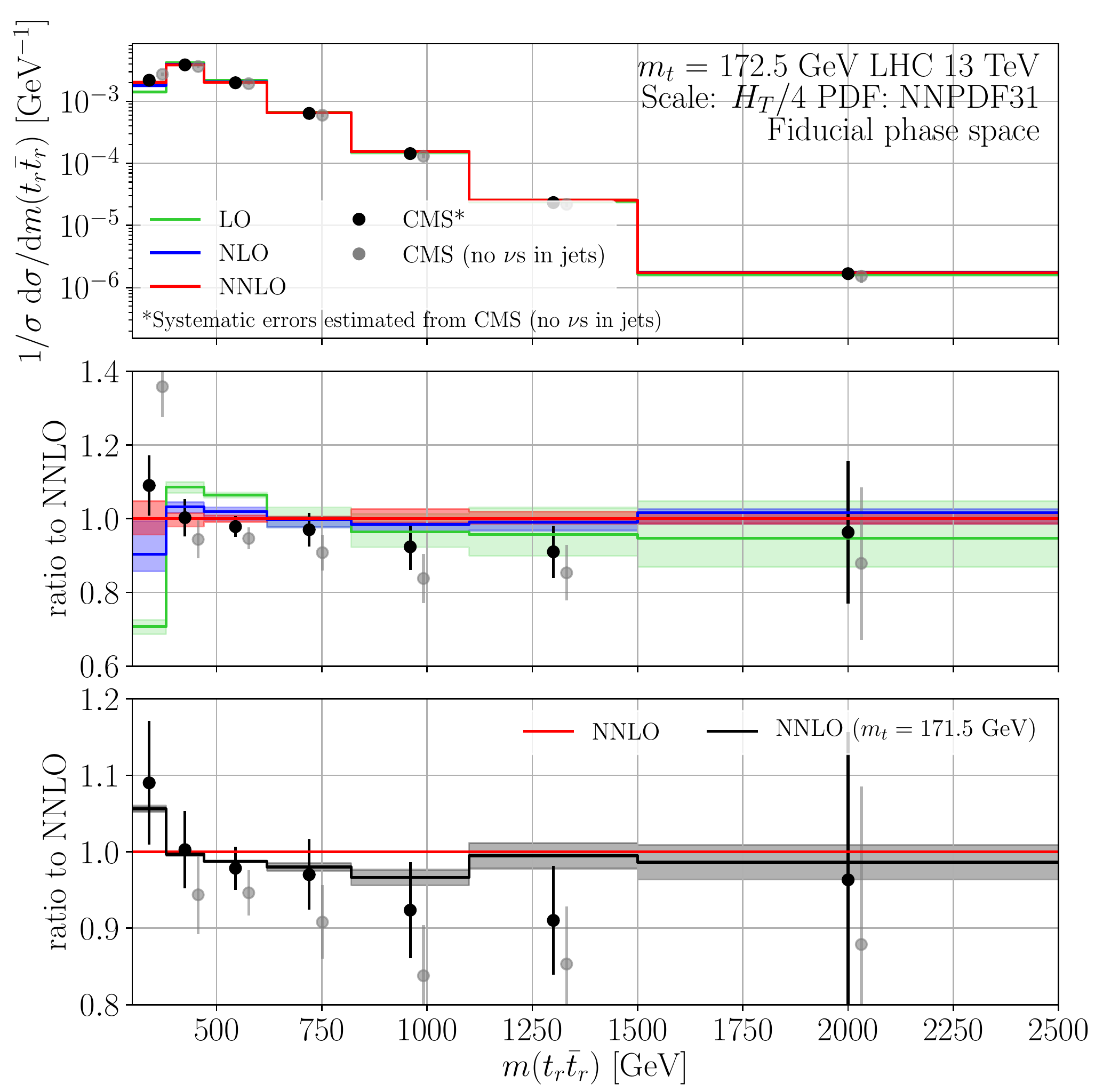}
\includegraphics[width=6.5cm]{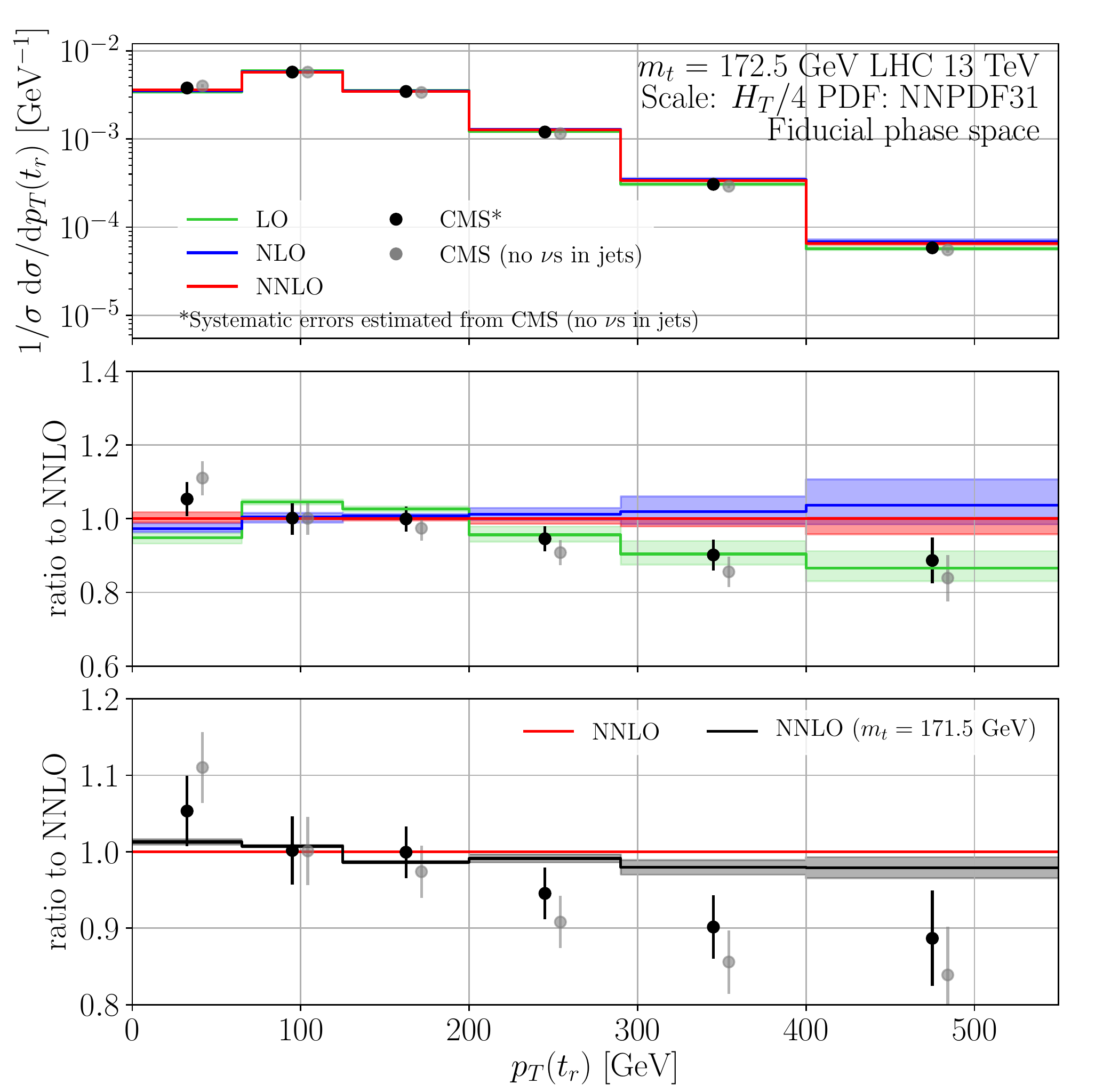}\\
\includegraphics[width=6.5cm]{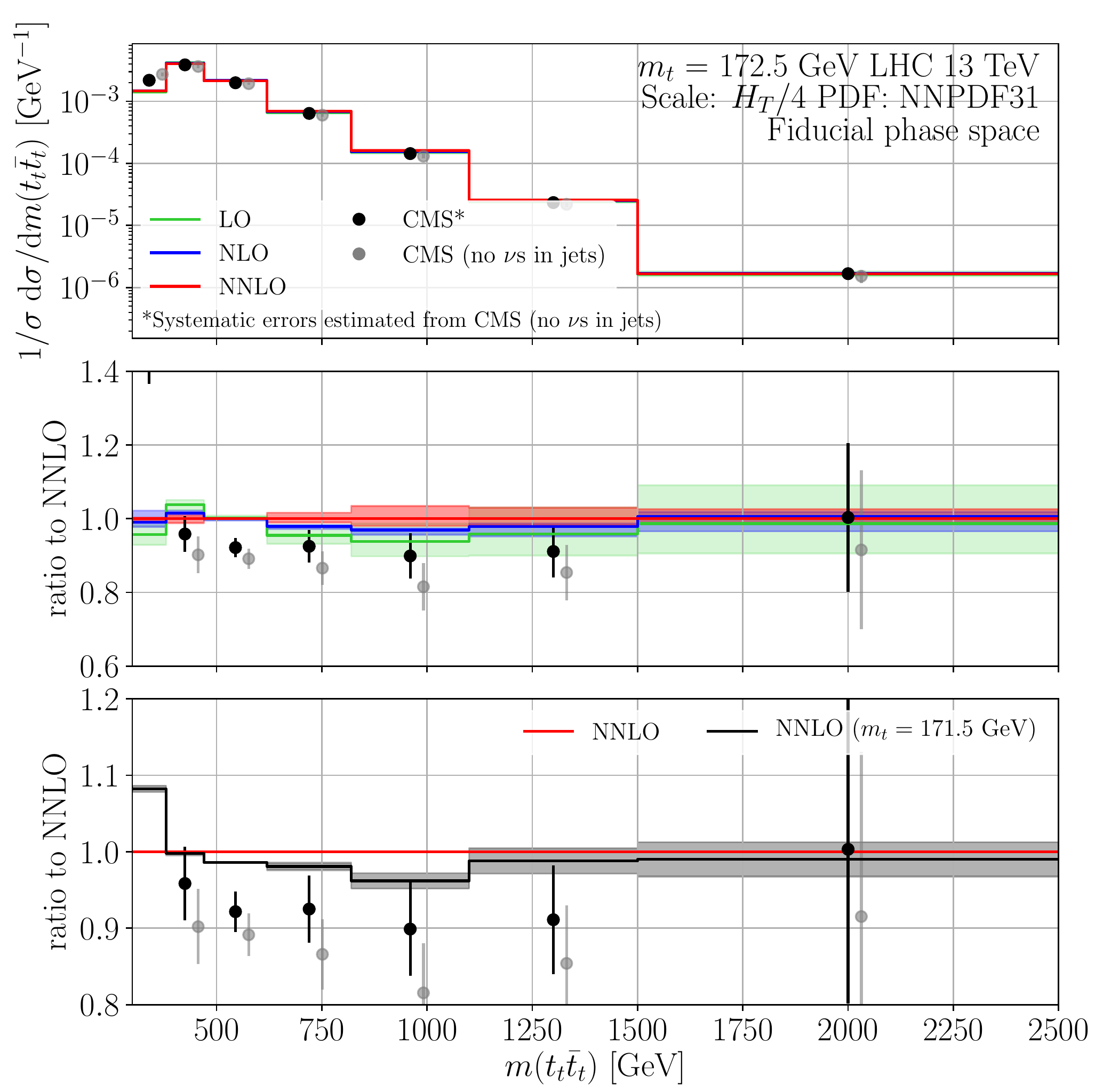}
\includegraphics[width=6.5cm]{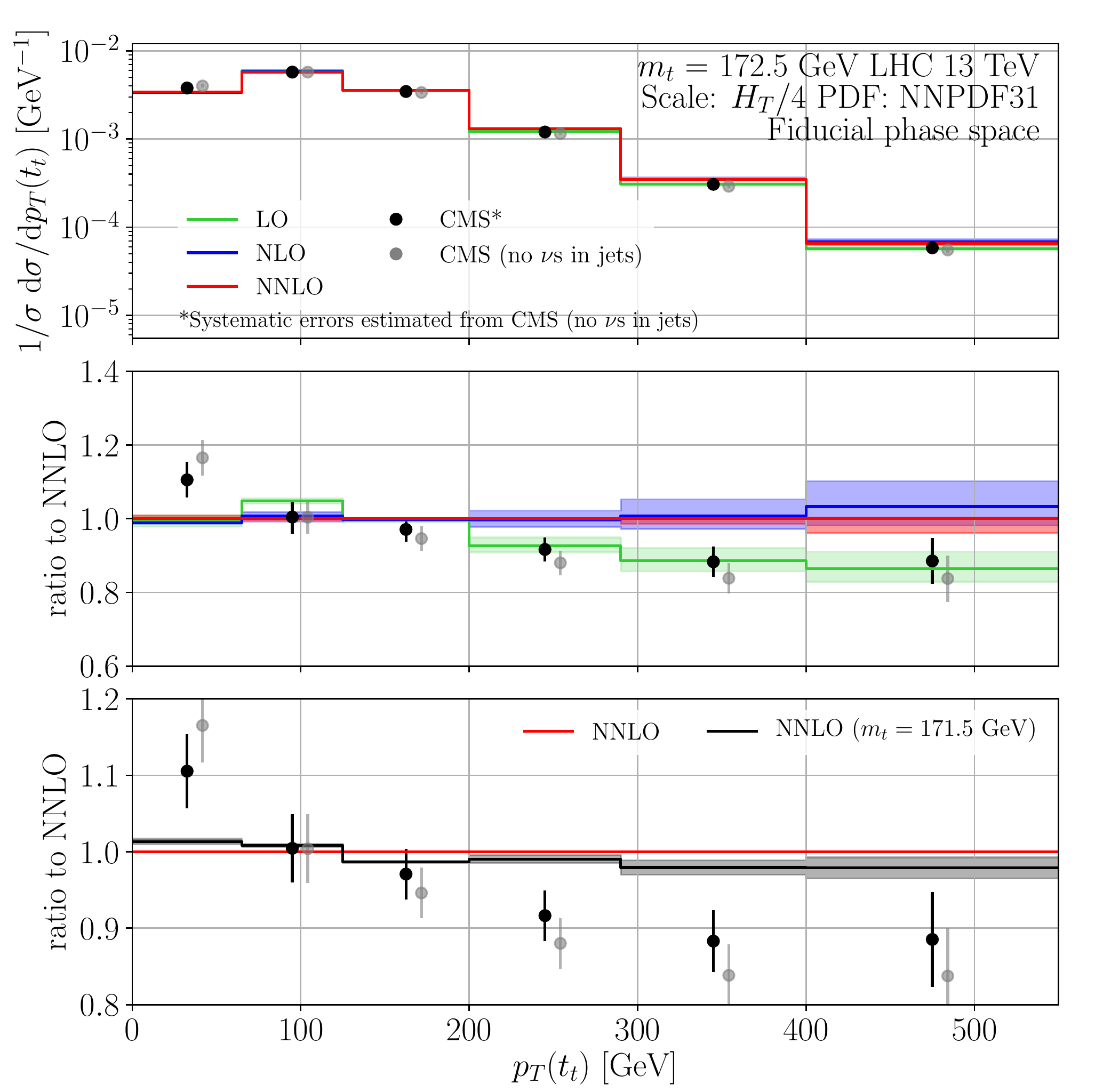}
\caption{As in fig.~\ref{fig:fid-top2-abs} but for the normalized distributions.}
\label{fig:fid-top2-norm}
\end{figure}

A number of interesting features can be observed in figs.~\ref{fig:fid-top1-abs},\ref{fig:fid-top1-norm},\ref{fig:fid-top2-abs},\ref{fig:fid-top2-norm}. The effect of the neutrinos from semileptonic decays on top-quark data is substantial and affects both the normalization and shapes of distributions. The size of the impact strongly depends on the distribution, or even on the specific bin, and varies from near perfect agreement to a difference larger than the experimental uncertainties. For example, the two rapidity distributions are affected only through their normalizations. In the kinematic range considered here the $p_T(t\bar{t})$ distribution is affected primarily through its normalization although shape difference arises at large $p_T(t\bar{t})$. The $p_T(t)$ distribution is affected about equally in terms of its normalization and shape. Interestingly, as can be seen in fig.~\ref{fig:fid-top2-norm} the slope of the top $p_T$ distribution changes significantly depending on how the neutrinos in the semileptonic decays are treated. Their inclusion in $b$-jets leads to a much softer top-quark high-$p_T$ tail. The modeling of this effect may have implication for the so-called top $p_T$ problem as well as for top quark mass measurements from the low-$p_T$ part of the spectrum. 

A particularly large sensitivity to the modeling of the neutrinos is observed in the $m(t\bar{t})$ distribution, especially towards small $m(t\bar{t})$ values which describe $t\bar t$ production close to absolute threshold. Both the absolute and normalized distributions are impacted. Notably, the effect on the first $m(t\bar{t})$ bin of the normalized distribution is much larger than the experimental uncertainties. Given the extreme sensitivity of this bin to the value of $m_t$ we caution that any theory-data comparison in this bin may be strongly impacted by such modeling and therefore have potentially significant impact on any $m_t$ extraction. The size of this effect may well be such that it eclipses other potential recently studied effects like the beyond-NNLO Coulomb corrections at and below threshold \cite{Ju:2020otc} or the impact from alternative definition of $m_t$ like, for example, in the $\overline{\rm MS}$-scheme at NNLO \cite{Catani:2020tko}.

\begin{figure}
\center
\includegraphics[width=7.4cm]{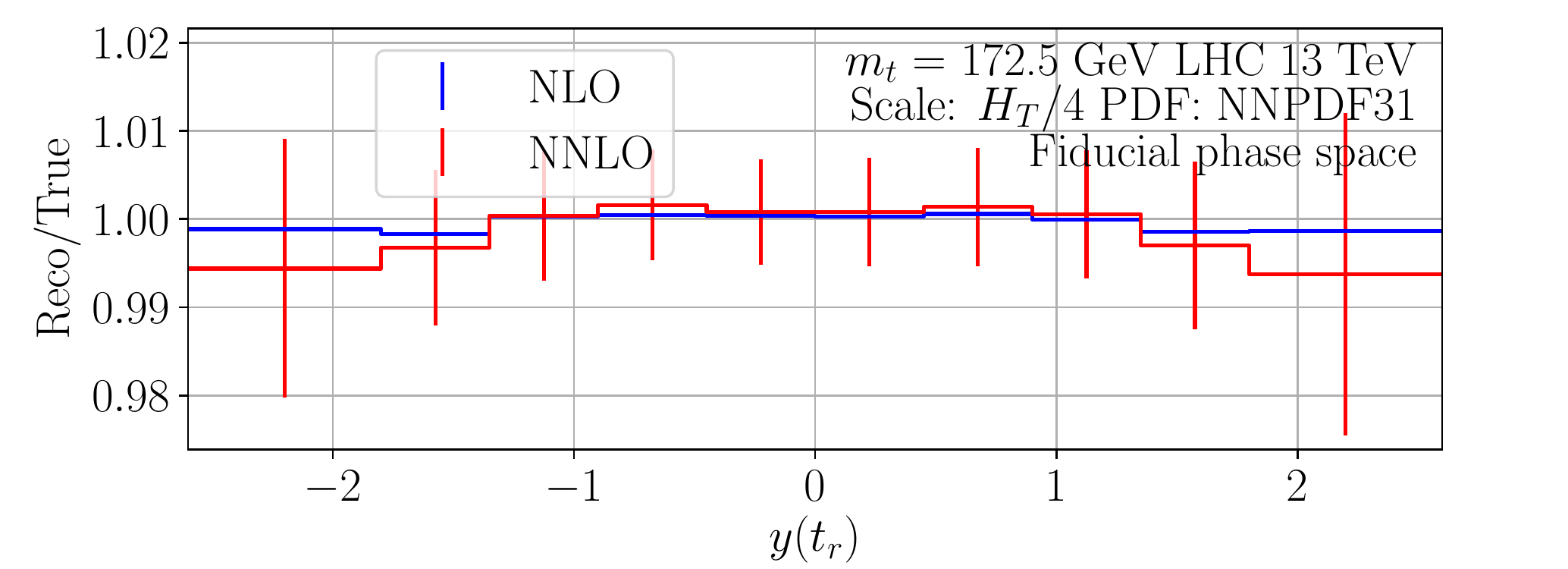}
\includegraphics[width=7.4cm]{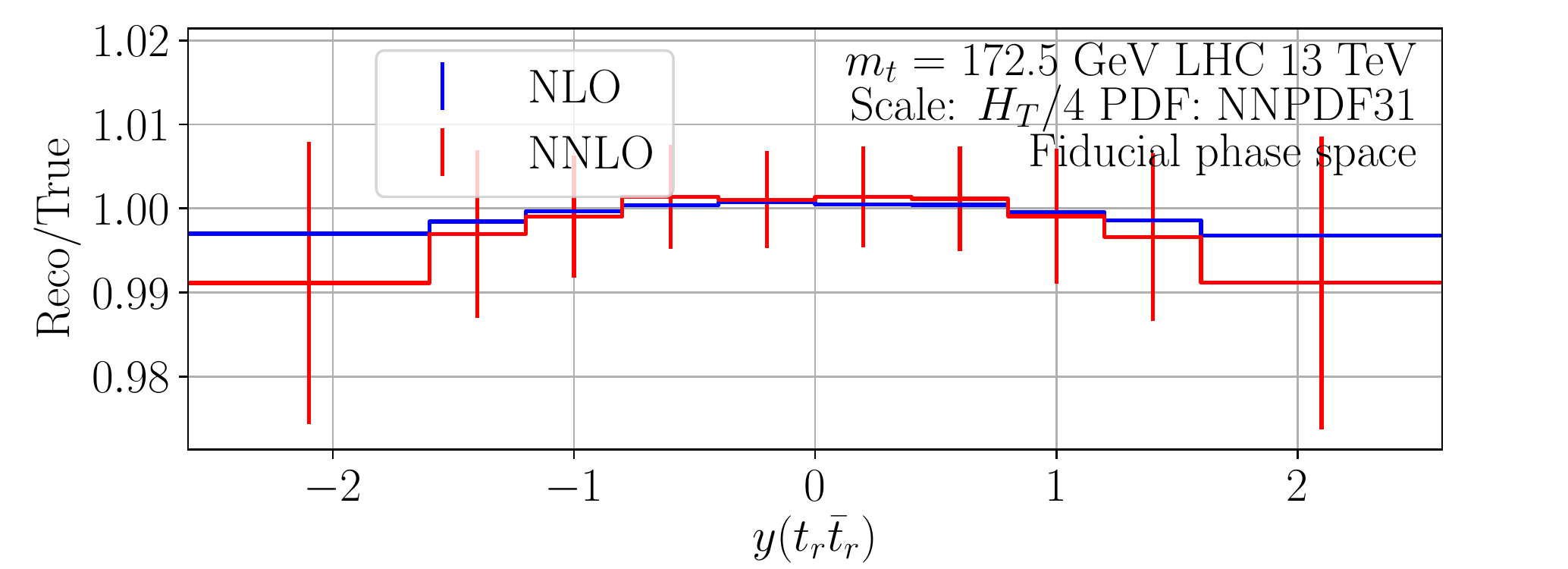}\\
\center
\includegraphics[width=7.4cm]{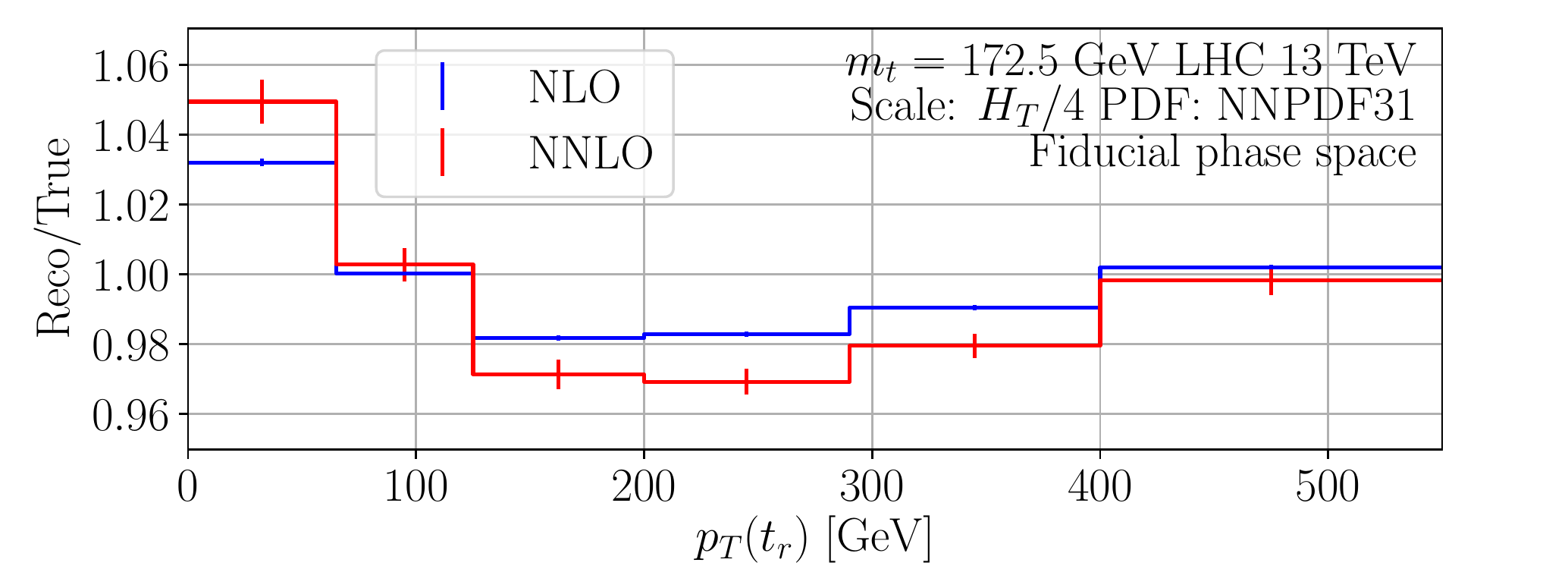}
\includegraphics[width=7.4cm]{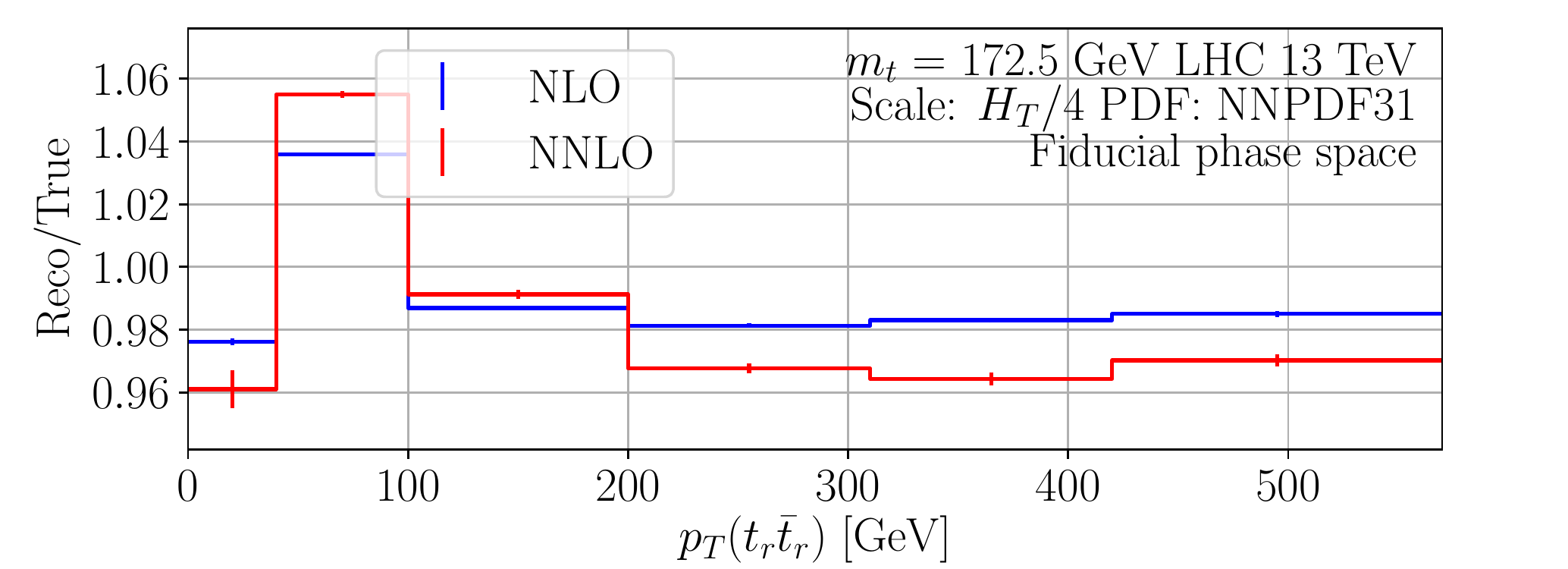}\\
\center
\includegraphics[width=7.4cm]{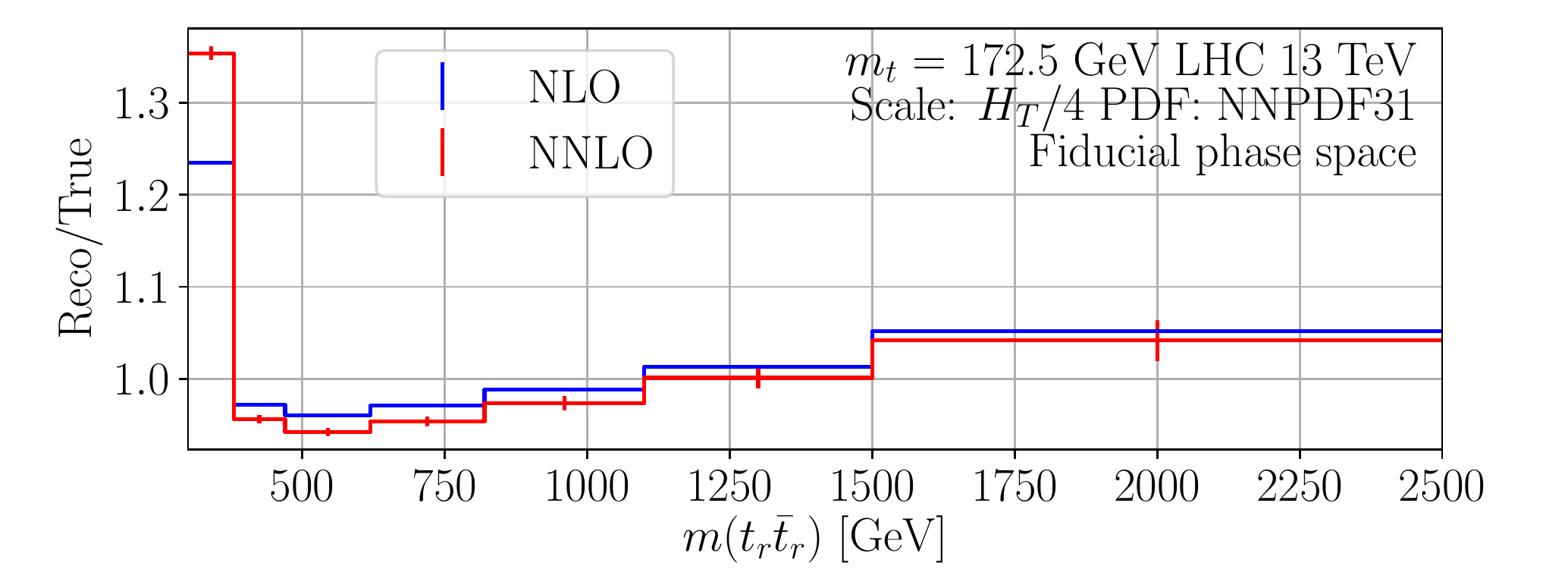}
\caption{Reconstructed versus true top quark at NLO (blue) and NNLO (red) QCD for each one of the top quark distributions considered here. The vertical bars show the MC error.}
\label{fig:reco-vs-true}
\end{figure}

The difference between reconstructed and true tops for each one of the distributions considered here can be found in fig.~\ref{fig:reco-vs-true}. The $t$ and $t\bar t$ rapidity distributions are least affected. At NLO the two predictions are very close while at NNLO, despite the much larger MC error, a clear shape-effect appears. The absolute size of the differences tends to be small - typically much below 1\% - which is much smaller than the experimental uncertainties. The most striking difference between true and reconstructed tops can be observed in the first bin of the $m(t\bar{t})$ distribution where it exceeds 20\% at NLO and 35\% at NNLO. The inclusion of the NNLO QCD corrections strongly impacts this difference. This means that this bin is particularly sensitive to the way measurements are translated to true top quarks. The rest of the $m(t\bar{t})$ distribution as well as the two $p_T$ distributions show only moderate sensitivity at the level of few percent to true-versus-reconstructed top. We believe it will be useful to extend the present study by adding predictions from MC event generators in order to gauge the true impact on top quark distributions from the top quark definition. 

For consistency, comparisons between theory and data should be done between the reconstructed tops and the data including neutrinos from semileptonic decays (i.e. the data in black). From figs.~\ref{fig:fid-top1-abs},\ref{fig:fid-top1-norm},\ref{fig:fid-top2-abs},\ref{fig:fid-top2-norm} we conclude that within uncertainties there is very good agreement between NNLO QCD prediction and data. This is the case for both absolute and normalized distributions. Notably, for all cases where the experimental uncertainties are small enough to make this comparison possible, one can see that NNLO QCD predictions describe data better than NLO QCD. In some cases this comparison becomes even sensitive to the value of $m_t$. For example, the shapes of the normalized $p_T(t)$ and $m(t\bar{t})$ distributions show small preference for the prediction based on $m_t=171.5$ GeV although the experimental uncertainties are still too large to allow for a detailed comparison. This may be an excellent opportunity for a future study.

The sensitivity of the fiducial top-quark differential distributions to $m_t$ roughly follow the well known pattern in inclusive distributions. Notably, an increased sensitivity can be observed towards high rapidities in both rapidity distributions. A more quantitative study of the $m_t$ dependence in fiducial top quark distributions will probably require a calculation with reduced MC errors. 

The pdf uncertainty in all distributions is generally rather small. It starts to be non-negligible only for the normalized $y(t\bar{t})$ distribution at large rapidity. It also becomes comparable to the NNLO scale variation for the $m(t\bar{t})$ distribution above about 1 TeV as well as for the normalized $p_T(t)$ distribution above about 400 GeV.

The pattern of higher-order corrections in the fiducial top-quark distributions considered in this section is similar to the one observed for the leptonic and $b$-jet fiducial distributions. NNLO QCD predictions have smaller scale variation than the NLO QCD ones, and the NNLO/NLO K-factors are small and most of the time negative. 

We conclude that, overall, high-precision description of LHC data by NNLO QCD is possible, especially for normalized distributions. At present the comparisons are dominated by the size of the experimental errors and future improvements in experimental LHC measurements can lead to very precise theory-data comparisons. While a number of potential subtleties have been highlighted here, it seems to us that there are no fundamental roadblocks to future high-precision comparisons at NNLO accuracy.

\section{Conclusions and Outlook}\label{sec:conclusions}

In this work we perform, for the first time, a comprehensive set of calculations for one- and two-dimensional distributions in $t\bar t$ production and decay to dilepton final states. The calculations are performed in the Narrow-Width Approximation and include all NNLO QCD corrections, both in top quark's production and decay stages. 

We give a complete technical account of our calculations, especially what concerns the implementation of NWA at NNLO in QCD. We then perform an extensive phenomenological analysis of a large set of observables measured at the LHC. 

We have considered $t\bar t$ spin correlations. We have provided a complete set of NNLO QCD prediction for the so-called spin-density matrix which completely parametrizes $t\bar t$ spin correlations. We find complete agreement with the measurements from CMS although the current experimental uncertainties are still too large to discriminate between the various perturbative orders. We also provide predictions for a set of measurements of spin-correlation-sensitive spectra published by both the CMS and ATLAS collaborations. The present calculations reinforce our findings from our previous work \cite{Behring:2019iiv} that fiducial-level comparisons between NNLO theory and data show complete agreement while comparisons at the fully inclusive level not always do. It is important to stress that this conclusion is based on rather different analyses in terms of the inclusiveness of their final states.

Besides spin correlations, we have also analyzed inclusive and fiducial differential distributions of leptons, $b$-jets and top quarks. With very few exceptions -- which can be understood as interplay between kinematics and fixed-order perturbation theory -- we observe very good agreement between NNLO QCD predictions and ATLAS and CMS measurements. Notably, the agreement holds for both inclusive and fiducial distributions. We believe that this is a particularly important result which demonstrates that fully differential calculations at the parton level at NNLO can successfully be compared to measurements. Such comparisons are not necessarily automatic and may require a dedication effort in the unfolding of data. Still, the comparisons carried out in this work are a proof of principle that such a program can be carried out successfully. 

For most distributions we have provided predictions for two values of the top quark mass. In many cases the experimental and theoretical precisions are high-enough so that one GeV change in $m_t$ can be discerned. This work will hopefully provide further guidance for future top quark measurements.

A number of interesting observations regarding the absolute threshold bin of the $m(t\bar t)$ distribution have also been made. We have found that this bin can be extremely sensitive to the reconstruction of the top quark from particle-level final states. In particular, we find very significant NNLO correction -- which is about 50\% larger than the NLO one -- to the value of this bin depending if true or reconstructed top quarks are used. This finding may have important implications for top physics, and top mass determinations in particular, since this bin is extremely sensitive to $m_t$, yet there may be potentially large beyond-NLO corrections that are missed in the translation of data to stable top quarks with current NLO-accurate event generators. Such potential missed corrections could, in turn, significantly affect measurements of $t\bar t$ at threshold. We would recommend a followup study where NNLO predictions are compared with state-of-the-art event generators with the hope of quantifying this potential effect. It is also worth mentioning that this bin is rather sensitive to the treatment of $b$-jets, specifically if neutrinos from semileptonic decays are included in the $b$-jets or not.

Finally, we are convinced that the findings of this work provide valuable insights in the context of extending fixed-order NNLO calculations like ours to future event generators with NNLO-precision.

\begin{acknowledgments}
We thank Mykola Savitskyi and the CMS Top and LHCTopWG conveners for extensive discussions. We also thank Malgorzata Worek for an insightful discussion about the behavior of the $m(b_1 b_2)$ distribution and Zahari Kassabov for providing us with reduced pdf sets. The work of M.C. was supported in part by a grant of the BMBF and by the Deutsche Forschungsgemeinschaft under grant 396021762 -- TRR 257. The work of A.M. and R.P. has received funding from the European Research Council (ERC) under the European Union's Horizon 2020 research and innovation programme (grant agreement No 683211). The work of A.M. was also supported by the UK STFC grants ST/L002760/1 and ST/K004883/1.
\end{acknowledgments}

\end{document}